\documentclass[a4paper,11pt]{phdthesis} 

\usepackage{fancyhdr}
\usepackage{calc}

\fancypagestyle{plain}{%
\fancyhead{} 
}

\usepackage[toc,page]{appendix}

\pdfoutput=1

\usepackage[italian,british,dutch,spanish]{babel}
\usepackage{graphicx}
\usepackage{epsfig}
\usepackage{amsfonts}
\usepackage{amssymb}
\usepackage{amsmath}
\usepackage{dsfont}
\usepackage{bbm}
\usepackage{multirow}
\usepackage[vcentermath]{youngtab}
\usepackage{latexsym}
\usepackage{verbatim}
\usepackage{mcite}
\usepackage{cite}
\usepackage{units}
\usepackage{cancel}
\usepackage[all]{xy}
\usepackage{eufrak}
\usepackage{mathcomp}
\usepackage{color}
\usepackage{bigints}
\usepackage{slashed}
\usepackage{arydshln}
\usepackage{pifont}
\usepackage{tikz}

\usetikzlibrary{shapes,snakes,shadows,arrows}
\def\ellip{(0,0) ellipse (12 and 6)}
\def\ellipp{(2,0) ellipse (6 and 3)}
\tikzstyle{P_3} = [draw,very thick]
\tikzstyle{P_4} = [draw,fill=blue!20,very thick]

\def\a{{\alpha}}
\def\b{{\beta}}
\def\g{{\gamma}}
\def\d{{\delta}}
\def\m{{\mu}}
\def\n{{\nu}}
\def\s{{\sigma}}
\def\t{{\tau}}
\def\r{{\rho}}
\def\eps{{\epsilon}}
\def\dm{{\dot{\mu}}}
\def\dn{{\dot{\nu}}}
\def\cC{{\mathcal{C}}}
\def\hi{{\hat{i}}}
\def\hj{{\hat{j}}}
\def\hk{{\hat{k}}}
\def\hl{{\hat{l}}}
\def\hm{{\hat{m}}}
\def\bg{{\bar{\gamma}}}

\DeclareMathAlphabet{\mathpzc}{OT1}{pzc}{m}{it}

\newcommand{\cN}{{\mathcal{N}}}
\newcommand{\cM}{{\cal M}}

\newcommand{\be}{\begin{equation}}
\newcommand{\ee}{\end{equation}}
\newcommand{\ben}{\begin{displaymath}}
\newcommand{\een}{\end{displaymath}}
\newcommand{\bea}{\begin{eqnarray}}
\newcommand{\eea}{\end{eqnarray}}
\newcommand{\nn}{\nonumber}

\newcommand{\bean}{\begin{eqnarray*}}
\newcommand{\eean}{\end{eqnarray*}}

\newcommand{\bbM}{{\mathbb{M}}}
\newcommand{\bbN}{{\mathbb{N}}}
\newcommand{\bbP}{{\mathbb{P}}}
\newcommand{\bbQ}{{\mathbb{Q}}}
\newcommand{\bbR}{{\mathbb{R}}}
\newcommand{\bbS}{{\mathbb{S}}}

\begin{document}
\selectlanguage{british}

\begin{titlepage}



\pagebreak
\thispagestyle{empty}
\vspace*{\stretch{1}}

\noindent{\hrule height .4mm} \vspace{0.5mm} \noindent{\hrule height .4mm} \vspace{1mm}

    \begin{center}
        \huge \textsf{Gauged Supergravities \\ and \\the Physics of Extra Dimensions }
    \end{center}

\noindent{\hrule height .4mm} \vspace{0.5mm} \noindent{\hrule height .4mm}

\vspace{6mm}

    \begin{flushright}
        \Large \textsc{Giuseppe Dibitetto}
    \end{flushright}

\vspace*{\stretch{3}}


\end{titlepage}

\newpage
\tableofcontents
\newpage

\chapter*{Introduction}
\markboth{Introduction}{Introduction}
\addcontentsline{toc}{chapter}{Introduction}
\label{introduction}
The aim of this thesis is to study gauged supergravities as effective descriptions for addressing the problem of moduli stabilisation in compactifications of string theory. The various formulations of string theory all point towards a unique theory (M-theory) which is generally thought to be a consistent proposal for a description of quantum gravity. Such a consistent theory of quantum gravity is something that theoretical physicists have been searching for for a long time. The reason behind all these difficulties is to be found in the intrinsic complications stemming from the attempt of combining together Quantum Field Theory (QFT) and General Relativity (GR) into a unique theory. QFT and GR are the bearing pillars of high-energy physics and we will try now to give a brief historical overview of them both.

QFT originates from the idea of merging together the physics of the very small (Quantum Mechanics) with Einstein's theory of Special Relativity describing objects travelling in proximity of the speed of light. In such a framework, elementary particles (like electrons, photons, etc.) are interpreted as \emph{quanta} of a propagating field which can be created and destroyed by means of interactions. The biggest triumph of QFT is often considered to be the prediction of very accurate experimental measurements such as the so-called $(g-2)_{e}$, \emph{i.e.} the gyromagnetic factor of the electron in the context of Quantum Electrodyanmics. 

Following this line in QFT, non-Abelian gauge theories have been used to describe the three fundamental interactions of nature (excluding gravity). These are the electromagnetic force, the weak nuclear force and the strong nuclear force. The idea of gauge theories is that of using symmetries as an organising principle in physics. In particular, a gauge symmetry is a local symmetry of a system and through the process of promoting a global symmetry to a local one, the description of interactions emerges in a natural way.

The best experimentally tested theory that describes the three fundamental interactions and includes all the elementary particles that we have observed so far, is called the \emph{Standard Model} (SM) and it consists of a QFT with gauge group $\textrm{SU}(3)\,\times\,\textrm{SU}(2)\,\times\,\textrm{U}(1)$, the first factor describing strong interactions and the other two the electroweak ones. Besides this internal symmetry, the SM also exhibits the Poincar\'e group (translations and Lorentz transformations) as spacetime symmetry required by Special Relativity. This very elegant construction of the SM crucially relies on the so-called \emph{Higgs mechanism} in order to give mass to all the elementary particles in a gauge-invariant way, that is, respecting the gauge symmetry of the theory. 

However, this mechanism should be driven by a scalar particle (the Higgs boson) that had not been detected by any particle accelerator before LHC (Large Hadron Collider), the new machine that is collecting data at present at CERN. Still, up to the electroweak scale ($\Lambda_{\textrm{ew}}\sim 250$ GeV), the SM seemed to be perfectly working according to all previous experiments. Detecting the Higgs boson was the first goal of the LHC and the analysis of 2012 has already shown the presence of a signal compatible with the Higgs at $m_{H}\sim 126$ GeV. During its second period of activity, LHC will register collisions involving centre-of-mass energies up to $\sqrt{s}=14$ TeV.

So far, the SM offers a valuable framework for describing three out of four fundamental interactions in nature, but still misses out gravity.
This is the object of study of the other building block of theoretical high-energy physics which is GR. This theory was proposed by Einstein in 1914 in order to classically describe gravity as a geometric effect. The main idea is that any source of energy (matter, etc.) curves the spacetime around it so that all the objects move along geodesics in a curved geometry as an effect of gravitational interaction, whether or not they have a mass. This feature makes gravity the dominant force at cosmological scales, where all the other interactions cease to be relevant.

GR has been widely tested at the experimental level and amongst its greatest successes we can mention \emph{e.g.} the prediction for the anomalous precession of Mercury's perihelion, or the explanation of the phenomenon of gravitational lensing, \emph{i.e.} the deflection of light beams in the vicinity of strong gravitational fields like those ones produced by galaxy clusters. 
From the formal perspective, the possibility of describing the same physics in any arbitrary reference frame can be viewed as the invariance under local reparametrisations and Lorentz transformations. This allows one to regard GR as a gauge theory where the symmetries that have been made local are then coordinate translations and Lorentz transformations.

Unfortunately, though, unlike the SM, GR happens to be a \emph{non-renormalisable} theory, \emph{i.e.} it is very sensitive to physics at higher energy scales. This completely spoils the predictive power of the theory beyond a certain scale. Thus, GR should be treated as an effective description which still needs a UV completion at high energies. Precisely because of its power-counting non-renormalisability, GR predicts the existence of spacetime singularities (black holes), which represent regions in spacetime where the curvature reaches infinity. Whenever one finds infinities in classical computations, the inevitable conclusion is that such a description should be abandoned in favour of the quantum theory. So, at the end, one can estimate that the typical scale at which quantum gravity is needed in order to understand physics is $M_{\textrm{Pl}}\sim 10^{19}$ GeV, which is normally referred to as the Planck scale.

Searching for a theory of quantum gravity implies, as we said, the combination of QFT and GR. On the other hand, this would provide a unification of all the four fundamental interactions in nature. Since we are now used to describing interactions by means of symmetries, this in some sense has as a first consequence the necessity of finding more fundamental symmetries which combine internal and spacetime symmetries in an elegant and simple universal formulation. Indeed, all the efforts of theoretical physicists since the last century have been focused on this aim.

Following the goal of unification, physicists started looking for gauge groups containing both the internal symmetries of the SM and the Poincar\'e group in a non-trivial way. By 'non-trivial' here we mean that the two parts should not commute in order to go beyond the direct product structure. This means that there should exist new conserved charges which do not commute with the Poincar\'e group, hence \emph{non-scalar} charges.

This attempt resulted in 1967 in a very important statement known in the scientific literature as the \emph{Coleman-Mandula Theorem} \cite{Coleman:1967ad}. This theorem states that it is impossible to construct a field theory in $D>2$ including tensorial conserved charges other than the Poincar\'e generators (4-momentum and angular momentum). The proof involves several technical assumptions which we do not discuss here. 

As a way out in order to circumvent the result by Coleman and Mandula, people thought of the possibility of having \emph{spinorial} conserved charges. Spinors are objects transforming in representations of the universal covering of rotation groups. As a consequence, this possibility led to a deeply novel sort of symmetries, which mix bosons and fermions. Such a symmetry is commonly referred to as \emph{supersymmetry} and its associated conserved charges are then called supercharges. The inclusion of supercharges in the algebra describing the symmetries of a given theory generalises the concept of Lie algebra to superalgebras.

Later on, supersymmetry was used in particle physics (see ref.~\cite{Martin:1997ns} to read more about this) to build supersymmetric extensions of the SM, like \emph{e.g.} the MSSM (Minimally Supersymmetric Standard Model). Such a model assumes the existence of supersymmetric partners for all the SM particles, whose masses could have been made higher (and hence not observable so far) by a soft supersymmetry breaking mechanism. The benefit of the MSSM is mainly that of solving the hierarchy problem of the SM by removing quadratic divergences in favour of logarithmic ones. This improved UV behaviour occurs thanks to supersymmetry and it reduces a lot the fine-tuning that one needs to introduce in order to overcome the aforementioned hierarchy.

From a phenomenological perspective, supersymmetry has several consequences that one might presently be able to test at LHC. Firstly, the MSSM would favour, at least in its maximally constrained version, a lighter Higgs boson ($m_{H}\lesssim 120$ GeV). The current peak which would be compatible with the Higgs at $126$ GeV would require a version of the MSSM with a less constrained parameter space. Secondly, for what is concerning flavour physics, supersymmetry is expected to significantly affect certain cross-sections at the TeV scale that we should be able to observe at LHC. This would happen via the appearence of powers of $\tan \beta$ in the expressions of the corresponding loop-induced MSSM cross-sections \cite{Altmannshofer:2010zt}.

Still during the 1960's and in a completely independent line of investigation, physicists started to study the possibility of constructing a theory at high energy scales by making the assumption that the fundamental objects are tiny vibrating strings. The general idea was that the vibrational modes of the string should correspond with observable particle states. In this way \emph{bosonic string theory} was first conceived as a way of describing strong interactions. Nevertheless, around 1973/'74, an alternative theory for strong interactions was developed, which goes under the name of Quantum Chromodynamics (QCD) and it became immediately clear that string theory was not the correct candidate for the description of strong interactions.

Subsequently, supersymmetry was employed to give birth to superstring theory and obtain a completely tachyon-free theory describing the dynamics of strings.  Only then string theory started to be considered as a possible candidate for describing quantum gravity, since it was found to contain the graviton (\emph{i.e.} the quantum of the gravitational field) in the spectrum and to reproduce ten-dimensional supergravities (supersymmetric versions of GR) in the low-energy limit. Furthermore it does not suffer from non-renormalisability like GR and supergravity. Moreover, people began to realise that it allows for gauge groups which are in principle big enough for containing the SM interactions as well. Following this line, one is naturally led to the hypothesis that string theory might be the unified theory that we are looking for.

Later, in the mid 1990's people started discovering the most peculiar and interesting feature of string theory, that is the presence of \emph{dualities}. These are essentially relations between different theories in different regimes which allow one to view them as different limits of the same theory. It was indeed realised that the five different formulations of string theory known and perturbatively investigated up to that moment were in fact related to one another by taking different limits of a unique theory (M-theory), which we already mentioned at the very beginning of this introduction.  

Another issue that string theory brings into the game is that of \emph{extra dimensions}. In fact, a consistent quantisation of the superstring requires the target-space to be ten-dimensional. The extra challenge for string theorists became then that of finding some mechanisms providing compactifications of string theory down to four dimensions in order to make contact with the evidences of our low-energy observations. Historically, the first compactifications which were studied were on particular Ricci-flat six-dimensional internal manifolds called Calabi-Yau manifolds. These have the nice feature of preserving some supersymmetry and of giving rise to Minkowski vacua in four dimensions.

\begin{figure}[h!]
\begin{center}
\includegraphics[scale=0.50]{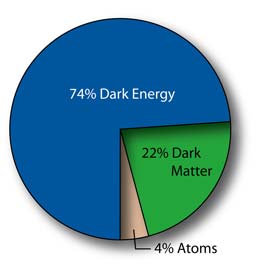}
\end{center}
\caption{{\it The concordance model of cosmology predicts that our universe has a cosmological constant $\Lambda>0$. In a recent phase of the history of the universe the vacuum energy took over and became the dominant energy content.}}
\label{figure:concordance}
\end{figure}

However, in the last fifteen years another fact came out of some cosmological observations: our universe contains \emph{dark energy}. This source of energy/matter in the universe satisfies an anomalous equation of state with respect to ordinary matter or radiation and it corresponds to the vacuum energy present in our universe. Dark energy can be accomodated inside GR by including an extra term to the Einstein equations which is often called cosmological constant and normally denoted by $\Lambda$. Combined measurements coming from supernovae \cite{Riess:1998cb, Perlmutter:1998np}, the Cosmic Microwave Background (CMB) radiation \cite{Jaffe:2000tx, Pryke:2001yz} and the Baryonic Acoustic Oscillations (BAO) \cite{Tegmark:2003ud, Eisenstein:2005su} concluded that we live in a universe with positive and small cosmological constant and gave rise to what we call nowadays the concordance model of cosmology. The energy/matter content giving the best fit is depicted in figure~\ref{figure:concordance}.

The cosmological constant drives an accelarated expansion of the universe which is described by de Sitter spacetime. This suggests that, after dark energy started to dominate, our universe started approaching a de Sitter vacuum rather than a Minkowski one. This implies that the suitable string compactifications for phenomenological purposes should give rise to de Sitter vacua. One can show that plain Calabi-Yau compactifications present the unfortunate feature of producing a large amount of massless scalar fields (a.k.a. moduli). Hence, in order to reproduce de Sitter vacua, one should go beyond these well-known compactifications.

An extra motivation for considering accelerated expanding universes in string theory is that of embedding \emph{inflationary models} within string theory. Inflation describes a phase of accelerated expansion of the universe right after the big bang. This was proposed to explain an almost perfect homogeneity and isotropy relating regions in the sky which had never been in causal contact with each other throughout the history.
Inflationary models are described by a quasi-de Sitter phase driven by a scalar field called the inflaton. 

The above issues provide two challenges for string theory compactifications related to de Sitter. The first one is finding de Sitter vacua in order to describe the late-time accelerating phase we are approaching now. The second one is embedding inflation in string theory by providing examples of compactifications in which quasi-de Sitter phases are possible with a very flat potential for the inflaton. These approaches in string theory result in what is often called \emph{string cosmology} and they have been extensively followed in several directions in the last decade. 

Concentrating for a moment on inflation, it is a particularly striking fact that string theory suggests some preferred classes of inflationary models, in which, for instance, no detectable tensor modes are present in the spectrum of cosmological perturbations of the early universe. This information, which is encoded in the CMB, can still be detected now, and is the result of frozen quantum fluctuations grown to observable size in the present universe. Precision measurements on the CMB carried out in the last decade by WMAP \cite{Efstathiou:2006ak, Spergel:2006hy, Komatsu:2008hk} already provided very precious data, although the existence of tensor perturbations still remains an open question. There is a possibility that the PLANCK satellite, which is currently collecting data, might tell us more about this. Such an experimental input would be a valuable opportunity for constraining models of inflation, among which there are stringy inflationary proposals.

Coming now back to the search for de Sitter vacua in string theory, right after the experimental detection of the cosmological constant, the existence of a huge 'zoo' of vacua \cite{Susskind:2003kw, Denef:2004ze} (about $10^{500}$ !!) was conjectured on the basis of statistical analysis. This enormous amount of different string vacua is often referred to as the \emph{landscape}. However, there has been more recently a lot of debate on this after the many failed attempts of finding classical (\emph{i.e.} at tree level) de Sitter solutions from string theory compactifications.  

Going beyond the search for classical solutions in string theory, people have considered the possibility of stabilising the moduli in an anti-de Sitter vacuum by means of quantum non-perturbative effects \cite{Kachru:2003aw} and subsequently providing an uplifting to de Sitter by means of several mechanisms. In ref.~\cite{Kachru:2003aw} such an uplifting was provided by additional extended sources breaking supersymmetry explicitely. Nevertheless, this mechanism completely ignores the backreaction of such sources and some recent analyses indicate that it might cause the arising of a singularity \cite{Bena:2012tx, Bena:2012bk} and possibly related instabilities \cite{Blaback:2012nf}. In ref.~\cite{Burgess:2003ic} the possibility of D-term uplifting was considered. However, later in refs~\cite{Choi:2005ge, deAlwis:2005tf} the inconsistency of this construction was pointed out due to the violation of gauge invariance occurring in a supergravity model with D-terms and yet vanishing F-terms. In ref.~\cite{Achucarro:2006zf} a valid proposal is given to overcome this inconsistency. The third possible type of uplifting mechanism is F-term uplifting, which was worked out \emph{e.g.} in ref.~\cite{Dudas:2006gr}.

A parallel but somehow related research line has regarded supergravity models as lower-dimensional effective descriptions coming from flux compactifications. In this context a lot of work has been done in the case of flux backgrounds preserving minimal supersymmetry in four dimensions. Some work has been done also in the context of compactifications preserving larger amount of supersymmetry. A very welcome ingredient (or even crucial in the case of (half-)maximal supergravities) for obtaining de Sitter solutions turns out to be given by \emph{non-geometric fluxes}. These objects appear as deformation parameters in the lower-dimensional effective description even though they do not have a clear higher-dimensional interpretation. Their appearence was first conjectured in ref.~\cite{Shelton:2005cf} based on duality covariance arguments. 

The aim of this thesis will be to follow this last research line, that is, to study gauged supergravities as effective descriptions arising from string compactifications. The final goal is to first formulate the complete dictionary between fluxes and deformation parameters of lower-dimensional supergravities. Subsequently, one can think of studying the landscape of vacua of particular classes of string compactifications through their effective gauged supergravity description. Finally, one could use the framework of gauged supergravities in order to understand the role of string dualities, since at that level they are realised as symmetries. The hope is that this could shed a light on the still unclear origin of non-geometric fluxes.

The thesis is organised as follows. In chapter~\ref{Strings} the various string theories and string dualities are reviewed. In chapter~\ref{Gauged_Sugra} supersymmetry is discussed and supergravities (low energy limits of string theory) and their deformations are introduced. In chapter~\ref{Fluxes} an overview of string compactifications is provided as mechanisms for generating a potential for moduli fields and subsequently some duality covariant proposals for describing non-geometric fluxes are introduced. In chapter~\ref{DFT} we discuss the orbit classification of gaugings of maximal and half-maximal supergravities in dimension seven and higher; subsequently we provide a Double Field Theory uplift for each orbit of theories. In chapter~\ref{Half_Max} we firstly introduce the dictionary between half-maximal gauged supergravities in four dimensions and orientifold reductions of type string theories with fluxes. Secondly, we study the landscape of vacua of geometric type IIA and IIB compactifications and furthermore give some example of locally geometric backgrounds in type IIB.
In chapter~\ref{Maximal} we show how to embed type II flux backgrounds without supersymmetry-breaking local sources inside maximal gauged supergravity in four dimensions and examine the full mass spectrum of a class of type IIA solutions. Finally, some additional material can be found in the appendices.

\chapter{String Theory and Dualities}
\markboth{String Theory and Dualities}{String Theory and Dualities}
\label{Strings}
In this chapter we will discuss some generalities about string theory as the main candidate for a description of quantum gravity. We will start from the simpler example of the bosonic string to move further to the discussion of the different formulations of string theory and dualities as a way of relating them together. Later on, we will briefly deal with the case of the superstring and argue that supergravities in ten dimensions can be obtained as low energy effective descriptions thereof. Finally, we will introduce the concept of branes and extended objects in string theory.

\section{The Bosonic String}
\label{sec:bosonic_string}

The original idea is that of writing an action for a $1$-dimensional object (\emph{string}) propagating in a $D$-dimensional background described by the coordinates $\left\{X^{\mu}\right\}$, with $\m=0,\,\dots,\,(D-1)$. During its motion, the string describes a surface (a $(1+1)$-dimensional submanifold described by the coordinates $(\tau,\sigma)\,\equiv\,\sigma^{\a}$, with $\a=0,1$) embedded in the background spacetime which is often called \emph{world-sheet}. From this perspective, the motion of the string is described by the dynamics of $D$ scalar fields $\left\{X^{\mu}(\tau,\sigma)\right\}$ which parametrise the worldsheet.  The free action describing the aforementioned system reads
\be
\label{world-sheet_S}
S\,=\,-\frac{T}{2}\,\int{d\tau\,d\sigma\,\sqrt{h}\,h^{\a\b}\,g_{\m\n}(X)\,\partial_{\a}X^{\m}\,\partial_{\b}X^{\n}}\ ,
\ee
where $h_{\a\b}$ is the world-sheet metric, $h\,\equiv\,|\det({h_{\a\b}})|$, $g_{\m\n}$ is the background metric and $T$ is the tension of the string (\emph{i.e.} mass / volume unit). 

The world-sheet metric $h_{\a\b}$ contains in principle only $1$ on-shell degree of freedom after gauge fixing (by making use of the diffeomorphism invariance of \eqref{world-sheet_S}). The peculiar fact about the above action is that it has another extra symmetry with respect to \emph{Weyl rescalings} of the form 
\be
\begin{array}{cccc}
h_{\a\b} & \longmapsto & \Lambda\,h_{\a\b} & ,
\end{array}
\ee
where $\Lambda$ is an arbitratry function of the world-sheet coordinates $(\tau,\sigma)$. Moreover, the theory described by \eqref{world-sheet_S} is renormalisable by power-counting.

We shall start studying the free propagation of a string in a Minkowski background, \emph{i.e.}
\be
g_{\m\n}\,=\,\textrm{diag}(-1,\,\underbrace{+1,\dots,+1}_{(D-1)})\ .
\ee
The action \eqref{world-sheet_S} has the following world-sheet symmetries
\be
\begin{array}{lc}
\left\{\begin{array}{cclc} \delta_{\xi} X^{\m} & = & \xi^{\a}\,\partial_{\a}X^{\m} & , \\ \delta_{\xi} h_{\a\b} & = & \xi^{\g}\,\partial_{\g}h_{\a\b}\,-\,2\,\partial_{(\a}\xi_{\b)} & ,\end{array}\right. & \textrm{(diffeomorphisms)} \\[4mm]
\left\{\begin{array}{cclc} \delta_{\Lambda} X^{\m} & = & 0 & , \\ \delta_{\Lambda} h_{\a\b} & = & \Lambda\,h_{\a\b} & ,\end{array}\right. & \textrm{(Weyl rescalings)}
\end{array}
\ee
together with the following global (target space) Poincar\'e symmetry
\be
\left\{\begin{array}{cclc} \delta X^{\m} & = & {a^{\m}}_{\n}\,X^{\n}\,+\,b^{\m} & , \\ \delta h_{\a\b} & = & 0 & ,\end{array}\right.
\ee
with $a_{\m\n}\,\equiv\,\eta_{\m\rho}\,{a^{\rho}}_{\nu}$ antisymmetric. By making use of two diffeomorphisms $\xi_{\a}$ and a Weyl rescaling $\Lambda$, one can always gauge away all the degrees of freedom of the world-sheet metric such that
\be
\label{gauge_string}
h_{\a\b}\,=\,\eta_{\a\b}\,=\,\textrm{diag}(-1,\,+1)\ .
\ee

To be more precise, under \emph{local} Weyl rescalings, the action \eqref{world-sheet_S} transforms as 
\be
\delta_{\Lambda}S\,=\,\frac{1}{2}\,\int{d^{2}\sigma\,\sqrt{h}\,h^{\a\b}\,T_{\a\b}\,\Lambda(\sigma)}\ ,
\ee
where $T_{\a\b}\,\equiv\,-\frac{2}{T}\,\frac{1}{\sqrt{h}}\,\frac{\delta S}{\delta h^{\a\b}}$ is the stress-energy tensor associated with the scalar fields $\left\{X^{\m}\right\}$.
This implies that, in order for the action to be invariant under local Weyl rescalings, we actually need to impose the following constraint
\be
h^{\a\b}\,T_{\a\b}\,=\,0 \ .
\ee
Moreover, the gauge choice \eqref{gauge_string} is only compatible with the equations of motion for $h_{\a\b}$ once the condition
\be
\label{constraint_string}
T_{\a\b}\,\overset{!}{=}\,0
\ee
is satisfied. 

Once the gauge choice \eqref{gauge_string} is made and the constraint \eqref{constraint_string} is imposed, one can derive the following equations of motion
\be
\label{EOM_string}
\begin{array}{cccccc}
\Box\,X^{\m} & = & \left(\partial^{2}_{\tau}\,-\,\partial^{2}_{\sigma}\right)\,X^{\mu} & = & 0 & .
\end{array}
\ee
These equations of motion have as a consequence that the stress-energy tensor $T_{\a\b}$ is conserved.
If we now look carefully at the variation of the action with respect to $X^{\m}$, we will see that it contains the following boudary terms
\be
-T\,\int{d\t\,\left[\partial_{\s}X^{\m}\,\d X_{\m}\,|_{\s=\pi}\,-\,\partial_{\s}X^{\m}\,\d X_{\m}\,|_{\s=0}\right]}\ ,
\ee
which can be set to zero by means of suitable \emph{boundary conditions} (b.c.). The physical interpretations of these is requirement that no energy-momentum flow occurs at the extrema of the string. The possible b.c. are
\begin{itemize}
\item $\begin{array}{lc}\partial_{\s}X^{\m}|_{\s=0,\,\pi}\,=\,0 \ ,  &  \phantom{Dirichlet}\textrm{(Neumann b.c. for opens strings)} \end{array}$  
\item $\begin{array}{lc} \d X^{\m}|_{\s=0,\,\pi}\,=\,0 \ , &   \,\,\phantom{Dirichlet}\textrm{ (Dirichlet b.c. for opens strings)} \end{array}$  
\item $\begin{array}{lc} X^{\m}(\tau,\s)\,=\,X^{\m}(\tau,\s+\pi) \ . &   \,\,\textrm{ (periodic b.c. for closed strings)} \end{array}$            
\end{itemize}

The general solution to the equations of motion \eqref{EOM_string} is easily written in light-cone world-sheet coordintes $\s^{\pm}\,\equiv\,\tau\,\pm\,\s$:
\be
\label{gen_sol_string}
X^{\m}(\s)\,=\,X^{\m}_{R}(\s^{+})\,+\,X^{\m}_{L}(\s^{-})\ ,
\ee
where the subscripts '$R$' and '$L$' stand for (right-)left-moving.

Let us now concentrate on the case of closed strings, for which one has to impose periodic b.c.; (the $R$ and $L$ part of) the solution generally given in \eqref{gen_sol_string} can be then expanded in \emph{Fourier modes} as follows
\be
\label{sol_R/L_mov}
\begin{array}{lclc}
X^{\m}_{R} & = & \frac{1}{2}\,x^{\m}\,+\,\frac{\ell_{s}^{2}}{2}\,p^{\m}\,(\tau-\s)\,+\,\frac{i\ell_{s}}{2}\,\sum\limits_{n\ne 0}\frac{1}{n}\,\a^{\m}_{n}\,e^{-2in\,(\tau-\s)} & , \\[3mm]
X^{\m}_{L} & = & \frac{1}{2}\,x^{\m}\,+\,\frac{\ell_{s}^{2}}{2}\,p^{\m}\,(\tau+\s)\,+\,\frac{i\ell_{s}}{2}\,\sum\limits_{n\ne 0}\frac{1}{n}\,\tilde{\a}^{\m}_{n}\,e^{-2in\,(\tau+\s)} & , 
\end{array}
\ee
where $\ell_{s}$ is the fundamental string length and  $\a^{\m}_{n}$ and $\tilde{\a}^{\m}_{n}$ are the Fourier components of the right-(left-)movers respectively. The reality condition of the solution \eqref{gen_sol_string} implies
\be
\begin{array}{lclclc}
x^{\m},\,p^{\m}\,\in\,\mathbb{R} & , & \a^{\m}_{n}\,=\,\left(\a^{\m}_{-n}\right)^{*} & \textrm{and} & \tilde{\a}^{\m}_{n}\,=\,\left(\tilde{\a}^{\m}_{-n}\right)^{*} & .
\end{array}
\ee
The physical interpretation of the constants $x^{\m}$ and $p^{\m}$ in the expressions \eqref{sol_R/L_mov} for $X^{\m}_{R}$ and $X^{\m}_{L}$ is that of postion and momentum of the centre of mass of the string.

By requiring that the coordinates $X^{\m}$ and the corresponding momenta satisfy canonical Poisson Brackets (PB) at equal times, one finds that the Fourier modes $\a$ and $\tilde{\a}$ have to satisfy the following PB\footnote{Please note that these PB are independent of the string tension and length after choosing $T\,=\,\frac{1}{\pi\,\ell^{2}_{s}}\,\equiv\,\frac{1}{2\pi\a^{\prime}}$.} 
\be
\begin{array}{lclc}
\left[\a^{\m}_{m},\,\a^{\n}_{n}\right]_{\textrm{PB}} & = & i\,m\,\eta^{\m\n}\,\d_{m+n,0} & , \\[1mm]
\left[\tilde{\a}^{\m}_{m},\,\tilde{\a}^{\n}_{n}\right]_{\textrm{PB}} & = & -i\,m\,\eta^{\m\n}\,\d_{m+n,0} & , \\[1mm]
\left[\a^{\m}_{m},\,\tilde{\a}^{\n}_{n}\right]_{\textrm{PB}} & = & 0 & , \\[1mm]
\end{array}
\ee
with the convention that $\a^{\m}_{0}\,=\,\tilde{\a}^{\m}_{0}\,=\,\frac{\ell_{s}}{2}\,p^{\m}$. After introducing the Fourier components of the stress-energy tensor $T_{\a\b}$
\be
\begin{array}{lclc}
L_{m}\,=\,\frac{1}{2}\,\sum\limits_{n\in \mathbb{Z}}\a_{m-n}\cdot\a_{n} & \textrm{and} & \tilde{L}_{m}\,=\,\frac{1}{2}\,\sum\limits_{n\in \mathbb{Z}}\tilde{\a}_{m-n}\cdot\tilde{\a}_{n} & ,
\end{array}
\ee
one finds that their PB describe a \emph{Virasoro algebra}
\be
\label{Virasoro}
\left[L_{m},\,L_{n}\right]_{\textrm{PB}}\,=\,i\,(m-n)\,L_{m+n}\ ,
\ee
and the same holds for $\tilde{L}$'s, whereas $[L,\,\tilde{L}]_{\textrm{PB}}=0\,$.

\subsection*{Aspects of the Quantum Theory}

Bosonic string theory can be quantised following different approaches yet giving rise to the same final result. The possible different approaches historically studied are the following
\begin{itemize}
\item {\bf Old Covariant Method:} inspired by the quantisation procedure \`a la Gupta-Bleuler followed in electrodynamics,
\item {\bf Modern Covariant Method:} type of BRST quantaisation based on the introduction of Faddeev-Popov ghosts,
\item {\bf Light-cone Gauge Quantisation:} solving explicitely the constraints on $T_{\a\b}$ by breaking covariance from the start. 
\end{itemize}
By following the preferred quantisation procedure, one will promote the PB previousely introduced to \emph{commutators} between operators. This leads to a central extension of the Virasoro algebra at a quantum level coming from the normal ordering prescription. 

One discovers that $\a$ and $\tilde{\a}$ suitably normalised behave as creators and annihilators. Hence, by making use of them, one can uniquely construct the space of physical states. Following, \emph{e.g.} the old covariant method, the general presence of ghosts (\emph{i.e.} negative squared norm states) arises from the Minkowskian signature of the metric. The spectrum of physical states only turns out to be free of ghosts for $D=26$. If one follows different quantisation procedures, this conclusion remains valid.

If we focus on the case of closed bosonic strings, we find out that there is a vacuum state $|0\rangle$ corresponding to a \emph{tachyonic} scalar, whose mass is given by $M^{2}=-4/\a^{\prime}$. The first excited states, instead, constitute the massless spectrum and include the following objects
\be
|\Omega^{ij}\rangle\,=\,\a^{i}_{-1}\,\a^{j}_{-1}\,|0\rangle\ ,
\ee
where the operators of the type $\a^{i}_{-1}$ denote transverse creation operators. Such an object lives then in the following representation of the little group SO($24$)
\be
\underbrace{\textbf{24}\,\otimes\,\textbf{24}}_{\tiny{\yng(1)}\,\otimes\,\tiny{\yng(1)}}\,=\,\textbf{1}\,\oplus\,\underbrace{\,\,\textbf{276}\,\,}_{\tiny{\yng(1,1)}}\,\oplus\,\underbrace{\,\,\textbf{299}\,\,}_{\tiny{\yng(2)}} \ .
\ee
The above irrep's describe the following massless fields
\begin{itemize}
\item the \emph{metric} $g_{\m\n}$,
\item a \emph{two-form} $b_{\m\n}$,
\item a scalar $\phi$, often called the \emph{dilaton}.
\end{itemize}
This field content is often referred to as the common sector of all string theories.

\section{Superstring Theory}
\label{superstrings}

In the previous section we have seen that bosonic string theory still suffers from the presence of a tachyon even in the closed string sector, which clearly would make our theory not unitary. Besides, there is no room for fermions in the spectrum of the bosonic string. In order to try to improve these unwanted features, we will supplement the action \eqref{world-sheet_S} with extra \emph{fermionic} world-sheet degrees of freedom called $\psi^{\m}$. For some further reading on the topic, we suggest to take a look at refs~\cite{Castellani:1991ev, GSW_book}.

Let us consider the action
\be
S_{1}\,=\,-\frac{1}{2\pi\a^{\prime}}\,\int{d^{2}\s\,\sqrt{h}\,\left(h^{\a\b}\,\partial_{\a}X^{\m}\,\partial_{\b}X_{\m}\,-\,i\,\overline{\psi}^{\m}\,\rho^{\a}\,\partial_{\a}\psi_{\m}\right)}\ ,
\ee
where $\rho^{\a}$ is a 2-dimensional realisation of gamma matrices (see section~\ref{susy} for the formal aspects of spinors and supersymmetry):
\be
\begin{array}{lclclc}
\rho^{0}\,=\,\left(\begin{array}{cc}0 & -i \\ i & 0\end{array}\right) & , & \rho^{1}\,=\,\left(\begin{array}{cc}0 & i \\ i & 0\end{array}\right) & \textrm{with} & \left\{\rho^{\a},\,\rho^{\b}\right\}\,=\,-2\eta^{\a\b} & . 
\end{array}
\nn\ee

By adding these two extra terms in the action
\be
S_{2}\,=\,-\frac{1}{\pi\a^{\prime}}\,\int{d^{2}\s\,\sqrt{h}\,\left(\overline{\chi}_{\a}\,\rho^{\b}\,\rho^{\a}\,\psi^{\m}\,\partial_{\b}X_{\m}\,+\,\frac{1}{4}\,(\overline{\psi}^{\m}\,\psi_{\m})\,(\overline{\chi}_{\a}\,\rho^{\b}\,\rho^{\a}\,\chi_{\b})\right)}\ ,
\ee
one finds that the full action $S\,=\,S_{1}\,+\,S_{2}$, apart from having a symmetry under Weyl rescalings that generalises the form presented in the purely bosonic case, has a completely novel type of symmetry
\be
\left\{\begin{array}{cclc} 
\d_{\eps}X^{\m} & = & \overline{\eps}\,\psi^{\m} & , \\ 
\d_{\eps}\psi^{\m} & = & -i\,\rho^{\a}\,\eps\,\left(\partial_{\a}X^{\m}\,-\,\overline{\psi}^{\m}\,\chi_{\a}\right) & , \\
\d_{\eps} h_{\a\b} & = & -2i\,\overline{\eps}\,\rho_{\a}\,\chi_{\b} & , \\
\d_{\eps} \chi_{\a} & = & \partial_{\a}\eps & ,
\end{array}\right.
\ee
where $\eps$ is an arbitrary Majorana spinor (see again section~\ref{susy}) in 2 dimensions. This symmetry relates bosonic fields ($X^{\m}$ and $h_{\a\b}$) to fermionic ones ($\psi^{\m}$ and $\chi_{\a}$) and is normally called \emph{supersymmetry}. Particular realisations of supersymmetry in field theory will be presented in the next chapter. 

Finally, there is an extra local symmetry transforming only the fermions $\chi_{\a}$ (and leaving all the other fields invariant) in the following way
\be
\d \chi_{\a}\,=\,i\,\rho_{\a}\,\eta \ ,
\ee
where $\eta$ is again a Majorana spinor.
Just as in the bosonic case, one can gauge away all the degrees of freedom inside the world-sheet metric by using local diffeomorphisms and Weyl rescalings and perform the gauge choice in \eqref{gauge_string}. Moreover, we can now make use of the two supersymmetries generated by $\eps$ and the two extra fermionic symmetries generated by $\eta$ in order to gauge away $\chi_{\a}$. Summarising, once we use all the symmetries at our disposal, we can always perform the following gauge choice
\be
\label{gauge_superstring}
\begin{array}{lclc}
h_{\a\b}\,=\,\eta_{\a\b} & , & \chi_{\a}\,=\,0 & .
\end{array}
\ee

At this point, we can vary the total action to get the following equations of motion for $X^{\m}$ and $\psi^{\m}$
\be
\label{EOM_superstring}
\begin{array}{lccclc}
\underbrace{\Box \, X^{\m} \,=\,0}_{\textrm{Wave eqn}} & & \textrm{and} & & \underbrace{\rho^{\a}\,\partial_{\a}\psi^{\m} \,=\,0}_{\textrm{Dirac eqn}} & ,
\end{array}
\ee
which have to be supplemented with the constraints that guarantee the consistency of the gauge choice performed in \eqref{gauge_superstring}. These translate into the vanishing of the currents associated with the symmetries we made use of, \emph{i.e.} the stress-energy tensor $T_{\a\b}$ and a supercurrent $\mathcal{J}_{\a}$
\be
\begin{array}{lclc}
\mathcal{J}_{\a} & = & \dfrac{1}{2}\,\rho^{\b}\,\rho_{\a}\,\psi^{\m}\,\partial_{\b}X_{\m} \,\overset{!}{=}\,0 & , \\[2mm]
T_{\a\b} & = & \partial_{\a}X^{\m}\,\partial_{\b}X_{\m}\,+\,\dfrac{i}{2}\,\overline{\psi}^{\m}\,\rho_{(\a}\,\partial_{\b)}\psi_{\m} \,\overset{!}{=}\,0 & .
\end{array}
\ee

Similarly to the bosonic case, one can solve the equations of motion \eqref{EOM_superstring} by introducing the same right- and left-movers for $X^{\m}$ and doing something analogous for $\psi^{\m}$ in light-cone coordinates
\be
\psi^{\m}(\s)\,=\,\psi^{\m}_{R}(\s^{+})\,+\,\psi^{\m}_{L}(\s^{-})\ .
\ee
After choosing by convention $\psi_{R}(\tau,0)\,=\,+\psi_{L}(\tau,0)$, we have two inequivalent possilities for fixing the b.c. at $\s=\pi$:
\be
\begin{array}{llcl}
\bullet & \psi_{R}(\tau,\pi)\,=\,+\psi_{L}(\tau,\pi) & , & \textrm{(Ramond (R) b.c.)} \\[2mm]
\bullet & \psi_{R}(\tau,\pi)\,=\,-\psi_{L}(\tau,\pi) & . & \textrm{(Neveu-Schwarz (NS) b.c.)}
\end{array}
\ee
From the above b.c., \emph{e.g.} in the case of closed strings, we get the following mode expansions
\be
\begin{array}{lc}
\left\{\begin{array}{lclc} \psi_{L} & = & \frac{1}{\sqrt{2}}\,\sum\limits_{n\in\mathbb{Z}}d^{\m}_{n}\,e^{-in\,(\tau-\s)} & , \\[1mm] \psi_{R} & = & \frac{1}{\sqrt{2}}\,\sum\limits_{n\in\mathbb{Z}}\tilde{d}^{\m}_{n}\,e^{-in\,(\tau+\s)} & , \end{array}\right. & \textrm{(R)} \\[4mm]
\left\{\begin{array}{lclc} \psi_{L} & = & \frac{1}{\sqrt{2}}\,\sum\limits_{r\in\left(\mathbb{Z}+\frac{1}{2}\right)}b^{\m}_{r}\,e^{-ir\,(\tau-\s)} & , \\[1mm] \psi_{R} & = & \frac{1}{\sqrt{2}}\,\sum\limits_{r\in\left(\mathbb{Z}+\frac{1}{2}\right)}\tilde{b}^{\m}_{r}\,e^{-ir\,(\tau+\s)} & . \end{array}\right. & \textrm{(NS)} 
\end{array}
\ee

When quantising the theory, one can follow the same approaches briefly described at the end of section~\ref{sec:bosonic_string}. The starting point is again introducing canonical commutation relations (coming from classical PB) for $X^{\m}$ and canonical \emph{anti-commutation} relations for $\psi^{\m}$. This operation results in creation and annihilation operators: $(\a,\,\tilde{\a})$ type for the bosons, $(b,\,\tilde{b})$ type for fermions with NS b.c. and finally $(d,\,\tilde{d})$ type for fermions with R b.c.

Quadratic cobinations of creators and annihilators give now rise to the new Fourier components of $T_{\a\b}$ and $\mathcal{J}_{\a}$ and generate a graded extension (see def. of superalgebras in section~\ref{susy}) of the Virasoro algebra given in \eqref{Virasoro}, with central extensions again related with the process of quantisation. In the case of the superstring, the consistency of the quantum theory requires the critical dimension to be $D=10$. However, the resulting quantum theory still contains a tachyon in the NS sector just like in the bosonic case.

Supersymmetry requires the total number of physical degrees of freedom associated to bosons and fermions to be equal (see the theorem stated in section~\ref{susy}). This is achieved by the so-called Gliozzi-Scherk-Olive (GSO) projection \cite{Gliozzi:1976qd}, which defines a notions of fermionic parity and eliminates all the states in the spectrum being parity odd.  The GSO projection leaves one with a massless sector consisting of an NS vector and a R spinor, that is, in terms of irrep's\footnote{All throughout the text $\textbf{8}_{\textrm{V}}$, $\textbf{8}_{\textrm{S}}$ and $\textbf{8}_{\textrm{C}}$ denote the triality of SO($8$) irrep's of dimension $8$, \emph{i.e.} the vector and the spinors of the two chiralities.} of the little group SO($8$)
\be
\begin{array}{cccc}
\textbf{8}_{\textrm{V}}\,\oplus\,\textbf{8}_{\textrm{C}} & \longleftrightarrow & A_{\m}\,\,\oplus\,\,\lambda & ,
\end{array}
\ee
which exactly defines the field content of an $\cN=1$ vector multiplet (see table~\ref{table:vector_multiplet}).

The different choices for what is regarding the GSO projection give rise to \emph{inequivalent string theories}, which we summarise in this paragraph
\begin{itemize}
\item {\bf Type II String Theories:} $\psi_{R}$ and $\psi_{L}$ are treated independently when it comes to perform the GSO projection. This gives rise to two inequivalent choices, depending on whether it selects opposite or equal signs (called type IIA and IIB, respectively): 
%

\resizebox{.95\hsize}{!}{$
\begin{array}{llcc}
\textrm{IIA:} & \left(\textbf{8}_{\textrm{V}}\,\oplus\,\textbf{8}_{\textrm{C}}\right)\,\otimes\,\left(\textbf{8}_{\textrm{V}}\,\oplus\,\textbf{8}_{\textrm{S}}\right) & = & \left\{\begin{array}{c}\big((\textbf{1}\,\oplus\,\textbf{28}\,\oplus\,\textbf{35}_{\textrm{V}})_{\textrm{NS-NS}}\,\oplus\,(\textbf{8}_{\textrm{V}}\,\oplus\,\textbf{56}_{\textrm{V}})_{\textrm{R-R}}\big)_{\textrm{B}} \\[1mm] \oplus\,\big((\textbf{8}_{\textrm{S}}\,\oplus\,\textbf{56}_{\textrm{S}})_{\textrm{NS-R}}\,\oplus\,(\textbf{8}_{\textrm{C}}\,\oplus\,\textbf{56}_{\textrm{C}})_{\textrm{R-NS}}\big)_{\textrm{F}} \end{array}\right\}\ , \\[7mm]
\textrm{IIB:} & \left(\textbf{8}_{\textrm{V}}\,\oplus\,\textbf{8}_{\textrm{C}}\right)\,\otimes\,\left(\textbf{8}_{\textrm{V}}\,\oplus\,\textbf{8}_{\textrm{C}}\right) & = & \left\{\begin{array}{c}\big((\textbf{1}\,\oplus\,\textbf{28}\,\oplus\,\textbf{35}_{\textrm{V}})_{\textrm{NS-NS}}\,\oplus\,(\textbf{1}\,\oplus\,\textbf{28}\,\oplus\,\textbf{35}_{\textrm{C}})_{\textrm{R-R}}\big)_{\textrm{B}} \\[1mm] \oplus\,\big((\textbf{8}_{\textrm{S}}\,\oplus\,\textbf{56}_{\textrm{S}})_{\textrm{NS-R}}\,\oplus\,(\textbf{8}_{\textrm{S}}\,\oplus\,\textbf{56}_{\textrm{S}})_{\textrm{R-NS}}\big)_{\textrm{F}} \end{array}\right\}\ , \\
\phantom{a} & \phantom{a} & \phantom{a} & \phantom{a} 
\end{array}$}
which gives exactly the field content of type IIA and IIB supergravities in $D=10$, as we will see in \eqref{fields_IIA_IIB}.
\item {\bf Type I String Theory:} it consists of open and closed strings; it is $\cN=1$ supersymmetric and the universal field content (\emph{i.e.} NS-NS and NS-R) is supplemented by $496$ vector multiplets which describe an SO($32$) gauge theory. Such a theory can be as well obtained by modding out type IIB string theory with respect to a $\mathbb{Z}_{2}$ parity flipping the sign of the world-sheet coordinates.
\item {\bf Heterotic String Theories:} constructed by combining the bosonic left-moving sector with the fermionic right-moving one. The bosonic sector then has to be compactified from $D=26$ down to $D=10$, giving rise to internal gauge symmetries. Anomaly cancellation forces the only two consistent possibilities to be SO($32$) and E$_{8}\,\times\,$E$_{8}$. 
\end{itemize} 
The common sector of both type I and heterotic string theories exactly matches the field content of $\cN=1$ supergravity in $D=10$. This common sector is then coupled to $496$ vector multiplets which describe a gauge theory with gauge group SO($32$) or E$_{8}\,\times\,$E$_{8}$.

Summarising, we have seen that the low-energy spectrum of all the five consistent string theories recovers the field content of the possible supergravities in ten dimensions. Moreover, these degrees of freedom also turn out to be described by a low energy effective action which is that of ten-dimensional supergravities (coupled to vector multiplets in the case of $\cN=1$).

\section{Beyond Ordinary Field Theory: Dualities}
\label{sec:dualities}

In the previous section we have seen that five different string theories in ten dimensions can be constructed perturbatively. We would like to stress that string theory contains two deeply different types of perturabtive expansions: the first one is in terms of the string coupling $g_{s}$ which is equal to $e^{\phi}$ for backgrounds with constant dilaton and it plays the role of $\hbar$ in loop expansions; this is a quantum theory defined on the spacetime. $g_{s}$ corrections take us away from the supergravity limit but simply by completing it with quantum corrections.
Moreover, there is a second and dramatically different expansion defined on the world-sheet which is carried out with respect to $\a^{\prime}$. $\a^{\prime}$ corrections take us away from the field theory description and hence they have a purely \emph{stringy} nature and do not have any analogue in QFT.

However, in general perturbation theory is insufficient to completely understand the physics described by a given quantum theory. In QFT, for instance, one often needs the so-called \emph{path-integral} formulation of the theory in order to capture possible non-perturbative effects.
Unfortunately, no analogue of the path-integral formulation is known in string theory. Still, there is one interesting feature of string theories that can be seen as an opportunity to understand some physical features thereof. One is able to prove that the five string theories are related amongst them via \emph{dualities}. Duality relates equivalent descriptions of a theory in which perturbation theory is done around different points, that is, dual descriptions are different ways of taking the limit \emph{i.e.} $\hbar\rightarrow 0$. The general advantage of dual descriptions is that, whenever a description enters the strong coupling regime and hence pertution theory is not to be trusted anymore, its dual description will conversely be in the semi-classical limit. To read more about dualities and non-perturbative aspects of string theory, we recommend refs~\cite{Giveon:1994fu, Sen:1998kr}.

\begin{figure}[h!]
\begin{center}
\includegraphics[scale=0.40]{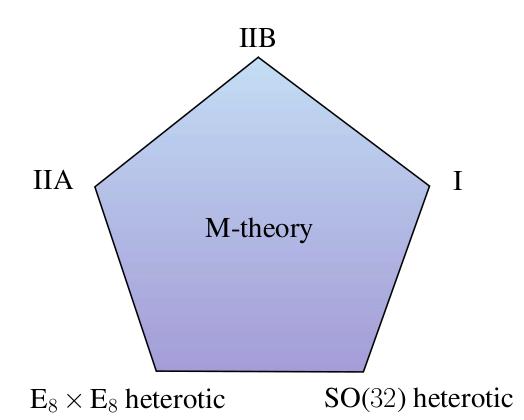}
\end{center}
\caption{{\it The different string theories in ten dimensions as different perturbative limits of an eleven-dimensional theory called M-theory.}}
\label{figure:string_theories}
\end{figure}

The situation depicted in figure~\ref{figure:string_theories} precisely shows how all the different string theories simply are different perturabtive expansions of the so-called M-theory, which is often regarded as the best candidate to a unified description of gravity and gauge theories. This theory has the peculiarity of not having a coupling like $g_{s}$ anymore and its low-energy limit is given by eleven-dimensional supergravity (see section~\ref{sec:supergravities}). 

As an example, let us now examine in detail the explicit nature of the dualities relating some string theories in figure~\ref{figure:string_theories}. We will start observing that type IIA and type IIB string theories are related by what we call \emph{T-duality}. This duality has perturbative nature, \emph{i.e.} it can be proven order by order in ($g_{s}$) perturbation theory. Its origin in this case is the fact that the two aforementioned string theories become the same theory in $D=9$ when reducing them on a circle $S^{1}$. The dictionary between the IIA and IIB side of the duality is constructed by
\be
\label{T_IIA_IIB_R}
R_{A}\,=\,\frac{\a^{\prime}}{R_{B}}\ ,
\ee
where $R_{A,B}$ are the radii of the circle $S^{1}$ in the two compactifications. The \eqref{T_IIA_IIB_R} implies that T-duality interchanges the role of momentum and winding modes in the spectrum.
From the world-sheet perspective, the above T-duality acts as
\be
\left\{\begin{array}{cccc}
X_{R}^{9} & \longmapsto & -X_{R}^{9} & , \\[2mm]
\psi^{9}  & \longmapsto & -\psi^{9} & ,
\end{array}\right.
\ee
where the direction labelled here by '$9$' is the compact one. The action of T-duality on the massless NS-NS sector fields $g_{\m\n}$, $b_{\m\n}$ and $\phi$ is known in the literature as the \emph{Buscher rules} \cite{Buscher:1987sk}. This duality manifests itself at the level of the nine-dimensional theory as an SO($1,1;\mathbb{Z}$) symmetry. As we will see later in table~\ref{table:string_dualities}, this gets generalised to SO($d,d;\mathbb{Z}$) when reducing type II (A or B it does not matter!) on a torus $T^{d}$.

As a further example, we want to illustrate the deeply different nature of \emph{S-duality}, which is yet non-perturbative and hence intrinsically difficult to prove. In constrast with the previous case of T-duality, where the theories can be compared order by order in $g_{s}$ and different contributions from different orders never mix, here such a mixing will occur. This makes it meaningless to compare the spectra state by state on the two sides of the duality.
Indeed, such non-perturbative dualities are normally conjectured and subsequently tested. The instruments at our disposal in order to test S-duality are those objects which are \emph{protected by supersymmetry} like
\begin{itemize}
\item the spectrum of BPS (\emph{i.e.} partially supersymmetric) states (see non-renormalisation theorems which protect supersymmetric objects from quantum corrections),
\item the low-energy effective Lagrangian (constrained by supersymmetry to match the supergravity action).
\end{itemize}
S-duality turns out to transform type IIB string theory into itself, the bosonic massless sector transforming as described in table~\ref{table:S_IIB}.
\begin{table}[h!]
\begin{center}
\scalebox{1}[1]{
\begin{tabular}{| c | c | c |}
\hline
sector & IIB fields & S-duals \\[1mm]
\hline \hline
\multirow{3}{*}{NS-NS} & $g_{\m\n}$ & $g_{\m\n}$ \\[1mm]
\cline{2-3}  & $b_{\m\n}$ & $C^{(2)}_{\m\n}$ \\[1mm]
\cline{2-3}  & $\phi$ & $-\phi$ \\[1mm]
\hline \hline 
\multirow{3}{*}{R-R} &  $C^{(0)}$ & $C^{(0)}$ \\[1mm]
\cline{2-3}  & $C^{(2)}_{\m\n}$  & $-b_{\m\n}$ \\[1mm]
\cline{2-3}  & $C^{(4)}_{\m\n\rho\s}$  & $C^{(4)}_{\m\n\rho\s}$ \\[1mm]
\hline
\end{tabular}
}
\end{center}
\caption{{\it The transformation law of the massless IIB fields under S-duality. For backgrounds with constant dilaton, such a duality takes $g_{s}\,\mapsto\,\frac{1}{g_{s}}$, thus interchanging in the spectrum the role of perturbative objects and solitons.}}
\label{table:S_IIB}
\end{table}
One can actually show that such S-duality can be completed to form a larger discrete group of non-pertirbative dualities given by SL($2,\mathbb{Z}$), of which S-duality represents the element $\left(\begin{array}{cc} 0 & -1 \\ 1 & 0\end{array}\right)$. This duality generalises the concept of electromagnetic duality for Maxwell theories \cite{Harvey:1996ur}.

\begin{figure}[h!]
\begin{center}
\includegraphics[scale=0.75]{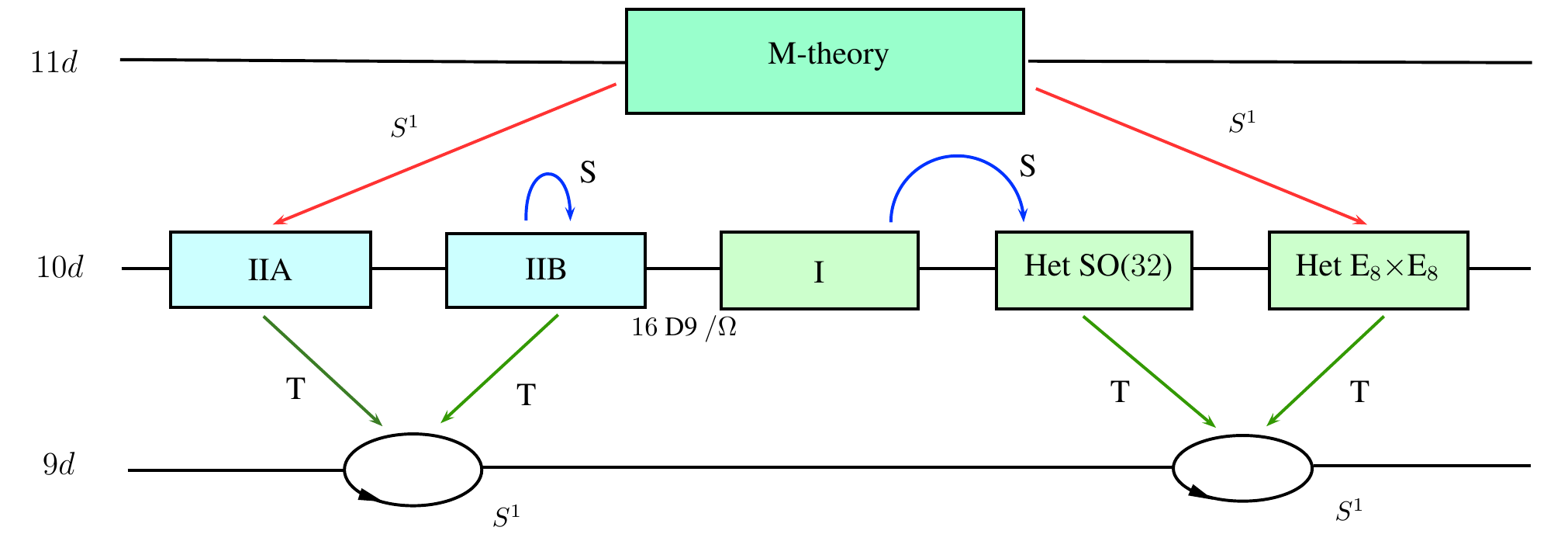}
\end{center}
\caption{{\it The net of all the dualities connecting the five string theories in ten dimensions and their links with M-theory. Please note that type II theories (on the left of the diagram) have $32$ supercharges, whereas, after modding out by the discrete symmetry $\Omega$, supersymmetry is broken such in a way that the other theories on the right (type I and heterotic) only retain $16$ supercharges.}}
\label{figure:string_dualities}
\end{figure}

The full net of dualities relating the different five string theories is presented in figure~\ref{figure:string_dualities}.

In type II theories, one can think of combining perturbative and non-perturbative dualities by applying a chain of S- and T-dualities. In such a way, one realises that they contain an enhanced duality group called consisting of more general dualities usually called \emph{U-dualities}. When reducing M-theory on a torus $T^{n}$, such a duality manifests itself as an E$_{n(n)}(\mathbb{Z})$ symmetry \cite{Hull:1994ys}. The duality groups of the compactified type II theories are summarised in table~\ref{table:string_dualities}. 
\begin{table}[h!]
\begin{center}
\begin{tabular}{| c || c | c |}
\hline
$D$ & T-duality & U-duality \\[1mm]
\hline \hline 
$9$ & O($1,1;\mathbb{Z}$) & SL($2,\mathbb{Z}$) \\[1mm]
\hline
$8$ & SL($2,\mathbb{Z})\,\times\,$SL($2,\mathbb{Z}$) & SL($2,\mathbb{Z})\,\times\,$SL($3,\mathbb{Z}$) \\[1mm]
\hline
$7$ & SL($4,\mathbb{Z}$) & SL($5,\mathbb{Z}$) \\[1mm]
\hline
$6$ & O($4,4;\mathbb{Z}$) & O($5,5;\mathbb{Z}$) \\[1mm]
\hline
$5$ & O($5,5;\mathbb{Z}$) & E$_{6(6)}(\mathbb{Z})$ \\[1mm]
\hline
$4$ & O($6,6;\mathbb{Z}$) & E$_{7(7)}(\mathbb{Z})$ \\[1mm]
\hline
\end{tabular}
\end{center}
\caption{{\it The various T-(U-)duality groups emerging as symmetries of type II (M-) theories on a $T^{10-D}$ ($T^{11-D}$). Please note that the corresponding supergravity theories enjoy the full continuous symmetries, as we will see in the next chapter. Quantum effects in string theory break the duality groups to the discrete subgroups shown here.}}
\label{table:string_dualities}
\end{table}

\section{Branes and Sources}

In the previous section we have seen that dualities are a very peculiar feature of string theory and that they generally relate descriptions in weakly and strongly coupled regime to each other. We also saw that non-perturbative dualities are very difficult to test and that analysing the spectrum of BPS states can be an important instrument in this sense.

In the spectrum of the various string theories, not only can we find states representing excitations of the so-called fundamental string itself, but by making use of dualities, also extended objects called \emph{branes} appear as solitonic states in the spectrum. These extended objects have a \emph{world-volume} action which is very similar to the world-sheet action of a string. Upon imposing certain b.c., the aforementioned branes can define backgrounds in string theory which preserve partial amounts of supersymmetry (BPS branes).

For example, D$p$-branes are extended objects whose world-volume is $(p+1)$-dimensional and Neumann b.c. are fixed on it, whereas in all the transverse directions, Dirichlet b.c. are chosen. D$p$-branes can be charged electrically under the R-R gauge potential $C^{(p+1)}$ or magnetically under $C^{(7-p)}$. 
Identically one could imagine to have branes which are electrically (magnetically) charged under NS-NS gauge potentials. In this case, though, we only have the $b$-field at our disposal, thus resulting in the fundamental string (which we denote by NS1) and the NS5-brane. The tension (\emph{i.e.} mass per volume unit) of the various objects named above is in general a function of the couplings $g_{s}$ and $\ell_{s}$ in string theory and solitonic objects have a tension which scales as negative powers of the couplings such that it becomes very high in the weakly coupled regime. This information is collected in table~\ref{table:brane_tensions}.

The extended objects introduced above all have positive tension. Furthermore, one can introduce \emph{orientifold planes} (O$p$), which are objects with negative tension located at the fixed points of some discrete involution. The main difference with respect to \emph{e.g.} D-branes is that O-planes are strictly speaking no dynamical objects, in the sense that, as we just saw, their position in the target space is not dynamically determined. Besides, on D-branes one can construct a gauge theory upon the introduction of extra matter content.

Other extended objects in string theory are the so-called \emph{KK monopoles} (after Kaluza-Klein). These objects are highly non-perturbative and they are charged under mixed symmetry fields like the dual graviton. In $D=10$, they are sometimes referred to as KK5-branes even though, strictly speaking they are only \emph{pre-branes}, in the sense that they become branes upon T-dualisation. The conjecture is that, since the KK monopole is T-dual to an NS5-brane, its tension should still scale as $g_{s}^{-2}$.
We suggest refs~\cite{Hull:1995xh, Dabholkar:1997zd} to find more about dualities in string backgrounds containing branes and orientifold planes.  
\begin{table}[h!]
\begin{center}
\begin{tabular}{| c | c |}
\hline 
Branes & Tension \\[1mm]
\hline \hline 
NS1 & $\ell_{s}^{-2}$\\[1mm]
\hline
NS5 & $g_{s}^{-2}\,\ell_{s}^{-6}$\\[1mm]
\hline \hline
D$p$ & $g_{s}^{-1}\,\ell_{s}^{-(p+1)}$\\[1mm]
\hline
\end{tabular}
\end{center}
\caption{{\it The tension (mass per volume unit) of several extended objects in string theory as a function of $g_{s}$ and $\ell_{s}$.}}
\label{table:brane_tensions}
\end{table}

Let us go back to type IIB string theory in order to see which branes can be coupled to the massless fields of the theory. In the NS-NS sector, the only gauge potential is the Kalb-Ramond $2$-form $b$ and hence we can have fundamental strings NS1 and NS5-branes, which are respectively electrically and magnetically charged under $b$. In the R-R sector, instead, we have $C^{(0)}$, $C^{(2)}$ and $C^{(4)}$; with respect to these fields D($-$1), D1 and D3 are electrically charged, whereas D7, D5 and again D3 are magnetically charged. Please note that the D($-$1) has the peculiarity of being localised both in space and time, thus it is a particular type of \emph{instanton}.
In table~\ref{table:IIB_self-S-duality} we summarise how S-duality acts on the BPS objects of the IIB spectrum. As we said previously, these provide a very import opportunity for testing S-duality.
\begin{table}[h!]
\begin{center}
\begin{tabular}{| c || c |}
\hline 
IIB & IIB$'$ \\[1mm]
\hline \hline 
$\left(g_{s},\,\ell_{s}\right)$ & $\left(g_{s}^{-1},\,\ell_{s}\,g_{s}^{1/2}\right)$ \\[1mm]
\hline
NS1 & D1 \\[1mm]
\hline
NS5 & D5 \\[1mm]
\hline
D1 & NS1 \\[1mm]
\hline
D3 & D3 \\[1mm]
\hline
D5 & NS5 \\[1mm]
\hline
D7 &  $\widetilde{\textrm{D7}}$ \\[1mm]
\hline
\end{tabular}
\end{center}
\caption{{\it The action of S-duality on type IIB branes. By acting with the full SL($2,\mathbb{Z}$), we can show that D3-branes are singlets, whereas (NS1, D1) and (NS5, D5) are doublets and hence we could obtain the dyonic $(p,q)$-string or 5-brane. D7-branes belong to a triplet of 7-branes labelled by the integers $(p,q,r)$. However, acting with SL($2,\mathbb{Z}$) on the D7, one only has access to a 2-dimensional conjugacy  class spanned by D7 and $\widetilde{\textrm{D7}}$ \protect\cite{Bergshoeff:2006ic}.}}
\label{table:IIB_self-S-duality}
\end{table}

\chapter{Gauged Supergravities}
\markboth{Gauged Supergravities}{Gauged Supergravities}
\label{Gauged_Sugra}
As we concluded in the previous chapter, supergravity theories in ten and eleven dimensions give a low-energy effective description of string theory and M-theory respectively. Upon toroidal reduction, these supergravities are related to supergravities in $D<10$. 
In this chapter we will briefly review how supergravities in various dimensions can be obtained by supersymmetrising a gravity theory. We will refer to them as \emph{ungauged supergravities}.
Furthermore we will show how to introduce deformations in supergravities to give rise to gauged supergravity theories.
In order to arrive there, we will need to first discuss supersymmetry and its relation to gravity.

\section{Supersymmetry}
\label{susy}

As already sketched in the introduction, supersymmetry is the result of the search for a fundamental symmetry unifying spacetime and internal symmetries in a non-trivial way. This is realised by certain spinorial conserved charges $Q$ called \emph{supercharges}. These are such that
\be
\textrm{bosons}\qquad\overset{Q}{\longrightarrow}\qquad\textrm{fermions}\ , \nn
\ee
and they square to bosonic transformations, such as translations, Lorentz transformations, etc.

Spinors in $1+(D-1)$ dimensions are the building blocks of fermionic representations of SO($1,D-1$). These are precisely the objects that one needs in order to discuss supersymmetry in given spacetime dimensions and signatures. In the so-called Dirac representation, the Lorentz generators are given by $\frac{1}{4}\,\gamma_{\mu\nu}\,\equiv\,\frac{1}{8}\,\left[\gamma_{\mu},\gamma_{\nu}\right]$, where the Dirac matrices $\left\{\gamma_{\mu}\right\}_{\mu=0,\dots,D-1}$ satisfy the Clifford algebra
\be
\left\{\gamma_{\mu},\gamma_{\nu}\right\}\,=\,2\,\eta_{\mu\nu}\ ,
\ee
where $\eta_{\mu\nu}\,=\,$diag$(-1,\underbrace{+1,\cdots,+1}_{D-1})$ is the Minkowski metric. Such a representation has real dimension equal to $2^{\textrm{\textlbrackdbl} D/2 \textrm{\textrbrackdbl}+1}$, where \textlbrackdbl$x$\textrbrackdbl$\,\,\,$denotes the integer part of $x$. However, depending on different spacetime dimensions and signatures, the components of a Dirac spinor might not all transform amongst themselves, yet they might contain different irreducible pieces, which are obtained by imposing some Lorentz-invariant constraint on a Dirac spinor.
For instance, in any even dimension, one can have \emph{chiral} spinors. These are obtained by imposing a chirality condition on a given Dirac spinor $\chi$
\be
\label{chiral}
\gamma_{*}\,\chi\,\equiv\,\left[\frac{(-i)^{D/2+1}}{D!}\,\epsilon^{\mu_{1}\cdots\mu_{D}}\,\gamma_{\mu_{1}}\,\cdots\,\gamma_{\mu_{D}}\right]\,\chi\,=\,\pm\,\chi\ ,
\ee
where $\epsilon^{\mu_{1}\cdots\mu_{D}}$ represents the Levi-Civita symbol in $D$ dimensions and the $+$ and the $-$ refer to right- and left-handed spinors respectively. Chiral spinors have therefore only $2^{\textrm{\textlbrackdbl} D/2 \textrm{\textrbrackdbl}}$ independent real components.

Another possible projection is a \emph{reality condition} giving rise to Majorana spinors. These irreducible spinors are objects satisfying the following constraint

\be
\label{reality}
\chi^{T}\,\mathcal{C}=\,\chi^{\dagger}\,\gamma_{0}\ ,
\ee
which reduces indeed to a reality condition for the components of $\chi$ whenever the charge-conjugation matrix $\mathcal{C}$ is chosen to be equal to $\gamma_{0}$. In any other case, \eqref{reality} plays only the formal role of a reality condition without being it in a strict sense. 

In general, whenever one decomposes a Dirac spinor $\chi$ in terms of its chiral components, these will violate the Majorana condition \eqref{reality}. Nevertheless, there are some special cases in which the conditions \eqref{chiral} and \eqref{reality} can be satisfied simultaneously by some irreducible spinors having only $2^{\textrm{\textlbrackdbl} D/2 \textrm{\textrbrackdbl}-1}$ real independent components. These spinors are called \emph{Majorana-Weyl} (MW) spinors. 

The complete details about spinors in various spacetime dimensions and signatures have been worked out in large detail in ref.~\cite{VanProeyen:1999ni}. For the sake of simplicity, we refrain from the full discussion and summarise some relevant information in table~\ref{table:spinors}.

\begin{table}[h!]
\begin{center}
\begin{tabular}{| c | c | c |}
\hline
$D$ (mod $8$) & Spinor irrep's & Real components \\[1mm]
\hline \hline
$1$, $3$ & M & $2^{(D-1)/2}$ \\[1mm]
$2$ & MW & $2^{D/2-1}$ \\[1mm]
$4$, $8$ & M & $2^{D/2}$ \\[1mm]
$5$, $7$ & D & $2^{(D+1)/2}$ \\[1mm]
$6$ & W & $2^{D/2}$ \\[1mm]
\hline
\end{tabular}
\end{center}
\caption{{\it The different irreducible spinors in various dimensions, but always with one single time direction. D stands for 'Dirac', M stands for 'Majorana', W for 'Weyl' and MW for 'Majorana-Weyl' spinors.}}
\label{table:spinors}
\end{table}
In a theory, the amount of supercharges has to be a multiple $\mathcal{N}$ of the number of real components of an irreducible spinor in $D$ dimensions\footnote{This is required by Lorentz invariance, since the components of an irreducible spinor all transform into each other under an SO($1,D-1$) transformation.}. The supercharges are then objects of the form ${Q^{i}}_{\alpha}$, where $i=1,\dots,\mathcal{N}$ and $\alpha$ is an irreducible spinor index. 

Each supercharge relates two fields whose helicity differs by $\frac{1}{2}$, thus filling the so-called \emph{supermultiplets} (representations of supersymmetry) with fields of increasing helicity. Because of this, there is an upper bound \cite{Nahm:1977tg} on the maximal number of supercharges that a theory can have if we do not want our supermultiplets to contain fields with spin higher than two. This requirement is related to the difficulties encountered in constructing an interacting Lagrangian for higher-spin particles even at a classical level\footnote{This statement refers to the assumption of higher-spin fields in a Minkowski background.}.
In particular, in theories with global supersymmetry, one cannot have more than 16 supercharges in the game in order to avoid gravitational degrees of freedom which would require gauged supersymmetry. In theories with local supersymmetry (supergravities), one is allowed to include up to spin 2 degrees of freedom. This enhances the maximal amount of supercharges to 32. We will refer to these theories as maximal supergravities. We would like to stress that this general analysis can be done by discarding the possibility of including higher-spin fields in the theory. However, the study of the dynamics of higher-spin fields has been studied over the years in the literature \cite{Fubini:1967zz, Fradkin:1987ks, Bergshoeff:1988jm, Vasiliev:1995dn} and it has recently received new attention \cite{Vasiliev:2004qz, Vasiliev:2004cm, Sorokin:2008tf, Bergshoeff:2009tb, Bergshoeff:2011pm}.

Summarising, the introduction of supersymmetry provides a unification of spacetime and internal symmetries by promoting ordinary Lie algebras to \emph{superalgebras} \cite{Kac:1977em}, objects in which the supercharges ${Q^{i}}_{\alpha}$ appear as fermionic generators. A superalgebra $\mathfrak{S}$ is defined as follows:
\begin{itemize}

\item $\mathfrak{S}$ is a graded vector space, \emph{i.e.} it admits a map
\be
\textrm{gr : }\quad\mathfrak{S}\quad\longrightarrow\quad\mathbb{Z}_{2}\ ,
\ee
which decomposes $\mathfrak{S}$ into $\mathfrak{S}^{(0)}\,\oplus\,\mathfrak{S}^{(1)}$ such that
\be
\begin{array}{lclc}
\textrm{gr}(B)=0\textrm{ mod }2 & , & \forall\,B\,\in\,\mathfrak{S}^{(0)} & , \\
\textrm{gr}(F)=1\textrm{ mod }2 & , & \forall\,F\,\in\,\mathfrak{S}^{(1)} & , 
\end{array}
\ee
which define bosonic ($B$) and fermionic ($F$) generators respectively,

\item there exists a bilinear and supercommutative internal composition law $\{\,,\,]$ 
\be
\left\{A,\,B\right]\,=\,(-1)^{1+\textrm{gr}(A)\textrm{gr}(B)}\,\left\{B,\,A\right]\ ,
\ee
such that
\begin{itemize}
\item $\{\,,\,]$ is additive with respect to ~gr,
\be
\textrm{gr}(\{A,\,B])\,=\,\textrm{gr}(A)\,+\,\textrm{gr}(B)\ ,
\ee

\item the super-Jacobi identities are satisfied for any $A,\,B,\,C\,\in\,\mathfrak{S}$,
\be
(-1)^{1+\textrm{gr}(A)\textrm{gr}(C)}\,\big\{\{A,\,B],\,C\big]\,+\,\textrm{(cyclic perm.)}\,=\,0\ .
\ee

\end{itemize}

\end{itemize}
A classification of superalgebras can be found in ref.~\cite{Castellani:1991et}; among the physically relevant superalgebras we find \emph{e.g.} the orthosymplectic superalgebra Osp($4|\cN$) \cite{Fabbri:1999ay}, which has as bosonic Lie algebra SO($3,2)\,\times\,$SO($\cN$) and corresponds to the AdS superalgebra. Another important superalgebra is the \emph{superconfromal} one SU($2,2|\cN$), having as Lie algebra SO($4,2)\,\times\,$SU($\cN)\,\times\,$U($1$).

Basically, a superalgebra defines an extension of an ordinary Lie algebra generated by a set of bosonic generators by the addition of a set of fermionic generators, for which the commutation relations with the bosonic symmetries and the anti-commutation relations among themselves are specified. Together with this 'fermionic' extension, a superalgebra includes a new bosonic symmetry called $R$-\emph{symmetry}, which is defined as the largest subgroup of the automorphism group of the supersymmetry algebra that commutes with Lorentz transformations. Therefore, $R$-symmetry transforms the internal index $i=1,\dots,\mathcal{N}$ carried by the supercharges.
For more details about the origin of supersymmetry and superalgebras we refer to \cite{West_book}.

\subsection*{The Different Supermultiplets}
\label{subsec:supermultiplets}

In any supersymmetric theory, all the fields must be arranged into \emph{supermultiplets}, which are representations of supersymmetry grouping together all the different degrees of freedom that are related to each other by supersymmetry (\emph{i.e.} superpartners). A possible approach to construct different supermultiplets is that of using the superfield formalism. Superfields are objects defined on the so-called superspace, which is an extension of ordinary spacetime obtained by supplementing it with a number of Grassmann (\emph{i.e.} anticommuting) coordinates depending on the value of $\mathcal{N}$. However, we are not going to discuss this approach here in detail. The most common supermultiplets encountered in supergravity are

\begin{itemize}

\item {\bf Gravity multiplets:} It is the minimal multiplet containing the graviton. It contains all the fields that represent the supersymmetry algebra on-shell. The explicit field content of these multiplets is given in table~\ref{table:gravity_multiplet} for $D=4$.

\begin{table}[h!]
\begin{center}
\begin{tabular}{| c | c || c | c | c | c | c | c | c |}
\hline
field & $s$ & $\cN=1$ & $\cN=2$ & $\cN=3$ & $\cN=4$ & $\cN=5$ & $\cN=6$ & $\cN=8$ \\[1mm]
\hline \hline 
$g_{\mu\nu}$ & $2$ & $1$ & $1$ & $1$ & $1$ & $1$ & $1$ & $1$ \\[1mm]
\hline
$\psi_{\mu}$ & $\frac{3}{2}$ & $1$ & $2$ & $3$ & $4$ & $5$ & $6$ & $8$ \\[1mm]
\hline
$A_{\mu}$ & $1$ &  & $1$ & $3$ & $6$ & $10$ & $16$ & $28$ \\[1mm]
\hline
$\lambda$ & $\frac{1}{2}$ &  &  & $1$ & $4$ & $11$ & $26$ & $56$ \\[1mm]
\hline
$\phi$ & $0$ &  &  &  & $2$ & $10$ & $30$ & $70$ \\[1mm]
\hline
\end{tabular}
\end{center}
\caption{{\it The field content of the gravity multiplets in $D=4$ for the various supergravity theories with different values of $\cN$. The number of on-shell degrees of freedom are to be multiplied by $2$ for every state with $s>0$. Please note that the $\cN=7$ analysis gives the same field content as in the $\cN=8$ case. Adapted from ref.~\protect\cite{SUGRABOOK}.}}
\label{table:gravity_multiplet}
\end{table}

\item {\bf Vector multiplets:} These multiplets contain only states with spin up to $1$ and they exist only for $\cN \le 4$. As it happens in type I string theory, the gauge fields of these multiplets can gauge an extra Yang-Mills-like group which is not part of the superalgebra. The explicit field content of these multiplets is given in table~\ref{table:vector_multiplet} for $D=4$.

\begin{table}[h!]
\begin{center}
\begin{tabular}{| c | c || c | c | c |}
\hline
field & $s$ & $\cN=1$ & $\cN=2$ & $\cN=4$ \\[1mm]
\hline \hline 
$A_{\mu}$ & $1$ & $1$ & $1$ & $1$ \\[1mm]
\hline
$\lambda$ & $\frac{1}{2}$ & $1$ & $2$ & $4$ \\[1mm]
\hline
$\phi$ & $0$ &  & $2$ & $6$ \\[1mm]
\hline
\end{tabular}
\end{center}
\caption{{\it The field content of the vector multiplets in $D=4$ for the various supergravity theories with different values of $\cN$. The number of on-shell degrees of freedom are to be multiplied by $2$ for every state with $s>0$. Please note that the $\cN=3$ analysis gives the same field content as in the $\cN=4$ case. Adapted from ref.~\protect\cite{SUGRABOOK}.}}
\label{table:vector_multiplet}
\end{table}

\item {\bf Chiral multiplets:} These are multiplets which only contain states with spin $0$ and $1/2$. In four dimensions, they only exist in $\cN=1$ theories. Supersymmetry requires the scalars to span a \emph{K\"ahler-Hodge} manifold, as we will see in more detail in section~\ref{section:N=1W}. The field content of chiral multiplets in $D=4$ is presented in table~\ref{table:matter_multiplet} together with that one of hypermultiplets.

\item {\bf Hypermultiplets:} They are the analog of chiral multiplets for $\cN=2$ theories and they also only contain states with spin $0$ and $1/2$. $\cN=2$ supersymmetry restricts the scalar to span a so-called \emph{Quaternionic K\"ahler} (QK) manifold. 

\begin{table}[h!]
\begin{center}
\begin{tabular}{| c | c || c | c |}
\hline
field & $s$ & $\cN=1$ & $\cN=2$ \\[1mm]
\hline \hline 
$\lambda$ & $\frac{1}{2}$ & $1$ & $2$ \\[1mm]
\hline
$\phi$ & $0$ & $2$ & $4$ \\[1mm]
\hline
\end{tabular}
\end{center}
\caption{{\it The field content of chiral ($\cN=1$) and hypermultiplets ($\cN=2$) in $D=4$. The number of on-shell degrees of freedom are to be multiplied by $2$ for every state with $s>0$. Adapted from ref.~\protect\cite{SUGRABOOK}.}}
\label{table:matter_multiplet}
\end{table}

\item {\bf Tensor multiplets:} These multiplets include the presence of antisymmetric tensors $T_{\mu\nu}$. However, in dimensions four and five, such tensors can be dualised to scalars and vectors respectively\footnote{We would like to stress that, in $D=5$, the presence of 2-forms still causes important physical differences in the gauged theory (see the line referring to $D=5$ in table~\ref{table:half-max}).}. In $D=6$, instead they can have (anti-)selfduality properties and hence tensor multiplets have a completely new physical content. Tensor multiplets can appear in $D=6$ $\cN=(2,0)$ supergravity (iib) (see table~\ref{table:diff_sugra's}).

\end{itemize}

Generically, supersymmetry is realised on-shell (\emph{i.e.} only when the equations of motion are satisfied), in the sense that the supersymmetry algebra closes only up to terms which are zero when evaluated at a solution of the equations of motion. In order to construct an off-shell realisation of supersymmetry, one typically needs to introduce a bunch of \emph{auxiliary fields} whose variation under supersymmetry transformations precisely cancels the contributions coming from other fields which prevent the superalgebra from closing off-shell. These auxiliary fields are, however, non-dynamical since there is no kinetic term associated to them in the Lagrangian. Besides, they are very difficult to interpret physically since their dimensionality is larger than $(D-1)/2$.

Once the field content of a supersymmetric theory is determined, the following theorem always turns out to hold:

\emph{The number of fermionic degrees of freedom always matches the number of bosonic ones in any realisation of supersymmetry whenever the right-hand side of the anticommutation relation between two supersymmetries is an invertible operator.}

This relation between bosonic and fermionic degrees of freedom is exactly what makes a supersymmetric theory much more constrained on the one hand, but, on the other hand, much better-behaved in the UV, since there are certain physical quantities computable from the theory which are protected by supersymmetry.

\section{Ungauged Supergravities}
\label{sec:supergravities}

A relativistic gravity theory in $1+3$ dimensions such as Einstein's general relativity (GR) describes all the objects as sources of the energy-momentum tensor curving spacetime around them, thus rendering gravity a geometric effect. However, one can always describe spacetime by means of a so-called locally inertial frame, in which spacetime looks locally flat and it is only when moving away from a given point that one can see the spacetime curvature as an effect of gravitational interaction. This locally inertial frame corresponds to the choice of a certain \emph{tetrad} (\emph{i.e.} vierbien) ${e_{\mu}}^{a}$ which can be arbitrarily rotated at every point and is subject to local diffeomorphisms.

This manifestly shows that GR is invariant under local Lorentz transformations and translations; these objects generate the Poincar\'e algebra $\mathfrak{iso}(1,3)$
\be
\begin{array}{lclclc}
\left[M_{\mu\nu},M^{\rho\sigma}\right]\,=\,-2\,\delta^{[\rho}_{[\mu}\,{M_{\nu]}}^{\sigma]} & , & \left[P_{\mu},M_{\nu\rho}\right]\,=\,\eta_{\mu[\nu}\,P_{\rho]} & , & \left[P_{\mu},P_{\nu}\right]\,=\,0 & ,
\end{array}
\ee
where $M_{\mu\nu}\,=\,M_{[\mu\nu]}$ represent the SO($1,3$) Lorentz generators, $P_{\mu}$ denote spacetime translations and all the spacetime indices can be raised and lowered by using the metric $\eta$.

GR can be obtained in a very elegant way by gauging the Poincar\'e algebra \cite{Utiyama:1956sy,Kibble:1961ba} given above, where the vierbein ${e_{\mu}}^{a}$ and the spin connection ${\omega_{\mu}}^{ab}$ are regarded as independent gauge fields. This construction is called \emph{first order formalism} \cite{MacDowell:1977jt} and in general it gives rise to a description of gravity with torsion. This formalism turns out to play an important role in supergravity, in that the fermionic fields induce a torsion as a consequence of the equations of motion.

Supergravities in various dimensions are supersymmetric extensions of GR. Their local symmetries are certain superalgebras extending the bosonic spacetime symmetries, \emph{e.g.} the super-Poincar\'e algebra, which reads
\be
\hspace{-4mm}
\begin{array}{lll}
\left[M_{\mu\nu},\,M^{\rho\sigma}\right]\,=\,-2\,\delta^{[\rho}_{[\mu}\,{M_{\nu]}}^{\sigma]}\ , & \left[P_{\mu},\,M_{\nu\rho}\right]\,=\,\eta_{\mu[\nu}\,P_{\rho]}\ , & \left[P_{\mu},\,P_{\nu}\right]\,=\,0\ , \\[2mm]
\left[M_{\mu\nu},\,{Q^{i}}_{\alpha}\right]\,=\,-\frac{1}{4}\,{\left(\gamma_{\mu\nu}\right)_{\alpha}}^{\beta}\,{Q^{i}}_{\beta} \ , & \left[P_{\mu},\,{Q^{i}}_{\alpha}\right]\,=\,0\ , & \hspace{-3mm}\left\{{Q^{i}}_{\alpha},\,{Q^{j}}_{\beta}\right\}\,=\,\left(\gamma^{\mu}\,\mathcal{C}^{-1}\right)_{\alpha\beta}\,P_{\mu}\,\delta^{ij}\ ,
\end{array}
\ee
where $i,\,j\,=\,1,\dots\,\mathcal{N}$ run over the number of supersymmetries. Since Majorana spinors in $1+3$ dimensions have $4$ independent real components, the maximal theory corresponds to $\mathcal{N}=8$. In the rest of the thesis, we shall refer to the case $\mathcal{N}=1$ in four dimensions as minimal supersymmetry, whereas any other case with $\mathcal{N}>1$ will be called extended supersymmetry.

Other superalgebras called \emph{superconformal} are obtained by performing the same supersymmetric extension of SO($2,d$) algebras, which describe the symmetries of a conformal field theory (CFT) in $d$ dimensions. Superconformal algebras have been used in the past in order to construct supergravity theories in different dimensions and with different amounts of supersymmetry. See for instance refs~\cite{Deser:1976eh, VanNieuwenhuizen:1981ae} for the construction of minimal supergravity in four dimensions and refs~\cite{Bergshoeff:1980is, Bergshoeff:1983qk} for the case of extended supergravities. Furthermore, the reader can find the topic presented in a more pedagogical approach in ref.~\cite{VanProeyen:2000}.
We would like to stress that so far superconformal algebras have been used merely as a tool for constructing supergravities and, even though there are some indications that they might play a more fundamental role (\emph{e.g.} in the context of supersymmetric charged black holes \cite{Dibitetto:2010sp}), their relevance in supergravity still remains unclear.

\subsection*{The Different Supergravity Theories}

In section~\ref{susy} we showed the different irreducible spinors in various dimensions and we also presented a bound on the total number of supercharges that a supergravity theory can have. If we combine these two pieces of information, we are able to see which are the values of $\mathcal{N}$ which are possible for different values of $D$. Given $\mathcal{N}>1$, according to whether chirality is defined in $D$ dimensions, one might have different possibilities in the choice for the chirality of the different supersymmetry generators (see \emph{e.g.} the case of $\mathcal{N}=2$ supergravities in $D=10$ $\rightarrow\,$IIA and IIB). This gives rise to the 'zoo' of all possible supergravity theories in various dimensions. The different supergravities for different values of $D$ are summarised in table~\ref{table:diff_sugra's}. 

Theories with $32$ supercharges are often called \emph{maximal supergravities}, whereas those ones with $16$ are called \emph{half-maximal supergravities}. $\mathcal{N}=1$ theories in any $D$ are often referred to as \emph{minimal supergravities}, but the corresponding number of supercharges increases with $D$, up to $D=11$ where the minimal and the maximal theory coincide. More details about the possible supergravities in different dimensions can be found in refs~\cite{Salam:1989fm, deWit:1997sz, Tanii:1998px, deWit:2002vz, VanProeyen:2003zj}.
\begin{table}[h!]
\begin{center}
\begin{tabular}{| c | c | c |}
\hline
$D$ & Supergravities ($\mathcal{N}$) & N$^{\underline{\circ}}$ of supercharges \\[1mm]
\hline \hline
$11$ & $1$ & $32$ \\[1mm]
\hline
$10$ & $\begin{array}{ccc} (1,0)\equiv\,\textrm{I}\,, & (1,1)\equiv\,\textrm{IIA}\,, & (2,0)\equiv\,\textrm{IIB} \end{array}$ & $16$, $32$, $32$ \\[1mm]
\hline
$9$, $8$, $7$ & $1$, $2$ & $16$, $32$ \\[1mm]
\hline
$6$ & $\begin{array}{cc} (1,0)\equiv\,\textrm{i}\,, &  (1,1)\equiv\,\textrm{iia}\,,\quad (2,0)\equiv\,\textrm{iib} \\ (2,1)\,,\quad (3,0)^{*}\,, & (2,2)\,,\quad (3,1)^{*}\,,\quad (4,0)^{*} \end{array}$ & $\begin{array}{cc} 8, & 16,\, 16 \\ 24,\, 24^{*}, & 32,\, 32^{*}, \, 32^{*} \end{array}$ \\[1mm]
\hline
$5$ & $1$, $2$, $3$, $4$ & $8$, $16$, $24$, $32$ \\[1mm]
\hline
$4$ & $1$, $2$, $3$, $4$, $5$, $6$, $8$ & $4$, $8$, $12$, $16$, $20$, $24$, $32$ \\[1mm]
\hline
\end{tabular}
\end{center}
\caption{{\it The possible supergravities in different dimensions labelled by the number $\mathcal{N}$ of supersymmetries. Theories with $16$ or more supercharges can have different gaugings. Theories with up to $16$ supercharges can be coupled to matter multiplets (vector, tensor or hypermultiplets, depending on the case). Please note that the theories in $D=6$ marked with a $^{*}$ \protect\cite{Townsend:1983xt, Hull:2000zn} cannot be constructed in terms of a metric tensor but, instead, in terms of a more complicated irrep of the Poincar\'e group. So, strictly speaking, they are not supergravity theories.}}
\label{table:diff_sugra's}
\end{table}

In section~\ref{Strings}, we have mentioned that supergravity theories emerge as low energy effective descriptions of string and M-theory. Given this as a starting point, the most natural question that one can ask after looking at table~\ref{table:diff_sugra's} is whether all the aforementioned supergravities have their origin from string theory. This issue goes under the name of \emph{universality} of supergravities and it  has been discussed from different perspectives in the literature.

Starting from $D=11$, we see that there a unique supergravity theory and it corresponds exactly with the low energy limit of M-theory. In $D=10$, there are two inequivalent maximal supergravities, \emph{i.e.} type IIA and type IIB which are in perfect agreement with the corresponding superstring theories discussed in section~\ref{superstrings}. As for $\mathcal{N}=1$, there is a unique possibility, even though we have not yet specified the possible $496$-dimensional gauge groups. Recently \cite{Adams:2010zy} it has been proven that the only consistent (\emph{i.e.} anomaly free) gaugings at a quantum level are E$_{8}\,\times\,$E$_{8}$ and SO($32$), which exactly match the two possible heterotic string theories. 

Unfortunately, we are still unable to complete the picture: the more we go down with $D$ and $\mathcal{N}$, the more possibilities open up and it is not obvious how to generate all the lower-$\mathcal{N}$ supergravities from some dimensional reduction of string theory. There are some cases in which this uplift still remains an open problem. For example, the $\mathcal{N}=(1,0)$ supergravity in $D=6$ with gauge group E$_{6}\,\times\,$E$_{7}\,\times\,$U($1)_{R}$, for which still no link with string theory is known. However, still in $D=6$, there have been interesting recent developments in the context of universality \cite{Kumar:2009us}. 

One of the main goals of the present work is address the problem of universality in the context of gauged half-maximal and maximal supergravities in various dimensions. This issue will be analysed mainly in section~\ref{DFT}, even though other chapters contain sections which are not completely unrelated to it.

\subsection*{The Scalar Cosets $G_{0}/H$ in Extended Supergravities}
\label{G0/H}

For the main purpose of this work, let us concentrate on half-maximal and maximal supergravities. The fields of these supergravity theories transform in certain irrep's of the global symmetry group $G_{0}$. In particular, the scalars span the adjoint representation of $G_{0}$; however, all the scalar modes corresponding to compact $G_{0}$ generators are not physical, in the sense that they can always be rotated away. As a result, the physical degrees of freedom span a coset $G_{0}/H$, where $H$ is the maximal compact subgroup of $G_{0}$. 

Therefore, the scalars can be represented by a vielbein $\mathcal{V}$ which transforms under global $G_{0}$ transformations from the left and local $H$ transformations from the right
\be
\mathcal{V}\quad\longrightarrow\quad L \, \mathcal{V} \, h(x)\ ,
\ee
where $L\in G_{0}$ and $h(x)\in H$. The role of the vielbein $\mathcal{V}$ is going to be crucial both in the ungauged and in the gauged theory in order to construct couplings between $p$-form gauge fields and fermions. This is due to the fact that fermions only transform with respect to $H$ but not with respect to $G_{0}$ and hence one needs the scalar coset representative $\mathcal{V}$ to mediate all the interactions between fermions and bosons by converting local $H$ indices into global $G_{0}$ ones and vice versa.

The total number of physical scalars is then equal to the dimension of the coset space $G_{0}/H$. In every $D$, these numbers are presented in table~\ref{table:max} and \ref{table:half-max} of section~\ref{section:Theta}. These scalars are divided into \emph{dilatons} and \emph{axions}. The number of dilatons can be easily derived from the eleven-dimensional origin; after a reduction $11\,\rightarrow\,D=11-d$, one has a dilaton for any reduced dimension, thus $d$ in total. This exactly corresponds to the number of Cartan generators inside $G_{0}$ (the rank of $\mathfrak{g}_{0}$). All the other scalars are axions and their number turns out to be equal to the number of positive roots of $G_{0}$.

As the above analysis shows, the global symmetry group $G_{0}$ is generally bigger than simply SL($d$), which is what the eleven-dimensional origin of maximal supergravities would suggest. This is the reason why historically $G_{0}$ was called 'hidden symmetry' \cite{Cremmer:1978ds, Cremmer:1980gs}. Nevertheless, quite recently  a new formalism has been developed which allows us to understand this hidden symmetry from Kac-Moody algebras. When compactifying eleven-dimensional supergravity on a $T^{11}$, one finds a duality symmetry described by the infinite-dimensional algebra E$_{11}\,=\,$E$_{8}^{+++}$. A non-linear realisation of E$_{11}$ was conjectured in ref.~\cite{West:2001as} to describe an extension of eleven-dimensional supergravity. Subsequently, in ref.~\cite{Schnakenburg:2001ya}, non-linear realisations of E$_{11}$ were also shown to give rise to extensions of type IIA and type IIB supergravity. In general, the duality group of maximal supergravity in $D$ dimensions comes from the decomposition of E$_{11}$ \cite{Riccioni:2007au} into A$_{D-1}\,\times\,G_{D}$
\be
\,\underbrace{\,\,G_{D}\,\,}_{\textrm{duality
group}}\,\times\,\underbrace{\,\,\textrm{A}_{D-1}\,\,}_{\textrm{gravity line}}\ ,
\ee
where A$_{D-1}\,=\,$SL($D$) represents the diffeomorphism group in $D$ dimensions (\emph{i.e.} spacetime symmetry) and $G_{D}$ is such that the product A$_{D-1}\,\times\,G_{D}$ is a maximal subgroup of E$_{11}$ and it represents the duality group of maximal supergravity in $D$ dimensions. Further work in the same line was done in refs~\cite{Bergshoeff:2007qi, Bergshoeff:2008qd, Bergshoeff:2008xv}.

\begin{figure}[h!]
\begin{center}
\includegraphics[scale=0.05]{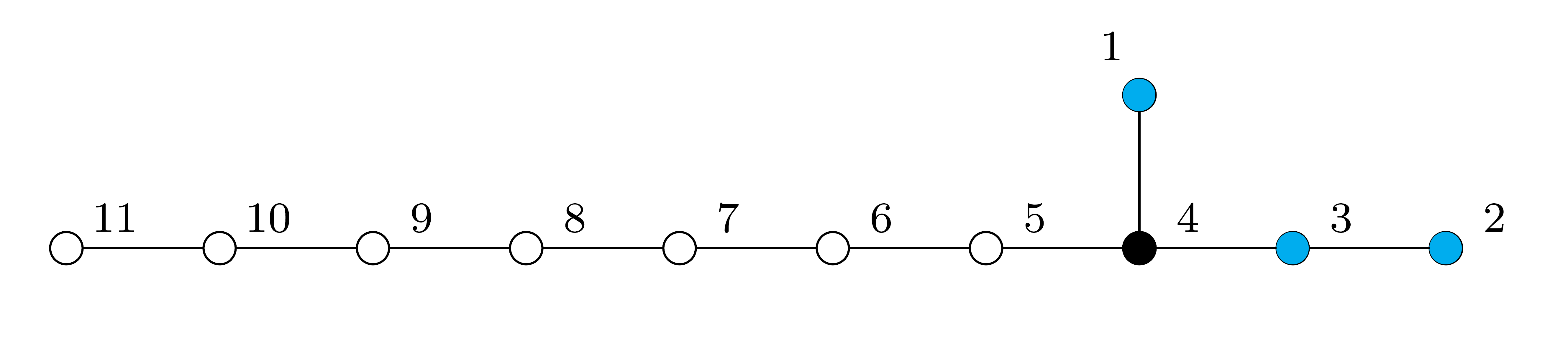}
\caption{{\it Decomposition of E$_{11}$ in terms of $\,(\textrm{A}_{1}\times\textrm{A}_{2})\,\times\,\textrm{A}_{7}$. All the black nodes neighbouring the gravity line are disabled and the cyan nodes represent the duality symmetry of the maximal theory in eight dimensions.}}\label{E11}
\end{center}
\end{figure}

\begin{figure}[h!]
\begin{center}
\includegraphics[scale=0.075]{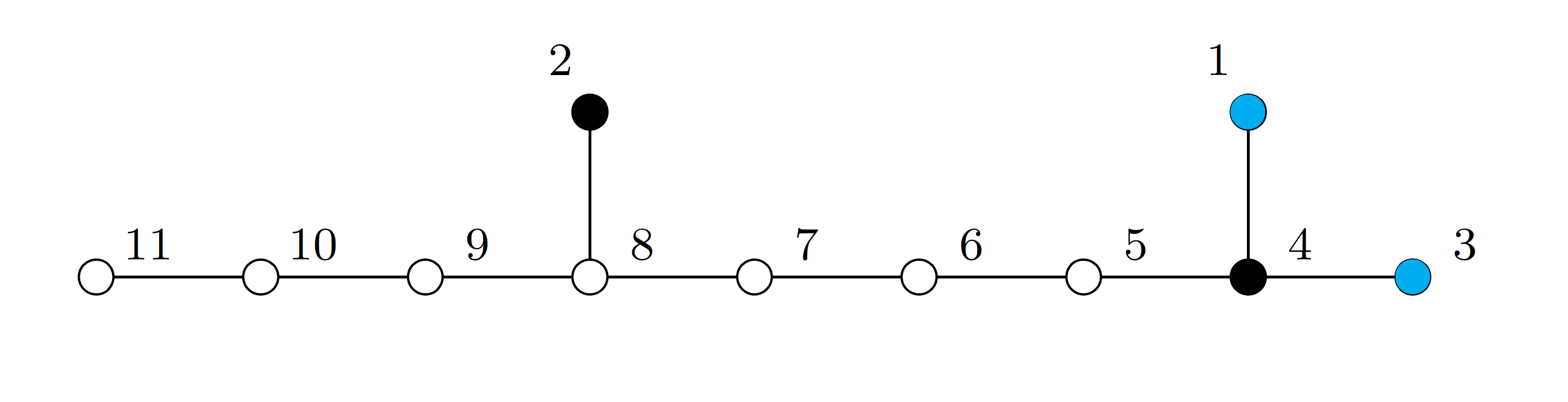}
\caption{{\it Decomposition of D$_{8}^{+++}$ in terms of $\,(\textrm{A}_{1}\times\textrm{A}_{1})\,\times\,\textrm{A}_{7}$. Again, all the black nodes neighbouring the gravity line are disabled and the cyan nodes represent the duality symmetry of the half-maximal theory in eight dimensions.}}\label{D8+++}
\end{center}
\end{figure}

This construction can be reproduced in the context of half-maximal supergravities \cite{Bergshoeff:2007vb} by using different Kac-Moody decompositions. The first example is the rank-11 algebra D$_{8}^{+++}$, which works in the case of $10-D$ vector multiplets; in other cases, different Kac-Moody algebras have been used (see \emph{e.g.} B$_{7}^{+++}$ and B$_{8}^{+++}$ in ref.~\cite{Bergshoeff:2007vb}). 
The Kac-Moody approach to (half-)maximal supergravities consists then in disintegrating the preferred Kac-Moody algebra into the gravity line (A$_{D-1}$) times the duality group. In this way, the full spectrum of the theory and its deformations can be determined. The general idea is sketched in the examples in figures~\ref{E11} and \ref{D8+++}. More details on this approach can be found in ref.~\cite{Teake:2010}. For similar analyses in theories with $8$ supercharges see ref.~\cite{Gomis:2007gb, Riccioni:2008jz}.
We have briefly presented the Kac-Moody approach as a valid method for deriving the duality symmetries, the spectra and the consistent deformations of extended supergravities, but we would like to stress that it still remains unclear whether Kac-Moody symmetries play a more fundamental role in supergravities and string theory. The case of E$_{11}$ has recently received a lot of attention in the literature \cite{West:2010ev, Steele:2010tk, Berman:2011jh, West:2011mm}, but there is still no final answer to the question whether E$_{11}$ can provide an organising principle for understanding the symmetries of eleven-dimensional supergravity.

\subsubsection*{Maximal Supergravities in $D=11$}

In $D=11$, a Majorana spinor has $32$ real components and hence the only possible supersymmetric theory that one can have is maximal supergravity, which, in this case, corresponds to $\mathcal{N}=1$. Maximal supersymmetry restricts the field content to one massless supermultiplet, \emph{i.e.} the gravity multiplet. These massless degrees of freedom are classified in terms of SO($9$) irrep's, where SO($9$) is the little group. They are divided into
\be
D=11: \qquad (\textbf{44}\,\oplus\,\textbf{84})_{\textrm{B}}\,\oplus\,(\textbf{128})_{\textrm{F}} \qquad \leftrightarrow \qquad \left\{{e_{\mu}}^{a},\,C_{\mu\nu\rho};\,\psi_{\mu}\right\}\ ,
\ee
which represent the vielbein, a 3-form gauge potential and a Majorana gravitino respectively.
The full action reads \cite{Cremmer:1978km}
\bea
S &=& \frac{1}{2\,\kappa^{2}}\int d^{11}x\,e\,\left[e^{a\mu}\,e^{b\nu}\,R_{\mu\nu ab}(\omega)\,-\,\overline{\psi}_{\mu}\,\gamma^{\mu\nu\rho}\,D_{\nu}\left(\frac{1}{2}\,(\omega\,+\,\check{\omega})\right)\,\psi_{\rho}\,-\,\frac{1}{24}\,G^{\mu\nu\rho\sigma}\,G_{\mu\nu\rho\sigma} \right. \nn\\
& &  -\,\frac{\sqrt{2}}{192}\,\overline{\psi}_{\nu}\,\left(\gamma^{\alpha\beta\gamma\delta\nu\rho}\,+\,12\,\gamma^{\alpha\beta}\,g^{\gamma\nu}\,g^{\delta\rho}\right)\,\psi_{\rho}\,\left(G_{\alpha\beta\gamma\delta}\,+\,\check{G}_{\alpha\beta\gamma\delta}\right) \nn\\
& & \left. -\,\frac{2\,\sqrt{2}}{(144)^{2}}\,e^{-1}\,\epsilon^{\alpha'\beta'\gamma'\delta'\alpha\beta\gamma\delta\mu\nu\rho}\,G_{\alpha'\beta'\gamma'\delta'}\,G_{\alpha\beta\gamma\delta}\,C_{\mu\nu\rho}\right]\ , \label{Lagrangian_11}
\eea
where 
\bea
\omega_{\mu ab} &=& \omega_{\mu ab}(e) \,+\, K_{\mu ab} \ , \nn\\
\check{\omega}_{\mu ab} &=& \omega_{\mu ab}(e) \,-\, \frac{1}{4}\,\left(\overline{\psi}_{\mu}\,\gamma_{b}\,\psi_{a}\,-\,\overline{\psi}_{a}\,\gamma_{\mu}\,\psi_{b}\,+\,\overline{\psi}_{b}\,\gamma_{a}\,\psi_{\mu}\right) \ , \nn\\
K_{\mu ab} &=& -\,\frac{1}{4}\,\left(\overline{\psi}_{\mu}\,\gamma_{b}\,\psi_{a}\,-\,\overline{\psi}_{a}\,\gamma_{\mu}\,\psi_{b}\,+\,\overline{\psi}_{b}\,\gamma_{a}\,\psi_{\mu}\right)\,+\,\frac{1}{8}\,\overline{\psi}_{\nu}\,{\gamma^{\nu\rho}}_{\mu ab}\,\psi_{\rho}\ , \nn\\
\check{G}_{\mu\nu\rho\sigma} &=& 4\,\partial_{[\mu}C_{\nu\rho\sigma]}\,+\,\frac{3}{2}\,\sqrt{2}\,\overline{\psi}_{[\mu}\,\gamma_{\nu\rho}\,\psi_{\sigma]}\ ,
\eea
and the covariant derivative $D$ acts on spinors as usual $D_{\nu}\psi_{\rho}\equiv\partial_{\nu}\psi_{\rho}\,+\,\frac{1}{4}\,\omega_{\nu ab}\,\gamma^{ab}\,\psi_{\rho}$ .

This theory has an on-shell $\mathbb{R}^{+}$ symmetry acting as
\be
\begin{array}{lclclc}
g_{\mu\nu}\,\rightarrow\,\lambda^{2}\,g_{\mu\nu} & , & C_{\mu\nu\rho}\,\rightarrow\,\lambda^{3}\,C_{\mu\nu\rho} & , & \psi_{\mu}\,\rightarrow\,\lambda^{1/2}\,\psi_{\mu}\ ,
\end{array}\label{lambda_11}
\ee
where $\lambda\,\in\,\mathbb{R}^{+}$. However, the Lagrangian \eqref{Lagrangian_11} has a non-trivial weight under the rescaling \eqref{lambda_11}. This implies that this $\mathbb{R}^{+}$ cannot be promoted to an off-shell symmetry. Such a symmetry is often referred to in the literature as \emph{trombone symmetry} \cite{Cremmer:1997xj}. The presence of the trombone symmetry is a general feature of all ungauged supergravities in any $D$.

\subsubsection*{Maximal Supergravities in $D=10$} 

In $D=10$ with one time direction, MW fermions are the irreducible spinors. The maximal theories correspond to $\mathcal{N}=2$; since the two supersymmetry generators in the theory are real and chiral, there are two discrete inequivalent possibilities (see also table~\ref{table:diff_sugra's}): $\mathcal{N}=(1,1)$ (opposite chiralities) and $\mathcal{N}=(2,0)$ (same chirality). These correspond to type IIA and type IIB respectively. The consistence of the corresponding superalgebra in IIA and IIB implies the possibility of extension by including gauge symmetries of different rank. This translates into the fact that the two inequivalent supergravities have different types of gauge fields. 

In this subsection we will explicitly follow the conventions of ref~\cite{Roest:2004pk}. Again because of maximal supersymmetry, only the gravity multiplet is allowed; its on-shell degrees of freedom rearranged in terms of SO($8$) irrep's read
\be
\begin{array}{cc}
\textrm{IIA}: & \big((\textbf{1}\,\oplus\,\textbf{28}\,\oplus\,\textbf{35}_{\textrm{V}})\,\oplus\,(\textbf{8}_{\textrm{V}}\,\oplus\,\textbf{56}_{\textrm{V}})\big)_{\textrm{B}}\,\oplus\,\big((\textbf{8}_{\textrm{S}}\,\oplus\,\textbf{56}_{\textrm{S}})\,\oplus\,(\textbf{8}_{\textrm{C}}\,\oplus\,\textbf{56}_{\textrm{C}})\big)_{\textrm{F}}\ ,\\[2mm]
\textrm{IIB}: & \big((\textbf{1}\,\oplus\,\textbf{28}\,\oplus\,\textbf{35}_{\textrm{V}})\,\oplus\,(\textbf{1}\,\oplus\,\textbf{28}\,\oplus\,\textbf{35}_{\textrm{C}})\big)_{\textrm{B}}\,\oplus\,\big(2\,\cdot\,(\textbf{8}_{\textrm{S}}\,\oplus\,\textbf{56}_{\textrm{S}})\big)_{\textrm{F}}\ .
\end{array}\label{IIA_IIB_sugra}
\ee
The degrees of freedom in \eqref{IIA_IIB_sugra} can be translated into the following field contents
\be
\begin{array}{cc}
\textrm{IIA}: & \left\{g_{\mu\nu},\,B_{\mu\nu},\,\phi,\,C^{(1)}_{\mu},\,C^{(3)}_{\mu\nu\rho};\,\psi_{\mu},\,\chi\right\}\ ,\\[2mm]
\textrm{IIB}: & \left\{g_{\mu\nu},\,B_{\mu\nu},\,\phi,\,C^{(0)},\,C^{(2)}_{\mu\nu},\,C^{(4)}_{\mu\nu\rho\sigma}|_{\textrm{SD}};\,\psi_{\mu},\,\chi\right\}\ ,
\end{array}\label{fields_IIA_IIB}
\ee
where the subscript SD on $C^{(4)}$ stands for self-dual and the fermions $\psi_{\mu}$ and $\chi$ are chosen in IIA to be real and containg two irreducible spinors of both chiralities, whereas in IIB, they are complex and containg two irreducible spinors of only one chirality.

After introducing the modified field strenghts for the $p$-form potentials
\be
G^{(d+1)}\,\equiv\,dC^{(d)}\,-\,dB\,\wedge\,C^{(d-2)}\ ,
\ee
one can define the duality relation between a $(d+1)$-form and a $(9-d)$-form
\be
\star\,G^{(9-d)}\,=\,(-1)^{(d+1)/2}\,e^{(4-d)\,\phi/2}\,G^{(d+1)}\ ,
\ee
which turns out to give rise to a self-duality (SD) condition for $G^{(5)}$, which is the field strength of $C^{(4)}$ appearing in \eqref{fields_IIA_IIB}. 

The bosonic part of the Lagrangian of type IIA supergravity reads
\bea
\mathcal{L}_{\textrm{IIA}} &=& \sqrt{-g}\,\bigg[R\,-\,\frac{1}{2}\,(\partial \phi)^{2}\,-\,\frac{1}{2}\,e^{-\phi}\,|H|^{2}\,-\,\frac{1}{2}\,\sum_{d=1,3}e^{(4-d)\,\phi/2}\,|G^{(d+1)}|^{2} \nn\\
& &  -\,\frac{1}{2}\,\star\, \left(dC^{(3)}\,\wedge\,dC^{(3)}\,\wedge\,B\right)\bigg] \ ,
\eea
where $H$ is the field strength associated to the NS-NS 2-form $B$. Type IIA supergravity has two different $\mathbb{R}^{+}$ symmetries: the first one is the trombone symmetry, analog to the one already encountered in $D=11$, whereas the second one is a proper symmetry of the Lagrangian and it acts on the fields in the following way
\be
\begin{array}{cccc}
e^{\phi}\,\rightarrow\,\lambda\,e^{\phi}\ , & B\,\rightarrow\,\lambda^{1/2}\,B\ , & C^{(1)}\,\rightarrow\,\lambda^{1/2}\,C^{(1)}\ , & C^{(3)}\,\rightarrow\,\lambda^{-1/4}\,C^{(3)}\ ,
\end{array}
\ee
and leaves the rest of the fields invariant.

The bosonic part of the Lagrangian of type IIB supergravity reads
\bea
\mathcal{L}_{\textrm{IIB}} &=& \sqrt{-g}\,\bigg[R\,-\,\frac{1}{2}\,(\partial \phi)^{2}\,-\,\frac{1}{2}\,e^{-\phi}\,|H|^{2}\,-\,\frac{1}{2}\,\sum_{d=0,2,4}e^{(4-d)\,\phi/2}\,|G^{(d+1)}|^{2} \nn\\
& &  -\,\frac{1}{2}\,\star\, \left(C^{(4)}\,\wedge\,dC^{(2)}\,\wedge\,B\right)\bigg] \ , \label{L_IIB}
\eea
which has to be supplemented by the SD condition for $G^{(5)}$. Since there is no way of having an off-shell formulation of type IIB supergravity which already takes this condition into account, \eqref{L_IIB} defines what is often called a pseudo-action.
Type IIB supergravity has two different symmetries: a trombone symmetry (which is, as always, only realised on-shell) and an SL($2$) symmetry. Any element 
\be
{\Lambda^{\alpha}}_{\beta}\,\equiv\,\left(\begin{array}{cc}a & b \\ c & d \end{array}\right)\,\in\,\textrm{SL}(2)
\ee 
acts on the fields in the second row of \eqref{fields_IIA_IIB} in the following way
\be
\begin{array}{ccc}
\tau\,\rightarrow\,\dfrac{a\,\tau\,+\,b}{c\,\tau\,+\,d}\ , & B^{\alpha}\,\rightarrow\,{\left(\Lambda^{-1}\right)_{\beta}}^{\alpha}\,B^{\beta}\ , & C^{(4)}\,\rightarrow\,C^{(4)}\ ,
\end{array}\nn
\ee
\be
\begin{array}{cc}
\psi_{\mu}\,\rightarrow\,\left(\dfrac{c\,\tau^{*}\,+\,d}{c\,\tau\,+\,d}\right)^{1/4}\,\psi_{\mu}\ ,\quad \chi\,\rightarrow\,\left(\dfrac{c\,\tau^{*}\,+\,d}{c\,\tau\,+\,d}\right)^{3/4}\,\chi \ ,
\end{array}\nn
\ee
where, for convenience, we have defined $\tau\,\equiv\,C^{(0)}\,+\,i\,e^{-\phi}$ and $B^{\alpha}\,\equiv\,\left(-B,\,C^{(2)}\right)$. In ref.~\cite{Bergshoeff:2005ac} the SL($2$) covariant reformulation of type IIB supergravity can be found. Type IIB string theory breaks SL($2$) into its discrete subgroup SL($2,\mathbb{Z}$). This group contains the so-called S-duality transformation which flips the sign of the dilaton $\phi$ in a background with vanishing axion $C^{(0)}$. Because of its very definition, S-duality turns out to be a non-perturbative duality relating the strong- and weak-coupling regimes. 

\section{The Embedding Tensor Formalism}
\label{section:Theta}

Any ungauged supergravity in any dimension can be deformed (\emph{i.e.} gauged) by promoting a certain subgroup of its global bosonic symmetry to a local one. In the last decade, a very powerful formalism has been developed in the context of extended supergravities in order to give an exhaustive formulation of the consistent gaugings of supergravity. This is called \emph{embedding tensor formalism} \cite{Nicolai:2000sc, deWit:2002vt, deWit:2005hv}. In this section we will briefly present a general discussion in the case of (half-)maximal supergravities in various dimensions; for more details on this part, we refer to \cite{Weidner:2006rp}. An analogous formalism may be developed also in the minimally extended case (\emph{i.e.} $\mathcal{N}=2$ in four dimensions, see for instance ref.~\cite{deWit:2011gk}), but this goes beyond the aim of this thesis. 

The global symmetry group of the (half-)maximal theory, which is fixed by supersymmetry\footnote{Actually, in the half-maximal case, it is only fixed after choosing the number of vector multiplets $n$ that one wants to couple to gravity, whereas in the maximal theory it is really fixed since maximal supersymmetry does not allow for extra matter content.}, turns out to rigidly determine and organise all the possible deformations, which can therefore be described in a universal covariant formulation. As we will see later on more explicitly, the global symmetries of these theories can be interpreted as the remnant of dualities relating the different string theories from which they originate.

From now on, we will denote the global symmetry group of our ungauged supergravity theory by $G_{0}$. The gauging procedure promotes a subgroup $G\,\subset\,G_{0}$ to a local symmetry. This procedure breaks the symmetry of the gauged theory from $G_{0}$ to $G$. However, there is a way of promoting the structure constant of the gauge algebra to an embedding tensor $\Theta$ which transforms under the full $G_{0}$. To summarise this point, as long as one considers $\Theta$ as a tensor, the full $G_{0}$ covariance of the theory is recovered.

After gauging $G$, one needs to introduce minimal couplings of the vector gauge fields in order to preserve gauge invariance. This implies replacing ordinary derivatives $\partial_{\mu}$ with covariant ones $D_{\mu}$ in the Lagrangian. The algebra $\mathfrak{g}_{0}=\mathfrak{lie}(G_{0})$ is generated by $\{t_{\alpha}\}_{\alpha=1,\dots,\textrm{dim}(\mathfrak{g}_{0})}$, where $\alpha$ is an adjoint index. They satisfy
\be
\left[t_{\alpha},\,t_{\beta}\right]\,=\,{f_{\alpha\beta}}^{\gamma}\,t_{\gamma}\ .
\ee

Let us denote by $V$ the representation in which the vectors ${A_{\mu}}^{M}$ of the theory in exam transform (for examples see tables~\ref{table:max} and \ref{table:half-max}).  These vectors will now transform under both global $G_{0}$ transformations $L^{\alpha}$ and local $G$ transformations $\Lambda^{M}(x)$
\be
\begin{array}{lclc}
\delta_{L}{A_{\mu}}^{M}\,=\,-L^{\alpha}\,{\left[t_{\alpha}\right]_{N}}^{M}\,{A_{\mu}}^{N} & , & \delta_{\Lambda}{A_{\mu}}^{M}\,=\,\partial_{\mu}\Lambda^{M} & .
\end{array}
\ee
In order to construct the covariant derivative $D_{\mu}$ we need to relate indices of $V$ ($M,N,\dots$) to adjoint indices ($\alpha,\beta,\dots$); this will allow us to write down a minimal coupling for the vector gauge fields. 
This is explicitly done by a linear map
\be
\label{ET_map}
\Theta\quad : \quad V \quad \longrightarrow \quad \mathfrak{g}_{0} \ ,
\ee
called \emph{embedding tensor} which precisely specifies how the vectors enter the gauging procedure, hence completely specifying the gauged theory. The map defined in \eqref{ET_map} allows us to write down the gauge-covariant derivative as
\be
D_{\mu}\,=\,\partial_{\mu}\,-\,g\,{A_{\mu}}^{M}\,{\Theta_{M}}^{\alpha}\,t_{\alpha}\ ,
\ee
where $g$ denotes the gauge coupling.

The embedding tensor $\Theta$ also explicitly specifies the generators $X_{M}$ of the gauge group 
\be
X_{M}\,=\,{\Theta_{M}}^{\alpha}\,t_{\alpha}\ .
\ee
As a consequence of \eqref{ET_map}, the embedding tensor will in general transform in the tensor product between the conjugate representation of $V$ (which we will denote by $V^{\prime}$) and the adjoint representation $\mathfrak{g}_{0}$ of $G_{0}$. This will in general contain several irrep's $\theta_{i}$, with $i=1,\dots\,n$
\be
\label{V'xg0}
\Theta\,\in\,V^{\prime}\,\otimes\,\mathfrak{g}_{0}\,=\,\theta_{1}\,\oplus\,\theta_{2}\,\oplus\,\cdots\,\oplus\theta_{n}\ .
\ee
However, consistency and supersymmetry restrict $\Theta$ to only live in a subset of all the possible irrep's in the r.h.s. of \eqref{V'xg0}. This goes under the name of \emph{linear constraint} (LC); the procedure of imposing the LC can be regarded as projecting out all the embedding tensor irrep's which are forbidden by consistency.
It is worth mentioning that, after imposing gauge invariance of the vectors and the higher-rank tensor fields, supersymmetry will in general still impose further restrictions. This is why it is normally stated that the LC is eventually demanded by supersymmetry, even though we would like to stress that, except for very few counterexamples, bosonic consistency already requires the LC in most of the cases\footnote{In any case, the reason why this turns out to be the generic situation still remains obscure.}. The set of consistent deformations of (half-)maximal supergravities in various dimensions is shown in tables~\ref{table:max} and \ref{table:half-max}.

\begin{table}[h!]
\renewcommand{\arraystretch}{1.25}
\begin{center}
\scalebox{0.9}[0.9]{
\begin{tabular}{|c|c|c|c|c|c|}
\hline
$D$ & $G_{0}$ & $H$ & \# scalars & vectors & $\Theta$ \\
\hline \hline
$9$ &  $\mathbb{R}^{+}\,\times\,$SL($2$) & SO($2$) & $3$ & $\textbf{1}_{(+4)}\,\oplus\,\textbf{2}_{(-3)}$ & $\textbf{2}_{(+3)}\,\oplus\,\textbf{3}_{(-4)}$\\
\hline
$8$ &  SL($2)\,\times\,$SL($3$) & SO($2)\,\times\,$SO($3$) & $7$ & $(\textbf{2},\,\textbf{3}^\prime)$ & $(\textbf{2},\,\textbf{3})\,\oplus\,(\textbf{2},\,\textbf{6}^\prime)$\\
\hline
$7$ &  SL($5$) & SO($5$) & $14$ & $\textbf{10}^\prime$ & $\textbf{15}\,\oplus\,\textbf{40}^\prime$\\
\hline
$6$ &  SO($5,5$) & SO($5)\,\times\,$SO($5$) & $25$ & $\textbf{16}$ & $\textbf{144}$\\
\hline
$5$ &  E$_{6(6)}$ & USp($8$) & $42$ & $\textbf{27}^\prime$ & $\textbf{351}$\\
\hline
$4$ &  E$_{7(7)}$ & SU($8$) & $70$ & $\textbf{56}$ & $\textbf{912}$\\
\hline
\end{tabular}
}
\end{center}
\caption{{\it Summary of some important facts about maximal gauged supergravities in various $D$. As one can see here, the number of physical scalar degrees of freedom and the number of embedding tensor components increase rather fast when moving to lower dimensions.}} \label{table:max}
\end{table}
\begin{table}[h!]
\renewcommand{\arraystretch}{1.25}
\begin{center}
\scalebox{0.87}[0.87]{
\begin{tabular}{|c|c|c|c|c|c|}
\hline
$D$ & $G_{0}$ & $H$ & \# scalars & vectors & $\Theta$ \\
\hline \hline
$9$ &  $\mathbb{R}^{+}\,\times\,$SO($1,n$) & SO($n$) & $1+n$ & $\textbf{(1+n)}_{(+1)}$ & $\tiny{\yng(1)}\,\oplus\,\tiny{\yng(1,1,1)}$\\[2mm]
\hline
$8$ &   $\mathbb{R}^{+}\,\times\,$SO($2,n$) & SO($2)\,\times\,$SO($n$) & $1+2n$ & $\textbf{(2+n)}_{(+1)}$ & $\tiny{\yng(1)}\,\oplus\,\tiny{\yng(1,1,1)}$\\[2mm]
\hline
$7$ &   $\mathbb{R}^{+}\,\times\,$SO($3,n$) & SO($3)\,\times\,$SO($n$) & $1+3n$ & $\textbf{(3+n)}_{(+1)}$ & $\textbf{1}\,\oplus\,\tiny{\yng(1)}\,\oplus\,\tiny{\yng(1,1,1)}$\\[2mm]
\hline
$6$a &   $\mathbb{R}^{+}\,\times\,$SO($4,n$) & SO($4)\,\times\,$SO($n$) & $1+4n$ & $\textbf{(4+n)}_{(+1)}$ & $\tiny{\yng(1)}\,\oplus\,\tiny{\yng(1,1,1)}$\\[2mm]
\hline
$6$b &  SO($5,n$) & SO($5)\,\times\,$SO($n$) & $5n$ & none & none \\[1mm]
\hline
$5$ &   $\mathbb{R}^{+}\,\times\,$SO($5,n$) & SO($5)\,\times\,$SO($n$) & $1+5n$ & $\textbf{(5+n)}_{(+1)}\,\oplus\,\textbf{1}_{(-2)}$ & $\tiny{\yng(1)}\,\oplus\,\tiny{\yng(1,1)}\,\oplus\,\tiny{\yng(1,1,1)}$\\[2mm]
\hline
$4$ &   SL($2)\,\times\,$SO($6,n$) & SO($2)\,\times\,$SO($6)\,\times\,$SO($n$) & $2+6n$ & $(\textbf{2},\,\textbf{6+n})$ & $\left(\textbf{2},\,\tiny{\yng(1)}\right)\,\oplus\,\left(\textbf{2},\,\tiny{\yng(1,1,1)}\,\right)$\\[2mm]
\hline
\end{tabular}
}
\end{center}
\caption{{\it Summary of some important facts about half-maximal gauged supergravities in various $D$. The free parameter $n$ represents the number of extra vector multiplets that can be coupled to the gravity sector. Only in the case $D=6$b the theory does not contain any vectors; in this case $n$ represents the number of self-dual tensor multiplets.}} \label{table:half-max}
\end{table}

Gauge transformations $\Lambda$ act in the following way on $\Theta$
\be
\delta_{\Lambda}{\Theta_{N}}^{\alpha}\,=\,g\,\Lambda^{M}\,{\Theta_{M}}^{\beta}\,\left({\left[t_{\beta}\right]_{N}}^{P}\,{\Theta_{P}}^{\alpha}\,-\,{f_{\beta\gamma}}^{\alpha}\,{\Theta_{N}}^{\gamma}\right)\ .
\ee
Requiring gauge invariance of the embedding tensor implies a set of \emph{quadratic constraints} (QC) which are then needed for consistency. These QC can be rewritten in terms of the gauge generators ${\left(X_{M}\right)_{N}}^{P}\,\equiv\,{X_{MN}}^{P}$ expressed in the $V$ representation. This yields
\be
\label{Closure}
\left[X_{M},\,X_{N}\right]\,=\,-{X_{MN}}^{P}\,X_{P}\ ,
\ee
which translates immediately into the closure of the gauge algebra. This set of QC \eqref{Closure} contains both a symmetric and an antisymmetric part in $M\leftrightarrow N$, which are respectively interpreted as a condition imposing the antisymmetry of the brackets and the Jacobi identities. We would like to stress that the tensor ${X_{MN}}^{P}$ representing the generalised structure constants of the gauge group is in general not antisymmetric in $M\leftrightarrow N$ and, indeed, consistency only requires that ${X_{(MN)}}^{P}$ is killed whenever contracted with $X_{P}$. Precisely because of this, the embedding tensor formulation of a gauged supergravity requires the introduction of higher-rank form potentials in the theory.

Any embedding tensor configuration $\Theta$ satisfying the LC and QC, \emph{i.e.} schematically
\bea \mathbb{P}_1(\Theta)&=&0\qquad\text{ (LC) ,} \label{LC}\\
\mathbb{P}_2(\Theta\otimes\Theta)&=&0\qquad\text{ (QC) ,}\label{QC} \eea
defines a consistent gauged theory, where, in \eqref{LC} and \eqref{QC}, $\mathbb{P}_{1}$ and $\mathbb{P}_{2}$ represent suitable projectors selecting the forbidden linear and quadratic irreducible pieces that $\Theta$ could in principle generate. 

In many explicit cases in various $D$, gauged supergravities have been worked out in detail in the literature by making use of the embedding tensor formalism. For further details on this topic, we want to refer to \cite{deWit:2004nw, Samtleben:2005bp, deWit:2005hv, Samtleben:2007an, Bergshoeff:2007ef, FernandezMelgarejo:2011wx} and \cite{Schon:2006kz} for gauged maximal and half-maximal supergravities respectively.

\subsubsection*{Fermions and Supersymmetry}

So far we have seen that the ungauged theory defines how deformation parameters modify the terms in the Lagrangian for $p$-form potentials in order for the gauged action to be gauge-invariant. This is basically done by means of the minimal substitution $\partial\,\rightarrow\,D$ and by defining gauge-covariant field strenghts. Nevertheless, this does not yet guarantee invariance under supersymmetry.

The aim of this section is to understand how to deform the Lagrangian in order to obtain a well-defined, gauge-invariant action which, on top of this, preserves supersymmetry. We shall see that the addition of couplings between the fermions and a scalar potential, both driven by the embedding tensor, together with an extra gauge-covariant topological term, restores supersymmetry. As a consequence, one also needs to modify the supersymmetry transformations for the fermions and therefore the Killing spinor equations.
\be
\mathcal{L}_{\textrm{gauged}}\,=\,\mathcal{L}_{\textrm{ungauged}}[\partial\rightarrow D]\,+\,\mathcal{L}_{\textrm{top}}\,+\,\mathcal{L}_{\textrm{fermi mass}}\,+\,\mathcal{L}_{\textrm{pot}}\ .
\label{gauged_L}
\ee

We already saw that scalars in (half-)maximal supergravities span the coset geometry $G_{0}/H$, where $G_{0}$ is the global symmetry group and $H$ its maximal compact subgroup. All the fermions only tranform non-trivially under local $H$ transformations and, as we already observed in section~\ref{G0/H}, the scalar coset representative $\mathcal{V}$ needs to mediate all the interactions with the $p$-forms. Moreover, in the gauged theory, $\mathcal{V}$ is also needed in order to couple the fermions to the embedding tensor.

In particular, the fermions always transform under the $R$-symmetry (see def. in section~\ref{susy}) group, which is in general only a subgroup of $H$. It only coincides with the full $H$ in maximal theories, whereas in half-maximal theories it is strictly a proper subgroup of $H$. The gravity multiplet contains two different types of fermions: the gravitino $\psi_{\mu}$ (helicity $3/2$) and the dilatino $\chi$ (helicity $1/2$). The inclusion of $n$ extra vector multiplets (only possible in the half-maximal case): the gaugino $\lambda$ (again helicity $1/2$). The way these different fermions transform with respect to $H$ is given in tables~\ref{table:R-symm} and \ref{table:R-symm_half}.
\begin{table}[h!]
\renewcommand{\arraystretch}{1.25}
\begin{center}
\scalebox{0.95}[0.95]{
\begin{tabular}{|c|c|c|c|c|}
\hline
$D$ & $H$ & $H_{R}$ & $\psi_{\mu}$ irrep's & $\chi$ irrep's  \\
\hline \hline
$9$ & SO($2$) & U($1$) & $\textbf{1}_{(+1)}\,\oplus\,\textbf{1}_{(-1)}$ & $2\,\cdot\,\textbf{1}_{(+1)}\,\oplus\,2\,\cdot\,\textbf{1}_{(-1)}$ \\[1mm]
$8$ & SO($2)\,\times\,$SO($3$) & U($2$) & $\textbf{2}_{(+1)}\,\oplus\,\overline{\textbf{2}}_{(-1)}$ & $\textbf{2}_{(+1)}\,\oplus\,\overline{\textbf{2}}_{(-1)}\,\oplus\,\textbf{4}_{(+3)}\,\oplus\,\overline{\textbf{4}}_{(-3)}$\\[1mm]
$7$ & SO($5$) & USp($4$) & \textbf{4} & \textbf{16} \\[1mm]
$6$ & SO($5)\,\times\,$SO($5$) & USp($4)\,\times\,$USp($4$) & $(\textbf{1},\,\textbf{4})\,\oplus\,(\textbf{4},\,\textbf{1})$ & $(\textbf{5},\,\textbf{4})\,\oplus\,(\textbf{4},\,\textbf{5})$ \\[1mm]
$5$ & USp($8$) & USp($8$) & $\textbf{8}$ & $\textbf{48}$ \\[1mm]
$4$ & SU($8$) & SU($8$) & $\textbf{8}\,\oplus\,\overline{\textbf{8}}$ & $\textbf{56}\,\oplus\,\overline{\textbf{56}}$\\[1mm]
\hline
\end{tabular}
}
\end{center}
\caption{{\it $R$-symmetry groups of the various maximal supergravities in $D>3$. The different fermions' irrep's are given with respect to $H_{R}$.}} \label{table:R-symm}
\end{table}
\begin{table}[h!]
\renewcommand{\arraystretch}{1.25}
\begin{center}
\scalebox{0.95}[0.95]{
\begin{tabular}{|c|c|c|c|c|c|}
\hline
$D$ & $H$ & $H_{R}$ & $\psi_{\mu}$ irrep's & $\chi$ irrep's & $\lambda$ irrep's\\
\hline \hline
$9$ & SO($n$) & none  & & & \\[1mm]
$8$ & SO($2)\,\times\,$SO($n$) & U($1$)  & $\textbf{1}_{(+1)}\,\oplus\,\textbf{1}_{(-1)}$ & $\textbf{1}_{(+1)}\,\oplus\,\textbf{1}_{(-1)}$ & $\textbf{1}_{(+1)}\,\oplus\,\textbf{1}_{(-1)}$ \\[1mm]
$7$ & SO($3)\,\times\,$SO($n$) & SU($2$)  & $\textbf{2}$ & $\textbf{2}$ & $\textbf{2}$ \\[1mm]
$6$ & SO($4)\,\times\,$SO($n$) & SU($2)\,\times\,$SU($2$) & $\left(\textbf{2},\,\textbf{1}\right)\,\oplus\,\left(\textbf{1},\,\textbf{2}\right)$ & $\left(\textbf{2},\,\textbf{1}\right)\,\oplus\,\left(\textbf{1},\,\textbf{2}\right)$ & $\left(\textbf{2},\,\textbf{1}\right)\,\oplus\,\left(\textbf{1},\,\textbf{2}\right)$ \\[1mm]
$5$ & SO($5)\,\times\,$SO($n$) & USp($4$) & $\textbf{4}$ & $\textbf{4}$ & $\textbf{4}$ \\[1mm]
$4$ & SO($6)\,\times\,$SO($n$) & U($4$) & $\textbf{4}_{(+1)}\,\oplus\,\overline{\textbf{4}}_{(-1)}$ & $\textbf{4}_{(+1)}\,\oplus\,\overline{\textbf{4}}_{(-1)}$ & $\textbf{4}_{(+1)}\,\oplus\,\overline{\textbf{4}}_{(-1)}$ \\[1mm]
\hline
\end{tabular}
}
\end{center}
\caption{{\it $R$-symmetry groups of the various half-maximal supergravities in $D>3$. The different fermions' irrep's are given with respect to $H_{R}$.}} \label{table:R-symm_half}
\end{table}
Let us now examine the term $\mathcal{L}_{\textrm{fermi mass}}$ in \eqref{gauged_L} in the simpler case of maximal supergravities\footnote{The maximal case is simpler because $H\,=\,H_{R}$. In half-maximal theories, $H\,=\,H_{R}\,\times\,$SO($n$) and fermions in the gravity multiplet ($\psi_{\mu}$ and $\chi$) are SO($n$) singlets. The gaugini $\lambda$, instead, transform non-trivially under SO($n$) and one must introduce new fermionic couplings (giving rise to new $T$-tensor irrep's) which transform non-trivially under SO($n$) as well.}. By fermionic mass terms we mean bilinear couplings between the fermions without derivatives. These couplings, as we commented before, must be mediated by the scalar fields through $\mathcal{V}$. Schematically, they are the form
\be
e^{-1}\,\mathcal{L}_{\textrm{fermi mass}}\,=\,g\,\left(A_{1}\,\overline{\psi}_{\mu}\,\gamma^{\mu\nu}\,\psi_{\nu} \,+\, A_{2}\,\overline{\chi}\,\gamma^{\mu}\,\psi_{\mu} \,+\, A_{3}\,\overline{\chi}\,\chi \right) \,+\, \textrm{h.c.} \ ,
\label{fermi_mass}
\ee
where $e$ is the determinant of the spacetime vielbein, $g$ the gauge coupling and $A_{1}$, $A_{2}$ and $A_{3}$ represent some tensors linear in the embedding tensor and depending on the scalars. These objects are usually called \emph{fermionic shifts}, for a reason which will become clear in a moment.

Part of the terms in \eqref{fermi_mass} are needed to cancel the supersymmetry variation of the new couplings in the Lagrangian between the gauge fields. The rest of them, need a modification of the Killing spinor equations induced by the gauging
\bea
\left(\delta\,\psi_{\mu}\right)_{\textrm{gauged}} &=& \left(\delta\,\psi_{\mu}\right)_{\textrm{ungauged}} \,+\, g\, A_{1}\,\epsilon \ ,\label{susy_psi}\\
\left(\delta\,\chi\right)_{\textrm{gauged}} &=& \left(\delta\,\chi\right)_{\textrm{ungauged}} \,+\, g\, A_{2}\,\epsilon \ , \label{susy_chi}
\eea
where $\epsilon$ parametrises a local supersymmetry transformation and $A_{1}$ and $A_{2}$ are the same objects appearing in \eqref{fermi_mass}. As we already anticipated, the modification of \eqref{susy_psi} and \eqref{susy_chi} implies the presence of a new potential term in \eqref{gauged_L} of the form
\be
e^{-1}\,\mathcal{L}_{\textrm{pot}} \,=\, -g^{2}\,V \,=\, 2\,g^{2}\,\left(|A_{1}|^{2}-|A_{2}|^{2}\right)\ ,
\label{L_pot}
\ee
where $A_{1}$ and $A_{2}$ depend on the scalars.

\subsubsection*{The $T$-tensor}

In the last section we formally introduced the fermionic shifts $A_{1}$, $A_{2}$ and $A_{3}$. We already saw that these are certain combinations of the embedding tensor and scalars which only transform with respect to local $H$ transformations. These objects are obtained from $\Theta$ by acting on any of its fundamental indices of $G_{0}$ with $\mathcal{V}$ from the right. What we then get is often called the \emph{T-tensor} \cite{deWit:1981eq}. Group-theoretically, it corresponds with the branching of the embedding tensor ($G_{0}$) irrep's with respect to its maximal compact subgroup $H$:
\be
\Theta\,=\,\theta_{1}\,\oplus\,\theta_{2}\,\oplus\,\cdots \quad \overset{H\,\subset\,G_{0}}{\longrightarrow} \quad T\,=\,\left(t_{11}\,\oplus\,t_{12}\,\oplus\,\cdots\right)\,\oplus\,\left(t_{21}\,\oplus\,t_{22}\,\oplus\,\cdots\right)\,\oplus\,\cdots \ .
\ee
The irreducible components $t_{ij}$ of the $T$-tensor are precisely the building-blocks of the fermionic shifts. We will give the explicit construction in due course in the following chapters whenever examining a specific theory in detail. For the moment we summarise in table~\ref{T_tens_max} the decomposition in $T$-tensor irrep's in the case of maximal supergravities in various dimensions.

\begin{table}[h!]
\renewcommand{\arraystretch}{1.25}
\begin{center}
\scalebox{1}[1]{
\begin{tabular}{|c|c|c|}
\hline
$D$ & $H\,=\,H_{R}$ & $T$-tensor irrep's\\
\hline \hline
$9$ & U($1$) & $\textbf{1}_{(-2)}\,\oplus\,\textbf{1}_{(-1)}\,\oplus\,\textbf{1}_{(0)}\,\oplus\,\textbf{1}_{(+1)}\,\oplus\,\textbf{1}_{(+2)}$\\[1mm]
$8$ & U($1)\,\times\,$SU($2$) & $\textbf{1}_{(-1)}\,\oplus\,\textbf{1}_{(+1)}\,\oplus\,\textbf{3}_{(-1)}\,\oplus\,\textbf{3}_{(+1)}\,\oplus\,\textbf{5}_{(-1)}\,\oplus\,\textbf{5}_{(+1)}$\\[1mm]
$7$ & USp($4$) & $\textbf{1}\,\oplus\,\textbf{5}\,\oplus\,\textbf{14}\,\oplus\,\textbf{35}$ \\[1mm]
$6$ & USp($4)\,\times\,$USp($4$) & $(\textbf{4},\,\textbf{4})\,\oplus\,(\textbf{4},\,\textbf{16})\,\oplus\,(\textbf{16},\,\textbf{4})$ \\[1mm]
$5$ & USp($8$) & $\textbf{36}\,\oplus\,\textbf{315}$ \\[1mm]
$4$ & SU($8$) & $\textbf{36}\,\oplus\,\overline{\textbf{36}}\,\oplus\,\textbf{420}\,\oplus\,\overline{\textbf{420}}$ \\[1mm]
\hline
\end{tabular}
}
\end{center}
\caption{{\it The irreducible components of the $T$-tensor in gauged maximal supergravities in $D>3$. These irrep's contribute to the different fermionic quadratic couplings inside $\mathcal{L}_{\textrm{fermi mass}}$ driven by $A_{1}$, $A_{2}$ and $A_{3}$.}} \label{T_tens_max}
\end{table}

In general, we can say that the $T$-tensor analysis turns out to be very useful in order to study the problem of the stabilisation of the scalars into a maximally symmetric vacuum and computing the mass spectra for the scalars at the critical points. This is related to the fact that, for such an analysis, we need the formulation of the gauged supergravity theory in exam in a specific point of moduli space (\emph{i.e.} the scalar coset $G_{0}/H$, in this case). Whenever specifying the theory to a particular point, the full $G_{0}$ covariance is broken to its compact part $H$ and hence the $T$-tensor becomes the natural object to use.

In this section, we hope to have covered some relevant generalities about the embedding tensor and the $T$-tensor. The fact that the global symmetry group differs for any $D$ implies that the development of the formalism for a specific theory and the group-theoretical analysis behind it needs a case-by-case study. In the following chapters, some specific theories will be studied and examined in detail. This will allow us to use them for various physical purposes.

\section{Minimal $D=4$ Supergravities and Deformations}
\label{section:N=1W}

In the previous sections we have introduced maximal supergravities in $D=10$, $11$ and subsequently we have discussed the embedding tensor formalism as a tool for classifying deformations of extended supergravities. In this section, we will now discuss minimal supergravities in $D=4$ and a particular set of deformations which will be relevant for us later in this thesis.
 
The minimal amount of supercharges in four dimensions is $4$, corresponding to $\cN=1$ (see table~\ref{table:diff_sugra's}). $\cN=1$ supergravities in $D=4$ have a universal part coming from the gravity multiplet containing the graviton ${e_{\mu}}^{a}$ and a Majorana gravitino $\psi_{\mu}$. The universal part of the action is made out of an Einstein term plus a Rarita-Schwinger term describing the dynamics of a spin-$\frac{3}{2}$ particle coupled to gravity (see \emph{e.g.} refs~\cite{Castellani:1991eu, Bailin:1994qt, SUGRABOOK})
\be
\label{action_N=1}
S\,=\,\frac{1}{2\,\kappa^{2}}\,\int{d^{4}x\,e\,\left(e^{a\mu}\,e^{b\nu}\,R_{\mu\nu ab}(\omega)\,-\,\overline{\psi}_{\mu}\,\gamma^{\mu\nu\rho}\,D_{\nu}\psi_{\rho}\right)}\ ,
\ee
where $D_{\mu}\psi_{\nu}\,\equiv\,\partial_{\nu}\psi_{\rho}\,+\,\frac{1}{4}\,{\omega_{\mu}}^{ab}\,\gamma^{ab}\,\psi_{\nu}$. The above action is invariant under the following local supersymmetry transformations
\be
\label{susy_N=1}
\begin{array}{lcl}
\delta\,{e_{\mu}}^{a}\,=\,\frac{1}{2}\,\overline{\epsilon}\,\gamma^{a}\,\psi_{\mu} & , & \delta\,\psi_{\mu}\,=\,D_{\mu}\epsilon  
\end{array}
\ee
only at a linear level in $\psi$ if $\omega$ is the usual torsion-free connection $\omega(e)$. In order for the \eqref{action_N=1} to be invariant at all orders one needs to include interaction terms in the Lagrangian which are higher-order in the gravitino. These terms are obtained by correcting $\omega(e)$ with torsion terms quadratic in $\psi$. 

\subsection*{The Coupling to Chiral Multiplets}

As we saw in table~\ref{table:matter_multiplet}, $\cN=1$ theories allow for an arbitrary number $n$ of chiral multiplets. The total field content is described by
\be
\left\{z^{\a},\,\lambda^{\a}\right\}_{\a=1,\dots,n}\ ,
\ee
where each of the $z^{\a}$ is a complex scalar field and each of the $\lambda^{\a}$ is a chiral fermion. These fields describing physical degrees of freedom must be supplemented by an extra unphysical \emph{compensator multiplet} $\left\{z^{0},\,\lambda^{0}\right\}$. The above multiplets can be combined into $\left\{z^{I},\,\lambda^{I}\right\}_{I=0,\dots,n}$. In section~\ref{subsec:supermultiplets} we already saw that supersymmetry restricts the complex scalar fields to span a K\"ahler-Hodge manifold. A K\"ahler manifold $\mathcal{M}_{K}$ is a manifold of real dimension $2n+2$ described by local complex coordinates $z^{I}\,\equiv\,\phi^{I}\,+\,i\,\phi^{I+n}$, for $I=0,\dots,n$ such that
\begin{itemize}
\item $\mathcal{M}_{K}$ is \emph{Hermitian}, \emph{i.e.} there exists a coordinate system in which the metric takes the form 
\be
ds^{2}\,=\,2\,K_{I\bar{J}}\,dz^{I}\,\otimes\,d\bar{z}^{\bar{J}}\ ,\nn
\ee
\item the hermitian metric $K_{I\bar{J}}$ defines a closed \emph{K\"ahler form}
\be
J\,\equiv\,-2\,i\,K_{I\bar{J}}\,dz^{I}\,\wedge\,d\bar{z}^{\bar{J}}\ ,\nn
\ee 
with $d J\,=\,0$.
\end{itemize}
As a consequence, every K\"ahler manifold has a metric which, locally in every coordinate patch, can be written as 
\be
\label{Kahler_metric}
K_{I\bar{J}}\,=\,\partial_{I}\partial_{\bar{J}}\,K(z,\bar{z})\ ,
\ee
where the real function $K(z,\bar{z})$ is called \emph{K\"ahler potential}. Please note that $K$ is not uniquely determined by \eqref{Kahler_metric} since the following transformation (therefore called K\"ahler transformation)
\be
\begin{array}{ccc}
K(z,\bar{z}) & \longmapsto & K(z,\bar{z})\,+\,f(z)\,+\,\bar{f}(\bar{z})
\end{array}
\ee
leaves the metric $K_{I\bar{J}}$ invariant and therefore it is called K\"ahler symmetry. Such a symmetry transforms the fermions non-trivially and defines a U($1$) bundle on the K\"ahler manifold. This ensures that K\"ahler transformations are locally well-defined on intersections of charts.
In order for $\mathcal{M}_{K}$ to be K\"ahler-Hodge the following topological condition has be satisfied by the K\"ahler form $J$
\be
\label{KH_cond}
q\,\iint\,J\,=\,4\,\pi\,n\ ,
\ee
where $q$ is the charge defining the U($1$) covariant derivatives and $n$ a suitable integer\footnote{Mathematically, the condition \eqref{KH_cond} requires the existence of a line bundle $\begin{array}{ccc} L & \longrightarrow & \cM_{K}\end{array}$ such that \cite{SUGRABOOK} $c_{1}(L)=[J]$, where $c_{1}$ denotes the first Chern class and $[J]$ the cohomology class defined by the K\"ahler form.}.

Going back to our $\cN=1$ supergravity coupled to chiral multiplets, the $n$ complex scalars $\left\{z^{I}\right\}$ will span a K\"ahler manifold $\mathcal{M}_{K}$ and the K\"ahler metric will determine the kinetic Lagrangian for the scalar fields
\be
\label{kin_N=1}
\mathcal{L}_{\textrm{kin}}\,=\,-K_{I\bar{J}}\,\partial_{\mu}z^{I}\,\partial^{\mu}\bar{z}^{\bar{J}}\ .
\ee

\subsection*{Superpotential Deformations}

$\cN=1$ supergravity coupled to chiral multiplets can be deformed by adding an arbitrary \emph{holomorphic} function $W(z)$ called superpotential. $W(z)$ is invariant under coordinate transformations on $\mathcal{M}_{K}$ and we will see later how it transforms under K\"ahler transformations in order to keep physical quantities invariant. Superpotential deformations of $\cN=1$ supergravity have the following physical consequences:

\begin{itemize}

\item A scalar potential is induced for $\left\{z^{I}\right\}$ of the form
\be
\label{V_N=1}
V\,=\,e^{K}\left(-3\,|W|^{2}\,+\,K^{I\bar{J}}\,D_{I}W\,D_{\bar{J}}\overline{W}\right)\ ,
\ee
where $K^{I\bar{J}}$ is the inverse K\"ahler metric and the covariant derivative $D$ is defined as $D_{I}W\,\equiv\,\partial_{I}W\,+\,\partial_{I}K\,W$.

\item Quadratic fermionic couplings to the scalars (often called fermionic mass terms) arise of the form
\be
\mathcal{L}_{\textrm{fermi mass}}\,=\,\frac{1}{4}\,m_{3/2}\,\overline{\psi}_{\mu}\,(\mathds{1}-\gamma_{*})\,\gamma^{\mu\nu}\,\psi_{\nu}\,-\,\frac{1}{2}\,m_{IJ}\,\overline{\chi}^{I}\,\chi^{J}\,+\,m_{I}\,\overline{\psi}_{\mu}\,\gamma^{\mu}\,\chi^{I}\,+\,\textrm{h.c.}\ ,
\ee
where 
\be
\label{fermi_mass_N=1}
\begin{array}{lc} 
m_{3/2}\,=\,e^{K/2}\,W & , \\[1mm]
m_{IJ}\,=\,e^{K/2}\,D_{I}D_{J}W & , \\[1mm]
m_{I}\,=\,\dfrac{1}{\sqrt{2}}\,e^{K/2}\,D_{I}W & .
\end{array}
\ee
\item The fermions $\psi_{\mu}$ and $\chi^{I}$ have modified supersymmetry rules by terms in which the superpotential enters. These new scalar-dependent terms are
\be
\label{susy_W}
\begin{array}{lclc}
\left(\delta\,\psi_{\mu}\right)_{W}\,=\,\dfrac{1}{2}\,e^{K/2}\,W\,\gamma_{\mu} & , & \left(\delta\,\chi^{I}\right)_{W}\,=\,-\dfrac{1}{\sqrt{2}}\,e^{K/2}\,K^{I\bar{J}}\,D_{\bar{J}}\overline{W} & .
\end{array}
\ee
\end{itemize}

From \eqref{susy_W} we can see that, in a maximally symmetric vacuum with only non-zero scalar vev's, having $D_{I}W\,\ne\,0$ inevitably breaks supersymmetry, whereas having $W\,\ne\,0$ does not necessarily imply supersymmetry breaking since the supersymmetry variation of the gravitino always has the universal contribution in \eqref{susy_N=1} given by the covariant derivative of $\epsilon$ which can compensate the $W$ term. Looking at \eqref{V_N=1}, one realises that the negative definite contribution in the scalar potential is the only contribution associated with supersymmetric vacua, which therefore can only be Minkowski and anti-de Sitter (AdS). De Sitter (dS) vacua are only possible by turning on the positive contribution in \eqref{V_N=1} and hence can never be supersymmetric. Moreover, from \eqref{fermi_mass_N=1} one can infer that the cosmological constant in a supersymmetric vacuum can be interpreted as the gravitino mass.

\chapter{Flux Compactifications}
\markboth{Flux Compactifications}{Flux Compactifications}
\label{Fluxes}
Starting from the various string theories in ten dimensions described in chapter~\ref{Strings}, one needs to construct compactifications thereof in order to make contact with four-dimensional low energy descriptions. The first comapctifications studied in the literature were those ones on \emph{Ricci-flat} six-dimensional manifolds (\emph{e.g.} tori or Calabi-Yau (CY) manifolds). Unfortunately, as we mentioned in the introduction, these turned out to give to rise to a number of massless scalars (a.k.a. \emph{moduli}) which are in contradiction both with particle phenomenology and with the precision tests of GR.

In this chapter we will briefly review flux compactifications as a mechanism for generating a scalar potential for the moduli. The lower-dimensional effective description is generically a gauged supergravity and it enjoys a duality symmetry coming from the winding modes of strings along the compact internal directions. This naturally suggests the introduction of \emph{non-geometric fluxes}, whose name derives from their unclear string-theoretical origin. In the last section we discuss T-duality covariant proposals such as Generalised Complex Geometry (GCG) and Double Field Theory (DFT) as possible frameworks in which to address the problem of the higher-dimensional origin of non-geometric fluxes.

\section{Dimensional Reductions}
\label{Dim_Reductions}

In this section we want to analyse a class of well-understood (geometric) compactifications of string theory. This corresponds to studying the propagation of superstrings on a background of the form
\be
\mathcal{M}_{10}\,=\,\mathcal{M}_{(1,9-d)}\,\times\,\mathcal{M}_{d}\ ,
\ee
where $\mathcal{M}_{(1,9-d)}$ indicates the $(10-d)$-dimensional background spacetime and $\mathcal{M}_{d}$ is the compact internal $d$-dimensional manifold. The physical properties of the lower-dimensional effective theory obtained after the compactification procedure depend on the internal geometry of $\mathcal{M}_{d}$.

\subsection*{Kaluza-Klein Reductions}

The original construction of a compactification was done in the context of pure gravity theories. In refs~\cite{Kaluza:1921tu, Klein:1926tv} they considered the reduction of five-dimensional Einstein gravity on a circle $S^{1}$. This construction is called Kaluza-Klein (KK) reduction, after the authors of the above references. 
Starting from the Einstein Lagrangian in $D+1$ dimensions\footnote{Note that here all the hatted quantities indicate $(D+1)$-dimensional objects.}
\be
\label{Einstein_Lagrangian}
\hat{\mathcal{L}}\,=\,\sqrt{-\hat{g}}\,\hat{R}\ ,
\ee
one can compactify the $(D+1)$-th direction $y$ on $S^{1}$. The coordinates $\hat{x}^{M}$ are split into $\left(x^{\mu},\,y\right)$, where $\mu$ runs from $0$ to $D-1$. As a consequence of the periodicity in $y$, the metric admits a Fourier decomposition of the form
\be
\hat{g}_{MN}(x^{\mu},y)\,=\,\sum_{n\,\in\,\mathbb{Z}}\hat{g}_{MN}^{(n)}(x^{\mu})\,e^{i\,\frac{ny}{L}}\ ,
\label{FourierKK}\ee
where $L$ is the compactification radius. The KK reduction consists of the aforementioned compactification on a circle, together with the restriction to the massless sector of the lower-dimensional theory. This approach is justified by observing that all the non-zero modes in the expansion~\eqref{FourierKK} have a mass which is proportional to $1/L$ and hence relatively large if $L$ is chosen to be small enough.

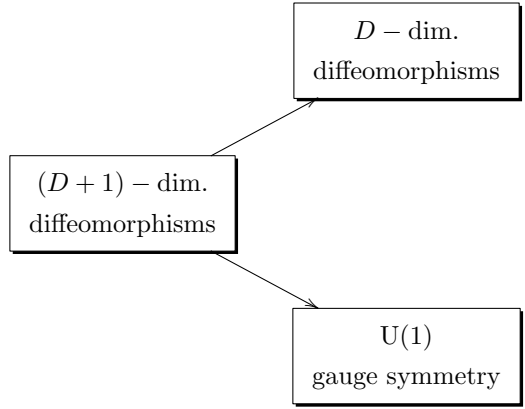
\begin{figure}
\label{fig:internal-external}
\begin{center}
\scalebox{0.9}[0.9]{\xymatrix{ & *+[F-,]{\begin{array}{c} D-\textrm{dim.}\\ \textrm{diffeomorphisms}\end{array}} \\ 
*+[F-,]{\begin{array}{c} (D+1)-\textrm{dim.}\\ \textrm{diffeomorphisms}\end{array}} \ar[ru]\ar[rd] & \\
 & *+[F-,]{\begin{array}{c} \textrm{U}(1)\\ \textrm{gauge symmetry}\end{array}} 
 }}
\end{center}
\caption{{\it The KK reduction from $D+1$ to $D$ breaks higher-dimensional diffeomorphisms into lower-dimensional diffeomorphisms plus U($1$) gauge transformations.}} \end{figure}

The reduced $D$-dimensional theory describes a vector $A_{\mu}$ and a scalar $\phi$ in addition to the metric $g_{\mu\nu}$. The reduction \emph{Ansatz} is the following
\be
\label{KKred_ans}
d\hat{s}^{2}\,=\,e^{2\,\alpha\,\phi}\,ds^{2}\,+\,e^{-2\,\alpha\,(D-2)\,\phi}\,\left(dy\,+\,A\right)^{2}\ ,
\ee
where $A\,\equiv\,A_{\mu}\,dx^{\mu}$ and $\alpha^{2}\,\equiv\,\frac{1}{2\,(D-1)\,(D-2)}$. After performing the reduction, the $D$-dimensional Lagrangian reads
\be
\mathcal{L}\,=\,\sqrt{-g}\,\left(R\,-\,\frac{1}{2}\,(\partial \phi)^{2}\,-\,\frac{1}{4}\,e^{-2\,(D-1)\,\alpha\,\phi}\,F_{\mu\nu}\,F^{\mu\nu}\right)\ ,
\label{LD_KK}\ee
where $F_{\mu\nu}\,\equiv\,2\,\partial_{[\mu}A_{\nu]}$ is the field strength associated to $A_{\mu}$. The Lagrangian~\eqref{LD_KK} describes the Einstein-Maxwell theory in $D$-dimensions coupled to the scalar field $\phi$ (often called \emph{dilaton}). The corresponding equations of motion read
\be
\begin{array}{cc}
G_{\mu\nu}\,=\,\frac{1}{2}\,\left(\partial_{\mu}\phi\,\partial_{\nu}\phi\,-\,\frac{1}{2}\,(\partial \phi)^{2}\,g_{\mu\nu}\right)\,+\,\frac{1}{2}\,e^{-2\,(D-1)\,\alpha\,\phi}\,\left(F_{\mu\rho}\,{F_{\nu}}^{\rho}\,-\,\frac{1}{4}\,F^{2}\,g_{\mu\nu}\right) & , \\[3mm]
\nabla_{\mu}\,\left(e^{-2\,(D-1)\,\alpha\,\phi}\,F^{\mu\nu}\right)\,=\,0 & , \\[2mm]
\Box\phi\,=\,-\frac{1}{2}\,(D-1)\,\alpha\,e^{-2\,(D-1)\,\alpha\,\phi}\,F^{2} & ,
\end{array}
\label{EOM_KK}\ee
where $G_{\mu\nu}$ is the Einstein tensor, $F^{2}\,\equiv\,F_{\mu\nu}\,F^{\mu\nu}$ and $\Box\,\equiv\,\partial_{\mu}\,\partial^{\mu}$. From the third equation of motion in \eqref{EOM_KK}, one immediately realises that setting $\phi\,=\,$const. is inconsistent, which means that the dilaton associated with the size of the compactification circle has to be dynamical. After such a reduction on a circle, part of what used to be spacetime (external) symmetry in the $(D+1)$-dimensional theory becomes gauge (internal) symmetry in the $D$-dimensional theory, as summarised in figure~\ref{fig:internal-external}.   

In a more general context, as we saw at the beginning of the section, one needs to compactify more than one dimension in order to relate four-dimensional (or any other $D<9$) supergravity to string theory. The straightforward generalisation of the above construction is a so-called \emph{toroidal reduction}, where $\mathcal{M}_{d}$ is chosen to be $T^{d}\,\equiv\,\underbrace{S^{1}\,\times\,\cdots\,\times\,S^{1}}_{d\textrm{ times }}$. The dimensional reduction of type II supergravity on $T^{d}$ (or equivalently of eleven-dimensional supergravity on $T^{d+1}$) gives rise to maximal (ungauged) supergravity in $10-d$ dimensions.
Instead, if one wants to go beyond the ungauged case to include a scalar potential, one has to go beyond tori and deal with internal manifolds with non-trivial geometry. However, there is no guarantee that such compactifications are consistent in the general case.

A KK reduction is called consistent \cite{Duff:1984hn} when all the gauge bosons of the isometry group $G$ of the compact manifold are retained in a truncation keeping only a finite number of lower-dimensional fields, with the essential requirement that setting the truncated fields to zero is consistent with their own equations of motion. Put in another way, the reduction ansatz is consistent if all the higher-dimensional equations of motion are satisfied as a consequence of the equations of motion for the retained lower-dimensional fields. It is only in very exceptional cases that such consistent KK reductions on compactifying spaces other than tori are possible.

A particular class of compactifications which has received attention in the literature is that of coset reductions, which include the relevant case of \emph{sphere reductions}\footnote{Please note that the sphere $S^{d}$ can be seen as the coset space $\frac{\textrm{SO}(d+1)}{\textrm{SO}(d)}$.}. These are particularly interesting since they preserve supersymmetry completely and they can be used for obtaining semisimple gaugings of maximal supergravities in various dimensions. Nevertheless, only in very few cases this procedure has been proven to be consistent. The known sphere compactifications giving rise to gauged maximal supergravities are summarised in table~\ref{table:spheres}. A further discussion on this point can be found in refs~\cite{Cvetic:2000dm, Duff:1986hr}.

\begin{table}[h!]
\begin{center}
\begin{tabular}{| c | c | c |}
\hline
$D$ & Gauging & Origin \\[1mm]
\hline \hline
$8$ & SO($3$) & IIA on $S^{2}$ \cite{Salam:1984ft}\\[1mm]
$7$ & SO($5$) & $11D$ on $S^{4}$ \cite{Nastase:1999cb, Nastase:1999kf}\\[1mm]
$6$ & SO($5$) & IIA on $S^{4}$ \cite{Cvetic:2000ah}\\[1mm]
$5$ & SO($6$) & IIB on $S^{5}$ \cite{Cvetic:2000nc}\\[1mm]
$4$ & SO($8$) & $11D$ on $S^{7}$ \cite{deWit:1986iy}\\[1mm]
\hline
\end{tabular}
\end{center}
\caption{{\it The semisimple compact gaugings of maximal supergravities obtained from sphere reductions of maximal supergravities in $D=10$ or $11$. For every case we give the refs where the consistency of the corresponding sphere reduction is discussed. We would like to stress that the $S^{5}$ reduction, which provided the first evidence for the AdS/CFT correspondence \protect\cite{Maldacena:1997re}, has not yet fully been proven to be consistent.}}
\label{table:spheres}
\end{table}

\subsection*{Twisted Reductions}

In the previous section we have seen how a toroidal reduction of a gravity theory can give rise to theories in lower dimensions including lower-spin degrees of freedom (\emph{i.e.} vectors and scalars). However, such reductions do not contain any mechanism to stabilise the scalar fields into a vacuum by giving them a mass. Subsequently, we saw that going beyond toroidal compactifications is not only needed, but in general very difficult to achieve because of consistency issues. As an example of this, we briefly discussed sphere reductions.

In this section, we will see how to exploit the global symmetry of a gravity theory in order to obtain a deformed lower-dimensional theory, \emph{i.e.} in which the field strenghts associated to the gauge fields are modified and a scalar potential appears. In this context we mean by twisted reduction the dimensional reduction over a \emph{group manifold} \cite{Scherk:1979zr} (SS). These reductions will turn out to be very useful since their consistence can proven in a very simple and general way.

A group manifold $G$ is a set equipped with both a group structure defined by a multiplication operation and a differentiable manifold structure. The extra compatibility condition between the two so far independent structures is that the group multiplication operation and the map defining the inverse of a group element are differentiable maps. This allows us to introduce a coordinate system $\left\{y^{m}\right\}_{m=1,\,\dots,\,\textrm{dim}(G)}$ in terms of which one is able to parametrise the general element $g(y)\,\in\,G$. Because of the definition of group manifold, one can always define the following two diffeomorphism (\emph{i.e.} differentiable coordinate transformations)
\be
\begin{array}{lclc}
g\quad\longmapsto\quad\Lambda_{L}(h)\,g\,\equiv\,h\,g & , & g\quad\longmapsto\quad\Lambda_{R}(h)\,g\,\equiv\,g\,h & ,
\end{array}
\ee
which are called left and right multiplication, respectively. However, $\Lambda_{L}$ and $\Lambda_{R}$ are not isometries of the metric in general. Still, one can always define a set of left-invariant 1-forms $\left\{\sigma^{m}\right\}$
\be
T_{m}\,\sigma^{m}\,=\,g^{-1}\,dg\ ,
\ee
where $\left\{T_{m}\right\}$ are the generators of $G$ and $d\,\equiv\,dy^{m}\,\frac{\partial}{\partial y^{m}}$. Expressed on the coordinate basis of 1-forms $\left\{dy^{m}\right\}$, the left-invariant forms read
\be
\sigma^{m}\,=\,{U^{m}}_{n}(y)\,dy^{n}\ ,
\ee
where ${U^{m}}_{n}(y)$ are suitable functions on $G$. It turns out that the quantities
\be
\label{fmnp_group}
{f_{mn}}^{p}\,=\,-2\,{\left(U^{-1}\right)^{q}}_{m}\,{\left(U^{-1}\right)^{r}}_{n}\,\partial_{[q}\,{U^{p}}_{r]}
\ee
do not depend on $y$ and exactly represent the structure constants of $G$.
The left-invariant 1-forms $\left\{\sigma^{m}\right\}$ define a class of metrics on $G$ for which $\Lambda_{L}$ is an isometry
\be
\label{L-invariant-metric}
ds_{G}^{2}\,=\,g_{mn}\,\sigma^{m}\,\otimes\,\sigma^{n}\ .
\ee
These isometries defined by $\Lambda_{L}$ are generated by a basis of Killing vectors $\left\{L_{m}\right\}$ which satisfy the so-called Maurer-Cartan equations
\be
\left[L_{m},\,L_{n}\right]\,=\,{f_{mn}}^{p}\,L_{p}\ ,
\ee
where ${f_{mn}}^{p}$ are given in~\eqref{fmnp_group}. Please note that there exist group manifolds for which ${f_{mn}}^{n}\,\ne\,0$, but strictly speaking they are not good compactifying internal manifolds \cite{Scherk:1979zr}, since their volume form is trivial in cohomlogy and hence it is not well-defined \cite{Grana:2006kf}.

\subsubsection*{Reductions of Gravity on Twisted Tori}

Starting from the \emph{Ansatz} for toroidal reductions (generalisation of \eqref{KKred_ans} for a $T^{d}$ with coordinates $\left(x^{\mu},\,y^{m}\right)$)
\be
\label{KKred_ans_gen}
d\hat{s}^{2}\,=\,e^{2\,\alpha\,\phi}\,ds^{2}\,+\,e^{2\,\beta\,\phi}\,M_{mn}\,\left(dy^{m}\,+\,{A^{m}}_{\mu}\,dx^{\mu}\right)\,\left(dy^{n}\,+\,{A^{n}}_{\nu}\,dx^{\nu}\right)\ ,
\ee
where $m,\,n\,=\,1,\,\dots,\,d$ and
\be
\begin{array}{lclc}
\alpha^{2}\,\equiv\,\dfrac{d}{2\,(D+d-2)\,(D-2)} & , & \beta\,\equiv\,-\dfrac{(D-2)\,\alpha}{d} & ,
\end{array}
\ee
we promote the internal part of the metric to the left-invariant metric on a $d$-dimensional group manifold defined in eqn.~\eqref{L-invariant-metric}. This procedure yields the following reduction \emph{Ansatz}
\be
d\hat{s}^{2}\,=\,e^{2\,\alpha\,\phi}\,ds^{2}\,+\,e^{2\,\beta\,\phi}\,M_{mn}\,\left(\sigma^{m}\,+\,{A^{m}}_{\mu}\,dx^{\mu}\right)\,\left(\sigma^{n}\,+\,{A^{n}}_{\nu}\,dx^{\nu}\right)\ ,
\ee
where $\sigma^{m}\,\equiv\,{U^{m}}_{n}\,dy^{n}$. The consistency of this construction is guaranteed by group theory, in the sense that the KK truncation retains as lower-dimensional fields only those ones which are left-invariant. An inconsistency in the reduction would imply the appearence of left-invariant fields in the equations of motion of the truncated fields. However, this is impossible because one can never build a quantity which is not left-invariant out of left-invariant fields.

As sketched in fig.~\ref{fig:internal-external}, the reduction \emph{Ansatz} \eqref{KKred_ans} breaks the $D$-dimensional diffeomorphisms into $(D-d)$-dimensional diffeomorphisms times an internal symmetry GL($d$). The general idea of these twisted reductions is to make use of the $y$-dependent GL($d$) transformation $U$ (called \emph{twist matrix}) in the compactification procedure. According to the \eqref{fmnp_group}, $U$ will turn on non-zero structure constants ${f_{mn}}^{p}$ which will appear in the lower-dimensional theory as deformation parameters linearly modifying the field strengths of the vectors ${A^{m}}_{\mu}$ and inducing a quadratic scalar potential.

By evaluating the Einstein Lagrangian \eqref{Einstein_Lagrangian} for the \emph{Ansatz} \eqref{KKred_ans_gen}, one finds
\be
\mathcal{L}\,=\,\sqrt{-g}\,\left[R\,+\,\frac{1}{4}\,\textrm{Tr}(D_{\mu}M\,D^{\mu}M^{-1})\,-\,\frac{1}{2}\,(\partial \phi)^{2}\,-\,\frac{1}{4}\,e^{2\,(\alpha-\beta)\,\phi}\,M_{mn}\,{F^{m}}_{\mu\nu}\,F^{n \mu\nu}\,-\,V\right]\ ,
\ee
with
\be
\begin{array}{lclc}
{F^{m}}_{\mu\nu}\,\equiv\,2\,\partial_{[\mu}\,{A^{m}}_{\nu]}\,-\,{f_{np}}^{m}\,{A^{n}}_{\mu}\,{A^{p}}_{\nu} & \textrm{and} & D_{\mu}M_{mn}\,\equiv\,\partial_{\mu}\,M_{mn}\,+\,2\,{f_{q(m}}^{p}\,{A^{q}}_{\mu}\,M_{n)p}\ ,
\end{array}
\ee
whereas the scalar potential $V$ reads
\be
V\,=\,\frac{1}{4}\,e^{2\,(\beta-\alpha)\,\phi}\,\left(2\,{f_{mn}}^{p}\,{f_{pq}}^{m}\,M^{nq}\,+\,{f_{mn}}^{p}\,{f_{qr}}^{s}\,M^{mq}\,M^{nr}\,M_{ps}\right)\ ,
\ee
where $M^{mn}$ represents the inverse of $M_{mn}$. 

The lower-dimensional deformed theory is exactly the one obtained by gauging a subgroup of GL($d$), which is the global symmetry group of the undeformed lower-dimensional theory obtained by means of a toroidal compactification. For compactifications of ten- or eleven-dimensional supergravities, the deformation parameters ${f_{mn}}^{p}$ are often called \emph{metric flux} (${\omega_{mn}}^{p}$) since they are associated with the spin connection of the internal manifold. 

Another ingredient that can be added to these compactifications are the so-called \emph{gauge fluxes}. These are possible whenever a $p$-form gauge potential appears in the $D$-dimensional theory coupled to gravity in a sort of Einstein-Maxwell theory. Gauge fluxes are then nothing but the components of the aforementioned gauge potential integrated along the internal directions. Such an integration gives rise to non-vanishing fluxes only when the internal manifold admits non-trivial $p$-cycles.

\section{String Compactifications with Metric and Gauge Fluxes}
\label{Geom_Flux_Comp}

In the previous section we have shown how compactifications of theories including gravity give rise to lower-dimensional theories with non-trivial internal symmetry, even though in general we saw that it might be difficult to find the internal geometry being able to describe the correct lower-dimensional effective description (\emph{e.g.} all the moduli stabilised in a dS vacuum). In this section we will review (geometric) flux compactifications in string theory with the inclusion of gauge and metric fluxes. 

As we saw more in general in the previous sections, toroidal reductions of string theory unfortunately give rise to a lower-dimensional ungauged supergravity, where therefore all the moduli are massless. What one needs is to land in a gauged supergravity, \emph{e.g.} in $D=4$, in order to effectively see a potential for the moduli fields. A way of generating a gauging in the effective theory is to include \emph{fluxes} in the compactification procedure. We extensively saw in the previous sections that curving the internal manifold is a source of potential that corresponds to adding metric flux. Now we will see which is the set of fluxes that have been studied in string compactifications and which kind of physics they give rise to. 

Since the turn of the millenium, a lot of progress has been made in the context of flux compactifications of string theory in order to obtain four-dimensional effective descriptions with a number of desired features. In particular, from a phenomenological point of view, one is interested in a vacuum with small but positive cosmological constant and spontaneously broken supersymmetry. This implies the necessity of finding de Sitter (dS) solutions from string theory compactifications. In addition to modelling dark energy, these are relevant for embedding descriptions of inflation in string theory. Moreover, Anti-de Sitter (AdS) solutions are employed in holographic applications in order to study physical systems which have a conformal symmetry realised in the UV. 
Interesting reviews in flux compactifications are refs~\cite{Grana:2005jc, Douglas:2006es}. 

\subsection*{Flux Compactifications in Type II String Theory}

Many type II string theory constructions related to flux backgrounds compatible with minimal supersymmetry have been studied so far. In particular, the mechanism of inducing an effective superpotential from fluxes \cite{Giddings:2001yu} has been extensively studied in the literature for those compactifications giving rise to a so-called $STU$-model as low energy description \cite{Derendinger:2004jn, DeWolfe:2004ns, Villadoro:2005cu, DeWolfe:2005uu, Dall'Agata:2009gv}. These theories arise from the $T^{6}/(\mathbb{Z}_{2}\,\times\,\mathbb{Z}_{2})$ orbifold compactifications of type IIB with O3/O7-planes (and duals thereof). 

\begin{figure}[h!]
\begin{center}
\scalebox{0.9}[0.9]{
\begin{tabular}{ccccc}
\includegraphics[scale=0.5,keepaspectratio=true]{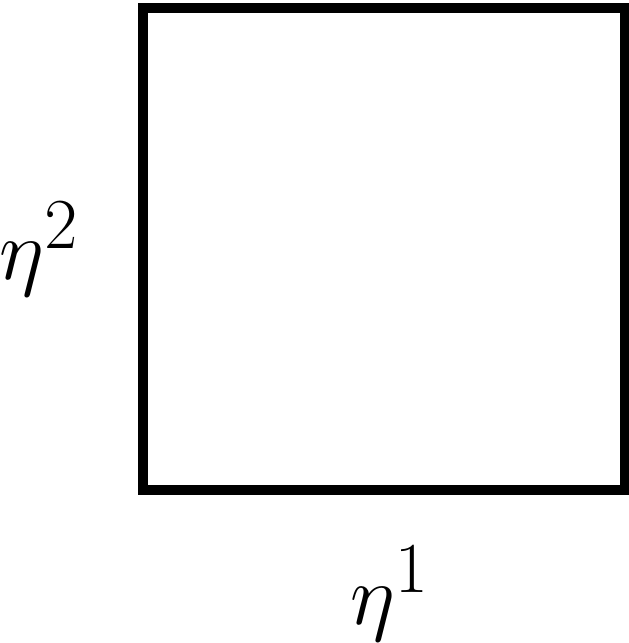} &    &  \includegraphics[scale=0.5,keepaspectratio=true]{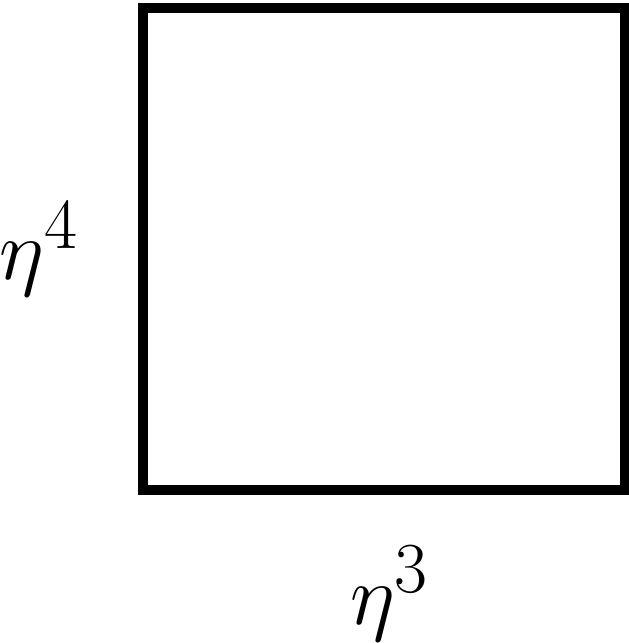}  &    & \includegraphics[scale=0.5,keepaspectratio=true]{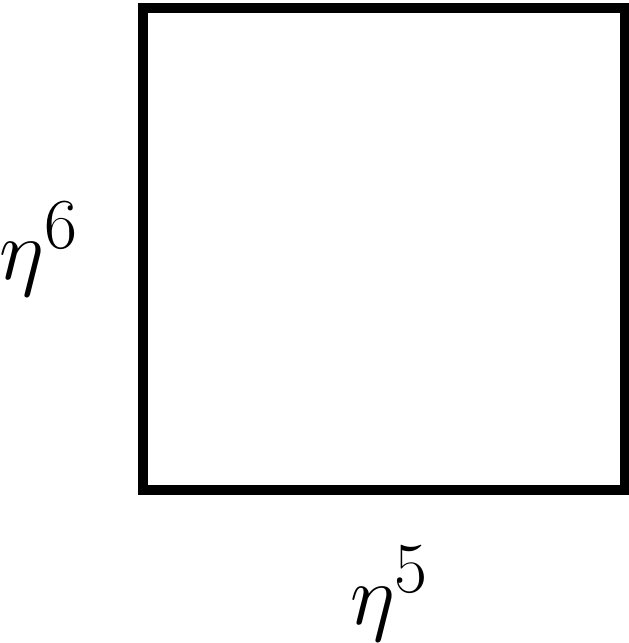} \\[-24mm]
     & \,\,\Large{$\times$} &        & \,\,\Large{$\times$} &
\end{tabular}
}
\end{center}
\vspace{9.5mm}
\caption{{\it $T^{6}\,= \,T^{2}_{1}\,\times\,T_{2}^{2}\,\times\,T_{3}^{2}$ torus factorisation and the coordinate basis.}}
\label{fig:Torus_Factor}
\end{figure}

After denoting by $\left\{\eta^{m}\right\}_{m=1,\dots,6}$ the basis of 1-forms on the torus, the $\mathbb{Z}_{2}\,\times\,\mathbb{Z}_{2}$ orbifold action is defined by\footnote{Here we follow the conventions in ref.~\cite{Guarino:2010zz}.} 
\be
\begin{array}{lclclc}
\theta_{1} & : & \left(\eta^{1},\,\eta^{2},\,\eta^{3},\,\eta^{4},\,\eta^{5},\,\eta^{6}\right) & \longmapsto & \left(\eta^{1},\,\eta^{2},\,-\eta^{3},\,-\eta^{4},\,-\eta^{5},\,-\eta^{6}\right) & , \\[1mm]
\theta_{2} & : & \left(\eta^{1},\,\eta^{2},\,\eta^{3},\,\eta^{4},\,\eta^{5},\,\eta^{6}\right) & \longmapsto & \left(-\eta^{1},\,-\eta^{2},\,\eta^{3},\,\eta^{4},\,-\eta^{5},\,-\eta^{6}\right) & , 
\end{array}
\ee
the full $\mathbb{Z}_{2}\,\times\,\mathbb{Z}_{2}$ being $\left\{1,\,\theta_{1},\,\theta_{2},\,\theta_{1}\theta_{2}\right\}$. Such an orbifold action forces the factorisation of $T^{6}$ shown in figure~\ref{fig:Torus_Factor}.
The above orbifold action has no invariant 1-forms (nor 5-forms), whereas it admits the following set of invariant 2-forms (and dual 4-forms)
\be
\label{H2M6Z}
\begin{array}{lclclc}
\omega_{1}\,=\,\eta^{12} & , & \omega_{2}\,=\,\eta^{34} & , & \omega_{3}\,=\,\eta^{56} & ,  \\[1mm]
\tilde{\omega}^{1}\,=\,\eta^{3456} & , & \tilde{\omega}^{2}\,=\,\eta^{1256} & , & \tilde{\omega}^{3}\,=\,\eta^{1234} & ,  
\end{array}
\ee
where $\eta^{12}\,\equiv\,\eta^{1}\,\wedge\,\eta^{2}$, and the following invariant 3-forms
\be
\label{H3M6Z}
\begin{array}{lclclclc}
\alpha_{0}\,=\,\eta^{135} & , & \alpha_{1}\,=\,\eta^{235} & , & \alpha_{2}\,=\,\eta^{451} & , & \alpha_{3}\,=\,\eta^{613} & , \\[1mm]
\beta^{0}\,=\,\eta^{246} & , & \beta^{1}\,=\,\eta^{146} & , & \beta^{2}\,=\,\eta^{362} & , & \beta^{3}\,=\,\eta^{524} & .
\end{array}
\ee
After choosing the normalisation
\be
\int_{\mathcal{M}_{6}}{\eta^{123456}}\,=\,\textrm{vol}_{6}\ ,
\ee
the invariant forms defined above satisfy
\be
\begin{array}{lclc}
\int_{\mathcal{M}_{6}}{\alpha_{0}\,\wedge\,\beta^{0}}\,=\,-\textrm{vol}_{6} & , & \int_{\mathcal{M}_{6}}{\alpha_{I}\,\wedge\,\beta^{J}}\,=\,\textrm{vol}_{6}\,\delta^{J}_{I} & ,
\end{array}
\ee
where $I,\,J\,=\,1,\,2,\,3$.

The K\"ahler 2-form $J$ can be expanded in terms of the $H^{2}(\mathcal{M}_{6},\mathbb{Z})$ basis elements introduced in \eqref{H2M6Z} as
\be
\label{moduli_A}
J\,=\,A_{1}\,\omega_{1}\,+\,A_{2}\,\omega_{2}\,+\,A_{3}\,\omega_{3}\ ,
\ee
where $\left\{A_{I}\right\}$ are real moduli (often called \emph{K\"ahler moduli}) measuring the area of the surface of $T^{2}_{I}$ for $I\,=\,1,\,2,\,3$.
The holomorphic 3-form $\Omega$ is, instead defined in terms of the $H^{3}(\mathcal{M}_{6},\mathbb{Z})$ basis elements introduced in \eqref{H3M6Z} as
\be
\label{moduli_z}
\Omega\,=\,\alpha_{0}\,+\,\alpha_{I}\,\tau^{I}\,+\,\beta^{I}\,\frac{\tau^{1}\,\tau^{2}\,\tau^{3}}{\tau^{I}}\,+\,\beta^{0}\,\tau^{1}\,\tau^{2}\,\tau^{3}\ ,
\ee
where $\left\{\tau^{I}\right\}$ are some extra complex moduli (often called \emph{complex structure moduli}). 

The six geometric moduli introduced in \eqref{moduli_A} and \eqref{moduli_z} come from the internal components of the ten-dimensional metric tensor and generate the moduli space parametrising the possible metrics on $T^{6}/\left(\mathbb{Z}_{2}\,\times\,\mathbb{Z}_{2}\right)$ away from singularities. These metrics can be explicitely written as
\be
ds^{2}_{6}\,=\,\sum_{I=1}^{3}{\frac{A_{I}}{\textrm{Im}(\tau^{I})}\,\left(|\tau^{I}|^{2}\,\left(\eta^{2I-1}\right)^{2}\,+\,\left(\eta^{2I}\right)^{2}\,-\,2\,\textrm{Re}(\tau^{I})\,\eta^{2I-1}\,\otimes\,\eta^{2I}\right)}\ .
\ee

Compactifications of type IIB string theory on a $T^{6}/\left(\mathbb{Z}_{2}\,\times\,\mathbb{Z}_{2}\right)$ orbifold break supersymmetry down to $\frac{1}{4}$ of the original amount, thus giving rise to effective four-dimensional descriptions preserving $\cN=2$ supersymmetry.

Subsequently we will further break supersymmetry to $\cN=1$ by means of the following $\mathbb{Z}_{2}$ action
\be
\label{O3_involution}
\begin{array}{lclclc}
\sigma & : & \left(\eta^{1},\,\eta^{2},\,\eta^{3},\,\eta^{4},\,\eta^{5},\,\eta^{6}\right) & \longmapsto & \left(-\eta^{1},\,-\eta^{2},\,-\eta^{3},\,-\eta^{4},\,-\eta^{5},\,-\eta^{6}\right) & .
\end{array}
\ee
The combined action of the orientifold $\sigma$ with the orbifold group generated by $\theta_{1}$ and $\theta_{2}$ gives rise to a $\mathbb{Z}_{2}^{3}$ parity describing O3- and O7-planes in the type IIB duality frame. The $T^{6}/\left(\mathbb{Z}_{2}\,\times\,\mathbb{Z}_{2}\right)$ orbifold and orientifolds thereof are very interesting setups since the internal manifold is its own mirror. The situation in CY compactifications is much more complicated and one needs \emph{mirror symmetry} \cite{Grana:2006hr} to relate different CY manifolds which are linked to each other via dualities. In the $\mathbb{Z}_{2}\,\times\,\mathbb{Z}_{2}$ orbifold, instead, one can have low-energy effective descriptions which are related by dualities and are still formally described by the same effective theory, where only the fields and the couplings have been transformed. In particular, this means that everything which we are here introducing in the context of type IIB compactifications with O3- and O7-planes can reformualted or reinterpreted,\emph{e.g.} in the language of type IIA with O6-planes.

The O3-planes are described by the involution $\sigma$ given in \eqref{O3_involution}, which are required to sit at each of the $4\,\times\,4\,\times\,4\,=\,64$ fixed points of $\sigma$. The O7-planes, instead, are in a triplet and are described by the involutions $\left\{\sigma\,\theta_{1},\,\sigma\,\theta_{2},\,\sigma\,\theta_{1}\,\theta_{2}\right\}$.

The moduli space of type IIB orientifolds with O3- and O7-planes is given by $\left\{S,\,T_{I},\,U_{I}\right\}$, where $S$ represents the so-called \emph{axiodilaton}, $T_{I}$ are the K\"ahler moduli and $U_{I}$ are the complex structure moduli
\be
\begin{array}{cclc}
S & = & C_{0}\,+\,i\,e^{-\phi} & , \\[1mm]
T_{I} & = & \dfrac{1}{\textrm{vol}_{6}}\,\bigintss\limits_{\mathcal{M}_{6}}{\left(C_{4}\,\wedge\,\omega_{I}\,+\,i\,e^{-\phi}\,A_{J}\,A_{K}\right)} & , \\[1mm]
U_{I} & = & \tau^{I} & , 
\end{array}
\ee
with $I\ne J\ne K$. These scalars span a coset manifold given by $\left(\dfrac{\textrm{SL}(2)}{\textrm{SO}(2)}\right)^{7}$. Please note that the expression of $T_{I}$ simplifies due to the absence of $C_{2}$ and $B_{2}$ which are projected out by the orientifold action. These compactifications, as anticipated above, preserve $\cN=1$ supersymmetry and the dynamics of the scalars in the case with no fluxes is totally encoded in the following K\"ahler potential
\be
\label{Kaehler_STU}
K\,=\,-\log\left(-i\,(S-\bar{S})\right)\,-\,\sum_{I=1}^{3}{\log\left(-i\,(T_{I}-\bar{T}_{I})\right)}\,-\,\sum_{I=1}^{3}{\log\left(-i\,(U_{I}-\bar{U}_{I})\right)}\ .
\ee
In the absence of fluxes, the Lagrangian for the above $7$ complex moduli is only given by the kinetic part given in terms of $K$ that we already introduced in \eqref{kin_N=1}. Now we will discuss how the introduction of fluxes induces a potential for the above scalars which can be written, as according to \eqref{V_N=1}, from a superpotential.

Gauge fluxes are constant non-zero background values that the field strengths of the $p$-form fields in the theory can aquire. In type IIB with O3- and O7-planes, the only allowed gauge fluxes are $\bar{H}_{3}$ and $\bar{F}_{3}$, which are associated with both the NS-NS $H_{3}$ and R-R $\tilde{F}_{3}$ 3-forms respectively. These field strengths are given by
\be
\label{H3F3fields}
\begin{array}{cclc}
\tilde{F}_{3} & = & F_{3}\,-\,H_{3}\,\wedge\,C_{0}\,+\,\bar{F}_{3} & , \\[1mm]
H_{3} & = & dB_{2}\,+\,\bar{H}_{3} & . 
\end{array}
\ee
The background fluxes introduced in \eqref{H3F3fields} can be now expanded on the basis of 3-forms introduced in \eqref{H3M6Z}, thus yielding
\be
\begin{array}{cclc}
\bar{H}_{3} & = & b_{3}\,\alpha_{0}\,+\,b_{2}^{(I)}\,\alpha_{I}\,+\,b_{1}^{(I)}\,\beta^{I}\,+\,b_{0}\,\beta^{0} & , \\[1mm]
\bar{F}_{3} & = & a_{3}\,\alpha_{0}\,+\,a_{2}^{(I)}\,\alpha_{I}\,+\,a_{1}^{(I)}\,\beta^{I}\,+\,a_{0}\,\beta^{0} & . 
\end{array}
\ee
The above fluxes induce a superpotential deformation in the effective $\cN=1$ description which was first studied in ref.~\cite{Gukov:1999ya}. The superpotential $W$ reads
\be
W\,=\,\int_{\mathcal{M}_{6}}{\left(\bar{F}_{3}\,-\,S\,\bar{H}_{3}\right)\,\wedge\,\Omega}\ .
\ee
After plugging here the expression of the holomorphic 3-form $\Omega$ given in \eqref{moduli_z}, one finds
\be
W\,=\,P_{1}(U_{I})\,+\,P_{2}(U_{I})\,S\ ,
\ee
where $P_{1}$ and $P_{2}$ are cubic polynomials in the complex structure moduli given by
\be
\begin{array}{cclc}
P_{1}(U_{I}) & = & a_{0}\,-\,\sum\limits_{I}{a_{1}^{(I)}\,U_{I}}\,+\,\sum\limits_{I}{a_{2}^{(I)}\,\dfrac{U_{1}\,U_{2}\,U_{3}}{U_{I}}}\,-\,a_{3}\,U_{1}\,U_{2}\,U_{3} & , \\[3mm]
P_{2}(U_{I}) & = & -b_{0}\,+\,\sum\limits_{I}{b_{1}^{(I)}\,U_{I}}\,-\,\sum\limits_{I}{b_{2}^{(I)}\,\dfrac{U_{1}\,U_{2}\,U_{3}}{U_{I}}}\,+\,b_{3}\,U_{1}\,U_{2}\,U_{3} & . 
\end{array}
\ee
The $\cN=1$ supergravity defined by the above superpotential has a \emph{no-scale} feature due to the absence of the moduli $T_{I}$. This implies that they appear as completely flat directions in the scalar manifold. The line of including some non-perturbative effects such as gaugino condensation \cite{Font:1990nt} has been considered as a possible mechanism to further stabilise the K\"ahler moduli (see \emph{e.g.} refs~\cite{Witten:1996bn, Achucarro:2006zf}).

From now on, for simplicity, we will restrict ourselves to the so-called \emph{isotropic limit} of the $T^{6}/\left(\mathbb{Z}_{2}\,\times\,\mathbb{Z}_{2}\right)$ orbifold, which basically reduces to $3$ the number of independent compex moduli vevs by imposing
\be
\begin{array}{lclc}
T_{1}\,=\,T_{2}\,=\,T_{3}\,\equiv\,T & \textrm{ and } & U_{1}\,=\,U_{2}\,=\,U_{3}\,\equiv\,U & ,
\end{array}
\ee
which span the scalar coset $\left(\dfrac{\textrm{SL}(2)}{\textrm{SO}(2)}\right)^{3}$. As far as the fluxes are concerned, the isotropic limit implies the following identifications
\be
\begin{array}{lclc}
a_{1}^{(1)}\,=\,a_{1}^{(2)}\,=\,a_{1}^{(3)}\,\equiv\,a_{1} & \textrm{ and } & b_{1}^{(1)}\,=\,b_{1}^{(2)}\,=\,b_{1}^{(3)}\,\equiv\,b_{1} & .
\end{array}
\ee
The theories described by this scalar content and the superpotential deformations thereof are often referred to in the literature as $STU$-models. Apart from being simpler, we will see in chapter~\ref{Half_Max} that these $\cN=1$ theories have an interesting relation with particular truncations of $\cN=4$ gauged supergravities. The most general isotropic superpotential for these supergravity models is up to linear in $S$ and up to cubic in $T$ and $U$, and therefore it  admits $2\,\times\,4\,\times\,4\,=\,32$ couplings. These couplings, as we will see in the next section, are related to the complete set of \emph{generalised fluxes}.

As we were anticipating previously, the simplicity of the $\mathbb{Z}_{2}\,\times\,\mathbb{Z}_{2}$ orbifold allows us to interpret the superpotential appearing in the effective $\cN=1$ description as arising from any compactification in the preferred duality frame. After choosing the duality frame, what will change is the flux label that one assigns to every superpotential coupling within the $STU$-model. In particular, also the set of allowed geometric fluxes will give rise to different superpotentials depending on the choice of duality frame.

The most general geometric (\emph{i.e.} gauge and metric fluxes) set of fluxes in type IIB with O3- and O7-planes only consists of $F_{3}$ and $H_{3}$ fluxes, since metric flux turns out to be projected out by the orientifold involution. The corresponding superpotential reads\footnote{We put the label GKP on this superpotential after the authors of ref.~\cite{Giddings:2001yu} who first considered this setup and analysed the vacua structure thereof.}
\be
\label{W_GKP}
W_{\textrm{GKP}}\,=\,P_{F}(U)\,+\,S\,P_{H}(U)\ ,
\ee
with $P_{F}(U)\,\equiv\,a_{0}\,-\,3\,a_{1}\,U\,+\,3\,a_{2}\,U^{2}\,-\,a_{3}\,U^{3}$ and $P_{H}(U)\,\equiv\,b_{0}\,-\,3\,b_{1}\,U\,+\,3\,b_{2}\,U^{2}\,-\,b_{3}\,U^{3}$.
In type IIA with O6-planes \cite{Derendinger:2004jn}, the set of geomteric fluxes includes some metric flux, even though part of it is still projected out by the orientifold involution, only half of the components of the possible NS-NS gauge flux $H$ and R-R gauge fluxes. The precise dictionary will be explained in detail later.

As originally argued in ref.~\cite{Kachru:2002sk}, applying a T-duality transformation to a background given by the NS-NS flux $H_{abc}$ along the $a$ direction gives rise to the metric flux ${\omega_{bc}}^{a}$, which, as we saw previously, can be interpreted as the structure constants of the isometry algebra of the internal space.
Furthermore, since type IIB is invariant under S-duality, it makes sense to construct a set of fluxes closed under S-duality. It turns out that such a duality transformation corresponds to an inversion of the axiodilaton $S$ and hence it interchanges all the superpotential couplings linear in $S$ with those ones independent of $S$. Thus, within type IIB geometric compactifications, $F_{3}$ and $H_{3}$ are a doublet under S-duality.

The $\cN=1$ effective descriptions presented above are a very useful playground to understand the mechanism of moduli stabilisation with fluxes. 
With respect to the string theory interpretation of the theories at hand, progress in this direction has been (partially) motivated by the search for dS solutions. Firstly, a \emph{no-go} result was proven which rules out the possibility of having dS solutions in the presence of only gauge fluxes \cite{Hertzberg:2007wc}. Further generalisations have investigated the possibility to circumvent this no-go theorem by including metric fluxes, see \emph{e.g.} refs~\cite{Silverstein:2007ac, Flauger:2008ad, Haque:2008jz, Caviezel:2008tf, Danielsson:2009ff, deCarlos:2009fq, Caviezel:2009tu}. 

However, by taking a look at the landscape of these geometric compactifications of type II theories, classical dS vacua with no tachyonic directions (\emph{i.e.} fully stable) have not been found so far. In particular, if one analyses the subset of $\cN=1$ constructions admitting an uplift to extended supergravities, the situation becomes even harder and we will see that dS solutions (even unstable) are ruled out. This, together with the purpose of constructing a duality-invariant completion to geometric compactifications, will lead us towards \emph{non-geometric fluxes} in a very natural way. 

\subsection*{Flux Compactifications in Heterotic String Theory}

A different context to discuss flux compactifications is that of heterotic string theory. These compactifications preserve half-maximal supersymmetry and let O($d,d$) symmetry emerge very naturally as T-duality group. In this section we will see how covariance with respect to T-duality in the NS-NS sector of heterotic compactifications already suggests the concept of non-geometric fluxes and the link with other constructions like Generalised Geometry and Double Field Theory.

In ref.~\cite{Kaloper:1999yr} twisted reductions of heterotic string theory on a $T^{d}$ have been considered. The undeformed case (\emph{i.e.} pure toroidal reduction with no twist) gives rise to half-maximal supergravity in $10-d$ dimensions coupled to $16$ vector multiplets, which enjoys an O($d,d+16$) as global symmetry. They found that the twist parameters induce a gauging of the effective half-maximal supergravity and transform as tensors under O($d,d+16$) dualities (exactly as the embedding tensor $\Theta$ should do). 

We start from the (bosonic) low energy action of heterotic string theory
\be
S\,=\,\int{d^{10}x\,\sqrt{-G}\,e^{-\Phi}\,\left(R\,+\,(\partial \Phi)^{2}\,-\,\frac{1}{12}\,H_{\mu\nu\rho}\,H^{\mu\nu\rho}\,-\,\frac{1}{4}\,\sum_{I=1}^{16}{{F^{I}}_{\mu\nu}\,F^{I\mu\nu}}\right)}\ ,
\ee
where $H_{\mu\nu\rho}\,\equiv\,3\,\left(\partial_{[\mu}B_{\nu\rho]}\,-\,\frac{1}{2}\,\sum\limits_{I=1}^{16}{{A^{I}}_{[\mu}\,{F^{I}}_{\nu\rho]}}\right)$ is the modified field strength of the NS-NS 2-form $B_{\mu\nu}$ and ${F^{I}}_{\mu\nu}\,\equiv\,2\,\partial_{[\mu}{A^{I}}_{\nu]}$ is the field strength of ${A^{I}}_{\mu}$.
After reduction, the internal field components of $G$, $B$ and $A$ ($G_{MN}$, $B_{MN}$ and ${A^{I}}_{M}$ with $M=1,\dots,d$) combine into a scalar matrix which spans the coset
\be
\frac{\textrm{O}(d,d+16)}{\textrm{O}(d)\,\times\,\textrm{O}(d+16)}\ .\nn
\ee
The twisted reduction is done by means of the following \emph{Ansatz} for the \emph{Zehnbein} 
\be
{E^{a}}_{m}\,=\,\left(\begin{array}{cc} {e^{\alpha}}_{\mu} & {E^{A}}_{N}\,{V^{N}}_{\mu} \\ 0 & {E^{A}}_{M}\end{array}\right)\ ,
\ee
where $a\,=\,(\alpha,A)$ and $m\,=\,(\mu,M)$. This construction turned out to formally reproduce the scalar potential coming from part of the electric sector of half-maximal supergravity in $D=4$ coupled to $16$ vector muliplets. In the following part we are going to ignore the presence of the extra vector multiplets and concentrate on the common sector of the theory, in which all the fields and deformations are arranged into irrep's of O($d,d$). Some concrete constructions of twisted reductions can be found in the context of heterotic \cite{Hull:2005hk} and M-theory \cite{Hull:2006tp} compactifications respectively.

\begin{center}
\begin{figure}[h!]
\includegraphics[scale=0.40]{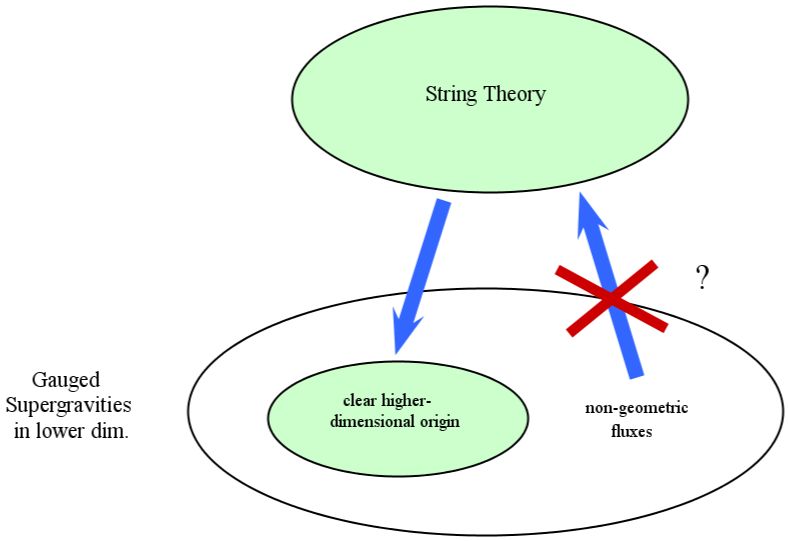}
\caption{{\it The complete set of embedding tensor deformations of the lower-dimensional theory is split into those ones which have a clear higher-dimensional origin and hence have a one-to-one mapping with geometric fluxes, and those ones for which such a higher-dimensional origin is not known.}}\label{figure:higher-dim}
\end{figure}
\end{center}

In the previous section, we saw that $\omega$ fluxes in the NS-NS sector can be obtained by applying a T-duality along an isometry direction to a background generated by $H_{3}$ flux. This implies that the equivalence between the two backgrounds is valid within the supergravity approximation as established by the so-called \emph{Buscher rules} \cite{Buscher:1987sk} (see section~\ref{sec:dualities}). However, the full embedding tensor deformations of the $(10-d)$-dimensional theory admit as a universal piece a full 3-form of O($d,d$) $f_{MNP}$. When decomposing $f_{MNP}$ into irrep's of the diffeomorphism subgroup GL($d)\,\subset\,$O($d,d$), one realises that the geometric set of fluxes (\emph{i.e.} $\left\{H_{mnp},\,{\omega_{mn}}^{p}\right\}$ where the fundamental index of O($d,d$) $M$ splits into $\left(_{m},\,^{m}\right)$ of GL($d$)) only accounts for \emph{half} of the total number of components of the 3-form. This has two related consequences: 
\begin{itemize}
\item There are some embedding tensor deformations of the effective half-maximal supergravity which cannot be obtained by means of geometric compactification and hence do not have a clear higher-dimensional origin. These correspond to possible extra ingredients in the compactification procedure that we do not fully understand. Therefore we call them non-geometric fluxes. This situation is sketched in figure~\ref{figure:higher-dim}.

\item Even starting from a perfectly geometric background described by $H_{3}$ and $\omega$ fluxes, a general T-duality transformation would take it to an effective description including non-geometric fluxes \cite{Shelton:2005cf}. The challenge of flux compactifications becomes that of establishing whether new physics (\emph{e.g.} dS vacua, stability, etc.) related to the presence of non-geometric fluxes can occur. If this turned out to be the case, then the second aim would be that of finding an uplift for theories including non-geometric fluxes. 
\end{itemize}

\section{Non-geometric Fluxes}
\label{sec:non-geom_fluxes}

As we saw in the previous sections, the study of non-geometric backgrounds is mainly motivated by duality covariance arguments \cite{Shelton:2005cf}, but also by the search for classical vacua in string theory \cite{Shelton:2006fd} with a special regard for dS solutions \cite{deCarlos:2009fq}. In this section we will first briefly introduce the set of generalised fluxes in heterotic compactifications and subsequently go back to $STU$-models arising from type II compactifications and introduce there the complete duality covariant flux-induced superpotential. 

\subsection*{T-duality Invariant Heterotic Fluxes}

Previously we argued that the geometric set of heterotic fluxes only reproduces half of the 3-form deformations that every half-maximal supergravity allows for. In this subsection we shall concentrate on the $D=4$ case, even though we would like to stress that none of the things presented here will crucially depend on this. To be precise, in fact, the four-dimensional case has the special feature of the SL($2$) electromagnetic duality enhancing the 3-form $f_{MNP}$ to an SL($2$) doublet $f_{\alpha MNP}$ of 3-forms, where $\alpha=(+,-)$. Nevertheless, a twisted toroidal reduction of heterotic string theory on a $T^{6}$, only gives rise to the purely electric (\emph{i.e.} only the $+$ components are non-zero) sector of $\cN=D=4$ supergravity. Restricting to $f_{+ MNP}$ makes the $D=4$ case perfectly analogous to all the others where a single O($d,d$) 3-form $f_{MNP}$ is allowed as deformation\footnote{Please note that in all half-maximal supergravities (see table~\ref{table:half-max}) other deformations are possible, but their higher-dimensional origin is not so clear and hence we restrict to gaugings in the 3-form which at least contain a geometric subset, non-geometric fluxes only representing the T-duality completion thereof.}.

From the four-dimensional point of view, O($6,6$) is a symmetry that relates equivalent effective descriptions with different higher-dimensional origins. Geometric twisted compactifications turn out to only be invariant under the GL($6$) subgroup of O($6,6$) describing diffeomorphisms and gauge transformations of the $B$-field on the 6-torus. In order to generate the full $\textbf{220}$ of O($6,6$) (\emph{i.e.} the 3-form representation), one needs to generalise the prescriptions of T-duality beyond the Buscher rules by considering the possibility of performing it along directions in which no isometries are present \cite{Hull:2006qs}. This allows us to complete the duality chain
\be
\label{Tduality}
\begin{array}{ccccccc}
H_{mnp} & \overset{T_{m}}{\longleftrightarrow} & {\omega_{np}}^{m} & \overset{T_{n}}{\longleftrightarrow} & {Q_{p}}^{mn} & \overset{T_{p}}{\longleftrightarrow} & R^{mnp}\ ,
\end{array}
\ee
where the second T-duality $T_{n}$ is still legitimate in the supergravity sense but it produces the so-called $Q$ flux which describes a locally geometric background for which, though, a global description is not possible; the third T-duality $T_{p}$ to obtain the so-called $R$ flux, on the contrary, is done in a direction with no isometries and hence it describes a background which does not even allow for a local description.

As already anticipated above, the NS-NS (non-)geometric fluxes introduced in the duality chain \eqref{Tduality} turn out to exactly fill out all the GL($6$) irrep's coming from the decomposition of the 3-form of O($6,6$):
\be
\label{Het_Fluxes}
\begin{array}{ccccccccc}
\textbf{220} & \overset{\textrm{GL}(6)\,\subset\,\textrm{O}(6,6)}{\longrightarrow} & \textbf{20} & \oplus & (\textbf{6}\,\oplus\,\textbf{84}) & \oplus & (\textbf{6}^{\prime}\,\oplus\,\textbf{84}^{\prime}) & \oplus & \textbf{20}^{\prime} \\[2mm]
f_{MNP} &  & H_{mnp} & & {\omega_{np}}^{m} & & {Q_{p}}^{mn} & & R^{mnp}
\end{array}\ .
\ee
Please note that the $\textbf{6}$ and $\textbf{6}^{\prime}$ irrep's present in \eqref{Het_Fluxes} correspond with the traces of $\omega$ and $Q$ respectively and they must be included in the counting in order to obtain the full $\textbf{220}$, even though there might be some consistency subtleties related to these backgrounds. For instance, in the case of a background with only metric flux, if this is not traceless, the volume form cannot be correctly defined because of cohomology issues. This would make it a problematic compactification from the mathematical viewpoint.

By including the complete set of generalised fluxes given in \eqref{Het_Fluxes}, one obtains the full electric sector of gauged $\cN=4$ supergravity with gaugings purely in the $\textbf{220}$. The fluxes turn out to reproduce the structure constants of the underlying $12$-dimensional gauge algebra generated by $\left\{Z_{m},\,X^{m}\right\}_{m=1,\dots,6}$. $Z_{m}$ are the six KK generators and correspond to the internal coordinate transformations $\delta x^{m}=\lambda^{m}$, whereas $X^{m}$ represent the generators associated with the internal gauge transformations of the $B$-field $\delta B_{mn}=\partial_{[m}\lambda_{n]}$. The brackets read 
\be
\label{Het_Gauge_Alg}
\begin{array}{cclc}
\left[Z_{m},\,Z_{n}\right] & = & {\omega_{mn}}^{p}\,Z_{p}\,+\,H_{mnp}\,X^{p} & , \\[1mm]
\left[Z_{m},\,X^{n}\right] & = & -{\omega_{mp}}^{n}\,X^{p}\,+\,{Q_{m}}^{np}\,Z_{p} & , \\[1mm]
\left[X^{m},\,X^{n}\right] & = & {Q_{p}}^{mn}\,X^{p}\,+\,R^{mnp}\,Z_{p} & . 
\end{array}
\ee
These first relations between heterotic compactifications with non-geometric fluxes and gauged $\cN=4$ supergravities already suggest the importance of half-maximal supergravities in understanding the role of T-duality as an organising principle for string compactifications. 

At this point there are two relevant questions that one could try to address. Firstly, one could wonder whether including non-geometric fluxes in our effective description always gives rise to new physics. By this we mean that it would be very important to be able to classify fluxes in terms of T-duality orbits to see which backgrounds are genuinely non-geometric and which other can be dualised to geometric ones. The second question could be, suppose one finds truly non-geometric flux backgrounds, how can one construct an uplift to string/M-theory? The origin of non-geometric fluxes by itself already suggests that, in order to describe those backgrounds, a construction or framework might be needed in which T-duality is promoted to a fundamental symmetry rather than just being a symmetry of the compactified versions of string theory. Examples of such constructions are Generalised Complex Geometry and Double Field Theory. 
In chapter~\ref{DFT} we will try to address both of these issues in the context of Double Field Theory. 

\subsection*{Generalised Fluxes in Type IIB Compactifications}

In section~\ref{Geom_Flux_Comp} we have introduced the $\mathbb{Z}_{2}\,\times\,\mathbb{Z}_{2}$ orbifold and shown that orientifolds thereof admit an effective $\cN=1$ description with a superpotential for the moduli induced by the presence of fluxes. Subsequently, we concentrated on the isotropic case, which is described by an $STU$-model. There we saw that, in type IIB with O3- and O7-planes, the only geometric background fluxes allowed are gauge fluxes, both NS-NS ($H_{3}$) and R-R ($F_{3}$). The superpotential induced by these was given in \eqref{W_GKP} and, as we will see, happens not to be invariant under duality transformations.

The aim of this subsection will be that of studying which kind of dualities these theories possess and how these transform the effective description. The following step, then will be writing down the fully duality invariant superpotential and interpreting its new couplings introduced for duality arguments as non-geometric fluxes. We will call the fluxes forming the complete set obtained in this way \emph{generalised fluxes}. This approach was first followed in refs~\cite{Aldazabal:2006up, Aldazabal:2008zza}. For a complete review of supergravity models induced by generalised fluxes in the $\mathbb{Z}_{2}\,\times\,\mathbb{Z}_{2}$ orbifold, we recommend refs~\cite{Wecht:2007wu, Guarino:2010zz}.

An $STU$-model enjoys a global symmetry of the form
\be
G_{0}\,=\,\textrm{SL}(2)_{S}\,\times\,\textrm{SL}(2)_{T}\,\times\,\textrm{SL}(2)_{U}\ ,
\ee
where the first factor can be interpreted as S-duality, while the rest generates a combination of S- and T-dualities. Note that the supergravity theory has the full continuous symmetry, whereas when all quantum corrections are included in string theory the above groups are broken into discrete SL($2,\mathbb{Z}$) factors (see table~\ref{table:string_dualities}).

As we already saw in chapter~\ref{Strings}, the general $\Lambda\,=\,\left(\begin{array}{cc} a & b \\ c & d \end{array}\right)\,\in\,\textrm{SL}(2)_{S}$ act on the axiodilaton $S$ as
\be
\label{S_duality}
\begin{array}{cccc}
S & \longmapsto & \dfrac{a\,S\,+\,b}{c\,S\,+\,d} & ,
\end{array}
\ee
and the same do $\textrm{SL}(2)_{T}$ and $\textrm{SL}(2)_{U}$ on $T$ and $U$, respectively. In order for the effective theory to be invariant under \eqref{S_duality}, the superpotential $W$ needs to transform as
\be
\begin{array}{cccc}
W(S) & \longmapsto & \dfrac{1}{c\,S\,+\,d}\,W(S) & .
\end{array}
\ee
This fact implies that the superpotential couplings (\emph{i.e.} the fluxes) must transform as well under SL($2)_{S}$ \cite{Shelton:2005cf} and the same for T-dualities. In particular, all the fluxes are paired into irrep's of $G_{0}$. For instance, $F_{3}$ and $H_{3}$ have to transform as a doublet of SL($2)_{S}$
\be
\begin{array}{cccc}
\left(\begin{array}{c} F_{3} \\ H_{3}\end{array}\right)& \longmapsto & \left(\begin{array}{cc} a & b \\ c & d\end{array}\right)\,\left(\begin{array}{c} F_{3} \\ H_{3}\end{array}\right) & .
\end{array}
\ee
As a consequence, the fully duality invariant superpotential must contain \emph{all} the couplings up to linear in $S$ and up to cubic in $T$ and $U$. 

Within the NS-NS sector, starting from $H$ flux and following the chain in \eqref{Tduality}, one could in principle generate $\omega$, $Q$ and $R$ just as in the heterotic case. However, the orientifold projection only allows for non-vanishing $Q$ flux. 
Towards the complete set of generalised fluxes, one finds the necessity of introducing the so-called $P$-flux building an S-duality doublet together with $Q$ \cite{Aldazabal:2006up}. This completes the S-duality invariant set of fluxes which still admit a locally geometric description. The corresponding superpotential is given by
\be
W^{(\textrm{loc.~geom.})}\,=\,(P_{F}\,+\,P_{H} \, S ) \,+\, 3 \, T \, (P_{Q} \,+\, P_{P} \, S )\ ,
\ee
with 
\be
\begin{array}{lcl}
\label{Poly_unprim1}
P_{F} = a_0 - 3 \, a_1 \, U + 3 \, a_2 \, U^2 - a_3 \, U^3 & \hspace{5mm},\hspace{5mm} & P_{H} = b_0 - 3 \, b_1 \, U + 3 \, b_2 \, U^2 - b_3 \, U^3\,,  \\[2mm]
P_{Q} = c_0 + C_{1} \, U - C_{2} \, U^2 - c_3 \, U^3 &
\hspace{5mm},\hspace{5mm} & P_{P} = d_0 + D_{1} \, U - D_{2}
\, U^2 - d_3 \, U^3\ ,
\end{array}
\ee
where, for the sake of convenience, we have introduced the flux combinations $\,C_{i} \equiv 2 \, c_i - \tilde{c}_{i}\,$, $\,D_{i} \equiv 2 \, d_i - \tilde{d}_{i}\,$ entering the superpotential, and hence the scalar potential and any other physical quantity.

Nevertheless, T-duality covariance requires yet new fluxes (called \emph{primed fluxes} in ref.~\cite{Aldazabal:2008zza}) to complete all the $STU$ polynomials. These fluxes do not have any interpretation even in the context of Generalised Geometry or Doubled Geometry, which were all naturally developed in the heterotic duality frame where these fluxes are absent. 

The complete set of generalised fluxes is presented in tables~\ref{table:unprimed_fluxes} and \ref{table:primed_fluxes} and the fully duality covariant induced superpotential $W$ reads
\be
\label{W_dual_fluxes}
W\,=\,(P_{F} +
P_{H} \, S ) + 3 \, T \, (P_{Q} + P_{P} \, S ) + 3 \, T^2 \, (P_{Q'}
+ P_{P'} \, S ) + T^3 \, (P_{F'} + P_{H'} \, S ) \ ,
\ee
where the unprimed sector was defined in \eqref{Poly_unprim1} and the primed sector is given by
\be
\begin{array}{lcl}
\label{Poly_prim1}
P_{F'} = a_3' + 3 \, a_2' \, U + 3 \, a_1' \, U^2 + a_0' \, U^3 & \hspace{5mm},\hspace{5mm} &P_{H'} = b_3' + 3 \, b_2' \, U + 3 \, b_1' \, U^2 + b_0' \, U^3\,,  \\[2mm]
P_{Q'} = -c_3' +  C'_{2} \, U + C'_{1} \, U^2 - c_0' \, U^3 &
\hspace{5mm},\hspace{5mm} & P_{P'} = -d_3' + D'_{2} \, U +
D'_{1} \, U^2 - d_0' \, U^3\ ,
\end{array}
\ee
with $\,C'_{i} \equiv 2 \, c'_i - \tilde{c}'_{i}\,$ and $\,D'_{i} \equiv 2 \, d'_i - \tilde{d}'_{i}\,$.

\begin{table}[h!]
\renewcommand{\arraystretch}{1.25}
\begin{center}
\scalebox{0.92}[0.92]{
\begin{tabular}{ | c || c | c | c |}
\hline
couplings & Type IIB & Type IIA & fluxes\\
\hline
\hline
$1 $&  $ {F}_{ ijk} $& $F_{aibjck}$ & $  a_0 $\\
\hline
$U $&  ${F}_{ ij c} $& $F_{aibj}$ & $   a_1 $\\
\hline
$U^2 $& ${F}_{i b c} $& $F_{ai}$ & $  a_2 $\\
\hline
$U^3 $& ${F}_{a b c} $& $F_{0}$ & $  a_3 $\\
\hline
\hline
$S $& $ {H}_{ijk} $& $ {H}_{ijk} $  & $  -b_0$\\
\hline
$S \, U $& ${H}_{ij c} $& ${{\omega}_{ij}}^{c}$ & $  -b_1 $\\
\hline
$S \, U^2 $&  ${H}_{ i b c}$ & $ {{Q}_{ i }}^{ b c}$  & $  -b_2 $\\
\hline
$S \, U^3 $& $ {H}_{a b c} $& $ {R}^{a b c} $ & $  -b_3 $\\
\hline
\hline
$T $& $  {Q_k}^{a b} $&$ H_{a b k} $ & $  c_0 $\\
\hline
$T \, U $& $ {Q_k} ^{a j} = {Q_k}^{i b} \,\,\,,\,\,\, {Q_a}^{b c} $& $ {\omega_{k a}}^{j} = {\omega_{b k}}^{i} \,\,\,,\,\,\, {\omega_{b c}}^a $  & $c_1 \,\,\,,\,\,\, \tilde {c}_1 $\\
\hline
$T \, U^2 $& $ {Q_c}^{ib} = {Q_c}^{a j} \,\,\,,\,\,\, {Q_k}^{ij} $& $ {Q_b}^{ci} = {Q_a}^{j c} \,\,\,,\,\,\, {Q_k}^{ij} $ & $c_2 \,\,\,,\,\,\,\tilde{c}_2 $\\
\hline
$T \, U^3 $& $  {Q_{c}}^{ij} $& $  R^{ijc} $ & $c_3 $\\
\hline
\hline
$S \, T $& $ {P_k}^{a b}$ & & $  -d_0 $\\
\hline
$S \, T \, U $& $ {P_k}^{a j} = {P_k}^{i b} \,\,\,,\,\,\, {P_a}^{b c} $&  & $-d_1 \,\,\,,\,\,\, -\tilde{d}_1 $\\
\hline
$S \, T \, U^2 $& $ {P_c}^{ib}= {P_c}^{a j} \,\,\,,\,\,\, {P_k}^{ij} $&  & $-d_2 \,\,\,,\,\,\,-\tilde{d}_2 $\\
\hline
$S \, T \, U^3 $& $  {P_{c}}^{ij} $&  & $-d_3 $\\
\hline
\end{tabular}
}
\end{center}
\caption{{\it Mapping between unprimed fluxes and couplings in the superpotential both in type IIB with O3 and O7 and in type IIA with O6. The six internal directions depicted in figure~\protect\ref{fig:Torus_Factor} are split into $\,``-"$ labelled by $i=1,3,5$ and $\,``\,|\,"$ labelled by $a=2,4,6$. Note that the empty boxes in type IIA are related to the presence of dual fluxes analogous to the 'primed' notation in type IIB.}}
\label{table:unprimed_fluxes}
\end{table}

\begin{table}[h!]
\renewcommand{\arraystretch}{1.25}
\begin{center}
\scalebox{0.92}[0.92]{
\begin{tabular}{ | c || c | c | c |}
\hline
couplings &  Type IIB &  Type IIA & fluxes\\
\hline
\hline
$T^3 \, U^3 $& $ {F'}^{ijk} $&  & $  a_0' $\\
\hline
$T^3 \, U^2 $& ${F'}^{ ij c} $& &$   a_1' $\\
\hline
$T^3 \, U $& ${F'}^{i b c} $& &$  a_2' $\\
\hline
$ T^3 $& ${F'}^{a b c} $& &$  a_3' $\\
\hline
\hline
$S \, T^3 \, U^3 $& $ {H'}^{ ijk} $& &$  -b_0'$\\
\hline
$S \, T^3 \, U^2 $& $ {H'}^{i jc} $& &$ - b_1' $\\
\hline
$S \, T^3 \, U $& $ {H'}^{ i b c} $& & $  -b_2' $\\
\hline
$S  \, T^3 $& $ {H'}^{a b c} $& &$  -b_3' $\\
\hline
\hline
$T^2 \, U^3 $& $  {{Q'}_{a b}}^k $& &$  c_0' $\\
\hline
$T^2 \, U^2 $& $ {{Q'}_{a j}}^k = {{Q'}_{i b}}^k \,\,\,,\,\,\, {{Q'}_{b c}}^a $& &$c_1' \,\,\,,\,\,\, \tilde{c}_1' $\\
\hline
$T^2 \, U $& $ {{Q'}_{ib}}^c = {{Q'}_{a j}}^c \,\,\,,\,\,\, {{Q'}_{ij}}^k $& &$c_2' \,\,\,,\,\,\,\tilde{c}_2' $\\
\hline
$T^2 $& $ {{Q'}_{ij}}^{c} $& &$c_3' $\\
\hline
\hline
$S \, T^2 \, U^3$& $  {{P'}_{a b}}^k $& &$  -d_0' $\\
\hline
$S \, T^2 \, U^2 $& $ {{P'}_{a j}}^k = {{P'}_{i b}}^k \,\,\,,\,\,\, {{P'}_{b c}}^a $& &$-d_1' \,\,\,,\,\,\, -\tilde{d}_1' $\\
\hline
$S \, T^2 \, U $& $ {{P'}_{ib}}^c = {{P'}_{a j}}^c \,\,\,,\,\,\, {{P'}_{ij}}^k $& &$-d_2' \,\,\,,\,\,\,-\tilde{d}_2' $\\
\hline
$S \, T^2  $& $  {{P'}_{ij}}^{c} $& & $-d_3' $\\
\hline
\end{tabular}
}
\end{center}
\caption{{\it Mapping between primed fluxes and couplings in the superpotential. The conventions are the as in table~\protect\ref{table:unprimed_fluxes} and again, just as there, the empty column should be filled in with extra dual fluxes.}}
\label{table:primed_fluxes}
\end{table}

We will see in chapter~\ref{Half_Max} that a particular subset of all the theories described by the duality invariant superpotential \eqref{W_dual_fluxes} originate from a truncation of $\cN=4$ supergravity. This will be the case whenever the couplings in $W$ satisfy the $\cN=4$ QC required for the consistency of the gauging. On the other hand, from a stringy viewpoint, these QC should be interpreted as the requirement that all the supersymmetry-breaking objects (branes and dual branes) are absent.

\section{T-duality Covariant Constructions}

In the previous section we have seen that the existence of non-geometric fluxes was first conjectured in order for the low energy effective theory to be duality covariant and subsequently they turned out to be a crucial ingredient for dS extrema.
Dualities are correctly encoded in the global symmetry of the underlying gauged supergravity in four dimensions \cite{Samtleben:2008pe}. In this sense, T-duality singles out the important role of half-maximal supergravities, whereas, in order to supplement it with non-perturbative dualities to generate the full U-duality group, one has to consider maximal supergravity. 

Concentrating in particular on T-duality \cite{Alvarez:1989ad, Lauer:1989ax}, different ways have been investigated in the literature in order to implement T-duality covariance at a more fundamental level in order to gain a better understanding of how non-geometric fluxes change the compactification prescription, thus appearing in the effective theory \cite{Andriot:2011uh, Andriot:2012wx, Andriot:2012an}. One direction to follow is Generalised Complex Geometry \cite{Grana:2006hr, Grana:2008yw, Coimbra:2011nw}, in which the internal manifold is given a particular bundle structure in which the gauge fields now span the full T-duality group. Another possibility is that of doubling the internal coordinates \cite{Hull:2006va, Hull:2007jy, Hull:2009sg} by supplementing them with the corresponding duals to winding modes and viewing a non-geometric flux background as something created by means of a twisted double torus compactification \cite{Dall'Agata:2007sr}.

Recently, this second approach has been further developed into the so-called Double Field Theory (DFT) \cite{Hull:2009mi, Hohm:2010jy, Hohm:2010pp, Jeon:2011cn}, which aims to promote T-duality to a fundamental symmetry even independently of whether spacetime directions are compact or not \cite{Duff:1989tf, Duff:1990hn}. This theory in $10+10$ dimensions is formulated in terms of a generalised metric, whose action can be constructed to be fully O($10,10$) invariant. Moreover, there are some evidences that the gaugings of $\cN=4$ supergravity might follow from DFT reductions \cite{Aldazabal:2011nj, Geissbuhler:2011mx, Grana:2012rr}. This would provide a higher-dimensional origin for non-geometric flux backgrounds, even though the concrete construction leading to the most general background still needs to be accomplished.

\subsection*{Doubled Geometry}

Beyond conventional geometric string backgrounds consisting of a manifold equipped with a metric and gauge fields, one can consistently define string theory on a background in which all the local patches are geometric but these patches need to be glued together by special transition functions which include not only diffeomorphisms and gauge transformations, but also T-duality transformations \cite{Dabholkar:2005ve}.
Such an object describing non-geometric backgrounds is called T-fold and historically it represented the first T-duality covariant proposal for understanding non-geometric fluxes beyond the effective lower-dimensional supergravity description.

A T-fold can be introduced by means of the so-called \emph{doubled formalism} \cite{Hull:2004in} developed in the context of toroidal fibrations of the form $U\,\times\,T^{d}\,\times\,\tilde{T}^{d}$, where $U$ is an open set in the internal manifold $\mathcal{M}_{6}$, whereas the doubled torus $T^{d}\,\times\,\tilde{T}^{d}$ is the tangent space and is described by the internal coordinates $y^{m}$ paired up with their corresponding winding coordinates $\tilde{y}_{m}$ to give rise to the doubled coordinates $\mathbb{Y}^{M}\,\equiv\,(y^{m},\tilde{y}_{m})$ transforming in the fundamental representation of O$(d,d)$. 

Let $\left\{U_{\alpha}\right\}$ be an open cover of the internal manifold $\mathcal{M}_{6}$, \emph{i.e.} a collection of open sets such that $\mathcal{M}_{6}\,=\,\bigcup\limits_{\alpha}\,U_{\alpha}$. Then a T-fold is constructed as the collection of patches $U\,\times\,T^{d}$, in each of which a metric $g_{\alpha}$ and a two-form $b_{\alpha}$ are well-defined and, on overlapping patches $U_{\alpha}\,\cap\,U_{\beta}$, they are glued together in the ordinary geometric way by means of diffeomorphisms and $b$-transformations. The rest of the geometric features of a T-fold lies in the moduli fields $(g+b)_{mn}$ and in some U$(1)^{2d}$ connections $A_{\alpha}$ and $\tilde{A}_{\alpha}$. These objects are glued together over overlaps $U_{\alpha}\,\cap\,U_{\beta}$ by transition functions in $\textrm{O}(d,d)\,\times\,\textrm{U}(1)^{2d}$.

A background defined in the above way turns out to be geometric whenever the structure group $\Gamma_{d}$ reduces a subgroup of GL($d$). In any other case, T-folds describe non-geometric backgrounds. The aforementioned construction can be recast in terms of a doubled fibration $T^{d}\,\times\,\tilde{T}^{d}$, in which $\textrm{O}(d,d)\,\ltimes\,\textrm{U}(1)^{2d}$ acts geometrically. Given a doubled background, one has to perform a so-called \emph{choice of polarisation} in order to reproject the tangent bundle of the internal space back to a single torus $T^{d}$ by choosing $d$ physical coordinates out of the $2d$ coordinates given by $(y^{m},\tilde{y}_{m})$.

After fixing the O($d,d$) metric in light-cone coordinates to be
\be
\eta_{MN}\,=\,\left(\begin{array}{cc} 0 & \mathds{1}_{d} \\ \mathds{1}_{d} & 0 \end{array}\right)\ ,
\ee
one can introduce the so-called \emph{generalised metric} ${\cal
H}_{MN}$ on such a doubled torus, which is a symmetric O($d,d$) tensor defined as
\be
\label{gen_H_DG}
{\cal
H}_{MN}\,=\,\left(\begin{array}{cc} g^{-1} & -g^{-1}\,b \\ b\,g^{-1} & g\,-\,b\,g^{-1}\,g\end{array}\right)\ ,
\ee
where $g$ and $b$ are respectively the internal components of the metric and of the NS-NS two-form present in the theory. The introduction of the generalised metric allows one to describe the non-linear transformation of $\,(g+b)_{mn}$ under O($d,d$) by means of the tensorial transformation given by
\be
\begin{array}{ccc}
{\cal H} & \longmapsto & \Lambda^{T}\,{\cal H}\,\Lambda\ ,
\end{array}
\ee
where $\Lambda\,\in\,\textrm{O}(d,d)$.  

\subsubsection*{Twisted Reductions on Doubled Tori}

The doubled formalism has been used in the context of heterotic flux compactifications in order to produce effective supergravity descriptions including $Q$ and $R$ fluxes \cite{Hull:2009sg}. The dynamics can be formulated in terms of ${\cal H}$ which describes the metric on the doubled torus
\be
ds^{2}\,=\,{\cal H}_{MN}\,d\mathbb{Y}^{M}\,\otimes\,d\mathbb{Y}^{N}\ .
\ee
In ref.~\cite{Hull:2006va} they considered reductions of O($d,d$) invariant theories on a circle spanned by $z\,\sim\,z+1$ with the inclusion of an O($d,d$) duality twist. Such twist is specified by an O($d,d$) algebra element ${N^{M}}_{N}$, whose corresponding $z$-dependent group element is given by $\exp(N\,z)$. This duality twist reduction takes a $(10-d)$-dimensional theory exhibitng a $\textrm{U}(1)^{2d}$ gauge symmetry to a $(9-d)$-dimensional theory with a non-abelian gauge symmetry. If we denote by $\left\{T_{M}\right\}_{M=1,\dots,2d}$ the $\textrm{U}(1)^{2d}$ generators, after the twisted reduction they get supplemented by two extra generators $\left\{Z_{z},X^{z}\right\}$ corresponding, respectively, to shits in $z$ and $b$-transformations with one leg along the $z$ direction. The new non-abelian gauge algebra of the lower-dimensional theory reads
\be
\begin{array}{lclc}
\left[Z_{z},\,T_{M}\right]\,=\,-{N^{N}}_{M}\,T_{N} & , & \left[T_{M},\,T_{N}\right]\,=\,-N_{MN}\,X^{z} & ,
\end{array} 
\ee
where $N_{MN}\,\equiv\,\eta_{MP}\,{N^{P}}_{N}\,=\,-N_{NM}$ all the other commutators vanish.

From the viewpoint of flux compactifications, the twist matrix $N$ describes the following flux configuration
\be
{N^{M}}_{N}\,=\,\left(\begin{array}{cc} {\omega_{z m}}^{n} & {Q_{z}}^{mn} \\ H_{z m n} & -{\omega_{z n}}^{m}\end{array}\right)
\ee
and it describes a locally geometric background. At this stage, a further T-duality bringing $R$ flux into the game was only conjectured.

In ref.~\cite{Dall'Agata:2007sr}, instead, twisted doubled tori reductions \`a la Scherk and Schwarz (SS) have been considered in order to formally reproduce gaugings of half-maximal supergravity in $D=4$. 
In this case, one can define the following set of left-invariant 1-forms (see $\sigma^{m}$ in section~\ref{Dim_Reductions})
\be
\mathbb{E}^{M}\,\equiv\,{U^{M}}_{N}(\mathbb{Y})\,d\mathbb{Y}^{N}\ ,
\ee
where $U$ represents the twist matrix depending on the doubled coordinates $\mathbb{Y}^{M}\,\equiv\,(y^{m},\tilde{y}_{m})$. These 1-forms satisfy
\be
d\mathbb{E}^{M}\,=\,-\frac{1}{2}\,{f_{NP}}^{M}\,\mathbb{E}^{N}\,\wedge\,\mathbb{E}^{P}\ ,
\ee
for constant $f$. These constants can be seen as metric flux on a doubled space and they contain all the set of possible generalised fluxes according to the splitting of the fundamental O($6,6$) index $M$ into upper and lower GL($6$) indices. In ref.~\cite{Dall'Agata:2007sr} some explicit example of $U$ matrices are given which reproduce some particular heterotic backgrounds with non-geometric fluxes turned on. Further work in this direction can be found in ref.~\cite{ReidEdwards:2009nu}.

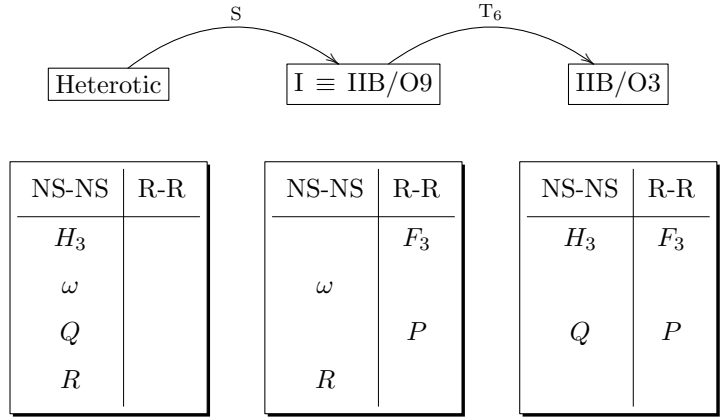
\begin{figure}
\label{fig:DG_NG}
\begin{center}
\scalebox{0.9}[0.9]{\xymatrix{ *+[F-:<3pt>]{\textrm{Heterotic}} \ar@/^2pc/[r]^{\textrm{S}} & *+[F-:<3pt>]{\textrm{I}\,\equiv\,\textrm{IIB}/\textrm{O9}} \ar@/^2pc/[r]^{\textrm{T}_{6}} & *+[F-:<3pt>]{\textrm{IIB}/\textrm{O3}} \\
*+[F-,]{\begin{tabular}{ c | c }
NS-NS & R-R \\[1mm]
\hline
$H_{3}$ & \\[1mm]
$\omega$ & \\[1mm]
$Q$ & \\[1mm]
$R$ & \\[1mm]
\end{tabular} } & *+[F-,]{\begin{tabular}{ c | c }
NS-NS & R-R \\[1mm]
\hline
 & $F_{3}$\\[1mm]
$\omega$ & \\[1mm]
 & $P$\\[1mm]
$R$ & \\[1mm]
\end{tabular} } & *+[F-,]{\begin{tabular}{ c | c }
NS-NS & R-R \\[1mm]
\hline
$H_{3}$ & $F_{3}$\\[1mm]
 & \\[1mm]
$Q$ & $P$\\[1mm]
 & \\[1mm]
\end{tabular} }
 }}
\end{center}
\caption{{\it The set of generalised fluxes which are allowed in different duality frames. Please note that doubled geometry can only describe backgrounds containing purely NS-NS fluxes. This is the reason why only in heterotic compactifications doubled geometry and ``non-geometry'' happen to coincide (see remark in ref.~\protect\cite{Dibitetto:2010rg}).}} \end{figure}

The concluding remark of this subsection is then precisely that, unlike in the heterotic case, compactifications on doubled twisted tori in different duality frames (\emph{i.e.} reductions of other string theories with or without orientifolds), do not reproduce the same generalisation of geometric flux compactification suggested by the introduction of the first non-geometric fluxes. This difference is depicted in figure~\ref{fig:DG_NG}.

\subsection*{Generalised Complex Geometry}

Generalised Complex Geometry (GCG) is a formalism introduced in ref.~\cite{Hitchin:2004ut} that interpolates between complex and symplectic manifolds. The basic idea is to treat tangent and cotangent space of the internal manifold $\mathcal{M}_{d}$ on equal footing by merging them together into a new bundle structure whose elements are formal sums of tangent vectors and 1-forms
\be
\begin{array}{lclc}
X\,+\,\xi & \in & T\mathcal{M}_{d}\,\oplus\,T^{*}\mathcal{M}_{d} & .
\end{array}
\ee
The physical relevance of this construction is to be found again in (non-)geometric flux compactifications since T-duality suggests that complex and symplectic geometry are each other's mirrors in string theory. Here we will only scketch some basic features of this formalism, but we indicate refs~\cite{Gualtieri:2003dx, Lindstrom:2004iw, Zabzine:2006uz, Hitchin:2010qz} as very interesting and complete reviews.

From a formal point of view, a \emph{generalised} almost-complex structure on such a bundle is an endomorphism ${\cal J}$ of $T\mathcal{M}_{d}\,\oplus\,T^{*}\mathcal{M}_{d}$ that squares to $-\mathds{1}_{2d}$, thus generalising the usual almost-complex structure on $T\mathcal{M}_{d}$. This generalised bundle amdits a natural metric ${\cal I}$ defined by
\be
{\cal I}(X\,+\,\xi,\,Y\,+\,\eta)\,\equiv\,\frac{1}{2}\,\left(\imath_{Y}\xi\,+\,\imath_{X}\eta\right)\ ,
\ee
where $\imath_{Y}\xi\,\equiv\,Y^{m}\,\xi_{m}$. In the coordinate basis $\left(\partial_{m},\,dx^{m}\right)$, it turns out to be proportional to the light-cone metric
\be
{\cal I}\,=\,\frac{1}{2}\,\left(\begin{array}{cc} 0 & \mathds{1}_{d} \\ \mathds{1}_{d} & 0 \end{array}\right)\ ,
\ee
which reduces the structure group to be O($d,d$).

A generalised almost-complex structure ${\cal J}$ is then a map
\be
\begin{array}{lclclc}
{\cal J} & : & T\mathcal{M}_{d}\,\oplus\,T^{*}\mathcal{M}_{d} & \longrightarrow & T\mathcal{M}_{d}\,\oplus\,T^{*}\mathcal{M}_{d} & ,
\end{array}
\ee
such that
\be
\label{alg_cond_J}
\begin{array}{cccc}
{\cal J}^{2}\,=\,-\mathds{1}_{2d} & \textrm{and} & {\cal J}^{T}\,{\cal I}\,{\cal J}\,=\,{\cal I} & .
\end{array}
\ee
This further reduces the structure group to U$(\frac{d}{2},\frac{d}{2})$.

Moreover, it turns out to be possible to define a generalisation of the Lie bracket, often called \emph{Courant bracket}, in the following way
\be
\label{Courant_bracket}
\left[X\,+\,\xi,\,Y\,+\,\eta\right]_{C}\,\equiv\,\left[X,\,Y\right]\,+\,{\cal L}_{X}\eta\,-\,{\cal L}_{Y}\xi\,-\,\frac{1}{2}\,d\left(\imath_{X}\eta\,-\,\imath_{Y}\xi\right)\ ,
\ee
where $[\,,\,\,]$ represent the ordinary Lie brackets and 
\be
{\cal L}_{X}\eta\,\equiv\,\left(X^{m}\,\partial_{m}\eta_{n}\right)\,dx^{n}\,+\,\left(\eta_{n}\,\partial_{m}X^{n}\right)\,dx^{m}\ .
\ee
The brackets defined in \eqref{Courant_bracket} are still anti-symmetric and admit a non-trivial automorphism defined by a closed two-form $b$
\be
e^{b}\,(X\,+\,\xi)\,\equiv\,X\,+\,\xi+\imath_{X}b\ .
\ee
This could be physically interpreted as a transformation of the $b$-field in a string background in the more general case in which $b$ is not closed anymore.

The Courant bracket defines a set of integrability conditions for ${\cal J}$, which are differential constraints that ${\cal J}$ has to satisfy together with the algebraic ones given in \eqref{alg_cond_J} in order to be called a generalised complex structure.

\subsubsection*{$G$ Structures and Flux Compactifications}

The formalism of GCG has been used in the literature \cite{Grana:2004bg, Grana:2005sn, Grana:2006kf} to study generalised flux compactifications preserving \emph{e.g.} $\cN=1$ supersymmetry in four dimensions. These analyses have been carried out by using the language of $G$ structures, with particular relevance of SU($3$) structures in type IIA compactifications \cite{Caviezel:2008tf} and SU($2$) structures in type IIB compactifications \cite{Caviezel:2009tu}.

Whenever it is possible to have two commuting almost complex structures whose product defines a positive-definite metric on $T\mathcal{M}_{d}\,\oplus\,T^{*}\mathcal{M}_{d}$, the structure group is broken to the compact group \cite{Grana:2006kf} $\textrm{U}(\frac{d}{2})\,\times\,\textrm{U}(\frac{d}{2})$. It has been shown that a $\textrm{U}(\frac{d}{2})\,\times\,\textrm{U}(\frac{d}{2})$ automatically provides a metric and a $b$-field through the natural T-duality covariant combination which is known as the generalised metric in DFT.

Specifying ourselves to the case $d=6$, one finds that the structure group is in fact further restricted to SU($3)\,\times\,$SU($3$). By making use of the mapping between spinors and polyforms, it is possible to construct the objects which define the geometry of the manifold in terms of two so-called \emph{pure spinors} $\Phi_{+}$ and $\Phi_{-}$. These can be in turn rewritten as bilinears of SO($d$) spinors. Pure spinors are never vanishing and globally defined SO($d,d$) spinors $\Phi$ mapped to bilinears $\eta_{1}\,\otimes\,\eta_{2}^{\dagger}$ such that 
\be
\overline{\eta}_{1}\,\gamma^{m_{1}\cdots m_{k}}\,\eta_{2}\,=\,0
\ee
for any $k\,\le\,d/2$. This definition turns out to be equivalent \cite{Berkovits:2000fe} to the one given in ref.~\cite{Hitchin:2010qz} in terms of Dirac structures.

For SU($3$) structures, these are a real $(1,1)$-type tensor $J$ and a holomorphic $(3,0)$-type tensor $\Omega$ such that
\be
\label{J&Omega}
\begin{array}{lccclc}
J\,\wedge\,\Omega\,=\,0 & & \textrm{and} & & i\,\Omega\,\wedge\,\bar{\Omega}\,=\,\dfrac{4}{3}\,J^{3} & ,
\end{array}
\ee
where now, in constrast with the case of a CY manifold (\emph{i.e.} K\"ahler manifold with SU($3$) holonomy), $J$ and $\Omega$ do \emph{not} need to be closed in general. $J$ and $\Omega$ in \eqref{J&Omega} are given in terms of the above pure spinors by
\be
\label{J&Omega_Phi}
\begin{array}{lccclc}
\Phi_{+}\,=\,\dfrac{1}{8}\,e^{-i\,J} & & \textrm{and} & & \Phi_{-}\,=\,-\dfrac{i}{8}\,\Omega & .
\end{array}
\ee
Supersymmetry equations for a given $\cN=1$ background can be written in terms of the spinors in \eqref{J&Omega_Phi} and subsequently, through the mapping between spinors and forms, one can rewrite those conditions in the language of $p$-forms, which are much easier to handle.

When analysing the supersymmetry equation of the following (warped) background
\be
ds^{2}\,=\,e^{2A}\,\eta_{\mu\nu}\,dx^{\mu}\,dx^{\nu}\,+\,ds^{2}_{6}\ ,
\ee
one schematically finds
\be
\begin{array}{lclc}
(d\,-\,H\,\wedge)\,\left(e^{2A-\phi}\,\Phi_{1}\right) & = & 0 & , \\[1mm]
(d\,-\,H\,\wedge)\,\left(e^{2A-\phi}\,\Phi_{2}\right) & = & e^{2A-\phi}\,dA\,\wedge\,\overline{\Phi}_{2}\,+\, \big\{\textrm{R-R Fluxes}\big\} & , 
\end{array}
\ee
where $\phi$ is the type IIA (IIB) dilaton and $\Phi_{1}\,=\,\Phi_{+}$ and $\Phi_{2}\,=\,\Phi_{-}$ for type IIA and vice versa for type IIB.
 
Another interesting bridge between GCG and (non-)geometric fluxes is represented by the possibility of reinterpreting the twisted Courant bracket defining the integrable generalised complex structure of the internal manifold as a gauge algebra of the compactified theory. Work in this direction has been done in refs~\cite{Grana:2008yw, Halmagyi:2008dr, Halmagyi:2009te}. 
In this sense, the generalised geometric structure of the internal manifold itself provides a local definition of the non-geometric NS-NS fluxes $H$, $\omega$, $Q$ and $R$. In terms of the following gauge generators
\be
\begin{array}{lclc}
{\cal X}^{a}\,\equiv\,{{\cal A}^{a}}_{m}\,dx^{m}\,+\,{\cal B}^{am}\,\partial_{m} & \textrm{and} & {\cal X}_{a}\,\equiv\,{\cal C}_{am}\,dx^{m}\,+\,{{\cal D}_{a}}^{m}\,\partial_{m} & ,
\end{array}
\ee
where ${\cal A}$, ${\cal B}$, ${\cal C}$ and ${\cal D}$ parametrise the full O($d,d$), the flux-induced gauge algebra reads
\be
\label{Het_Gauge_Alg_GCG}
\begin{array}{cclc}
\left[{\cal X}_{a},\,{\cal X}_{b}\right] & = & {\omega_{ab}}^{c}\,{\cal X}_{c}\,+\,H_{abc}\,{\cal X}^{c} & , \\[1mm]
\left[{\cal X}_{a},\,{\cal X}^{b}\right] & = & -{\omega_{ac}}^{b}\,{\cal X}^{c}\,+\,{Q_{a}}^{bc}\,{\cal X}_{c} & , \\[1mm]
\left[{\cal X}^{a},\,{\cal X}^{b}\right] & = & {Q_{c}}^{ab}\,{\cal X}^{c}\,+\,R^{abc}\,{\cal X}_{c} & , 
\end{array}
\ee
where $H$, $\omega$, $Q$ and $R$ are given in terms of derivatives of O($d,d$) gauge fields. Please note that these commutation relations exactly coincide with those ones given in \eqref{Het_Gauge_Alg}.

\subsection*{Double Field Theory}

DFT is a recent proposal that promotes T-duality to a symmetry in field theory \cite{Hull:2009mi, Hull:2009zb}, and is currently defined in terms of a background independent action \cite{Hohm:2010jy, Hohm:2010pp}. The theory is constructed on a double space \cite{Hull:2004in}, and its original version was created to describe the dynamics of closed strings on tori, the dual coordinates being associated to the winding modes of the strings. However, the
background independent action allows for more general spaces, and SS compactifications of DFT were shown to formally reproduce the bosonic (electric) sector of half-maximal gauged supergravities \cite{Aldazabal:2011nj, Geissbuhler:2011mx}. This already suggests its relation to non-geometric flux compactifications.

Detailed  reviews of DFT can be found in refs~\cite{Hohm:2011gs, Zwiebach:2011rg}. Here we will only provide a discussion of the minimal ingredients with the corresponding references to make contact with the results of the analysis of the following chapter. Many other interesting works on the subject towards the implementation of T-duality covariance can be found in refs~\cite{Rocen:2010bk, West:2010ev, Berman:2010is, Jeon:2010rw, Berman:2011cg, Berman:2011jh, Hohm:2011ex, Coimbra:2011ky, West:2011mm, Coimbra:2011nw, Berman:2011pe}.

As we saw in chapter~\ref{Strings}, T-duality relates a string background on a $T^{d}$ with radius $R$ with the one on $T^{d}$ with radius $\alpha^{\prime}/R$ via the non-compact duality group O($d,d$). To give a short review of DFT, we will here follow the conventions of ref.~\cite{Hohm:2011gs}. Starting from the string theory world-sheet action given by
\be
S\,=\,\int{d^{2}\sigma\, {\cal E}_{ij}(X)\,\partial_{+}X^{i}\,\partial_{-}X^{j}}\ ,
\ee
where $+$ and $-$ indicate light-come directions on the world-sheet and ${\cal E}_{ij}\,\equiv\,\left(G_{ij}+B_{ij}\right)$, one can derive a linear fractional transformation on ${\cal E}$. If the starting background admits commuting isometries, such a transformation relates equivalent backgrounds. However, this cannot be viewed as a symmetry of the world-sheet theory, since it involves non-trivial transformations of the couplings. 

From the string field theory viewpoint, this implies the appearence of \emph{winding modes} in the spectrum in addition to momentum modes. Therefore, the natural step towards understanding T-duality at a more fundamental level, became that of geometrising it by doubling the spacetime \cite{Kugo:1992md} thruogh the inclusion of ``winding-type" coordinates $\tilde{x}_{i}$ and allowing for novel spacetime rotation mixing ordinary coordinates with winding-type coordinates.

The low-energy spacetime action of the common sector of string theory
\be
\label{action_common}
S\,=\,\int{d^{10}x\,\sqrt{-g}\,e^{-2\,\phi}\,\left(R\,+\,4\,(\partial\phi)^{2}\,-\,\frac{1}{12}\,|H|^{2}\right)}\ ,
\ee
where $H\,\equiv\,db$, is not backgound independent, since the background ${\cal E}$ appears explicitely and hence O($10,10$) is not an actual symmetry of \eqref{action_common}.

Even though a background independent formulation of closed string theory is not known, recently such a construction has been performed for DFT. DFT is a field theory with manifest invariance under the O$(10,10)$ T-duality group, and therefore captures stringy features. The coordinates are combined to form fundamental vectors $\mathbb{X}^M =(\tilde x_{i}, x^{i})$, containing $10$ spacetime coordinates $x^i$ and $10$ dual coordinates $\tilde x_i$, $i=1,...,10$. The field content is that of the NS-NS sector, but defined on a doubled space. The metric of the global symmetry group
\be
\eta_{MN} = \begin{pmatrix}0^{ij} & \delta^i{}_j \\ \delta_i{}^j & 0_{ij}\end{pmatrix} \label{lightcone}
\ee
raises and lowers the indices of all the O($10,10$) tensors, thus relating spacetime components to winding-type ones. On the other hand, the dilaton $\phi$ is combined with the determinant of $g$ in a T-invariant way $e^{-2d}\,\equiv\,\sqrt{-g}\,e^{-2\phi}$.

After introducing the following derivative opretors
\be
\begin{array}{lclc}
{\cal D}_{i}\,\equiv\,\dfrac{\partial}{\partial x^{i}}\,-\,{\cal E}_{ik}\,\dfrac{\partial}{\partial \tilde{x}_{k}} & \textrm{and} & \bar{{\cal D}}_{i}\,\equiv\,\dfrac{\partial}{\partial x^{i}}\,+\,{\cal E}_{ik}\,\dfrac{\partial}{\partial \tilde{x}_{k}} & ,
\end{array}
\ee
one is able to write down the following background independent action
\be
\label{actionDFT_E}
\begin{array}{ccl}
S & = & \bigints{d^{10}x\,d^{10}\tilde{x}\,e^{-2\,d}\,\bigg[-\dfrac{1}{4}\,g^{ik}\,g^{jl}\,{\cal D}^{p}{\cal E}_{kl}{\cal D}_{p}{\cal E}_{ij}}\,+\,\dfrac{1}{4}\,g^{kl}\,\left({\cal D}^{j}{\cal E}_{ik}\,{\cal D}^{i}{\cal E}_{jl}\,+\,\bar{{\cal D}}^{j}{\cal E}_{ki}\,\bar{{\cal D}}^{i}{\cal E}_{lj}\right) \\[1mm]
 & & +\, \left({\cal D}^{i}d\,\bar{{\cal D}}^{j}{\cal E}_{ij}\,+\,\bar{{\cal D}}^{i}d\,{\cal D}^{j}{\cal E}_{ji} \right)\bigg] \ .
\end{array}
\ee
The action \eqref{actionDFT_E} is invariant under 
\be
\begin{array}{lccclc}
{\cal E}\quad\longmapsto\quad \left(a\,{\cal E}\,+\,b\right)\,\left(c\,{\cal E}\,+\,d\right)^{-1} & & \textrm{and} & & d\,\,\longmapsto\,\, d & ,
\end{array}
\ee
where $\left(\begin{array}{cc}a & b \\c & d\end{array}\right)\in\,$O($10,10$). Such a background independent action can be rewritten by making use of the generalised metric (see also definition in \eqref{gen_H_DG}), which is given in terms of the ordinary metric $g_{ij}$ and the Kalb-Ramond field $b_{ij}$ by the following symmetric element of O$(d,d)$
\be 
{\cal H}_{MN} = \begin{pmatrix} g^{ij} & -g^{ik}\,b_{kj} \\ b_{ik}\,g^{kj} & g_{ij}\,-\,b_{ik}\,g^{kl}\,b_{lj}\end{pmatrix}\ , 
\ee 
such that ${\cal H}_{MP} {\cal H}^{PN} = \delta^N_M$. In this formulation, the O($10,10$) invariant and background independent action \cite{Hohm:2010pp} can be written as
\be
\label{actionDFT_H}
\begin{array}{ccl}
S & = & \bigints d^{10}x\,d^{10}\tilde{x}\,e^{-2\,d}\,\bigg(\dfrac{1}{8}\,{\cal H}^{MN}\,\partial_{M}{\cal H}^{PQ}\,\partial_{N}{\cal H}_{PQ}\,-\,\dfrac{1}{2}\,{\cal H}^{MN}\,\partial_{N}{\cal H}^{PQ}\,\partial_{Q}{\cal H}_{MP} \\[1mm]
 & & -2\,\partial_{M}d\,\partial_{N}{\cal H}^{MN}\,+\,4\,{\cal H}^{MN}\,\partial_{M}d\,\partial_{N}d\bigg) \ .
\end{array}
\ee

The gauge invariance of DFT and closure of its gauge algebra gives rise to a set of constraints that restrict the coordinate dependence of the fields and gauge parameters. The original cubic formulation \cite{Hull:2009mi} turned out to require the so-called \emph{Weak Constraint} (WC), which imposes
\be
\partial_{M}\partial^{M}A\,=\,0\ ,
\ee
where $A$ represents a field or a gauge parameter. This constraint was found to arise from the level-matching condition in the sigma model, which has to be imposed in the massless sector of closed string theory (see also chapter~\ref{Strings})
\be
0\,\overset{!}{=}\,L_{0}\,-\,\overline{L}_{0}\,=\,-p_{i}\,w^{i}\ ,
\ee
where $p_{i}$ are the momenta and $w^{i}$ the winding number of a given state in the spectrum.

The final background independent formulation, on the other hand, was found to require yet another more restrictive constraint \cite{Hohm:2010jy}, which was therefore called \emph{Strong Constraint} (SC). This constraint not only requires the operator $\partial_{M}\partial^{M}$ to annhilate fields and gauge parameters, but also any product of them.

Later on though, in ref.~\cite{Grana:2012rr}, a more general formulation of DFT was proposed in which gauge invariance does not necessarily imply the SC. The first remark made there is that the very first formulation of DFT with the WC -- which was consistent up to cubic level -- does not really make sense as a complete theory, since the WC is not gauge invariant. The only subcase which gives rise to a well-defined theory is DFT with the SC, which once solved, gives us back ten-dimensional supergravity. In general, though, one finds that gauge invariance and the closure of the twisted Courant brackets require a much more involved \emph{slice constraint}, which even depends on several combinations of fields and gauge parameters and this makes it cumbersome to solve it in full generality.

A possible set of solutions to such a constraint is given by restricting the fields and gauge parameters to satisfy the SC. In such a situation, they can always be T-dualised to a frame in which the dependence on dual coordinates is cancelled. This restriction arises naturally in the context of toroidal compactifications. In such case, DFT provides an interesting framework in which ten-dimensional supergravity can be rotated to T-dual frames \cite{Andriot:2011uh, Andriot:2012wx, Andriot:2012an}. Different backgrounds violating the SC will be considered in the next chapter in the context of twisted reductions of DFT and we will see how these are related to non-geometric backgrounds in string theory.

\chapter{Duality Orbits and Double Field Theory}
\markboth{Duality Orbits and Double Field Theory}{Duality Orbits and Double Field Theory}
\label{DFT}
As we saw in the previous chapter, compactifications in duality covariant constructions such as GCG and DFT have proven to be suitable frameworks to reproduce gauged supergravities containing non-geometric fluxes. However, it is a priori unclear whether these approaches only provide a reformulation of old results, or also contain new physics. To address this question, we classify the T- and U-duality orbits of gaugings of (half-) maximal supergravities in dimensions seven and higher. It turns out that all orbits have a geometric supergravity origin in the maximal case, while there are non-geometric orbits in the half-maximal case. We show how the latter are obtained from compactifications of DFT. Some additional and technical material related to this chapter can be found in appendix~\ref{appendix:D=7,8}. Most of the results of this chapter were first obtained in refs~\cite{deRoo:2011fa, Dibitetto:2012rk}.

\section{Why Duality Orbits?}

In the context of half-maximal \cite{Schon:2006kz} and maximal \cite{deWit:2007mt} gauged supergravities, not only does supersymmetry tightly organise the ungauged theory, but also it strictly determines the set of possible deformations (\emph{i.e.} gaugings). The development of the so-called embedding tensor formalism (see section~\ref{section:Theta}) has enabled one to formally describe all the possible deformations in a single universal formulation, which therefore completely restores duality covariance. Unfortunately, not all the deformations have a clear higher-dimensional origin, in the sense that they can be obtained by means of a certain compactification of ten- or eleven-dimensional supergravity.

One of the most interesting open problems concerning flux compactifications is to reproduce, by means of a suitable flux configuration, a given lower-dimensional gauged supergravity theory. Although this was done in particular cases (see for example refs~\cite{Roest:2009dq, Dall'Agata:2009gv}), an exhaustive analysis remains to be done. This is due to fact that, on the one hand we lack a classification of the possible gauging configurations allowed in gauged supergravities and, on the other hand, only a limited set of compactification scenarios are known. Typically, to go beyond the simplest setups one appeals to dualities. The paradigmatic example \cite{Shelton:2005cf} starts by applying T-dualities to a simple toroidal background with a non-trivial two-form generating a single $H_{abc}$ flux. By T-dualising this setup, one can construct a chain
of T-dualities leading to new backgrounds (like twisted-tori or T-folds) and generating new (dual) fluxes, like the so-called ${Q_a}^{bc}$ and $R^{abc}$. We saw in the previous chapter that it is precisely by following duality covariance arguments in the lower-dimensional effective description that non-geometric fluxes were first introduced in order to explain the mismatch between particular flux compactifications and generic gauged supergravities. From the viewpoint of the lower-dimensional effective theory, it turns out that half-maximal and maximal gauged supergravities give descriptions which are explicitly covariant with respect to T- and U-duality respectively. This is schematically depicted in table~\ref{dualities_789}, even though only restricted to the cases we will address in this chapter.
\begin{table}[h]
\renewcommand{\arraystretch}{1.25}
\begin{center}
\scalebox{0.85}[0.85]{
\begin{tabular}{|c|c|c|}
\hline
$D$ & T-duality & U-duality \\
\hline \hline
$9$ &  O($1,1$) & $\mathbb{R}^{+}\,\times\,$SL($2$)\\
\hline
$8$ &  O($2,2)\,=\,$SL($2)\,\times\,$SL($2$) & SL($2)\,\times\,$SL($3$)\\
\hline
$7$ &  O($3,3)\,=\,$SL($4$) & SL($5$)\\
\hline
\end{tabular}
}
\end{center}
\caption{{\it The various T- and U-duality groups in $D>6$. These turn out to coincide with the global symmetry groups of half-maximal and maximal supergravities respectively (see tables~\protect\ref{table:max} and \protect\ref{table:half-max}).} \label{dualities_789}}
\end{table}

\begin{figure}[h!]
\begin{center}
\begin{tikzpicture}[scale=0.5,>=latex']
        \path[P_3] \ellip;
        \path[P_4] \ellipp;
        \draw[very thick] (-9,4) -- (-5,-5.45);
        \draw[very thick] (-4,5.65) -- (1,-6);
        \path (-3,2.25) node[left] {\textbf{A}}
            (-1.5,-3.5) node [right] {\textbf{B}};
        \path (-8,0) node[left] {orbit 1}
            (1,-4.5) node [right] {orbit 2};
        \path (16.5,-5) node[left] {\textbf{Flux configurations}}
            (0,.5) node [right] {\textbf{Geometric}};
        \path (-.5,-.75) node[right] {\textbf{configurations}};
\end{tikzpicture}
\caption{{\it The space of flux configurations sliced into duality orbits (vertical lines). Moving along a given orbit corresponds to applying dualities to a certain flux configuration and hence it does not imply any physical changes in the lower-dimensional effective description. Geometric fluxes only constitute a subset of the full configuration space. Given an orbit, the physically relevant question is whether (orbit 2 between A and B) or not (orbit 1) this intersects the geometric subspace. We refer to a given point in an orbit as a representative.}}\label{pic:orbits}
\end{center}
\end{figure}
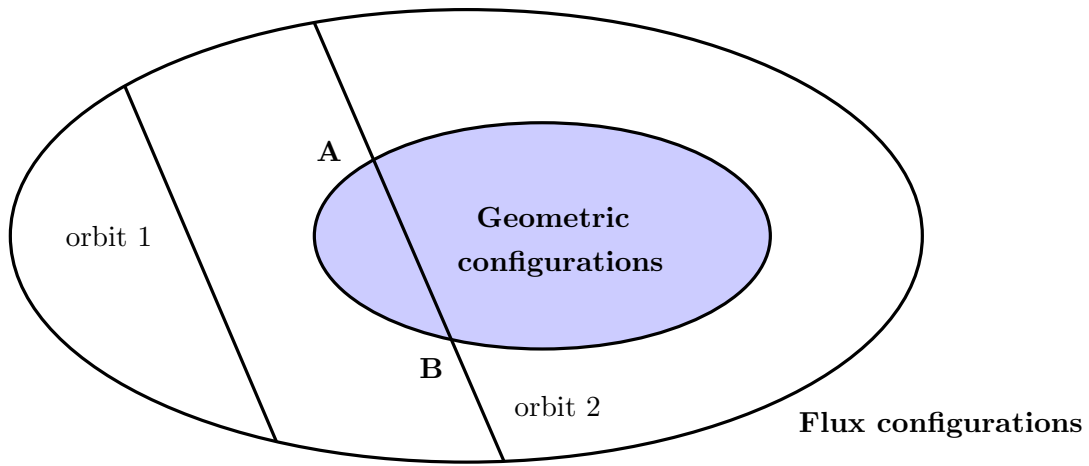

Here we would like to emphasise that all these (a priori) different T-duality connected flux configurations by definition lie in the same orbit of gaugings, and therefore give rise to the same lower-dimensional physics. In order to obtain a different gauged supergavity, one should consider more general configurations of fluxes, involving for example combinations of geometric and non-geometric fluxes, that can never be T-dualised to a frame in which the non-geometric fluxes are absent. For the sake of clarity, we depict this concept in figure~\ref{pic:orbits}.
Unfortunately, the original background independent formulation of DFT introduced in the previous chapter requires the SC for consistency and gauge invariance. This was found to imply that every consistent background (\emph{i.e.} satisfying the SC) can be rotated to a locally geometric one by means of an O($d,d$) transformation \cite{Hohm:2010jy}. In this scenario, DFT cannot possibly open up new T-duality orbits like orbit 1 in figure~\ref{pic:orbits}.

Subsequently, an indication has been given that gauge consistency of DFT does not need the WC and SC \cite{Grana:2012rr}. Following this direction, we could wonder whether relaxing these constraints can provide a higher-dimensional origin for all gaugings of extended supergravity through DFT. Our aim in the present work is to assess to what extent DFT can improve our description of non-geometric fluxes by giving a higher-dimensional origin to orbits which do not follow from standard supergravity compactifications. We will call such orbits of gaugings {\it non-geometric} (in figure~\ref{pic:orbits} they are represented by orbit 1).

As a starting point for this investigation, we will address the problem in the context of maximal and half-maximal gauged supergravities in dimension seven and higher, where the global symmetry groups are  small enough to allow for a general classification of orbits, without needing to consider truncated sectors. We will show that in the half-maximal supergravities in seven and higher dimensions, where the classifications of orbits can be done exhaustively, {\it all} the orbits (including geometric and non-geometric) admit an uplift to DFT, through SS \cite{Scherk:1979zr} compactifications on appropriate backgrounds. We provide explicit backgrounds for every orbit, and discuss their (un)doubled nature. The result is that truly doubled DFT provides the appropriate framework to deal with orbits that cannot be obtained from supergravity. In contrast, in maximal supergravities in eight and higher dimensions, all orbits are geometric and hence can be obtained without resorting to DFT.

\section{Twisted Reductions of DFT}
\label{DFT_twists}

While toroidal compactifications of DFT lead to half-maximal ungauged supergravities, SS compactifications on more general doubled spaces are effectively described by gauged supergravities like the ones we will analyse in the next sections. If the internal space is restricted in such a way that there always exists a frame without dual coordinate dependence, the only orbits allowed in the effective theory are those admitting representatives that can be obtained from compactifications of ten-dimensional supergravity. This is not the most general case, and we will show that some orbits require the compact space to be truly doubled, thus capturing information of both momentum and winding modes.

Recently in ref.~\cite{Grana:2012rr}, a new set of solutions to the constraints for DFT has been found. For these solutions the internal dependence of the fields is not dynamical, but fixed. The constraints of DFT restrict the dynamical external space to be undoubled, but allows for a doubling of the internal coordinates as long as the QC (see section~\ref{section:Theta}) for the gaugings are satisfied. Interestingly, these are exactly the constraints needed for consistency of gauged supergravity, so there is a priori no impediment to uplift any orbit to DFT in this situation. In fact, in the following sections we show that all the orbits in half-maximal $D = 7,8$ gauged supergravities can be reached from twisted double tori compactifications of DFT.

In the SS procedure described in section~\ref{Dim_Reductions}, the coordinates $X^M$ are split into  external directions $\mathbb{X}$ and compact internal $\mathbb{Y}$ coordinates. The former set contains pairs of O$(D,D)$ dual coordinates, while the latter one contains pairs of O$(n,n)$ dual coordinates, with $d = D + n$. This means that if a given coordinate is external (internal), its dual must also be external (internal), so the effective theory is formally a (gauged) DFT. The SS procedure is then defined in terms of a reduction ansatz, that specifies the dependence of the fields in $(\mathbb{X},\mathbb{Y})$ 
\be 
{\cal
H}_{MN} (\mathbb{X},\mathbb{Y}) = U(\mathbb{Y})^A{}_M \ \widehat
{\cal H}(\mathbb{X})_{AB}\ U(\mathbb{Y})^B{}_N
 \ , \ \ \ \ \ d(\mathbb{X},\mathbb{Y}) = \widehat d(\mathbb{X}) + \lambda(\mathbb{Y})\ .
\ee
Here the hatted fields $\widehat {\cal H}$ and $\widehat{d}$  are the dynamical fields in the effective theory, parameterising perturbations around the background, which is defined by $U(\mathbb{Y})$ and $\lambda(\mathbb{Y})$. The matrix $U$ is referred to as the \emph{twist matrix}, and must be an element of O$(n,n)$. It contains a DFT T-duality index $M$, and another index $A$ corresponding to the T-duality group of the effective theory. When DFT is evaluated on the reduction ansatz, the twists generate the gaugings of the effective theory 
\bea 
f_{ABC} &=& 3 \eta_{D[A}\
(U^{-1})^M{}_B (U^{-1})^N{}_{C]} \partial_M U^D{}_M \
,\label{f_from_U_}
\\
\xi_A &=& \partial_M (U^{-1})^M{}_A - 2 (U^{-1})^M{}_A \partial_M
\lambda \ ,\label{f_from_U}
\eea
where $f_{ABC}$ and $\xi_{A}$ build the generalised structure constants of the gauge group in the lower-dimensional theory. Note that these relations generalise eq.~\eqref{fmnp_group}.

Although $U$ and $\lambda$ are $\mathbb{Y}$ dependent quantities, the gaugings are forced to be constants in order to eliminate the $\mathbb{Y}$ dependence from the lower-dimensional theory.
When the external-internal splitting is performed, namely $d = D + n$, the dynamical fields are written in terms of their components which are a $D$-dimensional metric, a $D$-dimensional $2$-form, $2n$ $D$-dimensional vectors and $n^2$ scalars. These are the degrees of freedom of half-maximal supergravities. Since these fields are contracted with the gaugings, one must make sure that after the splitting the gaugings have vanishing Lorentzian indices, and this is achieved by stating that the twist matrix is only non-trivial in the internal directions. Therefore, although formally everything is covariantly written in terms of O$(d,d)$ indices $A,B,C,...$, the global symmetry group is actually broken to O$(n,n)$. We will not explicitly show how this splitting takes place, and refer to \cite{Aldazabal:2011nj} for more details. In this work,  for the sake of simplicity, we will restrict to the case $\xi_A  =0$, which should be viewed as a constraint for $\lambda$. Also we will restrict to O$(n,n)$ global symmetry groups, without additional vector fields.

There are two possible known ways to restrict the fields and gauge parameters in DFT, such that the action is gauge invariant and the gauge algebra closes. On the one hand, the so-called WC and SC can be imposed 
\be
\partial_M \partial^M A = 0\ , \ \ \ \ \partial_M A\ \partial^M B =0\ ,
\ee 
where $A$ and $B$ generically denote products  of (derivatives of) fields and gauge parameters. When this is the case, one can argue \cite{Hohm:2010jy} that there is always a frame in which the fields do not depend on the dual coordinates. On the other hand, in the SS compactification scenario, it is enough to impose the WC and SC only on the external space (\emph{i.e.}, on hatted quantities) 
\be
\partial_M \partial^M \widehat A = 0\ , \ \ \ \ \partial_M \widehat A\ \partial^M \widehat B = 0
\ ,
\ee 
and impose QC for the gaugings 
\be f_{E[AB} f^E{}_{C]D}
= 0 \ .
\ee 
This second option is more natural for our purposes, since these constraints exactly coincide with those of half-maximal gauged supergravities\footnote{We are working under the assumption that the structure constants not only specify the gauging, but all couplings of the theory. Reproducing the correct structure constants therefore implies reproducing the full theory correctly, as has been proven in $D=4$ and $D=10$ \cite{Aldazabal:2011nj, Geissbuhler:2011mx, Hohm:2011ex, Hohm:2010xe}.} (which are undoubled theories in the external space, and contain gaugings satisfying the QC).

Notice that if a given $U$ produces a solution to the QC, any T-dual $U$ will also. Therefore, it is natural to define the notion of {\it twist orbits}  as the sets of twist matrices connected through T-duality transformations. If a representative of a twist orbit generates a representative of an orbit of gaugings, one can claim that the twist orbit will generate the entire orbit of gaugings. Also, notice that if a twist matrix satisfies the WC and SC, any representative of its orbit will, so one can define the notions of undoubled and truly doubled twist orbits.

\subsection*{Non-geometry VS weak and strong constraint violation}

Any half-maximal supergravity can be uplifted to the maximal theory whenever the following constraint holds\footnote{$D=4$ half-maximal
supergravity is slightly different because its global symmetry group features an extra SL($2$) factor; for full details, see \cite{Aldazabal:2011yz, Dibitetto:2011eu}.}
\be 
f_{ABC}\,f^{ABC}\,=\,0\ . 
\label{Extra_f}
\ee
This constraint plays the role of an orthogonality condition between geometric and non-geometric fluxes. Interestingly, the constraint \eqref{Extra_f} evaluated in terms of the twist matrix $U$ and $\lambda$ can be rewritten as follows (by taking relations \eqref{f_from_U_} and \eqref{f_from_U} into account)
\be
f_{ABC}\,f^{ABC}\,=\,-3\,\partial_{D}{U^{A}}_{P}\,\partial^{D}{\left(U^{-1}\right)^{P}}_{A}-24\,\partial_{D}\lambda\,\partial^{D} \lambda\,+\,24\,\partial_{D}\partial^{D}\lambda\ . 
\label{Max_VS_Geom}
\ee
The RHS of this equation is zero whenever the background defined by $U$ and $\lambda$ satisfies the WC and SC. This immediately
implies that any background satisfying WC and SC defines a gauging which is upliftable to the maximal theory.
Conversely, if an orbit of gaugings in half-maximal supergravity does not satisfy the extra constraint \eqref{Extra_f}, the RHS of this equation must be non-vanishing, and then the WC and SC must be relaxed. In conclusion, the orbits of half-maximal supergravity that do not obey the QC of the maximal theory require truly doubled twist orbits, and are therefore genuinely non-geometric. This point provides a concrete criterion to label these orbits as non-geometric. Also, notice that these orbits will never be captured by non-geometric flux configurations obtained by T-dualising a geometric background\footnote{However, we would like to stress that, in general, it is not true that an orbit satisfying the QC constraints of maximal supergravity (\ref{Extra_f}) is necessarily generated by an undoubled twist orbit. An example can be found at the end of section 4.}.

For the sake of clarity, let us briefly review the definitions that we use. A twist orbit is non-geometric if it doesn't satisfy the WC/SC, and geometric if it does. Therefore, the notion of geometry that we consider is local, and we will not worry about global issues (given that the twist matrix is taken to be an element of the global symmetry group, the transition functions between coordinate patches are automatically elements of O$(n,n)$). On the other hand an orbit of gaugings is geometric if it contains a representative that can be obtained from ten-dimensional supergravity (or equivalently from a geometric twist orbit), and it is non-geometric it does not satisfy the constraints of maximal supergravity.

We have now described all the necessary ingredients to formally relate dimensional reductions of DFT and the orbits of half-maximal gauged supergravities. In particular, in what follows we will:
\begin{enumerate}
\item Provide a classification of all the orbits of gaugings in maximal and half-maximal supergravities in $D\geq7$.
\item Explore mechanisms to generate orbits of gaugings from twists, satisfying
\begin{itemize}
\item $U(\mathbb{Y}) \in \textrm{O}(n,n)$
\item Constant $f_{ABC}$
\item  $f_{E[AB} f^E{}_{C]D} = 0$
\end{itemize}
\item Show that in the half-maximal theories all the orbits of gaugings  can be obtained from twist orbits in DFT.
\item Show that in the half-maximal theories the orbits that satisfy the QC of maximal supergravity  admit a representative with a higher-dimensional supergravity origin. For these we provide concrete realisations in terms of unboubled backgrounds in DFT. Instead, the orbits that fail to satisfy (\ref{Extra_f}) require, as we argued, truly doubled twist orbits for which we also provide concrete examples.
\item Show that  there is a degeneracy in the space of twist orbits giving rise to the same orbit of gaugings. Interestingly, in some cases a given orbit can be obtained either from undoubled or truly doubled twist orbits.
\end{enumerate}

In the next sections we will classify all the orbits in (half-)maximal $D\geq 7$ supergravities, and provide the half-maximal ones with concrete uplifts to DFT, explicitly proving the above points.

\subsection*{Parametrisation of the duality twists}

Here we would like to introduce some notation that will turn out to be useful in the uplift of orbits to DFT. We start by noting the double internal  coordinates as $\mathbb{Y}^A = (\tilde y_a, y^a)$ with $a = 1,...,n$. As we saw, the SS compactification of DFT is defined by the twists $U(\mathbb{Y})$ and $\lambda(\mathbb{Y})$. The duality twist $U(\mathbb{Y})$ is not generic, but forced to be an element of O($n,n$), so we should provide suitable parameterisations. One option is the  {\it light-cone} parameterisation, where the metric of the (internal) global symmetry
group is taken to be of the form  
\be 
\eta_{AB} =
\begin{pmatrix} 0 & \mathds{1}_n \\ \mathds{1}_n & 0\end{pmatrix}\ . 
\ee 
The most general form of the twist matrix is then given by 
\be 
U(\mathbb{Y}) =
\left(\begin{matrix} e & 0 \\ 0 & e^{-T}\end{matrix}\right)
\,\left(\begin{matrix} \mathds{1}_n & 0 \\ -B &
\mathds{1}_n\end{matrix}\right)\, \left(\begin{matrix} \mathds{1}_n & \beta \\
0 & \mathds{1}_n\end{matrix}\right)\ , 
\ee 
with $e\,\in\,\textrm{GL}(n)$ and $B$ and $\beta$ are generic $n\times n$ antisymmetric matrices. When $\beta = 0$, $e = e(y^a)$ and $B = B(y^a)$, the matrix $e$ can be interpreted as a $n$-dimensional internal vielbein and $B$ as a background $2$-form for the $n$-dimensional internal Kalb-Ramond field $b$. Whenever the background is of this form, we will refer to it as geometric (notice that this still does not determine completely the background, which receives deformations from scalar fluctuations). In this case the gaugings take the simple form 
\bea
f_{abc} &=& 3 (e^{-1})^\alpha{}_{[a}(e^{-1})^\beta{}_{b}(e^{-1})^\gamma{}_{c]} \partial_{[\alpha} B_{\beta\gamma]}\ , \nn\\
f^a{}_{bc} &=& 2 (e^{-1})^\beta{}_{[b}(e^{-1})^\gamma{}_{c]}\partial_{\beta} e^a{}_{\gamma }\ , \nn\\
f^{ab}{}_c &=& f^{abc} = 0 \ .
\eea

If we also turn on a  $\beta (y^a)$, the relation of $e$, $B$ and $\beta$ with the internal $g$ and $b$ is less trivial, and typically the background will be globally well defined up to O$(n,n)$ transformations mixing the metric and the two-form (this is typically called a T-fold). In this case, we refer to the background as locally geometric but globally non-geometric, and this situation formally allows for non-vanishing
$f^{ab}{}_c$ and $f^{abc}$. Finally, if the twist matrix is a function of $\tilde y_a$, we refer to the background as locally non-geometric. Notice however, that if it satisfies the WC and SC, one would always be able to rotate it to a frame in which it is locally geometric, and would therefore belong to an undoubled orbit.

Alternatively, one could also define the {\it cartesian} parametrisation of the twist matrix, by taking the metric of the (internal) global symmetry group to be of the form
\be 
\eta_{AB} = \begin{pmatrix} \mathds{1}_n & 0 \\ 0 &
-\mathds{1}_n\end{pmatrix} \ .
\ee 
This formulation is related to the light-cone parametrisation through a SO$(2n)$ transformation, that must also rotate the coordinates. In this case the relation between the components of the twist matrix and the internal $g$ and $b$ is non-trivial. We will consider the O($n,n$) twist
matrix to contain a smaller O($n-1,n-1$) matrix in the directions $(y^2,...,y^n,\tilde y_2,...,\tilde y_n)$ fibred over the flat directions $(y^1,\tilde y_1)$. We have seen that this typically leads to constant gaugings.

Of course these are not the most general parameterisations and ansatz, but they will serve our purposes of uplifting all the orbits of half-maximal supergravity to DFT. Interesting works on how to generate gaugings from twists are \cite{Dall'Agata:2007sr, Andriot:2009fp}.

\section{U-duality Orbits of Maximal Supergravities}
\label{sec:U_Dualitites}

Following the previous discussion of DFT and its relevance for generating duality orbits, we turn to the actual classification of these. In particular, we start with orbits under U-duality of gaugings of maximal supergravity. Moreover, we will demonstrate that all such orbits do have a higher-dimensional supergravity origin.

Starting with the highest dimension for  maximal supergravity, $D=11$, no known deformation is possible here. Moreover, in $D=10$ maximal supergravities, the only possible deformation occurs in what is known as massive IIA supergravity\footnote{Throughout this chapter we will not consider the trombone gaugings giving rise to theories without an action principle, as discussed in \emph{e.g.} refs~\cite{Howe:1997qt, Bergshoeff:2002nv, LeDiffon:2008sh, LeDiffon:2011wt}.} \cite{Romans:1985tz}. It consists of a St\"uckelberg-like way of giving a mass to the 2-form $B_{2}$. Therefore, such a deformation cannot be interpreted as a gauging. The string theory origin of this so-called Romans' mass parameter is nowadays well understood as arising from D8-branes \cite{Polchinski:1995mt}. Furthermore, its DFT uplift has been constructed in ref.~\cite{Hohm:2011cp}. Naturally, the structure of possible orbits becomes richer when going to lower dimensions. In what follows we will perform
the explicit classification in dimensions nine and eight.

\subsection*{Orbits and origin of the $D=9$ maximal case}

\subsubsection*{Maximal $D=9$ gauged supergravity}

The maximal (ungauged) supergravity in $D=9$ \cite{Gates:1984kr} can be obtained by reducing either massless type IIA or type IIB supergravity in ten dimensions on a circle. The global symmetry group of this theory is
\be G_{0}\,=\,\mathbb{R}^{+}\,\times\,\textrm{SL}(2)\ . \notag
\ee
Note that $G_{0}$ is the global symmetry of the action and hence it is realised off-shell, whereas the on-shell symmetry has an extra
$\mathbb{R}^{+}$ with respect to which the Lagrangian has a non-trivial scaling weight. This is normally referred to as the \emph{trombone symmetry}. As a consequence, the on-shell symmetry contains three independent rescalings \cite{Bergshoeff:2002nv, Roest:2004pk}, which we summarise in table~\ref{rescalings}.
\begin{table}[h!]
\begin{center}
\scalebox{1}[1]{
\begin{tabular}{| c || c | c | c | c | c | c | c | c | c | c | c | c| c|}
\hline
\textrm{ID} & $e_{\mu}^{\phantom{\mu}a}$ & $A_{\mu}$ &  $A_{\mu}{}^{1}$ & $A_{\mu}{}^{2}$ & $B_{\mu\nu}{}^{1}$ & $B_{\mu\nu}{}^{2}$ & $C_{\mu\nu\rho}$ & $e^{\varphi}$ & $\chi$ & $e^{\phi}$ & $\psi_{\mu}$ & $\lambda\,,\,\tilde{\lambda}$ & $\mathcal{L}$  \\[1mm]
\hline \hline
$\alpha$ & $\frac{9}{7}$ & $3$ & $0$ & $0$ & $3$ & $3$ & $3$ & $\frac{6}{\sqrt{7}}$ & $0$ & $0$ & $\frac{9}{14}$ & $-\frac{9}{14}$ & $9$ \\[1mm]
\hline
$\beta$ & $0$ & $\frac{1}{2}$ & $-\frac{3}{4}$ & $0$ & $-\frac{1}{4}$ & $\frac{1}{2}$ & $-\frac{1}{4}$ & $\frac{\sqrt{7}}{4}$ & $-\frac{3}{4}$ & $\frac{3}{4}$ & $0$ & $0$ & $0$ \\[1mm]
\hline
$\gamma$ & $0$ & $0$ & $1$ & $-1$ & $1$ & $-1$ & $0$ & $0$ & $2$ & $-2$ & $0$ & $0$ & $0$ \\[1mm]
\hline
$\delta$ & $\frac{8}{7}$ & $0$ & $2$ & $2$ & $2$ & $2$ & $4$
& $-\frac{4}{\sqrt{7}}$ & $0$ & $0$ & $\frac{4}{7}$ & $-\frac{4}{7}$
&
$8$ \\[1mm] \hline
\end{tabular}
}
\end{center}
\caption{{\it The scaling weights of the nine-dimensional fields. As already anticipated, only three rescalings are independent since they are subject to the following constraint: $8\alpha-48\beta-18\gamma-9\delta=0$. As the scaling weight of the Lagrangian $\mathcal{L}$ shows, $\beta$ and $\gamma$ belong to the off-shell symmetries, whereas $\alpha$ and $\delta$ can be combined into a trombone symmetry and an off-shell symmetry.}}\label{rescalings}
\end{table}
The full field content consists of the following objects which arrange themselves into irrep's of $\mathbb{R}^{+}\,\times\,\textrm{SL}(2)$:
\be 
\textrm{9D :}\qquad \underbrace{e_{\mu}^{\phantom{\mu}a}\,,\,A_{\mu}\,,\,A_{\mu}{}^{i}\,,\,B_{\mu\nu}{}^{i}\,,\,C_{\mu\nu\rho}\,,\,\varphi\,,\,\tau=\,\chi\,+\,i\,e^{-\phi}}_{\textrm{bosonic dof's}}\,\,\,;\,\underbrace{\psi_\mu\,,\,\lambda\,,\,\tilde{\lambda}}_{\textrm{fermionic dof's}}\ ,\label{fields_9D} 
\ee
where $\mu,\nu,\cdots$ denote nine-dimensional curved spacetime, $a,b,\cdots$ nine-dimensional flat spacetime and $i,j,\cdots$ fundamental SL($2$) indices respectively.

The general deformations of this theory have been studied in detail in ref.~\cite{FernandezMelgarejo:2011wx}, where both embedding tensor deformations and gaugings of the trombone symmetry have been considered. For the present scope we shall restrict ourselves to the first ones. The latter ones would correspond to the additional mass parameters $m_{\textrm{IIB}}$ and $(m_{11},m_{\textrm{IIA}})$ in refs~\cite{Bergshoeff:2002nv, FernandezMelgarejo:2011wx}, which give rise to theories without an action principle.

The vectors of the theory $\{A_{\mu}\,,\,{A_{\mu}}^{i}\}$ transform in the $V^\prime\,=\,\textbf{1}_{(+4)}\,\oplus\,\textbf{2}_{(-3)}\,$ of $\mathbb{R}^{+}\,\times\,\textrm{SL}(2)\,$, where the $\mathbb{R}^{+}$ scaling weights are included as well\footnote{The $\mathbb{R}^{+}$ factor in the global symmetry is precisely the combination $\left(\frac{4}{3}\,\alpha\,-\,\frac{3}{2}\,\delta\right)\,$ of the different rescalings introduced in ref.~\cite{Bergshoeff:2002nv}.}. The resulting embedding tensor deformations live in the following tensor product
\be 
\mathfrak{g}_{0}\otimes V=\textbf{1}_{(-4)}\,\oplus\, 2\,\cdot\,\textbf{2}_{(+3)}\,\oplus\,\textbf{3}_{(-4)}\,\oplus\,\textbf{4}_{(+3)}\ . 
\ee
The LC projects out the $\textbf{4}_{(+3)}$, the $\textbf{1}_{(-4)}$ and one copy of the $\textbf{2}_{(+3)}$ since they would give rise to inconsistent deformations. As a consequence, the consistent gaugings are parameterised by embedding tensor components in the $\textbf{2}_{(+3)}\,\oplus\,\textbf{3}_{(-4)}$. We will denote these allowed deformations by $\theta^{i}$ and $\kappa^{(ij)}$.

The closure of the gauge algebra and the antisymmetry of the brackets impose the following QC
\bea
\epsilon _{ij}\,\theta^{i}\,\kappa^{jk} &=&0\ ,\qquad\qquad
\textbf{2}_{(-1)}\label{quadratic constraints in 9D1}\\
\theta^{(i}\,\kappa^{jk)} &=&0\ .\qquad\qquad\,
\textbf{4}_{(-1)}\label{quadratic constraints in 9D2}
\eea

\subsubsection*{The $\mathbb{R}^{+}\,\times\,$SL($2$) orbits of solutions to the QC}

The QC \eqref{quadratic constraints in 9D1} and \eqref{quadratic constraints in 9D2} turns out to be very simple to solve; after finding all the solutions, we studied the duality orbits, \emph{i.e.} classes of those solutions which are connected via a duality transformation. The resulting orbits of consistent gaugings in this case are presented in table~\ref{orbits_max9}.

\begin{table}[h!]
\begin{center}
\scalebox{1}[1]{
\begin{tabular}{| c | c | c | c |}
\hline
\textrm{ID} & $\theta^{i}$ & ${\kappa}^{ij}$ &  gauging \\[1mm]
\hline \hline
$1$ & \multirow{3}{*}{$(0,0)$} &  diag($1,1$) &  SO($2$) \\[1mm]
\cline{1-1}\cline{3-4} $2$ & & diag($1,-1$) & SO($1,1$) \\[1mm]
\cline{1-1}\cline{3-4} $3$ & & diag($1,0$) & $\mathbb{R}^{+}_{\gamma}$ \\[1mm]
\hline \hline $4$ & $(1,0)$ & diag($0,0$) & $\mathbb{R}^{+}_{\beta}$ \\[1mm]
\hline
\end{tabular}
}
\end{center}
\caption{{\it All the U-duality orbits of consistent gaugings in maximal
supergravity in $D=9$. For each of them, the simplest representative
is given. The subscripts $\beta$ and $\gamma$ refer to the
rescalings summarised in table~\ref{rescalings}.}}
\label{orbits_max9}
\end{table}

\subsubsection*{Higher-dimensional geometric origin}

The four different orbits of maximal $D=9$ theory have the following higher-dimensional origin in terms of geometric compactifications  \cite{Bergshoeff:2002mb}:

\begin{itemize}

\item \textbf{Orbits 1 -- 3:}  These come from reductions of type IIB supergravity on a circle with an SL($2$) twist.

\item \textbf{Orbit 4:}  This can be obtained from a reduction of type IIA supergravity on a circle with the inclusion of an $\mathbb{R}^{+}_{\beta}$ twist.

\end{itemize}

\subsection*{Orbits and origin of the $D=8$ maximal case}

\subsubsection*{Maximal $D=8$ gauged supergravity}

The maximal (ungauged) supergravity in $D=8$ \cite{Salam:1984ft} can be obtained by reducing eleven-dimensional supergravity on a $T^3$. The global symmetry group of this theory is
\be G_{0}\,=\,\textrm{SL}(2)\,\times\,\textrm{SL}(3)\ . \notag\ee
The full field content consists of the following objects which arrange themselves into irrep's of $\textrm{SL}(2)\,\times\,\textrm{SL}(3)$:
\be 
\textrm{8D :}\qquad \underbrace{e_{\mu}^{\phantom{\mu}a}\,,\,A_{\mu}{}^{\alpha m}\,,\,B_{\mu\nu m}\,,\,C_{\mu\nu\rho}\,,\,L_{m}^{\phantom{m}I}\,,\,\phi\,,\,\chi}_{\textrm{bosonic dof's}}\,\,\,;\,\underbrace{\psi_\mu\,,\,\chi_I}_{\textrm{fermionic dof's}}\ ,
\label{fields_8D} 
\ee
where $\mu,\nu,\cdots$ denote eight-dimensional curved spacetime, $a,b,\cdots$ eight-dimensional flat spacetime, $m,n,\cdots$ fundamental SL($3$), $I,J,\cdots$ fundamental SO($3$) and $\alpha,\beta,\cdots$ fundamental SL($2$) indices respectively. The six vector fields $A_{\mu}{}^{\alpha m}$ in \eqref{fields_8D} transform in the $V'=\left( \textbf{2},\textbf{3}^\prime\right)$.
There are eleven group generators, which can be expressed in the adjoint representation $\mathfrak{g}_{0}$.

The embedding tensor $\Theta $ then lives in the representation $\mathfrak{g} _{0}\,\otimes\,V$, which can be decomposed into irreducible representations as
\be 
\mathfrak{g}_{0}\otimes V=2\,\cdot\left(
\textbf{2},\textbf{3}\right) \oplus \left(
\textbf{2},\textbf{6}^\prime\right) \oplus  \left(
\textbf{2},\textbf{15}\right) \oplus \left(
\textbf{4},\textbf{3}\right)\,. 
\ee
The LC restricts the embedding tensor to the $\left( \textbf{2},\textbf{3}\right) \oplus \left(\textbf{2},\textbf{6}^\prime\right) $ \cite{Weidner:2006rp}. It is worth noticing that there are two copies of the $\left( \textbf{2},\textbf{3} \right) $ irrep in the above composition; the LC imposes a relation between them \cite{Samtleben:2008pe}. This shows that, for consistency, gauging some SL($2$) generators implies the necessity of gauging some SL($3$) generators as well. Let us denote the allowed embedding tensor irrep's by $\xi _{\alpha m}$ and $f_{\alpha}{}^{(mn)}$ respectively.

The QC then read \cite{Dani:2008, deRoo:2011fa}
\bea
\epsilon ^{\alpha\beta}\,\xi _{\alpha p}\xi _{\beta q} &=&0
\ ,\qquad\qquad
\left( \textbf{1},\textbf{3}^\prime\right)\label{quadratic constraints in 8D1}\\%
f_{(\alpha}{}^{np}\xi _{\beta)p} &=&0 \ , \qquad\qquad \left(
\textbf{3},\textbf{3}^\prime\right)\label{quadratic constraints in 8D2}\\%
\epsilon ^{\alpha\beta}\left(\epsilon
_{mqr}f_{\alpha}{}^{qn}f_{\beta}{}^{rp}+f_{\alpha}{}^{np}\xi _{\beta
m}\right)
&=&0\ . \qquad\left( \textbf{1},\textbf{3}^\prime\right)\oplus\left( \textbf{1},\textbf{15}\right)\label{quadratic constraints in 8D3} 
\eea
Any solution to the QC \eqref{quadratic constraints in 8D1}, \eqref{quadratic constraints in 8D2} and \eqref{quadratic constraints in 8D3} specifies a consistent gauging of a subgroup of SL($2)\,\times\,$SL($3$) where the corresponding generators are given by
\bea 
\label{gauge_gen_D=8} 
{\left(X_{\alpha
m}\right)_{\beta}}^{\gamma} &=& \delta_{\alpha}^{\gamma}\,\xi_{\beta
m}\,-\,\frac{1}{2}\,\delta_{\beta}^{\gamma}\,\xi_{\alpha m}\ ,\\
{\left(X_{\alpha m}\right)_{n}}^{p} &=&
\epsilon_{mnq}\,{f_{\alpha}}^{qp}\,-\,\frac{3}{4}\,\left(\delta_{m}^{p}\,\xi_{\alpha
n}\,-\,\frac{1}{3}\,\delta_{n}^{p}\,\xi_{\alpha m}\right)\ . 
\eea

\subsubsection*{The SL($2$)$\,\times\,$SL($3$) orbits of solutions to the QC}

We exploited an algebraic geometry tool called the Gianni-Trager-Zacharias (GTZ) algorithm \cite{GTZ}. This algorithm has been computationally implemented by the \textsc{\,Singular\,} project \cite{DGPS} and it consists in the primary decomposition of
ideals of polynomials (see chapter~\ref{Half_Max} for more details). After finding all the solutions to the QC by means of the algorithm mentioned above, one has to group together all the solutions which are connected through a duality transformation, thus obtaining a classification of such solutions in terms of duality orbits. The resulting orbits of consistent gaugings\footnote{Recently, also the possible vacua of the different
theories have been analysed \cite{deRoo:2011fa}. As we will explain later in this section, it was found that only  {\bf orbit 3} has maximally symmetric vacua.} in this case are
presented in table~\ref{orbits_max8}.

\begin{table}[h!]
\begin{center}
\scalebox{1}[1]{
\begin{tabular}{| c | c | c | c | c | c |}
\hline
\textrm{ID} & ${f_{+}}^{mn}$ & ${f_{-}}^{mn}$ & $\xi_{+m}$ & $\xi_{-m}$ &  gauging \\[1mm]
\hline \hline
$1$ & diag($1,1,1$) & \multirow{5}{*}{$\textrm{diag}(0,0,0)$} & \multirow{5}{*}{$(0,0,0)$} & \multirow{5}{*}{$(0,0,0)$} &  SO($3$) \\[1mm]
\cline{1-2}\cline{6-6} $2$ & diag($1,1,-1$) & & & &  SO($2,1$) \\[1mm]
\cline{1-2}\cline{6-6} $3$ & diag($1,1,0$) & & & &  ISO($2$) \\[1mm]
\cline{1-2}\cline{6-6} $4$ & diag($1,-1,0$) & & & &  ISO($1,1$) \\[1mm]
\cline{1-2}\cline{6-6} $5$ & diag($1,0,0$) & & & &  CSO($1,0,2$) \\[1mm]
\hline \hline $6$ & diag($0,0,0$) & diag($0,0,0$) & $(1,0,0)$ &
$(0,0,0)$ & Solv$_{2}\,\times\,$Solv$_{3}$\\[1mm]
\hline \hline $7$ & diag($1,1,0$) &
\multirow{3}{*}{$\textrm{diag}(0,0,0)$} & \multirow{3}{*}{$(0,0,1)$}
& \multirow{3}{*}{$(0,0,0)$} & \multirow{3}{*}{Solv$_{2}\,\times\,$Solv$_{3}$}\\[1mm]
\cline{1-2} $8$ & diag($1,-1,0$) & & & &   \\[1mm]
\cline{1-2} $9$ & diag($1,0,0$) & & & &   \\[1mm]
\hline \hline $10$ & diag($1,-1,0$) & \scalebox{0.7}[0.7]{$\left(\begin{array}{ccc}1 & 1 & 0\\
1 & 1 & 0\\ 0 & 0 & 0\end{array}\right)$} & $\frac{2}{9}(0,0,1)$ &
$(0,0,0)$ & Solv$_{2}\,\times\,$SO$(2)\,\ltimes\,$Nil$_{3}(2)$\\[1mm]

\hline
\end{tabular}
}
\end{center}
\caption{{\it All the U-duality orbits of consistent gaugings in maximal supergravity in $D=8$. For each of them, the simplest representative is given. We denote by  Solv$_{2}\,\subset\,$SL($2$) and Solv$_{3}\,\subset\,$SL($3$) a solvable algebra of dimension 2 and 3 respectively. To be more precise, Solv$_{2}$ identifies the Borel subgroup of SL($2$) consisting of $2\times 2$ upper-triangular matrices. Solv$_{3}$, instead, is a Bianchi type V algebra.}}
\label{orbits_max8}
\end{table}

\subsubsection*{Higher-dimensional geometric origin}

\begin{itemize}
\item \textbf{Orbits 1 -- 5:}  These stem from reductions of eleven-dimensional supergravity on a three-dimensional group manifold of type  A in the Bianchi classification \cite{Bergshoeff:2003ri}. The special case in orbit 1 corresponds to a reduction over an SO($3$) group manifold and it was already studied in ref.~\cite{Salam:1984ft}.

\item \textbf{Orbit 6:}  This can be obtained from a reduction of maximal nine-dimensional supergravity on a circle with the inclusion of an  $\mathbb{R}^{+}$ twist inside the global symmetry group.

\item \textbf{Orbits 7 -- 9:}  These can come from the same reduction from $D=9$ but upon inclusion of a more general $\mathbb{R}^{+}\,\times\,\textrm{SL}(2)$ twist.

\item \textbf{Orbit 10:}  This orbit seems at first sight more complicated to be obtained from a dimensional reduction owing to its non-trivial SL($2$) angles. Nevertheless, it turns out that one can land on this orbit by compactifying type IIB supergravity on a circle with an SL($2$) twist and then further reducing on another circle with $\mathbb{R}^{+}\,\times\,\textrm{SL}(2)$ twist given by the residual little group leaving invariant the intermediate nine-dimensional deformation.
\end{itemize}

\subsection*{Remarks on the $D=7$ maximal case}

The general deformations of the maximal theory in $D=7$ are constructed and presented in full detail in ref.~\cite{Samtleben:2005bp}. For the present aim we only summarise here a few relevant facts.

The global symmetry group of the theory is SL($5$). The vector fields $A_{\mu}{}^{MN}=A_{\mu}{}^{[MN]}$ transform in the \textbf{10}$^\prime$ of SL($5$), where we denote by $M$ a fundamental SL($5$) index. The embedding tensor $\Theta$ takes values in the following irreducible components
\be
\textbf{10}\,\otimes\,\textbf{24}\,=\,\textbf{10}\,\oplus\,\textbf{15}\,\oplus\,\textbf{40}^\prime\,\oplus\,\textbf{175}\,.
\ee
The LC restricts the embedding tensor to the $\textbf{15}\,\oplus\,\textbf{40}^\prime$, which can be parameterised by the following objects
\be 
Y_{(MN)}\,,\qquad\textrm{and}\qquad
Z^{[MN],P}\quad\textrm{with}\quad Z^{[MN,P]}=0\ . 
\ee
The generators of the gauge algebra can be written as follows
\be
{\left(X_{MN}\right)_P}^Q\,=\,\delta_{[M}^Q\,Y_{N]P}\,-\,2\,\epsilon_{MNPRS}\,Z^{RS,Q}\,,
\label{gen_max}\ee
or, identically, if one wants to express them in the $\textbf{10}$,
\be
{\left(X_{MN}\right)_{PQ}}^{RS}\,=\,2\,{\left(X_{MN}\right)_{[P}}^{[R}\,\,\delta_{Q]}^{S]}\,.
\label{gen_max10}\ee
The closure of the gauge algebra and the antisymmetry of the brackets imply the following QC
\be
Y_{MQ}\,Z^{QN,P}\,+\,2\,\epsilon_{MRSTU}\,Z^{RS,N}\,Z^{TU,P}\,=\,0\
, \label{QC_max7}
\ee
which have different irreducible pieces in the $\textbf{5}^\prime\,\oplus\,\textbf{45}^\prime\,\oplus\,\textbf{70}^\prime$.
Unfortunately, in this case, both the embedding tensor deformations and the QC reach a level of complexity that makes an exhaustive and general analysis difficult. Such analysis lies beyond the scope of our work.

\subsubsection*{The Critical Points of Maximal $D=8$ Supergravities}

Before going to the discussion of T-duality orbits of half-maximal theories, we would like in this paragraph to schematically present the study of critical points of maximal eight-dimensional supergravities. To this end, we will briefly summarise the results of ref.~\cite{deRoo:2011fa} where  the analysis of the scalar potential and its derivatives was preformed in full generality. The first result of the above reference, in fact, is the derivation of the scalar potential itself that we briefly sketch here. The scalar degrees of freedom present in \eqref{fields_8D} can rearranged into the following coset representatives 
\be
\label{scalarparamWM}
W_{\alpha\beta}\,=\,\left(
\begin{array}
{cc} e^{-\phi }+\chi ^{2}e^{\phi } & \chi e^{\phi } \\
\chi e^{\phi } & e^{\phi }
\end{array}
\right) \ ,\qquad
M_{mn}\,=\,L_{m}^{\phantom{m}i}L_{n}^{\phantom{m}j}\delta_{ij}\ .
\ee
of SL($2)/$SO($2$) and SL($3)/$SO($3$), respectively. 

The gravity/scalar part of the action reads \cite{AlonsoAlberca:2003jq}
\be 
S\,=\,\frac{1}{16\pi G_8}\int d^8x\,\,e\,\left(R\,+\,\frac{1}{4}\,\textrm{Tr}(\partial M\,\partial M^{-1})\,+\,\frac{1}{4}\,\textrm{Tr}(\partial W\,\partial W^{-1})\right)\,,
\label{action_max_D=8} 
\ee
where $e$ is the determinant of the vielbein. The full bosonic action, in addition to the terms in \eqref{action_max_D=8}, contains kinetic terms for the vector fields, the two- and three-forms and finally Chern-Simons terms.

The most general ansatz for the scalar potential induced by the gauging, consists of the following terms
\be
V=W^{\alpha\beta}\,\left[f_{\alpha}{}^{mn}f_{\beta}{}^{pq}\left(a\,M_{mp}M_{nq}+b\,M_{mn}M_{pq}\right)+c\,\xi_{\alpha m}\xi_{\beta n}M^{mn}\right]\ ,
\label{Ansatz 8D V}
\ee
$W^{\alpha\beta}$ and $M^{mn}$ denote the inverse matrices of $W_{\alpha\beta}$ and $M_{mn}$ appearing in \eqref{scalarparamWM}, and $a$, $b$ and $c$ are coefficients that are going to be determined. The way used for fixing these coefficients is to restrict them by means of the scalar potential in maximal $D=7$ supergravity, which we have just presented above.

In addition to the embedding tensor deformations extensively described in the previous paragraph, we need to introduce the scalar sector and scalar potential of maximal $D=7$ supergravities. Such scalar sector is described by the SL($5$)$/$SO($5$) coset geometry (see table~\ref{table:max}) parametrised by the symmetric matrix $\cM_{MN}$ with inverse $\cM^{MN}$. This divides the isometry group of the scalar
manifold SL($5$) into unphysical scalar degrees of freedom (generating the adjoint representation of SO($5$)) and physical scalar fields completing them to the $\textbf{24}$, \emph{i.e.} the adjoint representation of SL($5$). Maximal supersymmetry completely and uniquely determines the scalar potential to be of the form
\bea V&=&
\frac{1}{64}\bigg(2\cM^{MN}Y_{NP}\cM^{PQ}Y_{QM}-(\cM^{MN}Y_{MN})^2\bigg)+\nn\\
&+&Z^{MN,P}Z^{QR,S}\bigg(\cM_{MQ}\cM_{NR}\cM_{PS}-\cM_{MQ}\cM_{NP}\cM_{RS}\bigg)\,.
\label{7D V} 
\eea

Every gauging in $D=8$ must be an at most six-dimensional subgroup of the global symmetry group SL($2$)$\,\times\,$SL($3$). After dimensional reduction to $D=7$, the global symmetry group gets enhanced with respect to what one would naively expect\footnote{One would expect $\mathbb{R}^+\,\times\,\textrm{SL}(2)\,\times\,\textrm{SL}(3)$, whereas it turns out to be enlarged to an $\textrm{SL}(5)$.}; for this reason, one would certainly expect any consistent gauging of the eight-dimensional theory to be reduced to a consistent gauging of the seven-dimensional theory where the gauge group, though, undergoes an enlargement just in the same way as for the global symmetry group. This statement implies that the irreducible components of the embedding tensor in eight dimensions must be obtained as a truncation of the embedding tensor in $D=7$. This implies the possibility of deriving the scalar potential of maximal $D=8$ gauged supergravity from the expression of the
seven-dimensional scalar potential given in \eqref{7D V}, after understanding how the eight-dimensional degrees of freedom associated with internal symmetries sit inside SL($5$) irrep's. To this end, we need the branching of some relevant irrep's of SL($5$) with respect to irrep's of SL($2$)$\,\times\,$SL($3$), which is a maximal subgroup thereof. The embedding turns out to be unique and it gives rise to the following decompositions
\bea 
\textbf{5}\quad &\longrightarrow &\quad
(\textbf{2},\textbf{1})\,\oplus\,(\textbf{1},\textbf{3})\,,\label{br5}\\
\textbf{15}\quad &\longrightarrow &\quad
(\textbf{1},\textbf{6})\,\oplus\,(\textbf{2},\textbf{3})\,\oplus\,(\textbf{3},\textbf{1})\,,\label{br15}\\
\textbf{24}\quad &\longrightarrow &\quad
(\textbf{1},\textbf{1})\,\oplus\,(\textbf{1},\textbf{8})\,\oplus\,(\textbf{2},\textbf{3})\,\oplus\,(\textbf{2},\textbf{3}^{\prime})\,\oplus\,(\textbf{3},\textbf{1})\,,\label{br24}\\
\textbf{40}^{\prime}\quad &\longrightarrow &\quad
(\textbf{1},\textbf{3}^{\prime})\,\oplus\,(\textbf{1},\textbf{8})\,\oplus\,(\textbf{2},\textbf{1})\,\oplus\,(\textbf{2},\textbf{6}^{\prime})\,\oplus\,(\textbf{2},\textbf{3})\,\oplus\,(\textbf{3},\textbf{3}^{\prime})\,.\label{br40p}
\eea
The decomposition \eqref{br5} essentially tells that the fundamental SL($5$) index $M=1,2,3,4,5$ goes into $(\alpha\,;\,m)$, where $\alpha=+,-$ and $m=1,2,3$ represent fundamental SL($2$) and SL($3$) indices respectively. The decomposition \eqref{br24} tells us how the SL($2$)$\,\times\,$SL($3$) scalar degrees of freedom (living in the $(\textbf{1},\textbf{8})\,\oplus\,(\textbf{3},\textbf{1})$) are embedded in the adjoint of SL($5$). It is worth mentioning at this point that we are losing a Cartan generator in the branching procedure; such an abelian generator is realised as an extra $\mathbb{R}^+$ factor corresponding to a dilaton in the seven-dimensional theory, with respect to which any eight-dimensional object should have a scaling weight which we are omitting. This extra scalar exactly accounts for the $(\textbf{1},\textbf{1})$ irrep appearing in \eqref{br24}. The truncation that we need consists then in switching off all the off-diagonal axionic excitations (spanning the
$(\textbf{2},\textbf{3})$ and $(\textbf{2},\textbf{3}^{\prime})$ terms in \eqref{br24}), thus resulting in the following parametrisation
\be 
\cM_{MN}=\left(
\begin{array}{c|c}
e^{3\sigma}\,W_{IJ} & 0 \\[1mm]
\hline
\\[-4mm]
0 & e^{-2\sigma}\,M_{mn}
\end{array}
\right)\,,
\label{scalar coset decomposition}
\ee
where $\sigma$ is the extra dilaton corresponding to $\mathbb{R}^+$, whereas $W_{IJ}$ and $M_{mn}$ parametrise the SL($2$)$/$SO($2$) and SL($3$)$/$SO($3$) cosets respectively. It has been checked explicitly that the scaling weights of all the terms in the $D=8$ scalar potential with respect to the extra $\mathbb{R}^+$ are all equal such that it is perfectly consistent to set $\sigma=0$ in the rest of our derivation, since any other constant value can be seen as a change of normalisation of the potential energy in the Lagrangian.

The embedding tensor of the $D=8$ theory is embedded as follows inside the one of $D=7$ theory
\begin{eqnarray}
Z^{\alpha m,n}=-Z^{m\alpha,n} &=& \frac{1}{4}\,\epsilon^{\alpha\beta}f_{\beta}{}^{mn}-\frac{1}{16}\,\epsilon^{mnp}\epsilon^{\alpha\beta}\xi_{\beta p}\ ,\label{Z1}\\%
Z^{mn,\alpha}&=& \frac{1}{8}\,\epsilon^{mnp}\epsilon^{\alpha\beta}\xi_{\beta p}\ ,\label{Z2}\\%
Y_{\alpha m}=Y_{m\alpha}&=& \xi_{\alpha m}\ .\label{Y}%
\end{eqnarray}
One can check that substituting (\ref{Z1}), \eqref{Z2} and \eqref{Y} into the $D=7$ quadratic constraints \eqref{QC_max7} exactly leads to the ones in $D=8$ as shown in \eqref{quadratic constraints in 8D1}, \eqref{quadratic constraints in 8D2} and \eqref{quadratic constraints in 8D3}.

One can finally apply the decomposition rules (\ref{scalar coset decomposition}) and \eqref{Z1}, \eqref{Z2} and \eqref{Y} on the $D=7$ scalar potential (\ref{7D V}), so that the relative coefficients in (\ref{Ansatz 8D V}) can be determined, and by taking the normalisation of the action \eqref{action_max_D=8} into account one can further fix the overall factor of (\ref{Ansatz 8D V}). Then the $D=8$ scalar potential is fully derived:
\be
\label{8D V}
V= \frac{1}{2}\,W^{\alpha\beta}\,\left[f_{\alpha}{}^{mn}f_{\beta}{}^{pq}\,\left(2M_{mp}M_{nq}-M_{mn}M_{pq}\right)+\xi_{\alpha m}\xi_{\beta n}M^{mn}\right] \ .
\ee

In total there are 7 scalars for the coset $\frac{\textrm{SL}(2)}{\textrm{SO}(2)}\,\times\,\frac{\textrm{SL}(3)}{\textrm{SO}(3)}$. In \eqref{scalarparamWM} we already gave a parametrisation for the SL($2$) scalars; now we also specify a parametrisation of the
vielbein $L$ appearing in \eqref{scalarparamWM} containing the information about the SL($3$) scalars, which is given by
\be 
L_{m}^{\phantom{m}i} =\left(
\begin{array}{ccc}
e^{-\phi _{1}} & \chi _{1}e^{\frac{\phi _{1}-\phi _{2}}{2}} & \chi_{2}e^{\frac{\phi _{1}+\phi _{2}}{2}} \\%
0 & e^{\frac{\phi _{1}-\phi _{2}}{2}} & \chi _{3}e^{\frac{\phi _{1}+\phi _{2}}{2}} \\%
0 & 0 & e^{\frac{\phi _{1}+\phi _{2}}{2}}%
\end{array}
\right)
\ .  \label{scalar param. L}
\ee

Subsequently, by substituting such a parametrisation into the scalar potential (\ref{8D V}) and requiring that
\be
\frac{\delta V}{\delta \text{ (scalars)}}=0 \ ,
\ee
one obtains 7 equations which represent the extremality condition for the scalar potential. Since the full theory enjoys a global SL($2$)$\,\times\,$SL($3$) duality symmetry, one can choose to solve these equations in the origin of moduli space (setting all 7 scalars to zero\footnote{This translates into $W=\mathds{1}_2$ and $M=\mathds{1}_3$, from which it becomes manifest that the origin still presents a residual SO($2$)$\,\times\,$SO($3$) invariance.\label{footnote_origin}}.). As we will explain in more detail in chapter~\ref{Half_Max}, this can always be done without loss of generality by performing a non-compact duality transformation. This will translate the 7 equations of motion for the scalars into a set of 7 quadratic conditions in the embedding tensor components. Furthermore the quadratic constraints (\ref{quadratic constraints in 8D1}), (\ref{quadratic constraints in 8D2}) and (\ref{quadratic constraints in 8D3}) give another 30 equations in the embedding tensor components which need to be satisfied for the solution to be consistent. This set of 37 equations appears in the form an ideal consisting of homogeneous polynomial equations which can be solved for the components $\xi _{\alpha m}$ and $f_{\alpha}{}^{(mn)}$.

As explained in the footnote \ref{footnote_origin}, we still have compact duality transformations that we can use in order to simplify the general form of $\xi$ and $f$ without spoiling the choice of solving the equations of motion in the origin. For instance, we can make use of an SO($3$) transformation in order to diagonalise $f_{-}{}^{mn}$, whereas for the moment we do not need to exploit SO($2$) transformations.

By making again use of the GTZ algorythm \cite{GTZ} (see chapter~\ref{Half_Max}), we find in the end only one SO($2$)$\,\times\,$SO($3$) orbit of
solutions, in which the simplest representative is given by
\be 
f_{+}{}^{mn}= \left(
\begin{array}{ccc}
\lambda & 0 & 0\\
0 & \lambda & 0\\
0 & 0 & 0
\end{array}
\right)\text{ ,}
\,\,\,\,\,f_{-}{}^{mn}\,\,=\,\,\xi_{+m}\,\,=\,\,\xi_{-m}\,\,=\,\,0\text{
,}\label{solution_EOM_8}
\ee
where $\lambda$ represents an arbitrary real parameter, with $V=0$ (Minkowski). This theory belongs to \textbf{orbit 1} in table~\ref{orbits_max8}, whereas \emph{all} the other orbits of consistent theories listed there turn out to have \emph{no} critical points.

The solutions given in \eqref{solution_EOM_8} turn out to \emph{always non-supersymmetric} and have the following mass spectrum
\be
\begin{array}{cccc}
0\,\,(\times\,5)\,, & & & 8\,\lambda^2\,\,(\times\,2)\ ,
\end{array}
\ee
where, though, at least the SL($2$) dilaton $\phi$ can never be stabilised at any higher-order level since it corresponds to the overall $e^{\phi}$ behaviour typical of no-scale supergravities.
 
\section{T-duality Orbits of Half-maximal Supergravities}
\label{sec:T_Dualitites}

After the previous section on maximal supergravities, we turn our attention to theories with half-maximal supersymmetry. In particular, in this section we will classify the orbits under T-duality of all gaugings of half-maximal supergravity. We will only consider the theories with duality groups $\mathbb{R}^+\,\times\,\textrm{SO}(d,d)$ in $D= 10-d$, which places a restriction on the number of vector multiplets. For these theories we will classify all duality orbits, and find a number of non-geometric orbits. Furthermore, we demonstrate that DFT does yield a
higher-dimensional origin for all of them.

Starting from $D=10$ half-maximal supergravity without vector multiplets, it can be seen that there is no freedom to deform this theory, rendering this case trivial. In $D=9$, instead, we have the possibility of performing an Abelian gauging inside $\mathbb{R}^{+}\,\times\,$SO($1,1$), which will depend on one deformation parameter. However, this is precisely the parameter that one expects to generate by means of a twisted reduction from $D=10$.
This immediately tells us that non-geometric fluxes do not yet appear in this theory. In order to find the first non-trivial case, we will have to consider the $D=8$ case.

\subsection*{Orbits and origin of the $D=8$ half-maximal case}

\subsubsection*{Half-maximal $D=8$ gauged supergravity}

Half-maximal supergravity in $D=8$ is related to the maximal theory analysed in the previous section by means of a $\mathbb{Z}_{2}$ truncation. The action of such a $\mathbb{Z}_{2}$ breaks $\textrm{SL}(2)\,\times\,\textrm{SL}(3)$ into
$\mathbb{R}^{+}\,\times\,\textrm{SL}(2)\times\,\textrm{SL}(2)$, where $\textrm{SL}(2)\,\times\,\textrm{SL}(2)=\textrm{O}(2,2)$ can be interpreted as the T-duality group in $D=8$ as shown in table~\ref{dualities_789}. The embedding of $\mathbb{R}^{+}\,\times\,\textrm{SL}(2)$ inside SL($3$) is unique and it determines the following branching of the fundamental representation
\bea
\textbf{3}\,\,&\longrightarrow&\,\,\textbf{1}_{(+2)}\,\oplus\,\textbf{2}_{(-1)}\ ,\notag\\
m\,&\longrightarrow&\,\,(\bullet\,,\,i)\ ,\notag 
\eea
where the $\mathbb{R}^{+}$ direction labeled by $\bullet$ is parity even, whereas $i$ is parity odd, such as the other SL($2$) index $\alpha$. In the following we will omit all the $\mathbb{R}^{+}$ weights since they do not play any role in the truncation.

The embedding tensor of the maximal theory splits in the following way
\bea
(\textbf{2},\textbf{3})\,\,&\longrightarrow&\,\,\xcancel{(\textbf{2},\textbf{1})}\,\oplus\,(\textbf{2},\textbf{2})\ ,\notag\\
(\textbf{2},\textbf{6}^{\prime})\,\,&\longrightarrow&\,\,\xcancel{(\textbf{2},\textbf{1})}\,\oplus\,(\textbf{2},\textbf{2})\,\oplus\,\xcancel{(\textbf{2},\textbf{3})}\
,\notag 
\eea
where all the crossed irrep's are projected out because of $\mathbb{Z}_2$ parity. This implies that the consistent embedding tensor deformations of the half-maximal theory can be described by two objects which are doublets with respect to both SL($2$)'s. Let us denote them by $a_{\alpha i}$ and $b_{\alpha i}$. This statement is in perfect agreement with the Kac-Moody analysis performed in ref.~\cite{Bergshoeff:2007vb}. The explicit way of embedding $a_{\alpha i}$ and $b_{\alpha i}$ inside $\xi _{\alpha m}$ and $f_{\alpha}{}^{mn}$ is given by
\bea
{f_{\alpha}}^{i\bullet}&=&{f_{\alpha}}^{\bullet i}\,=\,\epsilon^{ij}\,a_{\alpha j}\ , \label{ET_Half_Max81}\\[2mm]
\xi _{\alpha i}&=&4\,b _{\alpha i}\ .
\label{ET_Half_Max82} 
\eea

The QC given in \eqref{quadratic constraints in 8D1}, \eqref{quadratic constraints in 8D2} and \eqref{quadratic constraints in 8D3} are decomposed according to the following branching
\bea
(\textbf{1},\textbf{3}^{\prime})\,\,&\longrightarrow&\,\,(\textbf{1},\textbf{1})\,\oplus\,\xcancel{(\textbf{1},\textbf{2})}\ ,\notag\\
(\textbf{3},\textbf{3}^{\prime})\,\,&\longrightarrow&\,\,(\textbf{3},\textbf{1})\,\oplus\,\xcancel{(\textbf{3},\textbf{2})}\ ,\notag\\
(\textbf{1},\textbf{15})\,\,&\longrightarrow&\,\,(\textbf{1},\textbf{1})\,\oplus\,\xcancel{2\,\cdot\,(\textbf{1},\textbf{2})}\,\oplus\,2\,\cdot\,(\textbf{1},\textbf{3})\,\oplus\,\xcancel{(\textbf{1},\textbf{4})}\
.\notag
\eea
As a consequence, one expects the set of $\mathbb{Z}_{2}$ even QC to consist of 3 singlets, a $(\textbf{3},\textbf{1})$ and 2 copies of the $(\textbf{1},\textbf{3})$. By plugging \eqref{ET_Half_Max81} and \eqref{ET_Half_Max82} into \eqref{quadratic constraints in 8D1}, \eqref{quadratic constraints in 8D2} and \eqref{quadratic constraints in 8D3}, one finds
\bea 
\epsilon ^{\alpha\beta}\,\epsilon^{ij}\,b_{\alpha i}\,b_{\beta
j}&=&0\ ,\qquad\qquad
\left(\textbf{1},\textbf{1}\right)\label{QC_Half_Tot_81}\\
\epsilon ^{\alpha\beta}\,\epsilon^{ij}\,a_{\alpha i}\,b_{\beta
j}&=&0\ ,\qquad\qquad
\left(\textbf{1},\textbf{1}\right)\label{QC_Half_Tot_82}\\
\epsilon ^{\alpha\beta}\,\epsilon^{ij}\,a_{\alpha i}\,a_{\beta
j}&=&0\ ,\qquad\qquad
\left(\textbf{1},\textbf{1}\right)\label{QC_Half_Tot_83}\\
\epsilon^{ij}\,a_{(\alpha i}\,b_{\beta) j}&=&0\ ,\qquad\qquad
\left(\textbf{3},\textbf{1}\right)\label{QC_Half_Tot_84}\\
\epsilon ^{\alpha\beta}\,a_{\alpha (i}\,b_{\beta j)}&=&0\
.\qquad\qquad
\left(\textbf{1},\textbf{3}\right)\label{QC_Half_Tot_85} 
\eea
With respect to what we expected from group theory, we seem to be finding a $(\textbf{1},\textbf{3})$ less amongst the even QC. This could be due to the fact that $\mathbb{Z}_{2}$ even QC can be sourced by quadratic expressions in the odd embedding tensor components that we truncated away. After the procedure of turning off all of them, the two $(\textbf{1},\textbf{3})$'s probably collapse to the same constraint or one of them vanishes directly.

The above set of QC characterises the consistent gaugings of the half-maximal theory which are liftable to the maximal theory, and hence they are more restrictive than the pure consistency requirements of the half-maximal theory. In order to single out only these we need to write down the expression of the gauge generators and impose the closure of the algebra. The gauge generators in the $(\textbf{2},\textbf{2})$ read
\be 
\label{Gen_Half_max8} 
{\left(X_{\alpha i}\right)_{\beta
j}}^{\gamma k} =
\frac{1}{2}\,\delta^{\gamma}_{\beta}\,\epsilon_{ij}\,\epsilon^{kl}\,a_{\alpha
l} \,+\, \delta^{\gamma}_{\alpha}\,\delta^{k}_{j}\,b_{\beta i} \,-\,
\frac{3}{2}\,\delta^{\gamma}_{\beta}\,\delta^{k}_{i}\,b_{\alpha j}
\,+\,
\frac{1}{2}\,\delta^{\gamma}_{\beta}\,\delta^{k}_{j}\,b_{\alpha i}
\,+\, \epsilon_{\alpha \beta}\,\epsilon^{\gamma
\delta}\,\delta^{k}_{j}\,b_{\delta i}\ . 
\ee
The closure of the algebra generated by \eqref{Gen_Half_max8} implies the following QC
\bea 
\epsilon ^{\alpha\beta}\,\epsilon^{ij}\,\left(a_{\alpha
i}\,a_{\beta j}\,-\,b_{\alpha i}\,b_{\beta j}\right)&=&0\
,\qquad\qquad
\left(\textbf{1},\textbf{1}\right)\label{QC_Half_81}\\
\epsilon ^{\alpha\beta}\,\epsilon^{ij}\,\left(a_{\alpha i}\,b_{\beta
j}\,+\,b_{\alpha i}\,b_{\beta j}\right)&=&0\ ,\qquad\qquad
\left(\textbf{1},\textbf{1}\right)\label{QC_Half_82}\\
\epsilon^{ij}\,a_{(\alpha i}\,b_{\beta) j}&=&0\ ,\qquad\qquad
\left(\textbf{3},\textbf{1}\right)\label{QC_Half_83}\\
\epsilon ^{\alpha\beta}\,a_{\alpha (i}\,b_{\beta j)}&=&0\
.\qquad\qquad \left(\textbf{1},\textbf{3}\right)\label{QC_Half_84}
\eea

To facilitate the mapping of gaugings $a_{\alpha i}$ and $b_{\alpha i}$ with the more familiar $f_{ABC}$ and $\xi_A$ in the DFT language, we have written a special section in the appendix~\ref{Appendix_A_'tHooft}. The mapping is explicitly given in \eqref{X2f_D=8}.

\subsubsection*{The O($2,2$) orbits of solutions to the QC}

After solving the QC given in \eqref{QC_Half_81}, \eqref{QC_Half_82}, \eqref{QC_Half_83} and \eqref{QC_Half_84} again with the aid of \textsc{\,Singular\,}, we find a 1-parameter family of T-duality orbits plus two discrete ones. The results are all collected in table~\ref{orbits_half_max8}.

\begin{table}[h!]
\begin{center}
\scalebox{1}[1]{
\begin{tabular}{| c | c | c | c |}
\hline
\textrm{ID} & $a_{\alpha i}$ & $b_{\alpha i}$ & gauging \\[1mm]
\hline \hline
$1$ & diag($\,\cos\alpha,0$) & diag($\,\sin\alpha,0$) &  Solv$_{2}\,\times\,$SO($1,1$) \\[1mm]
\hline \hline
$2$ & diag($1,1$) & diag($-1,-1$) &  \multirow{2}{*}{SL$(2)\,\times\,$SO($1,1$)} \\[1mm]
\cline{1-3}$3$ & diag($1,-1$) & diag($-1,1$) &  \\[1mm]
\hline
\end{tabular}
}
\end{center}
\caption{{\it All the T-duality orbits of consistent gaugings in half-maximal supergravity in $D=8$. For each of them, the simplest representative is given. Solv$_{2}$ refers again to the solvable subgroup of SL($2$) as already explained in the caption of table~\ref{orbits_max8}.} \label{orbits_half_max8}}
\end{table}

\subsubsection*{Higher-dimensional geometric origin}

The possible higher-dimensional origin of the three different orbits is as follows:

\begin{itemize}

\item \textbf{Orbit 1:} This orbit can be obtained by performing a two-step reduction of type I supergravity. In the first step, by reducing a circle, we can generate an $\mathbb{R}^{+}\,\times\,$SO($1,1$) gauging of half-maximal $D=9$ supergravity. Subsequently, we reduce such a theory again on a circle with the inclusion of a new twist commuting with the previous deformation. Also, these orbits include a non-trivial $\xi_A$ gauging, so we will not address it from a DFT perspective.

\item \textbf{Orbits 2 -- 3:} These do not seem to have any obvious geometric higher-dimensional origin in supergravity. In fact, they do not
 satisfy the extra constraints \eqref{Extra_f}, so one can only hope to reproduce them from truly doubled twist orbits in DFT.

\end{itemize}
Therefore we find that, while the half-maximal orbits in $D = 9$ all have a known geometric higher-dimensional origin, this is not the case for the latter two orbits in $D = 8$. We have finally detected the first signals of non-geometric orbits.

\subsubsection*{Higher-dimensional DFT origin}

As mentioned, the {\bf orbits 2} and {\bf 3} lack of a clear higher-dimensional origin. Here we would like to provide a particular twist matrix giving rise to these gaugings. We chose to start in the cartesian framework, and propose the following form for the SO$(2,2)$ twist matrix 
\be 
U = \begin{pmatrix} 1& 0 & 0 & 0 \\ 0
&
\cosh (m\,y^1 + n \, \tilde y_1) & 0 & \sinh (m\,y^1 + n \, \tilde y_1) \\
0& 0& 1& 0\\ 0 & \sinh (m\,y^1 + n \, \tilde y_1) & 0 & \cosh (m\,y^1
+ n \, \tilde y_1) \end{pmatrix}\ .  \vspace{2mm}
\ee

\noindent This is in fact an element of $\textrm{SO}(1,1)$ lying in the directions ($\tilde y_2, y^2$), fibred over the double torus ($\tilde y_1 , y^1 $). Here, the coordinates are written in the cartesian formulation, so we must rotate this in order to make contact with the light-cone case.

For this twist matrix, the WC and SC in the light-cone formulation read $(m+n) (m-n) = 0$, while the QC are always satisfied. The gaugings are constant, and when written in terms of $a_{\alpha i}$ and $b_{\alpha i}$ we find 
\be 
a_{\alpha i} = - b_{\alpha i } ={\rm diag} \left(-\frac{m+n}{2\,\sqrt{2}} ,\
\frac{m-n}{2\,\sqrt{2}}\right)\ , 
\ee 
so {\bf orbit 2} is obtained by choosing $m = 0$, $n = -2\,\sqrt{2}$, and {\bf orbit 3} by choosing $m = - 2\,\sqrt{2}$, $n = 0$. Notice that in both cases the twist orbit is truly doubled, so we find the first example of an orbit of gaugings without a clear supergravity origin, that finds an uplift to DFT in a truly doubled background.

\subsection*{Orbits and origin of the $D=7$ half-maximal case}

\subsubsection*{Half-maximal $D=7$ gauged supergravity}

A subset of half-maximal gauged supergravities is obtained from the maximal theory introduced at the end of section~\ref{sec:U_Dualitites} by means of a $\mathbb{Z}_2$ truncation. Thus, we will in this section perform this truncation and carry out the orbit analysis in the half-maximal theory. As we already argued before, this case is not only simpler, but also much more insightful from the point of view of understanding T-duality in gauged supergravities and its relation to DFT.

The action of our $\mathbb{Z}_2$ breaks\footnote{The $\mathbb{Z}_2$ element with respect to which we are truncating is the following USp($4)\,=\,$SO($5$) element 
\be
\alpha\,=\,\left(\begin{array}{cc}\mathds{1}_{2} & 0\\ 0 &
-\mathds{1}_{2}\end{array}\right) \notag
\ee 
projecting out half of the supercharges.} SL($5$) into $\mathbb{R}^+\,\times\,$SL($4$). Its embedding inside SL($5$) is unique and it is such that the fundamental representation splits as follows
\be
\textbf{5}\,\,\longrightarrow\,\,\textbf{1}_{(+4)}\,\oplus\,\textbf{4}_{(-1)}\,.
\ee
After introducing the following notation for the indices in the $\mathbb{R}^+$ and in the SL($4$) directions
\be M\,\,\longrightarrow\,\,(\,\diamond\,,\,m)\,, \ee
we assign an even parity to the $\diamond$ direction and odd parity to $m$ directions.

The embedding tensor of the maximal theory splits according to
\bea
\textbf{15}&\longrightarrow & \textbf{1}\,\oplus\,\xcancel{\textbf{4}}\,\oplus\,\textbf{10}\,,\\[2mm]
\textbf{40}^\prime&\longrightarrow &\xcancel{\textbf{4}^\prime}\,\oplus\,\textbf{6}\,\oplus\,\textbf{10}^\prime\,\oplus\,\xcancel{\textbf{20}}\,,
\eea
where again, as in the $D=8$ case, all the crossed irrep's are projected out because of $\mathbb{Z}_2$ parity. This implies that the embedding tensor of the half-maximal theory lives in the $\textbf{1}\,\oplus\,\textbf{6}\,\oplus\,\textbf{10}\,\oplus\,\textbf{10}^\prime$
and hence it is described by the following objects
\be
\theta\,\,,\,\,\xi_{[mn]}\,\,,\,\,M_{(mn)}\,\,,\,\,\tilde{M}^{(mn)}\,.
\label{Theta_half} \ee
This set of deformations agrees with the decomposition $\textrm{D}_8^{+++}\,\rightarrow\,\textrm{A}_3\,\times\,\textrm{A}_6$ given in ref.~\cite{Bergshoeff:2007vb}. The objects in \eqref{Theta_half} are embedded in $Y$ and $Z$ in the following way
\bea
Y_{\diamond\,\diamond}&=&\theta\,, \label{ExprY1}\\[2mm]
Y_{mn}&=&\frac{1}{2}\,M_{mn}\,,\label{ExprY2}\\[3mm]
Z^{mn,\,\diamond}&=&\frac{1}{8}\,\xi^{mn}\,,\label{ExprZ1}\\[2mm]
Z^{m\,\diamond,n}&=&-Z^{\diamond\,m,n}\,=\,\frac{1}{16}\,\tilde{M}^{mn}\,+\,\frac{1}{16}\,\xi^{mn}\,,\label{ExprZ2}
\eea
where for convenience we defined $\xi^{mn}\,=\, \frac{1}{2}\,\epsilon^{mnpq}\,\xi_{pq}$.

Now we will obtain the expression of the gauge generators of the half-maximal theory by plugging the expressions \eqref{ExprY1} -- \eqref{ExprZ2} into \eqref{gen_max}. We find
\be
{\left(X_{mn}\right)_p}^q\,=\,\frac{1}{2}\,\delta_{[m}^q\,M_{n]p}\,-\,\frac{1}{4}\,\epsilon_{mnpr}\,\left(\tilde{M}\,+\,
\xi\right)^{rq}\, \,, \label{gen_half-max}\ee
which extends the expression given in  ref.~\cite{Roest:2009tt} by adding an antisymmetric part to $\tilde{M}$ proportional to $\xi$. Note that the $\xi$ term is also the only one responsible for the trace of the gauge generators which has to be non-vanishing in order to account for $\mathbb{R}^+$ gaugings.

The presence of such a term in the expression \eqref{gen_half-max} has another consequence: the associated structure constants that one writes by expressing the generators in the $\textbf{6}$ ${\left(X_{mn}\right)_{pq}}^{rs}$ will not be automatically antisymmetric in the exchange between $mn$ and $pq$. This implies the necessity of imposing the antisymmetry by means of some extra QC\footnote{The QC which
ensure the antisymmetry of the gauge brackets are given by \\ ${\left(X_{mn}\right)_{pq}}^{rs}\,X_{rs}\,+\,(mn\,\leftrightarrow\,pq)\,=\,0$,
where $X$ is given in an arbitrary representation.}.

The QC of the maximal theory are branched into
\bea
\textbf{5}^\prime&\longrightarrow & \textbf{1}\,\oplus\,\xcancel{\textbf{4}^\prime}\,,\\[2mm]
\textbf{45}^\prime&\longrightarrow &\xcancel{\textbf{4}}\,\oplus\,\textbf{6}\,\oplus\,\textbf{15}\,\oplus\,\xcancel{\textbf{20}}\,,\\[2mm]
\textbf{70}^\prime&\longrightarrow &\textbf{1}\,\oplus\,\xcancel{\textbf{4}}\,\oplus\,\xcancel{\textbf{4}^\prime}\,\oplus\,\textbf{10}^\prime\,\oplus\,\textbf{15}\,\oplus\,\xcancel{\textbf{36}^\prime}\,.
\eea
By substituting the expressions \eqref{ExprY1} -- \eqref{ExprZ2} into the QC \eqref{QC_max7}, one finds
\bea
\theta\,\xi_{mn}&=&0\,,\qquad \,\,\,\,\,\,(\textbf{6}) \label{theta_xi}\\[2mm]
\left(\tilde{M}^{mp}\,+\,  \xi^{mp}\right)\, M_{pq}&=&0\,,\qquad ( \textbf{1}\,\oplus\,\textbf{15})\label{Q_Qtilde}\\[2mm]
M_{mp}\,\xi^{pn}\,-\,\xi_{mp}\,\left(\tilde{M}^{pn}\,+\,\xi^{pn}\right)&=&0\,,\qquad ( \textbf{1}\,\oplus\,\textbf{15})\label{Q_xi}\\[2mm]
\theta\,\tilde{M}^{mn}&=&0\,.\qquad \,\,\,\,\,(\textbf{10}^\prime)
\label{theta_Qtilde} \eea
Based on the Kac-Moody analysis performed in ref.~\cite{Bergshoeff:2007vb}, the QC constraints of the half-maximal theory should only impose conditions living in the $\textbf{1}\,\oplus\,\textbf{6}\,\oplus\,\textbf{15}\,\oplus\,\textbf{15}$. The problem is then determining which constraint in the $\textbf{1}$ is already required by the half-maximal theory and which is not.

By looking more carefully at the constraints \eqref{theta_xi} -- \eqref{theta_Qtilde}, we realise that the traceless part of \eqref{Q_Qtilde} exactly corresponds to the Jacobi identities that one gets from the closure of the algebra spanned by the generators \eqref{gen_half-max}, whereas the full \eqref{Q_xi} has to be imposed to ensure antisymmetry of the gauge brackets. Since there is only one constraint in the  $\textbf{6}$, we do not have ambiguities there\footnote{We would like to stress that the parameter $\theta$ within the half-maximal theory is a consistent deformation, but it does not correspond to any gauging and hence QC involving it cannot be derived as Jacobi identities or other consistency constraints coming from the gauge algebra.}.

We are now able to write down the set of QC of the half-maximal theory:
\bea
\theta\,\xi_{mn}&=&0\,,\qquad \,(\textbf{6}) \label{QC1}\\[2mm]
\left(\tilde{M}^{mp}\,+\,  \xi^{mp}\right)\, M_{pq}\,-\,\frac{1}{4}\,\left(\tilde{M}^{np}\,M_{np}\right)\,\delta_q^m&=&0\,,\qquad (\textbf{15})\label{QC2}\\[2mm]
M_{mp}\,\xi^{pn}\,+\,\xi_{mp}\,\tilde{M}^{pn}&=&0\,,\qquad (\textbf{15})\label{QC3}\\[2mm]
\epsilon^{mnpq}\,\xi_{mn}\,\xi_{pq}&=&0\,.\qquad \,\,\,(
\textbf{1})\label{QC4} \eea
We are not really able to confirm whether (\ref{QC1}) is part of the QC of the half-maximal theory, in the sense that there appears a top-form in the \textbf{6} from the $\textrm{D}_{8}^{+++}$ decomposition but it could either be a tadpole or a QC. This will however not affect our further discussion, in that we only consider orbits of gaugings in which $\theta=0$. The extra QC required in order for the gauging to admit an uplift to maximal supergravity are
\bea
\tilde{M}^{mn}\,M_{mn}&=&0\,,\qquad \,(\textbf{1}) \label{extra1}\\[2mm]
\theta\,\tilde{M}^{mn}&=&0\,.\qquad (\textbf{10}^\prime)
\label{extra2} \eea

\subsubsection*{The O($3,3$) orbits of solutions to the QC in the $\textbf{10}\,\oplus\,\textbf{10}^\prime$}

The aim of this section is to solve the constraints summarised in \eqref{QC1}, \eqref{QC2}, \eqref{QC3} and \eqref{QC4}. We will start by considering the case of gaugings only involving the $\textbf{10}\,\oplus\,\textbf{10}^\prime$. This restriction is motivated by flux compactification, as we will try to argue later on.

The only non-trivial QC are the following
\be
\tilde{M}^{mp}\,M_{pn}-\frac{1}{4}\left(\tilde{M}^{pq}\,M_{pq}\right)\,\delta^m_n\,=\,0\ ,
\label{QCQQtilde}\ee
which basically implies that the matrix product between $M$ and $\tilde{M}$, which in principle lives in the $\textbf{1}\,\oplus\,\textbf{15}$, has to be pure trace. We made use of a GL($4$) transformation in order to reduce $M$ to pure signature; as a consequence, the QC \eqref{QCQQtilde} imply that $\tilde{M}$ is diagonal as well \cite{Dibitetto:2010rg}. This results in a set of eleven 1-parameter orbits\footnote{We would like
to point out that the extra discrete generator $\eta$ of O($3,3$) makes sure that, given a certain gauging with $M$ and $\tilde{M}$, it lies in the same orbit as its partner with the role of $M$ and $-\tilde{M}$ interchanged.} of solutions to the QC which are given in table~\ref{orbits_halfmax7}.

\begin{table}[h!]
\begin{center}
\scalebox{.95}[1]{
\begin{tabular}{| c | c | c | c | c |}
\hline
\textrm{ID} & $M_{mn}/\,\cos\alpha\,$ & $\tilde{M}^{mn}/\,\sin\alpha\,$ & range of $\alpha$ & gauging \\[1mm]
\hline \hline
$1$ & diag($1,1,1,1$) & diag($1,1,1,1$) & $-\frac{\pi}{4}\,<\,\alpha\,\le\,\frac{\pi}{4}$ & $\left\{\begin{array}{cc}\textrm{SO}($4$)\ , & \alpha\,\ne\,\frac{\pi}{4}\ ,\\ \textrm{SO}(3)\ , & \alpha\,=\,\frac{\pi}{4}\ .\end{array}\right.$\\[4mm]
\hline
$2$ & diag($1,1,1,-1$) & diag($1,1,1,-1$) & $-\frac{\pi}{4}\,<\,\alpha\,\le\,\frac{\pi}{4}$ & SO($3,1$)\\[1mm]
\hline
$3$ & diag($1,1,-1,-1$) & diag($1,1,-1,-1$) & $-\frac{\pi}{4}\,<\,\alpha\,\le\,\frac{\pi}{4}$ & $\left\{\begin{array}{cc}\textrm{SO}($2,2$)\ , & \alpha\,\ne\,\frac{\pi}{4}\ ,\\ \textrm{SO}(2,1)\ , & \alpha\,=\,\frac{\pi}{4}\ .\end{array}\right.$\\[2mm]
\hline \hline
$4$ & diag($1,1,1,0$) & diag($0,0,0,1$) & $-\frac{\pi}{2}\,<\,\alpha\,<\,\frac{\pi}{2}$ & ISO($3$)\\[1mm]
\hline
$5$ & diag($1,1,-1,0$) & diag($0,0,0,1$) & $-\frac{\pi}{2}\,<\,\alpha\,<\,\frac{\pi}{2}$ & ISO($2,1$)\\[1mm]
\hline \hline
$6$ & diag($1,1,0,0$) & diag($0,0,1,1$) & $-\frac{\pi}{4}\,<\,\alpha\,\le\,\frac{\pi}{4}$ & $\left\{\begin{array}{cc}\textrm{CSO}(2,0,2)\ , & \alpha\,\ne\,\frac{\pi}{4}\ ,\\ \mathfrak{f}_{1}\quad(\textrm{Solv}_{6}) \ , & \alpha\,=\,\frac{\pi}{4}\ .\end{array}\right.$\\[4mm]
\hline
$7$ & diag($1,1,0,0$) & diag($0,0,1,-1$) & $-\frac{\pi}{2}\,<\,\alpha\,<\,\frac{\pi}{2}$ & $\left\{\begin{array}{cc}\textrm{CSO}(2,0,2)\ , & |\alpha|\,<\,\frac{\pi}{4}\ ,\\ \textrm{CSO}(1,1,2)\ , & |\alpha|\,>\,\frac{\pi}{4}\ ,\\ \mathfrak{g}_{0}\quad(\textrm{Solv}_{6}) \ , & |\alpha|\,=\,\frac{\pi}{4}\ .\end{array}\right.$\\[4mm]
\hline
$8$ & diag($1,1,0,0$) & diag($0,0,0,1$) & $-\frac{\pi}{2}\,<\,\alpha\,<\,\frac{\pi}{2}$ & $\mathfrak{h}_{1}\quad(\textrm{Solv}_{6})$\\[1mm]
\hline
$9$ & diag($1,-1,0,0$) & diag($0,0,1,-1$) & $-\frac{\pi}{4}\,<\,\alpha\,\le\,\frac{\pi}{4}$ & $\left\{\begin{array}{cc}\textrm{CSO}(1,1,2)\ , & \alpha\,\ne\,\frac{\pi}{4}\ ,\\ \mathfrak{f}_{2}\quad(\textrm{Solv}_{6}) \ , & \alpha\,=\,\frac{\pi}{4}\ .\end{array}\right.$\\[4mm]
\hline
$10$ & diag($1,-1,0,0$) & diag($0,0,0,1$) & $-\frac{\pi}{2}\,<\,\alpha\,<\,\frac{\pi}{2}$ & $\mathfrak{h}_{2}\quad(\textrm{Solv}_{6})$\\[1mm]
\hline \hline $11$ & diag($1,0,0,0$) & diag($0,0,0,1$) &
$-\frac{\pi}{4}\,<\,\alpha\,\le\,\frac{\pi}{4}$ &
$\left\{\begin{array}{cc}\mathfrak{l}\quad(\textrm{Nil}_{6}(3)\,)\ , & \alpha\,\ne\,0\ ,\\
\textrm{CSO}(1,0,3)\ , &
\alpha\,=\,0\ .\end{array}\right.$\\[4mm]
\hline
\end{tabular}
}
\end{center}
\caption{{\it All the T-duality orbits of consistent gaugings in half-maximal supergravity in $D=7$. Any value of $\,\alpha\,$ parameterises inequivalent orbits. More details about the non-semisimple gauge algebras $\mathfrak{f}_{1}$, $\mathfrak{f}_{2}$, $\mathfrak{h}_{1}$, $\mathfrak{h}_{2}$, $\mathfrak{g}_{0}$ and $\mathfrak{l}$ are given in appendix~\ref{Appendix_A_Gaugings}.} \label{orbits_halfmax7}}
\end{table}

As we will see later, some of these consistent gaugings in general include non-zero non-geometric fluxes, but at least in some of these cases one will be able to dualise the given configuration to a perfectly geometric background.

\subsubsection*{Higher-dimensional geometric origin}

Ten-dimensional heterotic string theory compactified on a $T^3$ gives rise to a half-maximal supergravity in $D=7$ where the SL($4$)$\,=\,$SO($3,3$) factor in the global symmetry of this theory can be interpreted as the T-duality group. The set of generalised fluxes which can be turned on here is given by (see decomposition in \eqref{Het_Fluxes})
\be 
\left\{f_{abc},\,{f_{ab}}^c,\,{f_a}^{bc},\,f^{abc}\right\}\equiv \left\{H_{abc},\,{\omega_{ab}}^c,\,{Q_a}^{bc},\,R^{abc}\right\}\
, \label{Fluxes_Het_DFT}
\ee
where $a,b,c\,=\,1,2,3$.

These are exactly the objects that one obtains by decomposing a three-form of SO($3,3$) with respect to its GL($3$) subgroup. The number of independent components of the above fluxes (including traces of $\omega$ and $Q$) amounts to $1+9+9+1\,=\,20$, which is the number of independent components of a three-form of SO($3,3$).
Nevertheless, the three-form representation is not irreducible since the Hodge duality operator in 3+3 dimensions squares to 1. This implies that one can always decompose it in a self-dual (SD) and anti-self-dual (ASD) part
\be 
\textbf{10}\,\oplus\,\textbf{10}^\prime\quad\textrm{of
SL}(4)\quad\longleftrightarrow\quad\textbf{10}_{\textrm{SD}}\,\oplus\,\textbf{10}_{\textrm{ASD}}\quad\textrm{of
SO}(3,3)\ , 
\ee
such that the matching between the embedding tensor deformations $(M_{mn},\,\tilde{M}^{mn})$ and the generalised fluxes given in \eqref{Fluxes_Het_DFT} now perfectly works. The explicit mapping between vectors of SO($3,3$) expressed in light-cone coordinates and two-forms of SL($4$) can be worked out by means of the SO($3,3$) 't Hooft symbols $\left[G_A\right]^{mn}$ (see appendix~\ref{Appendix_A_'tHooft}). This gives rise to the following dictionary between the $M$ and $\tilde{M}$-components and the fluxes given in \eqref{Fluxes_Het_DFT}
\be
M\,=\,\textrm{diag}\,\left(H_{123},\,{Q_1}^{23},\,{Q_2}^{31},\,{Q_3}^{12}\right)\
,\quad
\tilde{M}\,=\,\textrm{diag}\,\left(R^{123},\,{\omega_{23}}^1,\,{\omega_{31}}^2,\,{\omega_{12}}^3\right)\
.\label{dictionary}
\ee

The QC given in equations \eqref{QC1}-\eqref{QC4} enjoy a symmetry in the exchange
\be 
(M,\,\xi)\,\overset{\eta}{\leftrightarrow}\,(-\tilde{M},\,-\xi)\
. \label{triple_duality}
\ee
The discrete $\mathbb{Z}_2$ transformation $\eta$ corresponds to the following O($3,3$) element with determinant $-1$
\be 
\eta\,=\,\left(
\begin{array}{cc}
0 & \mathds{1}_3\\
\mathds{1}_3 & 0
\end{array}\right)\ ,
\label{eta}
\ee
which can be interpreted as a triple T-duality exchanging the three compact coordinates $y^{a}$ with the corresponding winding coordinates $\tilde{y}_{a}$ in the language of DFT.

Now we have all the elements to analyze the higher-dimensional origin of the orbits classified in table~\ref{orbits_halfmax7}.

\begin{itemize}

\item \textbf{Orbits 1 -- 3:}  These gaugings are non-geometric for every $\alpha\ne 0$; for $\alpha =0$, they correspond to coset
reductions of heterotic string theory. See \emph{e.g.} the $S^{3}$ compactification in ref.~\cite{Cvetic:2000dm} giving rise to the SO($4$) gauging. This theory was previously obtained in ref.~\cite{Salam:1983fa} as $\mathcal{N}=2$ truncation of a maximal supergravity in $D=7$.

\item \textbf{Orbits 4 -- 5:} For any value of $\alpha$ we can always dualise these representatives to the one obtained by means of a
twisted $T^{3}$ reduction with $H$ and $\omega$ fluxes.

\item \textbf{Orbits 6 -- 7:} For any $\alpha\ne 0$ these orbits could be obtained from supergravity compactifications on locally-geometric T-folds, whereas for $\alpha=0$ it falls again in a special case of the reductions described for orbits 4 and 5.

\item \textbf{Orbits 8 -- 11:} For any value of $\alpha$, these orbits always contain a geometric representative involving less general $H$ and $\omega$ fluxes.

\end{itemize}

To summarise, in the half-maximal $D=7$ case, we encounter a number of orbits which do not have an obvious higher-dimensional origin. To be more precise, these are \textbf{orbits 1, 2} and \textbf{3} for $\alpha\ne 0$. The challenge in the next subsection will be to establish what DFT can do for us in order to give these orbits a higher-dimensional origin. Again, before reading the following subsections we refer to section~\ref{DFT_twists} for a discussion of what we mean by light-cone and cartesian formulations.

\subsubsection*{Higher-dimensional DFT origin}

First of all we would like to show here how to capture the gaugings that only involve (up to duality rotations) fluxes $H_{abc}$ and ${\omega_{ab}}^{c}$. For this, we start from the light-cone formulation, and propose the following Ansatz for a {\it globally geometric twist} (involving $e$ and $B$ and physical coordinates $y$)
\bea
e &=& \begin{pmatrix}1 & 0& \frac{\omega_1}{\omega_3} \sin (\omega_1\,\omega_3\,y^2) \\ 0 & \cos(\omega_2\,\omega_3\,y^1) & -\frac{\omega_2}{\omega_3} \cos (\omega_1\,\omega_3\,y^2) \sin(\omega_2\,\omega_3\,y^1) \\ 0 & \frac{\omega_3}{\omega_2} \sin(\omega_2\,\omega_3\,y^1) & \cos (\omega_1\,\omega_3\,y^2) \cos(\omega_2\,\omega_3\,y^1)\end{pmatrix}\ ,\\
B&=& \begin{pmatrix}0& 0 & 0\\ 0& 0& H\,y^1\,\cos
(\omega_1\,\omega_3\,y^2) \\ 0 & -H\,y^1\,\cos
(\omega_1\,\omega_3\,y^2)& 0\end{pmatrix} \ , \\
\lambda &=& - \frac 12 \log(\cos (\omega_1\omega_3 y^2)) \ .
\eea
This is far from being the most general ansatz, but it serves our purposes of reaching a large family of geometric orbits. The parameters $\omega_i$ can be real, vanishing or imaginary, since $U$ is real and well-behaved in these cases. The QC, WC and SC are all automatically satisfied, and the gaugings read
\be 
M = {\rm diag} (H\ ,\ 0\ ,\ 0\ ,\ 0) \ ,\ \ \ \ \tilde M = {\rm
diag} (0\ , \ \omega_1^2\ ,\ \omega_2^2\ ,\ \omega_3^2)\ . 
\ee 
From here, by choosing appropriate values of the parameters the {\bf orbits 4, 5, 8, 10} and {\bf 11} can be obtained. Indeed these are geometric as they only involve gauge and (geo)metric fluxes.

Secondly, in order to address the remaining orbits, we consider an SO($2,2$) twist $U_{4}$ embedded in O($3,3$) in the following way
\be
U =
\begin{pmatrix}1 & 0 & 0 & 0\\ 0& A & 0 & B\\ 0& 0 & 1 & 0 \\ 0& C &
0 & D
\end{pmatrix}\ , \ \ \ \ \ \ U_{4} = \begin{pmatrix}A& B \\ C &
D\end{pmatrix}\ ,  \ \ \ \ \ \ \lambda =0\ . 
\ee
This situation is analog to the SO$(1,1)$ twist considered in the $D = 8$ case, but with a more general twist. Working in the cartesian formulation, one can define the generators and elements of SO$(2,2)$ as 
\be [t_{IJ}]_K{}^L = \delta^L_{[I}
\eta_{J]K}\ , \ \ \ \ \  U_4 = \exp\left(t_{IJ} \phi^{IJ}\right)\ ,
\ee 
where the rotations are generated by $t_{12}$ and $t_{34}$, and the boosts by the other generators. Also, we take $\phi^{IJ} = \alpha^{IJ} y^1 + \beta^{IJ} \tilde y_1$ to be linear.

From the above $\textrm{SO}(2,2)$ duality element one can reproduce the following orbits employing a {\it locally geometric twist} (including $e$, $B$ and $\beta$ but only depending on $y$, usually referred to as a T-fold):
~
\begin{itemize}

\item {\bf Orbit 6}  can be obtained by taking
\be 
{\bf (6)} \ \ \ \alpha^{12} = - \beta^{12} = - \frac
1{\sqrt{2}}\,(\cos \alpha + \sin \alpha)   \ , \ \ \ \alpha^{34} =
-\beta^{34}= - \frac 1{\sqrt{2}}\,(\cos \alpha + \sin \alpha)\
.\nn
\ee
and all other vanishing.

\item \textbf{Orbits 7} and {\textbf 9}  can be obtained by the following particular identifications
\be
\begin{array}{lclclc}
\phi^{14} = \phi^{23} & , & \phi^{12} = \phi^{34} & \textrm{and} &
\phi^{13} = \phi^{24} & .\end{array}\nn\ee
\be {\bf (7)} \ \ \ \alpha^{14} = - \beta^{14} = -
\frac{1}{\sqrt{2}}\,\sin \alpha \ , \ \ \ \alpha^{12} = - \beta^{12}
= - \frac{1}{\sqrt{2}}\,\cos \alpha \ , \ \ \ \alpha^{13} =
\beta^{13} =0 \ ,\nn\ee
\be {\bf (9)} \ \ \ \alpha^{14} = - \beta^{14} = -
\frac{1}{\sqrt{2}}\,\sin \alpha \ , \ \ \ \alpha^{12} = \beta^{12} =
0 \ , \ \ \ \alpha^{13} = \beta^{13} = - \frac{1}{\sqrt{2}}\,\cos
\alpha \ .\nn\ee
\end{itemize}
All these backgrounds satisfy both the WC and the SC  and hence they admit a locally geometric description. This is in agreement with the fact that the simplest representative of \textbf{orbits 6, 7} and {\textbf 9} given in table~\ref{orbits_halfmax7} contains $H$, $\omega$ and $Q$ fluxes but no $R$ flux.

Finally, one can employ the same SO($2,2$) duality elements with different identifications to generate the remaining orbits with a {\it non-geometric twist} (involving both $y$ and $\tilde{y}$ coordinates): ~
\begin{itemize}

\item \textbf{Orbits 1, 3}  can be again obtained by considering an SO($2)\,\times\,$SO($2$) twist  with arbitrary $\phi^{12}$ and $\phi^{34}$:
\be 
{\bf (1)} \ \ \ \alpha^{12} = - 2\,\sqrt{2}\,(\cos \alpha + \sin
\alpha) \ , \ \ \ \beta^{34} = 2\,\sqrt{2}\,(\cos \alpha - \sin
\alpha)\ , \ \ \ \alpha^{34} = \beta^{12} = 0 \ ,\nn
\ee
\be 
{\bf (3)} \ \ \ \alpha^{34} = - 2\,\sqrt{2}\,(\cos \alpha + \sin
\alpha) \ , \ \ \ \beta^{12} = 2\,\sqrt{2}\,(\cos \alpha - \sin
\alpha)\ , \ \ \ \alpha^{12} = \beta^{34} = 0 \ .\nn
\ee
\item \textbf{Orbit 2} can be obtained by means of a different SO($2,2$) twist built out of the two rotations and two boosts subject to the following identification
\be
\begin{array}{lcl}
\phi^{14} = \phi^{23} & \textrm{, } & \phi^{12} = \phi^{34}\ .
\end{array}
\ee
\be 
{\bf (2)} \ \ \ \alpha^{14} = \beta^{12} = \frac
1{\sqrt{2}}\,(\cos \alpha - \sin \alpha) \ , \ \ \ \alpha^{12} = -
\beta^{14} = - \frac 1{\sqrt{2}}\,(\cos \alpha + \sin \alpha)\ .
\nn
\ee
\end{itemize}
These backgrounds violate both the WC and the SC for $\alpha \neq 0$. This implies that these backgrounds are truly doubled and they do not even admit a locally geometric description.

Finally, let us also give an example of degeneracy in twist orbits-space reproducing the same orbit of gaugings. The following twist
\be
\phi^{12} = \phi^{13} \ , \ \ \ \phi^{34} = \phi^{24} \ , \ \ \ \phi^{23} = \phi^{14} = 0
\ee
\be {\bf (6)} \ \ \ \alpha^{13} = - \frac{1}{\sqrt{2}}(\cos \alpha + \sin
\alpha) \ , \ \ \ \beta^{24} = \frac{1}{\sqrt{2}}(\cos \alpha - \sin
\alpha)\ , \ \ \ \alpha^{24} = \beta^{13} = 0 \ ,\nn\ee
also reproduces the {\bf orbit 6}, but in this case through a non-geometric twist. What happens in this case is that although the twist matrix does not satisfy the WC/SC, the contractions in (\ref{Max_VS_Geom}) cancel.

\subsection*{Concluding Remarks}

In this chapter we have provided a litmus test to the notion of  non-geometry, by classifying the explicit orbits of consistent gaugings of different supergravity theories, and considering the possible higher-dimensional origins of these. The results turn out to be fundamentally different for the cases of U-duality orbits of maximal supergravities, and T-duality orbits of half-maximal theories.

In the former case we have managed to explicitly classify all U-duality orbits in dimensions $8 \leq D \leq 11$. This led to zero, one, four and ten discrete orbits in dimensions $D=11, 10,9$ and $8$, respectively, with different associated gauge groups. Remarkably, we have found that all of these orbits have a higher-dimensional origin via some geometric compactification, be it twisted reductions or compactifications on group manifolds or coset spaces. In our parlance, we have therefore found that all U-duality orbits are geometric. The structure of U-duality orbits is therefore dramatically different from the sketch of figure~\ref{pic:orbits} in the introduction. Although a full classification of all orbits in lower-dimensional cases becomes increasingly cumbersome, we are not aware of any examples that are known to be non-geometric. It could therefore hold in full generality that all U-duality orbits are necessarily geometric.

This is certainly not the case for T-duality orbits of gaugings of half-maximal supergravities. In this case, we have provided the explicit classification in dimensions $7 \leq D \leq 10$ (where in $D=7$ we have only included three-form fluxes). The numbers of distinct families of orbits in this case are zero, one, three and eleven in dimensions $D=10,\,9,\,8$ and $7$, respectively, which includes both discrete and one-parameter orbits. A number of these orbits do not have a higher-dimensional origin in terms of a geometric compactification. Such cases are {\bf orbits 2} and {\bf 3} in $D=8$ and {\bf orbits 1, 2} and {\bf 3} in $D=7$ for $\alpha\neq 0$. Indeed, these are exactly the orbits that do not admit an uplift to the maximal theory. As proven in section~\ref{DFT_twists}, all such orbits necessarily violate the WC and/or SC, and therefore need truly doubled backgrounds. Thus, the structure of T-duality orbits is very reminiscent of figure~\ref{pic:orbits} in the introduction. Given the complications that already arise in these simpler higher-dimensional variants, one can anticipate that the situation will be similar in four-dimensional half-maximal supergravity.

Fortunately, the formalism of DFT seems tailor-made to generate additional T-duality orbits of half-maximal supergravity. Building on the recent generalisation of the definition of DFT \cite{Grana:2012rr}, we have demonstrated that all T-duality orbits, including the non-geometric ones in $D=7,\,8$, can be generated by a twisted reduction of DFT. We have explicitly provided duality twists for all orbits. For locally-geometric orbits the twists only depend on the physical coordinates $y$, while for the non-geometric orbits these
necessarily also include $\tilde y$. Again, based on our exhaustive analysis in higher-dimensions, one could conjecture that also in lower-dimensional theories, all T-duality orbits follow from this generalised notion of DFT.

At this point we would like to stress once more that a given orbit of gaugings can be generated from different twist orbits. Therefore, there is a degeneracy in the space of twist orbits giving rise to a particular orbit of gaugings. Interestingly, as it is the case of {\bf orbit 6} in $D=7$  for instance, one might find two different twist orbits reproducing the same orbit of gaugings, one  violating WC and SC, the other one satisfying both. Our notion of a locally geometric orbit of gaugings is related to the existence of at least one undoubled background giving rise to it. However, this ambiguity seems to be peculiar of gaugings containing $Q$ flux. These can, in principle, be independently obtained by either adding a $\beta$ but no $\tilde{y}$ dependence (locally geometric choice, usually called T-fold), or by including non-trivial $\tilde{y}$ dependence but no $\beta$ (non-geometric choice) \cite{Aldazabal:2011nj}.

Another remarkable degeneracy occurs for the case of semi-simple gaugings, corresponding to {\bf orbits 1 -- 3} in $D=7$. For the special case of $\alpha = 0$, we have two possible ways of generating such orbits from higher-dimensions: either a coset reduction over a sphere or analytic continuations thereof, or a duality twist involving non-geometric coordinate dependence. Therefore $d$-dimensional coset reductions seem to be equivalent to $2d$-dimensional twisted torus reductions (with the latter in fact being more general, as it leads to all values of $\alpha$).
Considering the complications that generally arise in proving the consistency of coset reductions, this is a remarkable reformulation that would be interesting to understand in more detail. Furthermore, when extending the notion of DFT to type II and M-theory, this relation could also shed new light on the consistency of the notoriously difficult four-, five- and seven-sphere reductions of these theories.

Our results mainly focus on SS compactifications leading to gauged supergravities with vanishing $\xi_M$ fluxes. In addition, we have restricted to the NS-NS sector and ignored $\alpha'$-effects.
Also, we stress once again that relaxing the WC and SC is crucial in part of our analysis. If we kept the WC, typically the Jacobi identities would lead to backgrounds satisfying also the SC \cite{Grana:2012rr}.
However, from a purely (double) field theoretical analysis the WC is not necessary. A sigma model analysis beyond tori would help us to clarify the relation between DFT without the WC and SC and string field theory on more general backgrounds.

\chapter{Orientifold Compactifications}
\markboth{Orientifold Compactifications}{Orientifold Compactifications}
\label{Half_Max}
Many string theory constructions related to flux backgrounds compatible with minimal supersymmetry have been studied so far. In
particular, as we saw in chapter~\ref{Fluxes}, the mechanism of inducing an effective superpotential from fluxes has been extensively studied in the literature \cite{Giddings:2001yu,  Derendinger:2004jn, DeWolfe:2005uu, Camara:2005dc, Aldazabal:2006up,  Aldazabal:2007sn, Aldazabal:2008zza, Dall'Agata:2009gv, deCarlos:2009fq} for those compactifications giving rise to a so-called $STU$-model as low energy description. 
In this chapter we will firstly give an overview of the recent progress in understanding the link between half-maximally supersymmetric string backgrounds and gaugings of $\mathcal{N}=4$ supergravity \cite{Roest:2009dq, Dall'Agata:2009gv, Dibitetto:2010rg} and secondly show how this machinery can be exploited as a powerful tool for addressing the same issue in the context of $\mathcal{N}=4$ compactifications.

Another interesting opportunity offered by the study of such flux compactifications and their relation to half-maximal
supergravity, is that of addressing the issue of stability without supersymmetry in extended supergravity. More precisely, for a long time it was believed that there are no stable vacua of maximal or half-maximal supergravity that spontaneously break all supersymmetry. Recently \cite{Fischbacher:2010ec}, however, an example of an AdS critical point which is both non-supersymmetric and stable has been found in maximal supergravity. This adds further motivation to look for new such extrema in the half-maximal case as well. Furthermore, the possible existence of stable dS vacua in this context still remains an open discussion point \cite{Borghese:2010ei}. Most of the results presented in this chapter were first obtained in refs~\cite{Dibitetto:2010rg, Dibitetto:2011gm}. Some additional material related to this chapter can be found in appendix~\ref{appendix:Fluxes}.

\section{Gauged $\cN=4\,$, $D=4$ Supergravities}
\label{Review_N=4}

In this section we present a brief introduction to half-maximal ($\cN=4$) supergravity theories in four dimensions. We mostly follow the notation and conventions of ref.~\cite{Schon:2006kz} to work out the $\cN=4$ supergravity theory invariant under the action of the $G_{0}$ = SL($2$) $\times$ SO($6,6$) duality group (see table~\ref{table:half-max}) in four dimensions\footnote{We focus on the theory coupled to $n=6$ vector multiplets, which, as we will see in the next chapter, can be regarded as a $\mathbb{Z}_{2}$ truncation of $\cN=8$ supergravity. This relates it to orientifold reductions of type II string theory.}.

The $24$ vectors of the theory transform in the $(\textbf{2}, \textbf{12})$ of $G_{0}$. The SL($2$) factor inside the symmetry group is interpreted as electromagnetic duality. Beyond $\textrm{SL}(2)\,\times\,\textrm{SO}(6,6)$, one can embed the vector representation labelled by the indices $\a M$, inside the fundamental representation of Sp($24,\mathbb{R}$), where $\,\a=(+,-)\,$ is a fundamental SL($2$) index and $\,M=1,...,12\,$ is the SO($6,6$) fundamental index. Such symplectic transformations change the Lagrangian non-trivially such that the description that we give here would not be valid anymore. Nevertheless, Lagrangians in different symplectic frames describe the same theory at the level of the equation of motion.

The LC restricts the embedding tensor to the following irrep's (see table~\ref{table:half-max})
\be
\begin{array}{cccccc}
\Theta & \in & \underbrace{(\textbf{2}, \textbf{12})}_{\xi_{\a M}} & \oplus & \underbrace{(\textbf{2}, \textbf{220})}_{f_{\a MNP}} & .
\end{array}
\ee

\subsection*{Quadratic Constraints and Scalar Potential}

The scalars of the theory span the coset geometry
\be
\label{coset}
\frac{\textrm{SL}(2)}{\textrm{SO}(2)} \times \frac{\textrm{SO}(6,6)}{\textrm{SO}(6) \times \textrm{SO}(6)} \ .
\ee

We will name $\,M_{\alpha\beta}\,$ the scalars parameterising the first factor and $\,M_{MN}\,$ those ones parameterising the second factor in (\ref{coset}). For the former we will use the following explicit parameterisation
\be
\label{chi,phi}
M_{\alpha \beta} = e^{\phi}
\left(
\begin{array}{cc}
\chi^2 + e^{-2 \phi} & \chi \\
\chi & 1
\end{array}
\right) \hspace{5mm} , \hspace{5mm} \alpha = (+,-) \ ,
\ee
where the SL$(2)$ indices are raised and lowered using $\,\epsilon_{\alpha \beta} = \epsilon^{\alpha \beta}\,$ with $\,\epsilon^{+-} = -\epsilon^{-+} = 1$. The matrix $\,M_{MN}$, can be determined by starting from a 'vielbein' denoted by $\,\mathcal{V}_{M}^{\phantom{M}\underline{A}}$, where $\,\underline{A}\,$ is an SO($6$) $\times$ SO($6$) index whereas $\,M\,$ is an SO($6,6$) one. This object is such that
\be
\label{M=VV}
M=\mathcal{V} \, \mathcal{V}^{T} .
\ee
Global SO$(6,6)$ transformations act on $\,\mathcal{V}\,$ from the left, whereas local $\textrm{SO}(6) \times \textrm{SO}(6)$ transformations act from the right. Even though $\,\mathcal{V}\,$ is not by itself invariant under local $\textrm{SO}(6) \times \textrm{SO}(6)$ transformations, the particular combinations constructed out of it which will appear in the scalar potential are. In particular, the matrix $\,M\,$ itself is invariant.

The non-vanishing embedding tensor components $\xi_{\a M}$ and $\,f_{\a MNP}\,$ have to satisfy the following QC
\bea
\label{QC41}
&\hspace{-15mm} i)  & \xi_{\a M}\,\xi_{\b}^{\phantom{a}M}=0 \ , \\[2mm]
\label{QC42}
&\hspace{-15mm} ii) & \xi_{(\a}^{\phantom{a}P}\,f_{\b)PMN}=0 \ , \\[2mm]
\label{QC43}
&\hspace{-15mm} iii)& 3\,f_{\a R[MN}\,f_{\b PQ]}^{\phantom{abcde}R}\,+\,2\,\xi_{(\a [M}\,f_{\b)NPQ]}=0 \ , \\[2mm]
\label{QC44}
&\hspace{-15mm} iv) & \epsilon^{\a \b}\left(\xi_{\a}^{\phantom{a}P}\,f_{\b PMN}\,+\,\xi_{\a M}\,\xi_{\b N}\right)=0 \ , \\[2mm]
\label{QC45}
&\hspace{-15mm} v) & \epsilon^{\a \b}\left(f_{\a MNR}\,f_{\b PQ}^{\phantom{abcde}R}\,-\,\xi_{\a}^{\phantom{a}R}\,f_{\b R[M[P}\,\eta_{Q]N]}\,-\,\xi_{\a [M}\,f_{\b N]PQ}\,+\,\xi_{\a [P}\,f_{\b Q]MN}\right)=0 \ ,
\eea
which correspond with the following irrep's of the $\,\textrm{SL}(2)\,\times\,\textrm{SO}(6,6)\,$ symmetry of half-maximal supergravity,
\be
\label{irrepQC41}
i) \quad (\textbf{3},\textbf{1})
\hspace{10mm}  \hspace{10mm}
ii) \quad (\textbf{3},\textbf{66})
\hspace{10mm}  \hspace{10mm}
iii) \quad (\textbf{3},\textbf{495})
\ee
\be
\label{irrepQC42}
iv) \quad (\textbf{1},\textbf{66})
\hspace{10mm}  \hspace{10mm}
v) \quad (\textbf{1},\textbf{66})\,\oplus\,(\textbf{1},\textbf{2079}) \ .
\ee
The combination of supersymmetry and gaugings then induces the following scalar potential\footnote{We have set the gauge coupling constant to $\,g=\frac{1}{2}\,$ with respect to the conventions in ref.~\cite{Schon:2006kz}.}
\begin{align}
  V & = \dfrac{1}{64} \, f_{\alpha MNP} \, f_{\beta QRS} M^{\alpha
  \beta} \left[ \dfrac{1}{3} \, M^{MQ} \, M^{NR} \, M^{PS} + \left(\dfrac{2}{3} \, \eta^{MQ} - M^{MQ} \right) \eta^{NR}
  \eta^{PS} \right] + \notag \\
  & - \dfrac{1}{144} \, f_{\alpha MNP} \, f_{\beta QRS} \, \epsilon^{\alpha
  \beta} \, M^{MNPQRS}\, + \, \dfrac{3}{64} \, \xi_{\a M} \, \xi_{\b N} \, M^{\a\b} \, M^{MN} \ , \label{V}
\end{align}
where
\be
\label{MV6}
M_{MNPQRS} \equiv \epsilon_{\underline{mnpqrs}}\mathcal{V}_{M}^{\phantom{M}\underline{m}}\mathcal{V}_{N}^{\phantom{M}\underline{n}}\mathcal{V}_{P}^{\phantom{M}\underline{p}}\mathcal{V}_{Q}^{\phantom{M}\underline{q}}\mathcal{V}_{R}^{\phantom{M}\underline{r}}\mathcal{V}_{S}^{\phantom{M}\underline{s}} \ .
\ee
The underlined indices here are time-like rather than light-like SO($6$) indices, and they are related by the change of basis
\be
\label{lc_to_cart}
R \equiv \frac{1}{\sqrt{2}} \left(
\begin{array}{cc}
 -\mathds{1}_{6} & \mathds{1}_{6} \\
  \mathds{1}_{6} & \mathds{1}_{6}
\end{array}
\right) \ .
\ee
Because of this distinction between time- and space-like indices of SO$(6,6)$, this completely antisymmetric tensor is
invariant under local $\textrm{SO}(6) \times \textrm{SO}(6)$ transformations. Despite this, though, one would need to compute $\,\mathcal{V}\,$ associated with $\,M_{MN}\,$ explicitly in order to obtain the full form of the scalar potential.

As we saw previously, the theory contains vector fields $A_{\mu}$ in four dimensions which transform in the fundamental representation of SL($2$) $\times$ SO($6,6$),
\be
\begin{array}{cc}
A_{\mu}= V_{\mu}^{\a M} \, T_{\a M} \ ,
\end{array}
\ee

\subsection*{The Gauge Algebra}

In the \textit{ungauged} theory, only a subgroup $G=\textrm{U}(1)^{12} \subset \textrm{SO}(6,6)$ is realised and the vector fields become abelian, \emph{i.e.} $\left[  T_{\a M} , T_{\b N} \right]=0$. However, this ungauged theory can be deformed away from the abelian structure without breaking the $\,\cN=4\,$ supersymmetry so that a non-abelian subgroup $G \subset \textrm{SO}(6,6)$ is realised. From now on\footnote{In the main part of this chapter we will consider a group-theoretical truncation of $\cN=4$ supergravity in which the $(\textbf{2},\textbf{12})$ embedding tensor irrep will be entirely projected out.} , we will restrict to the case $\xi_{\a M}\,=\,0$. The commutation relations defining the algebra $G$ then read
\be
\label{SW_algebra}
\begin{array}{ccc}
\left[  T_{\a M} , T_{\b N} \right] &=&  {f_{\a\,MN}}^{P} \,\,T_{\b P} \ ,
\end{array}
\ee
with $f_{\a\,MNP}={f_{\a\,MN}}^{Q} \, \eta_{QP} = f_{\a\,[MNP]}\,$ being the structure constants of $\,G\,$ and with
$\,\eta_{MN}\,$ the SO($6,6$) metric. This automatically implies that only the $G \subset \textrm{SO}($6,6$)$ subgroups admitting $\eta_{MN}$ as a non-degenerate bi-invariant metric can be realised as deformations of the ungauged theory. In other words, the adjoint representation of $G$ has to be embeddable within the fundamental representation of SO($6,6$). This embedding may not be unique, resulting in non-equivalent realisations of the same $G$ subgroup. From now on, we will use light-cone coordinates, so that an SO($6,6$) index is raised or lowered by using the SO($6,6$) light-cone metric
\be
\label{eta_lc}
\eta_{MN}=\eta^{MN}=\begin{pmatrix} 0 & \mathds{1}_{6}  \\ \mathds{1}_{6} & 0 \end{pmatrix} \ .
\ee

Let us perform the GL($6$) splitting of the fundamental SO($6,6$) index $\,M \equiv ({}_{m} \, , \, {}^{m})\equiv (m,\bar{m})\,$ with
$\,m=1,...,6\,$ and $\,\bar{m}=\bar{1},...,\bar{6}\,$. Then, the vectors split as $\,T_{\a M} \equiv (Z_{\a m} \, , \, {X_{\a}}^{m})\,$ alike, and the algebra in (\ref{SW_algebra}) can be rewritten as the set of brackets
\be
\label{SW_algebra_ZX}
\begin{array}{cccccc}
\left[  Z_{\a m} , Z_{\b n} \right] &=&  {f_{\a\,mn}}^{p} \,\,Z_{\b p} &+& f_{\a\,mnp} \,\,{X_{\b}}^{p} & , \\[1mm]
\left[  Z_{\a m} , {X_{\b}}^{n} \right] &=&  {f_{\a\,m}}^{np} \,\,Z_{\b p} &+& f_{\a\, m\phantom{n}\,p}^{\phantom{\a}\phantom{m}n\phantom{p}} \,\,{X_{\b}}^{p} & , \\[1mm]
\left[ {X_{\a}}^{m}, Z_{\b n} \right] &=&  f^{\phantom{\a}m\phantom{n}p}_{\a\,\phantom{m}n\phantom{p}} \,\,Z_{\b p} &+& f_{\a\, \phantom{m}np}^{\phantom{\a} m\phantom{np}} \,\,{X_{\b}}^{p} & , \\[1mm]
\left[  {X_{\a}}^{m} , {X_{\b}}^{n} \right] &=&  {f_{\a\,}}^{mnp} \,\,Z_{\b p} &+& f_{\a\,\phantom{mn}p}^{\phantom{\a}mn\phantom{p}} \,\,{X_{\b}}^{p} & .\\[1mm]
\end{array}
\ee
It is worth noticing that this is only apparently a twenty-four-dimensional gauge algebra, but in fact the actual gauging is twelve-dimensional after imposing the constraints
\begin{align}
  \epsilon^{\alpha \beta} \,f_{\alpha MNP} \, T_\beta{}^P  = 0 \,, \label{abstract-QC}
\end{align}
which ensure the anti-symmetry of the brackets in \eqref{SW_algebra}. This fact is related to the observation in ref.~\cite{Dall'Agata:2007sr}, \emph{i.e.} that only the algebra realised on the vectors can be embedded in Sp($24,\mathbb{R}$), whereas the proper gauge algebra is that one realised on the curvatures, which is obtained from the previous one after dividing out by the abelian ideal consisting of all generators acting trivially on the curvatures. To summarise, in order to identify the correct
gauging, one has to solve these constraints by expressing half of the generators in terms of the other ones and plug the solution into the brackets of \eqref{SW_algebra_ZX}.

\subsection*{Vacua Analysis and Supersymmetry}

In this section we present the strategy followed to find the
complete set of extrema of the scalar potential induced by the
gaugings and tools for analysing the mass spectrum and supersymmetry
breaking.

\subsubsection*{Combining dualities and algebraic geometry techniques}

The investigation of the full vacua structure of a
particular truncation\footnote{We will in this part specify to the SO($3$) invariant sector of $\cN=4$ supergravity, which enjoys an SL($2$) $\times$ SO($2,2$) duality symmetry and contains three complex scalars called $S$, $T$ and $U$. The whole truncation procedure will be studied in detail later in section~\ref{FTheta_Dictionary}} is carried out by making use of the following
two ingredients: 
\begin{itemize}
\item part of the \,SL($2$) $\times$ SO($2,2$)\, duality
group in order to reduce the extrema scanning to the origin of the
moduli space without loss of generality\footnote{This approach
differs from that followed in ref.~\cite{Font:2008vd} where the
invariance under the action of the duality group was used to remove
redundant flux configurations producing physically equivalent
solutions.}. 

\item specific algebraic geometry techniques which
permit an exhaustive identification of the flux backgrounds
producing such moduli solutions.
\end{itemize}

Provided a set of vacuum expectation values (VEVs) for the moduli fields 
\be
\Phi_{0}\equiv\left( S_{0},T_{0},U_{0}\right) \nn
\ee
that satisfies the extremisation conditions of the scalar potential, $\,\left.\partial_{\Phi}V \right|_{\Phi_{0}}=0\,$,\, it can always be brought to the origin of the moduli space, \emph{i.e.}
\be
\label{moduli_origin}
S_{0}=T_{0}=U_{0}=i \ ,
\ee
by subsequently applying a real shift together with rescaling upon each of the complex moduli fields. These transformations span the non-compact part,
\be
\label{Gnc}
G_{n.c.} = \frac{\mathrm{SL}(2) \times \mathrm{SO}(2,2)}{\mathrm{SO}(2)^{3}} \ ,
\ee
of the duality group. In the case of the modulus $\,S$, they belong to the electric-magnetic SL($2$) factor, while transformations on the moduli $T$ and $U$ belong to SO$(2,2)$. In consequence, the fluxes will also transform in such a way that they compensate the transformation of the moduli fields and leave the scalar potential invariant.

Because of the aforementioned argument, 

\noindent\textit{restricting the search of
extrema to the origin of the moduli space does not imply a lack of
generality as long as the considered set of flux components is
invariant under the action of the non-compact part of the duality group.}

This statement automatically leaves us with two complementary descriptions of the same problem: the field and the flux pictures. In the former, a consistent flux background is fixed and the problem reduces to the search of extrema of the scalar potential in the field space. In the latter, the point in field space is fixed (the origin) and the problem reduces to find the set of consistent flux backgrounds compatible with the origin being an extremum of the scalar potential.
\begin{figure}[h!]
\begin{center}
\scalebox{0.9}[0.9]{
\begin{tabular}{ccc}
\includegraphics[keepaspectratio=true]{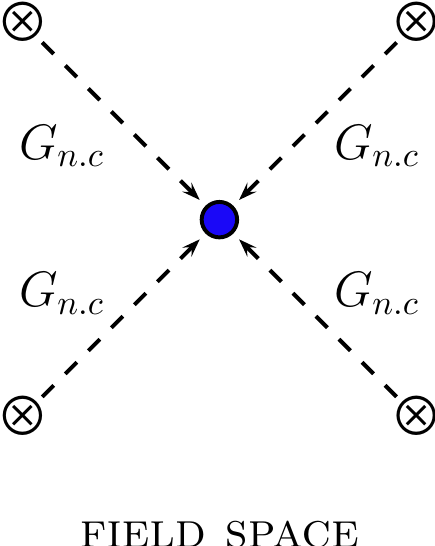} &  &
 \hspace{15mm}\includegraphics[keepaspectratio=true]{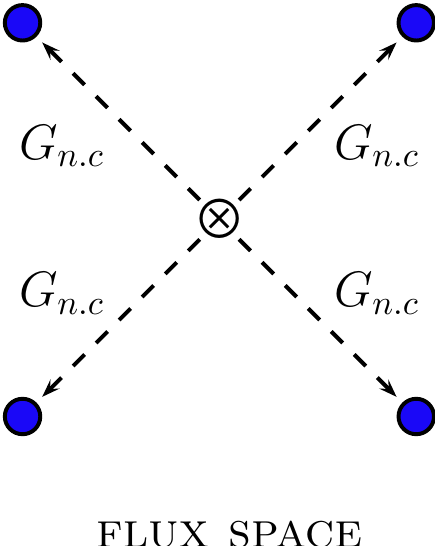} \\[-4cm]
&\hspace{15mm} \Large{$\longleftrightarrow$} &
\end{tabular}
}
\end{center}
\vspace{3cm}
\caption{{\it Sketch of the correspondence between the field picture (crossed dots) and the flux picture (filled dots). The left diagram represents moduli space, whereas the right diagram illustrates the space of fluxes.}}
\label{fig:field/flux_pic}
\end{figure}
The two descriptions are equivalent since dragging different moduli solutions down to the origin in the field space maps to a splitting of the corresponding flux background into various ones related by elements of $\,G_{n.c.}\,$ in the flux space. This correspondence is depicted in figure~\ref{fig:field/flux_pic}.

Using the flux picture results quite useful because, schematically, the
scalar potential induced by the gaugings takes the form of
\be
V = \sum_{\textrm{terms}} (\textrm{\small{fluxes}})^{2}
\,\cdotp\, (\textrm{\small{fields}})^{\textrm{high degree}} \ , 
\ee
hence being a sum of terms which are quadratic in the fluxes and
contain high degree couplings between the moduli fields. After
deriving the scalar potential with respect to the fields and going
to the origin of the moduli space, the extremum conditions reduce to
a set of quadratic conditions on the fluxes. Putting these
conditions together with the QC coming from the consistency of the gauging that will be later given for this specific case in (\ref{QCL}), we end up with a set of homogeneous polynomial equations, namely an ideal $\,I\,$ in the
ring $\,\mathbb{C}\left[a_{0},\dots,d'_{3} \right]\,$, involving the
different flux components as variables,
\be
\label{ideal_I}
I = \langle \,\left.\partial_{\Phi}V \right|_{\Phi_{0}} \,\,\, , \,\,\, \epsilon^{\alpha\beta}\,\Lambda_{\alpha\,
AB}^{\phantom{ABC\,}C}\,\Lambda_{\beta DEC} \,\,\, , \,\,\, \Lambda_{(\alpha\,
A[B}^{\phantom{ABC\,\,\,\,\,\,}C}\,\Lambda_{\beta)\, D]EC} \,  \rangle \ .
\ee
Nonetheless, only those solutions for which all the flux components turn out to be real are physically acceptable.

The study of non-trivial multivariate polynomial systems and their link to geometry is the subject of algebraic geometry \cite{CLO_book}. A powerful computer algebra system for polynomial computations is provided by the \textsc{\,Singular\,} project \cite{DGPS}. Moreover, a comprehensive introduction to the specifics of this software as well as to the algebraic geometry techniques implemented on it can be found in ref.~\cite{GP_book}. These techniques have been shown to be a successful approach to investigate the vacua structure of the effective supergravity theories coming from flux compactifications of string theory \cite{Gray:2006gn,Guarino:2008ik} and some extensions including both fluxes and non-perturbative effects\footnote{For a computational implementation of these algebraic geometry tools into a \textit{Mathematica} package exploring vacuum configurations, see ref.~\cite{Gray:2008zs}.} \cite{Gray:2007yq}.

Among the set of algebraic geometry tools implemented within \textsc{\,Singular}, in this work we will make extensive use of the Gianni-Trager-Zacharias (GTZ) algorithm \cite{GTZ} for primary decomposition into prime ideals (for more details on primary decomposition algorithms, see the appendix B of ref.~\cite{Gray:2006gn} and references therein). Specifically, we will apply this method to decompose the ideal $I$ of (\ref{ideal_I}) into a set of $\,n\,$ simpler prime ideals $\,J_{n}\,$,
\be
I= J_{1} \cap J_{2} \cap \ldots \cap J_{n} \ ,
\ee
which can be solved analytically. These prime ideals will only intersect in a finite number of disjoint points and, in general, they may have different dimension.

For the sake of simplicity, we are not running this decomposition in
the most general case in which all the forty embedding tensor
components (fluxes) allowed in the SO($3$) truncation are kept.
Instead, we are considering two examples of gauged supergravities
which have a well understood interpretation as type II string
compactifications in the presence of flux backgrounds: type IIA
compactifications with gauge and metric fluxes
\cite{Derendinger:2004jn,Villadoro:2005cu,Derendinger:2005ph,Dall'Agata:2009gv}
and type IIB compactifications with gauge fluxes
\cite{Giddings:2001yu,Kachru:2002he,DeWolfe:2004ns}.

Even though not all the fluxes are kept in these examples, the
previous argument for going to the origin of the moduli space
without loss of generality still holds since the transformation
needed to bring any moduli solution from its original location to
the origin (\emph{i.e.} an element of $G_{n.c.}$) does not turn on new flux components out of the initial
setup. We will restrict the analysis of more general flux
backgrounds for which a realisation in string theory is not known,
namely those including non-geometric fluxes, to a set of simple examples.

\subsubsection*{Supersymmetry breaking and full mass spectrum}

Two further important steps in the analysis of critical points are those of computing the amount of supersymmetry preserved at the extrema of the $\,\mathcal{N}=4\,$ theory and the mass spectrum of the scalar sector. As already pointed out in the introduction, carrying out such a computation for a whole set of vacua can help us shed further light on the relation between supersymmetry breaking and instability, which has recently been a crucial point of discussion in the context of extended supergravity. In order to do this, we will write down the fermionic mass terms involving the gravitini in the Lagrangian \cite{Schon:2006kz} induced by the irreducible components of the $T$-tensor (see section~\ref{section:Theta}). 

The fermionic sector of the theory is made out of $4$ gravitini $\psi_{\mu i}$, $4$ dilatini $\chi_{i}$ and $24$ gaugini $\lambda_{a i}$ coming from the matter sector, where $\,i=1,...,4\,$ is an SU($4)_{\textrm{time-like}}$ index, whereas $a=1,...,6\,$ is a fundamental SO$(6)_{\textrm{space-like}}$ index. The aforementioned mass terms then read
\be
\begin{array}{ccc}
e^{-1} \mathcal{L}_{\textrm{fermi mass}} & \supset \ & \dfrac{1}{6} \,A_{1}^{ij} \, \bar{\psi}_{\mu i} \, \Gamma^{\mu \nu} \, \psi_{\nu j}\,-\, \dfrac{i}{6} \,A_{2}^{ij} \, \bar{\psi}_{\mu i} \, \Gamma^{\mu} \, \chi_{j}\,+\,\dfrac{i}{2}\,{A_{2 a i}}^{j}\,{\bar{\psi}_{\mu}}{}^{i} \,\Gamma^{\mu} \,{\lambda^{a}}_{j} \,+\, \textrm{h.c.}\ ,
\end{array}
\ee
with $\,A_{1}^{ij}=A_{1}^{(ij)}$. The irreducible components of the $T$-tensor are
\be
\hspace{-4mm}
\begin{array}{lcl}
\left(\textbf{2},\textbf{12}\right)\,\oplus\,\left(\textbf{2},\textbf{220}\right) & \overset{\textrm{U}(4)_{R}\,\times\,\textrm{SU}(4)}{\longrightarrow} & 
\begin{array}{c}
2\,\cdot\,\left(\textbf{10},\textbf{1}\right)_{(-2)}\,\oplus\,2\,\cdot\,\left(\textbf{6},\textbf{1}\right)_{(+2)}\,\oplus\,2\,\cdot\,\left(\textbf{6},\textbf{15}\right)_{(+2)}\\[1mm]
2\,\cdot\,\left(\textbf{10},\textbf{1}\right)_{(+2)}\,\oplus\,2\,\cdot\,\left(\textbf{6},\textbf{1}\right)_{(-2)}\,\oplus\,2\,\cdot\,\left(\textbf{6},\textbf{15}\right)_{(-2)}
\end{array}
\,\oplus\,\textrm{c.c.} 
\end{array}
\ee
and they are given in terms of the complexified SL($2$) and SO($6,6$) vielbeins by
\be
\label{fermi_mass_N=4}
\begin{array}{cclc}
A_{1}^{ij} & = & \epsilon^{\alpha \beta} \, (\mathcal{V}_{\alpha})^{*} \, {\mathcal{V}^{M}}_{kl} \, \mathcal{V}^{N ik} \, \mathcal{V}^{P jl} \, f_{\beta MNP} & , \\[2mm]
A_{2}^{ij} & = & \epsilon^{\alpha \beta} \,\,\, \mathcal{V}_{\alpha} \,\,\,\,\,\, {\mathcal{V}^{M}}_{kl} \, \mathcal{V}^{N ik} \, \mathcal{V}^{P jl} \, f_{\beta MNP} \,+ \frac{3}{2} \, \epsilon^{\alpha \beta} \, \mathcal{V}_{\alpha} \, {\mathcal{V}}^{M ij} \, {\xi_{\beta M}} & , \\[2mm]
{A_{2\,a i}}^{j} & = & \epsilon^{\alpha \beta} \,\,\, \mathcal{V}_{\alpha} \,\,\,\,\,\, {\mathcal{V}^{M}}_{a} \, {\mathcal{V}^{N}}_{ik} \, \mathcal{V}^{P jk} \, f_{\beta MNP} - \frac{1}{4} \,\delta_{i}^{j} \, \epsilon^{\alpha \beta} \, \mathcal{V}_{\alpha} \, {\mathcal{V}^{M}}_{a} \, {\xi_{\beta M}} & .
\end{array}
\ee
The complexified SL($2$) vielbein $\,\mathcal{V}_{\alpha}\,$ is written as
\be 
\mathcal{V}_{\alpha} = e^{\phi/2} \, \left(
S \,,\, 1 \right) \hspace{10mm} , \hspace{10mm} \textrm{where}
\hspace{5mm} S= \chi + i \, e^{-\phi} \ , 
\ee
whereas the complexified (Lorentzian) SO($6,6$) vielbein $\,{\mathcal{V}_{M}}^{[ij]}\,$ is built from the $\,{\mathcal{V}_{M}}^{m}\,$ real vielbein by using the mapping
\be
{\mathcal{V}_{M}}^{ij}\,=\,{\mathcal{V}_{M}}^{m}\,\left[G_{m}\right]^{ij}\ ,
\ee
where $\left[G_{m}\right]^{ij}$ represent the time-like and ASD 't Hooft symbols given in \eqref{'tHooft_timelike} of appendix~\ref{appendix:Exceptional}. This choice is consistent with
\be
v_{ij} = (v^{ij})^{*} = -\dfrac{1}{2} \, \epsilon_{ijkl} \, v^{kl} \ ,
\ee
together with the normalisation
\be
- \, v^{m} \, \delta_{mn} \, v^{n} \, = \, \dfrac{1}{2} \, \epsilon_{ijkl} \, v^{ij} \, v^{kl}  \ .
\ee
Please note the difference with the conventions in ref.~\cite{Schon:2006kz}. Using the gravitino mass matrix $A_{1}^{ij}$, the Killing spinor equations determining the amount of supersymmetry at any extremum is translated into the eigenvalues equation
\be
\label{susy_cond}
A_{1}^{ij} \, q_{j} = \sqrt{-3 V_{0}} \, q^{i} \ ,
\ee
where $\,q^{i}\,$ is an SU($4$) vector and $\,V_{0}\,$ is the potential energy at either an AdS$_{4}$ or a Minkowski extremum.

Working in the SO($3$) truncation of the SO($6,6$) theory translates into an $\,A_{1}^{ij}\,$ gravitini mass matrix of the general form
\be
A_{1} =  \textrm{diag}\left( \, \kappa_{1} \,\,\,,\,\,\, \kappa_{2} \,,\,\kappa_{2} \,,\,\kappa_{2} \, \right)
\hspace{10mm} , \hspace{10mm}
\textrm{with} \hspace{5mm} \kappa_{1},\kappa_{2} \in
\mathbb{C} \ , \label{kappa-eigenvalues}
\ee
which reflects the splitting $\,\textbf{4}\rightarrow\,\textbf{1} \oplus \textbf{3}\,$ of the fundamental of SU($4$) under the action
of SO($3$). Consequently one expects that the amount of supersymmetry preserved would be
\begin{itemize}

\item[$i)$] $\mathcal{N}=4\,$ at those extrema where $\,|\kappa_{1}|=|\kappa_{2}|=\sqrt{-3 V_{0}}$.

\item[$ii)$] $\mathcal{N}=3\,$ at those extrema where $\,|\kappa_{1}| > |\kappa_{2}|\,$ with $\,|\kappa_{2}|=\sqrt{-3 V_{0}}$.

\item[$iii)$] $\mathcal{N}=1\,$ at those extrema where $\,|\kappa_{1}|< |\kappa_{2}|\,$ with $\,|\kappa_{1}|=\sqrt{-3 V_{0}}$.

\item[$iv)$] $\mathcal{N}=0\,$ at any other extremum.

\end{itemize}
The presented conditions for preserving supersymmetry only constrain the modulus of the eigenvalues of $A_1$ since the relation \eqref{susy_cond} exhibits a U($1$) $\times$ U($1$) covariance. The action of these transformations can be expressed in terms of the diagonal matrix $\text{diag}(\lambda\,, \,\mu\,,\,\mu\,,\,\mu)$, where $\lambda$, $\mu$ $\in$ U($1$).

Now it is worthwhile making a comment about the computation of the full mass spectrum of the scalar sector for a vacuum of the
$\mathcal{N}=4\,$ theory. To this purpose we applied the mass formula given in ref.~\cite{Borghese:2010ei}, where the scalar potential of the full $\mathcal{N}=4\,$ theory has been expanded up to second order around the origin in order to be able to read off the second derivatives of the potential with respect to all of the $38$ scalars of the theory evaluated in the origin of moduli space. The Hessian matrix evaluated in the origin is nevertheless not yet the physical mass matrix from where one can draw conclusions about stability of a solution. Suppose one has
\be
\label{Lag_can} e^{-1}
\mathcal{L}_{\textrm{canonic}}=\dfrac{1}{2} \,R -\dfrac{1}{2} \,
K_{ij} \, (\partial \phi^{i})(\partial \phi^{j}) - V \ ,
\ee
where $\,i=1,...,38\,$, then the covariant normalised mass$^2$ at an extremum $\phi_0$ of the scalar potential $V$ is then given by
\be
\label{mass}
{(\textrm{mass}^{2})^{i}}_{j} =\left.  \frac{1}{|V|} \, K^{ik} \,\,
\dfrac{\partial^{2} V}{\partial \phi^{k} \partial \phi^{j}}
\right|_{\phi=\phi_{0}} \ ,
\ee
where $\,K^{ij}\,$ denotes the inverse of the matrix $\,K_{ij}$ appearing in \eqref{Lag_can}. This (mass$^{2}$) matrix is known as
the canonically normalised mass matrix, which is consistent with taking the ``mostly plus'' signature for the space-time metric and its eigenvalues are to be read as the values for the squared mass in natural units\footnote{Every numerical value given in the following
sections for the energy and the mass is computed by setting the reduced Planck mass $m_p$ to $1$, whereas one needs to reinsert the value $\,m_{p}=\left( 8 \, \pi \, G \right)^{-1/2} = 2.43 \times 10^{18} \,\, \textrm{GeV} \,$ when expressing quantities in energy units.}. According to this definition of covariant mass, the Breitenlohner-Freedman (BF) bound for the stability of an AdS$_4$ moduli solution is given by \cite{Breitenlohner:1982jf}
\be
\label{BF_bound}
m^2\geq-\frac{3}{4}\ ,
\ee
where $m^2$ denotes the lightest eigenvalue of the mass matrix \eqref{mass} at the AdS$_4$ extremum. The mass formulae for the masses of the SL($2$) scalars, those ones of the SO($6,n$) sector and finally the mixing between them are given in ref.~\cite{Borghese:2010ei}. In the next sections, when presenting results, we shall give both a table with the values of the masses of the scalars in the SO($3$) truncation and the full mass spectrum for comparison's sake.

\section{Fluxes and Embedding Tensor: the Dictionary}
\label{FTheta_Dictionary}

Now our final aim in this chapter will be to apply the whole analysis presented in the previous section to $\cN=4$ gaugings coming from geometric type II string compactifications. To this end, let us first see what gauge algebras can be induced by flux compactifications. The starting point in this discussion are the results of Kaloper and Myers \cite{Kaloper:1999yr} (see the second part of section~\ref{Geom_Flux_Comp}). They found that the dimensional reduction of heterotic supergravity to four dimensions leads to a non-Abelian gauge algebra if one includes fluxes. In particular, they derived the four-dimensional effect of the following fluxes for the ten-dimensional field content consisting of the metric, a two-form and a dilaton\footnote{We will only include fluxes for the metric and the two-form. There is a similar possibility for the dilaton, which we will not consider, that leads to gauging with non-vanishing $\xi_{\alpha M}$ \cite{Derendinger:2007xp}.  Moreover, we will also not consider trombone gaugings of the type introduced in ref.~\cite{LeDiffon:2008sh} for the maximal theory.}.

As we saw previously, when reducing the metric from ten to four dimensions, one can generalise ordinary dimensional reduction by replacing the torus with a group manifold \cite{Scherk:1979zr}. A group manifold is specified by structure constants $\omega_{mn}{}^p$, where the indices run over the dimension of the group manifold. The four-dimensional effect of such so-called metric flux is to convert the gauge group U$(1)^6$, that corresponds to general coordinate transformations on the torus, to a non-Abelian group with commutation relations
\begin{align}
  \left[ Z_m, Z_n \right]\,=\, \omega_{mn}{}^p \, Z_p \,,
 \label{KKgenerators}
\end{align}
where $Z_m$ is the generator corresponding to the internal coordinate transformation $\delta x^m = \lambda^m$.

Due to the presence of the two-form gauge potential in the ten-dimensional theory, the four-dimensional gauge algebra is
actually larger. In particular, there is an additional U$(1)^6$ corresponding to internal gauge transformations of the form $\delta B_{mn} = \partial_{[m} \lambda_{n]}$. We will denote these generators by $X^p$. These commute amongst themselves, but form a representation of the group spanned by \eqref{KKgenerators}.
Furthermore, one can introduce gauge fluxes $H_{mnp}$ for this potential. The total algebra spanned by the six Kaluza-Klein and six gauge generators reads \cite{Kaloper:1999yr}
\be
\label{KM-algebra}
\begin{array}{cclc}
\left[Z_{m},\,Z_{n}\right] & = & {\omega_{mn}}^{p}\,Z_{p}\,+\,H_{mnp}\,X^{p} & , \\[1mm]
\left[Z_{m},\,X^{n}\right] & = & -{\omega_{mp}}^{n}\,X^{p}\, & , \\[1mm]
\left[X^{m},\,X^{n}\right] & = & 0 & . 
\end{array}
\ee
Note that the resulting algebra is purely electric. Furthermore, the gauge generators span an ideal of the algebra, and hence the full algebra is non-semi-simple.

In order to make contact with the $\textrm{SO}(6,6)$ notation of $\cN = 4$ supergravity, one needs to split up the $\textrm{SO}(6,6)$ index ${}^M =
({}_m, {}^m)$. The twelve doublets of generators then split up according to $X^{\alpha M} = (Z^\alpha{}_m, X^{\alpha m})$. The identification between the embedding tensor and the fluxes is then apparent:
\begin{align}
  f_{+mnp} = H_{mnp}  \ , \quad f_{+ mn}{}^p = \omega_{mn}{}^p \ , \label{KM-fluxes}
\end{align}
while the magnetic components vanish.

A natural question is how to generalise this to the case where one includes, in addition to gauge and metric flux, also the types of non-geometric fluxes introduced in ref.~\cite{Shelton:2005cf}. In section~\ref{sec:non-geom_fluxes} we saw that the full heterotic dictionary is given by \eqref{Het_Fluxes} and, following the T-duality chain \eqref{Tduality}, leads to the gauge algebra that we gave in \eqref{Het_Gauge_Alg}.
Note that this algebra, with all types of NS-NS fluxes, is still purely electric.

Subsequently one could reason that in the IIB duality frame with O3-planes one needs to mod out by the $\mathbb{Z}_2$ symmetry $(-)^{F_L} \Omega I_{4 \cdots 9}$. Under this symmetry, the only allowed fluxes are $H$ and $Q$. Therefore the algebra for these fluxes reads
\be
\label{STW-algebra2}
\begin{array}{cclc}
\left[Z_{m},\,Z_{n}\right] & = & H_{mnp}\,X^p & , \\[1mm]
\left[Z_{m},\,X^{n}\right] & = & Q_m{}^{np}\,Z_p & , \\[1mm]
\left[X^{m},\,X^{n}\right] & = & Q_p{}^{mn}\,X^p & . 
\end{array}
\ee
The relation between the embedding tensor and the fluxes can be easily read off from this algebra. Before we give it, let us introduce a slight generalisation by including S-duality related fluxes as well. For the two-form gauge potentials this is very natural, as we know that these form a doublet $(H,F)$ under S-duality. Similarly, it has been conjectured that there is a doublet of non-geometric fluxes $(Q,P)$ as well \cite{Aldazabal:2006up}. Including the two doublets of gauge and non-geometric fluxes, the relation to the embedding tensor that follows from \eqref{SW_algebra_ZX} is
\begin{alignat}{2}
  f_{+mnp} & = H_{mnp}  \,, \quad & f_{+ m}{}^{np} & = Q_{m}{}^{np} \ , \notag \\
  f_{-mnp} & = F_{mnp}  \,, \quad & f_{- m}{}^{np} & = P_{m}{}^{np} \ . \label{STW-ident}
\end{alignat}
The full algebra, including the commutation relations between electric and magnetic generators, then follows trivially from \eqref{SW_algebra}. Similarly, one can deduce the full set of constraints on the fluxes from the QC given in \eqref{QC41}--\eqref{QC45} specialised to the case\footnote{The only further subtlety is that the second set of QC in \eqref{QC_xi=0} can be obtained from \eqref{abstract-QC} by specifying it to the adjoint representation. Nevertheless, these sets of constraints are only equivalent if such adjoint representation is faithful, otherwise one has to take into account that the linear dependence relations between the $24$ generators have to be supplemented with the vanishing conditions for some of them.} $\xi_{\a M}=0$
\begin{align}
  f_{\alpha R[MN}\,f_{\beta P]Q}{}^R = 0 \ , \quad \epsilon^{\alpha \beta}\,f_{\alpha MNR}\,f_{\beta PQ}{}^R = 0 \ .
\label{QC_xi=0}
\end{align}
 
Note that the algebra \eqref{STW-algebra2} in general does not have any non-trivial ideals, and hence is not necessarily
non-semi-simple. This form of the algebra has been used in \emph{e.g.} \cite{deCarlos:2009fq} in their classification of the possible solutions
of the corresponding Jacobi identities. Indeed, they encountered simple and semi-simple possibilities. This poses a clear puzzle: we claim to have performed a number of dualities, under which the effective description should transform covariantly, and nevertheless the algebra \eqref{KM-algebra} of the starting point clearly differs from \eqref{STW-algebra2}. Indeed, one is necessarily non-semi-simple while the other is not. \emph{What has happened?} In our opinion, the confusion stems from the identification of the starting point.

The starting point of Kaloper and Myers corresponds to the heterotic string, and therefore contains an NS-NS two-form gauge potential.
However, in order to make contact with type II string theories with orientifold planes, \emph{e.g.}~the preferred duality frame of type IIB with O3-planes, one should first perform an S-duality. This takes one to type I string theory, or equivalently type IIB with
O9-planes. In this case the two-form is not NS-NS but rather R-R, which will be a crucial distinction when applying T-duality. As
mentioned before, in the NS-NS sector T-duality raises and lowers indices. In contrast, in the R-R sector the effect of T-duality is to create or annihilate indices:
\begin{align}
 \textrm{T}_p: \qquad
\begin{cases}
    F_{m_1 \cdots m_n} \rightarrow F_{m_1 \cdots m_n p} \ , \\
 F_{m_1 \cdots m_n p} \rightarrow F_{m_1 \cdots m_n} \ .
\end{cases}
\end{align}
In other words, a gauge potential remains a gauge potential but its rank changes.

The correct starting point for our purpose is
\be
\label{KM-algebra2}
\begin{array}{cclc}
\left[Z_{m},\,Z_{n}\right] & = & {\omega_{mn}}^{p}\,Z_{p}\,+\,F_{mnp}\,X^{p} & , \\[1mm]
\left[Z_{m},\,X^{n}\right] & = & -{\omega_{mp}}^{n}\,X^{p}\, & , \\[1mm]
\left[X^{m},\,X^{n}\right] & = & 0 & , 
\end{array}
\ee  
where $F_{mnp}$ is the R-R three-form flux. Upon a six-tuple T-duality to go to the type IIB duality frame with O3-planes, this transforms into
\be
\label{ACR-algebra1}
\begin{array}{cclc}
\left[Z_{m},\,Z_{n}\right] & = & 0 & , \\[1mm]
\left[Z_{m},\,X^{n}\right] & = & Q_{m}{}^{np}\,Z_p & , \\[1mm]
\left[X^{m},\,X^{n}\right] & = & Q_p{}^{mn}\,X^p \,+ \, {\tilde F}^{mnp}\,Z_p & , 
\end{array}
\ee
where ${\tilde F}^{mnp}\,\equiv\,\tfrac16 \epsilon^{mnpqrs} F_{qrs}$. This fixes the complete electric part of the gauge algebra. The remaining part follows straightforwardly once one has made the identification between the embedding tensor and the fluxes. Again we will give an
S-duality covariant set of fluxes, including the gauge doublet $(F,H)$ and the non-geometric doublet $(Q,P)$. With the algebra \eqref{ACR-algebra1} this identification reads
\begin{alignat}{2}
  f_{+}{}^{mnp} & = \tilde{F}^{mnp}  \,, \quad & f_{+ m}{}^{np} & = Q_{m}{}^{np} \,, \notag \\
  f_{-}{}^{mnp} & = \tilde{H}^{mnp}  \,, \quad & f_{- m}{}^{np} & = P_{m}{}^{np} \,. \label{ACR-ident}
\end{alignat}
The full algebra and corresponding QC then follow from \eqref{SW_algebra_ZX} and \eqref{QC_xi=0}. The latter read
\begin{align}
  & Q_r{}^{[mn}Q_q{}^{p]r} =   P_r{}^{[mn}P_q{}^{p]r}= 0 \,, \notag \\
  & P_r{}^{[mn}Q_q{}^{p]r} = Q_r{}^{mn}P_q{}^{pr} - P_r{}^{mn}Q_q{}^{pr} = 0  \,, \label{QC1}
\end{align}
involving only non-geometric flux, and
\begin{align}
  & \tilde{F}^{r[mn}Q_r{}^{pq]} =
  \tilde{H}^{r[mn}P_r{}^{pq]} = 0 \,, \notag \\
 &  \tilde{F}^{r[mn}P_r{}^{p]q}+ Q_r{}^{[mn}\tilde{H}^{p]qr}  = 0  \,, \label{QC2}
\end{align}
involving gauge fluxes as well. The fully anti-symmetric parts of the latter set of equations imply the absence of any 7-branes; these would break supersymmetry further to $\cN = 1$. The same form  of the algebra and QC was derived in the beautiful work\footnote{Due to different conventions regarding the $\textrm{SO}(6,6)$ and $\textrm{SL}(6)$ indices, our form of the identification
\eqref{ACR-ident} does not involve any non-trivial metrics, as in ref.~\cite{Aldazabal:2008zza}. Moreover, the QC given in ref.~\cite{Aldazabal:2008zza} are not all linearly independent, and hence can be written in a more economic way.} \cite{Aldazabal:2008zza} from a different starting point.

Note the differences between the two algebras\footnote{Most of the literature that uses \eqref{STW-algebra2} takes place in an $\cN = 1$ context, where the scalar potential is not given in terms of structure constants but rather a superpotential. Therefore our argument does not affect any of the results on $\cN =1$ moduli stabilisation etc.} \eqref{STW-algebra2} and \eqref{ACR-algebra1}. First of all, NS-NS fluxes induce a purely electric gauging in the former algebra \cite{Derendinger:2007xp}, while in the latter this involves magnetic generators as well. Moreover, the former can describe a (semi-)simple algebra \cite{Dall'Agata:2008qz, deCarlos:2009fq, Avramis:2009xi}, while the latter is always non-semi-simple algebra, as it should. This crucial difference between the two stems from the appearance of the Hodge dualised three-form $\tilde F$, instead of the three-form itself, in \eqref{ACR-algebra1}. This qualitative difference can be traced back to the different behaviour of NS-NS and R-R gauge potentials under T-duality.

Finally, the QC \eqref{QC2} are in general different for the two algebras. For instance, it can be seen from
the $\textrm{SL}(2)$ scaling weight that the last equation of \eqref{QC2} could never arise from \eqref{STW-ident}. However, in the truncation where one of the two non-geometric fluxes vanishes, \emph{e.g.} $P=0$, the QC bilinear in the NS-NS fluxes are in fact identical (provided $Q_m{}^{mn} = 0$). There is still a difference in the constraints bilinear in $Q$ and $F$: these are much stronger for the first identification \eqref{STW-ident} than those given in \eqref{QC2}.

\subsection*{What About De Sitter?}

All the gaugings that are known to give rise to dS solutions in $\cN = 4$ gauged supergravity \cite{deRoo:2002jf, deRoo:2003rm} are of the form
\begin{align}
  G = G_1 \times G_2 \times \cdots \,, \label{gaugegroup}
\end{align}
\emph{i.e.}~a direct product of a number of gauge factors. This is a solution to the QC \eqref{QC_xi=0} once the Jacobi identities are separately satisfied in the different factors. Moreover, in order to have a dS solution, the gauge group must contain electric and magnetic factors. Finally, the gauge factors have to be specific (semi-)simple groups. In particular, we will focus on the case of two gauge factors. Each factor is  of the form $\textrm{SO}(p,q)$ with $p+q=4$ and embedded in an $\textrm{SO}(3,3)$ factor. A number of examples of such gaugings with dS solutions was discussed in \cite{deRoo:2002jf, deRoo:2003rm}. Moreover, it was shown in ref.~\cite{deRoo:2006ms} that the contracted versions $\textrm{CSO}(p,q,r)$ with $p+q+r = 4$ of such gauge groups do not lead to any solutions with a positive scalar potential.  In this section we will assess to what extend one can obtain such gaugings from the flux compactifications considered earlier.

The direct product structure \eqref{gaugegroup} leads us to split $\textrm{SO}(6,6)$ into two $\textrm{SO}(3,3)$ factors in which to embed $G_{1}$ and $G_{2}$ respectively. Without loss of generality, we will take the first to be electric and lie in the directions $\{1,2,3,\bar 1 , \bar 2 , \bar 3 \}$, while the second is taken magnetic and lies in the complementary directions. We will discuss the embedding of the first factor in some detail; the discussion for the second factor is completely analogous. However, before we discuss $\textrm{SO}(4)$ embeddings in $\textrm{SO}(3,3) \simeq \textrm{SL}(4)$, we first generalise this to arbitrary $N$.

In general, the embedding of $\textrm{SO}(N)$ and its analytic continuations into SL$(N)$ can be written in terms of the following generators in
the fundamental representation
\begin{align}
\left(T_{ij}\right)^{k}_{\phantom{k}l}= 4 \delta^{k}_{\phantom{k}[i} M_{j]l}\ ,
\end{align}
in terms of a symmetric matrix $M$, that can always be diagonalised by a convenient choice of basis. It is in fact given by the identity
in the case of $\textrm{SO}(N)$. These generators labelled by antisymmetric pairs of indices satisfy the following commutation relations
\begin{align}
\left[T_{ij},T_{kl}\right]=f_{ij,kl}^{\phantom{ij,kl}mn} T_{mn}\ ,\quad
f_{ij,kl}^{\phantom{ij,kl}mn}= 8 \delta^{[m}_{\phantom{k}[i} M_{j][k}\delta^{n]}_{\phantom{k}l]}\ .
\end{align}
Analytic continuations of $\textrm{SO}(N)$  correspond to a number of minus signs in the $M$-matrix. Contractions thereof, denoted by
$\textrm{CSO}(p,q,r)$ with $p+q+r=N$ (see \emph{e.g.}~\cite{deRoo:2006ms}), can be understood in  this notation by replacing $r$ non-zero diagonal
entries of $M$ with zero entries.

However, the most general form of $\textrm{CSO}(p,q,r)$ structure constants for the special case of $N = 4$ is given in terms of two symmetric
matrices rather than one \cite{Roest:2009tt}, which we will denote by $M$ and $\tilde{M}$. The generators are then given by
\begin{align}
\left(T_{ij}\right)^{k}_{\phantom{k}l}= 4 \delta^{k}_{\phantom{k}[i} M_{j]l} - 2 \epsilon_{ijml}\tilde{M}^{mk}\,,
\end{align}
giving rise to the following general expression of the structure constants
\begin{align}
f_{ij,kl}^{\phantom{ij,kl}mn}= 8 \delta^{[m}_{\phantom{k}[i} M_{j][k}\delta^{n]}_{\phantom{k}l]} - \epsilon_{iji'j'}\epsilon_{klk'l'}\epsilon^{mni'l'} \tilde{M}^{j'k'} \ .
\end{align}
With such a form we need some extra consistency constraints in terms of $M$ and $\tilde{M}$, coming from imposing the Jacobi identities. These translate into
\begin{align}
M_{ij}\tilde{M}^{jk}-\tfrac{1}{4}\delta_i^{\phantom{a}k}M_{jl}\tilde{M}^{jl}=0\ .
\end{align}
If one still diagonalises $M$ by a convenient basis choice, the Jacobi identity imply $\tilde{M}$ to be diagonal as well. In this case the constraints reduce to
\begin{align}
M_{11}\tilde{M}^{11}=M_{22}\tilde{M}^{22}=M_{33}\tilde{M}^{33}=M_{44}\tilde{M}^{44}\,.
\end{align}

Let us now connect the adjoint representation in terms of $\textrm{SL}(4)$ indices to fundamental $\textrm{SO}(3,3)$ indices. This is exactly what we already showed in appendix~\ref{Appendix_A_'tHooft} and it leads to the following identification between the diagonal components of the two matrices $M$ and $\tilde M$, and the components of the embedding tensor $f_{\alpha MNP}$ in the first $\textrm{SO}(3,3)$ factor:
\begin{align}
  M & = \textrm{diag}(f_{+123}, f_{+1\bar2 \bar 3}, f_{+ \bar 1 2 \bar 3}, f_{+\bar 1 \bar 2 3}) \,, \notag \\
 \tilde{M} & =  \textrm{diag}(f_{+ \bar 1 \bar 2 \bar 3}, f_{+ \bar 1 2   3}, f_{+   1  \bar 2   3}, f_{+  1   2  \bar 3}) \,.
\end{align}
Other components of the embedding tensor in this $\textrm{SO}(3,3)$ factor, such as $f_{+ 1 \bar 1 2}$, correspond to off-diagonal components of
$M$ and $\tilde M$ and hence have been set equal to zero.

We have discussed in the previous sections how the embedding tensor can be sourced by different fluxes. In particular, we have discussed the two identifications \eqref{STW-ident} and \eqref{ACR-ident}. It will be illuminating to illustrate the different consequences of the two identifications in this context. Using the first identification, the matrices are given by
\begin{align}
  M & = \textrm{diag} (H_{123} , Q_{1}{}^{23} , Q_{2}{}^{31}, Q_{3}{}^{12}) \ ,  \notag \\
 \tilde{M}  & =  \textrm{diag}(0,0,0,0) \ .
\end{align}
In this case it would therefore be possible to use the different fluxes to generate a simple gauge factor. Given that the discussion in the second, magnetic factor is completely analogous, one could \emph{e.g.}~generate an $\textrm{SO}(4)_{\rm el}\,\times\,\textrm{SO}(4)_{\rm magn}$ gauge
group, which certainly leads to dS solutions. However, we have argued that this is not the correct identification; instead, one should use \eqref{ACR-ident}. In this case, the matrices read
\begin{align}
  M & = \textrm{diag} (0 , Q_{1}{}^{23} , Q_{2}{}^{31}, Q_{3}{}^{12}) \ ,  \notag \\
\tilde{M}  & =  \textrm{diag}(F_{456},0,0,0) \ .
\end{align}
The crucial point is that in this case the gauge flux does not enter in the $M$ matrix to make it non-singular; instead, it enters in the
other matrix $\tilde M$. These singular matrices only lead to non-semi-simple gauge groups. In particular, the matrix $M$ gives
rise to $\textrm{ISO}(3)$ and analytic continuations and contractions thereof. Provided the three components $Q_i{}^{jk}$ are non-zero,
the additional parameter $F_{456}$ does not modify the gauge group, but only describes different embeddings of it in $\textrm{SO}(3,3)$. Three of
these are inequivalent, corresponding to $F_{456}$ being positive, zero or negative. Exactly the same embeddings of $\textrm{ISO}(3)$ and
$\textrm{ISO}(2,1)$ were considered in ref.~\footnote{The relation to the notation of  \cite{deRoo:2006ms} is $\lambda^2 = (1-F_{456})/(1+F_{456})$.}\cite{deRoo:2006ms}, where it was found that such gauge groups do not give rise to scalar potentials with positive extrema.

Indeed, one can infer from the same reasoning that none of the gauge groups discussed in \cite{deRoo:2002jf, deRoo:2003rm} follows from a
flux compactification with the identification \eqref{ACR-ident}. The simple bottom line is that all the gauge groups necessarily consist
of (semi-)simple gauge factors, while one can only get non-semi-simple factors from flux compactifications.

In the following subsection, we will focus on deformations arising as consistent SO($3$) truncations of the general theory and will show that they admit a string theory realisation in terms of flux compactifications in the presence of generalised background fluxes. the underlying theories will be $STU$ models for which we will provide the fux-induced superpotential.

\subsection*{The SO($3$) Truncation}

Let us consider the SO($3$) truncation of the full theory enjoying an SL($2$) $\times$ SO($6,6$) global symmetry\footnote{This is the natural generalisation of the $\textrm{SL}(3)\,\times\,\textrm{SL}(3)$ truncation considered in ref.~\cite{Roest:2009dq}, and indeed will lead to a much richer landscape of vacua.}. In the following sections of this work we will be dealing with (non-) geometric flux compactifications of type II string theory having such a low-energy effective description. This truncation is performed by considering an SO($3$) subset in SO($6,6$) and keeping in the theory only the singlets with respect to this subgroup both in the scalar sector and in the embedding tensor part. Such a group theoretical
truncation is always guaranteed to be consistent in the sense that all of the non-singlet scalars can be consistently set to zero in
that their field equations can never be sourced by SO($3$) singlets. However, it by no means guarantees the stability of the non-singlets, and hence one must always explicitly check the mass spectrum of these fields as well.

\subsubsection*{The Scalar Sector}

The decomposition of the adjoint representation of $\textrm{SO}(6,6)$ contains six scalars
\be
{\bf 66} \rightarrow 6 \cdot ({\bf 1},{\bf 1}) \oplus {\text{non-singlet representations}} \ ,
\ee
amongst which two of them correspond to the product $\textrm{SO}(6)\times \textrm{SO}(6)$ and therefore they are pure gauge. This implies that the scalar coset associated with the matter multiplets is parameterised in terms of only four physical scalars: two dilatons ($\varphi_1$, $\varphi_2$) and two axions $(\chi_1, \chi_2)$. The scalar coset in this sector reduces in the following way under the SO(3) truncation
\be
\frac{\textrm{SO}(2,2)}{\textrm{SO}(2)\times \textrm{SO}(2)} \ .
\ee
The explicit parameterisation of $\,M_{M N}\,$ is defined in terms of a symmetric $\,G\,$ and an antisymmetric $\,B\,$ matrices as
\be
\label{M_MN}
M_{M N} \equiv \left(
\begin{array}{cc}
G^{-1}  & - G^{-1} \, B \\
B \, G^{-1}    &  G - B\,G^{-1}\,B
\end{array}
\right) \ ,
\ee
where $\,G\,$ and $\,B\,$ are given by
\be
\label{G&B}
G= e^{\varphi_2-\varphi_1}\left(
\begin{array}{cc}
 \chi_2^2 + e^{-2\varphi_2} & -\chi_2\\
 -\chi_2 & 1
\end{array}
\right) \otimes \mathds{1}_{3}
\hspace{10mm} , \hspace{10mm}
B= \left(
\begin{array}{cc}
 0 & \chi_1 \\
 -\chi_1 & 0
\end{array}
\right) \otimes \mathds{1}_{3} \ .
\ee
In consequence, we will choose the vielbein $\,\mathcal{V}\,$ in (\ref{M=VV}) to be
\be
\mathcal{V} \equiv \left(
\begin{array}{cc}
 \boldsymbol{e}^{T} & 0 \\
 B \, \boldsymbol{e}^{T} & \boldsymbol{e}^{-1}
\end{array}
\right)\,\otimes\,\mathds{1}_{3}
\hspace{5mm} , \hspace{5mm}
\boldsymbol{e} \equiv
e^{(\varphi_1+\varphi_2)/2} \left(
\begin{array}{cc}
 1 & \chi_2 \\
 0 & e^{-\varphi_2}
\end{array}
\right) \ ,
\ee
with $\,\boldsymbol{e}^{T}\,\boldsymbol{e} = G^{-1}$.

Using this parameterisation of the scalar sector in the truncated theory, the kinetic terms then reduce to
\bea
\label{L_kin}
\mathcal{L}_{\textrm{kin}}&=&\dfrac{1}{8} \, (\partial M_{\alpha \beta})(\partial M^{\alpha \beta}) + \dfrac{1}{16} \, (\partial M_{MN})(\partial M^{MN}) \\[2mm]
&=& -\dfrac{1}{4}\left[(\partial\phi)^2+e^{2\phi}(\partial\chi)^2+3
(\partial\varphi_1)^2+3 \, e^{2\varphi_1}(\partial\chi_1)^2+3
(\partial\varphi_2)^2+3 \, e^{2\varphi_2}(\partial\chi_2)^2\right] \nonumber \ .
\eea

\subsubsection*{The Quadratic Constraints}

First of all, the number of allowed embedding tensor components turns out to be 40, arranged into 20 SL(2) doublets, 20 being the number of SO($3$)-singlets contained in the decomposition of the ${\bf 220}$ of SO($6,6$):
\begin{align}
  ({\bf 2, 220}) \rightarrow 20 \cdot ({\bf 2},{\bf 1}) \oplus {\text{non-singlet representations}} \ .
\end{align}
A convenient way of describing these $20$ $\textrm{SO}(3)$-invariant doublets is described in ref.~\cite{Derendinger:2004jn}, where the relevant components of the embedding tensor are classified using the $\textrm{SO}(2,2)\times \textrm{SO}(3)$ subgroup of $\textrm{SO}(6,6)$ with embedding ${\bf 12}=({\bf 4},{\bf 3})$. In this case, one can rewrite every $\textrm{SO}(6,6)$ index $M$ as a pair $(A\,I)$, where $I=1,2,3$ is
a fundamental $\textrm{SO}(3)$ index, whereas $A=1,...,4$ is a fundamental $\textrm{SO}(2,2)$ index. Due to this decomposition, the structure constants of the gauge algebra can be factorised as follows
\be
\label{defL}
f_{\alpha MNP} = f_{\alpha \, AI \, BJ \, CK}=\Lambda_{\alpha ABC}\,\,\,\epsilon_{IJK}\ ,
\ee
from which one can infer that the $\textrm{SO}(2,2)$-tensor $\,\Lambda_{ABC}\,$ is \emph{completely symmetric}. This observation takes us back to the number of $20$ as expected from the group theoretical decomposition. What one can now do, is to write down the QC \eqref{QC_xi=0} in terms of the $\Lambda$ tensor. One obtains
\begin{align}
\epsilon^{\alpha\beta}\,\Lambda_{\alpha\,
AB}^{\phantom{ABC\,}C}\,\Lambda_{\beta DEC}=0\,,\qquad
\Lambda_{(\alpha\,
A[B}^{\phantom{ABC\,\,\,\,\,\,}C}\,\Lambda_{\beta)\, D]EC}=0\,,
\label{QCL}
\end{align}
where the extra indices $\alpha,\beta=(+,-)$ still represent the SL$(2)$ phase.

The first set of constraints in \eqref{QCL} takes values in the following representation of $\textrm{SL}(2)\,\times\,\textrm{SO}(2,2)$
\Yvcentermath1
\begin{align}\label{rep1} \left(\boldsymbol{1}, \left(\tiny{\yng(2)\,\otimes\,\yng(2)}\right)_{a}\right)\ ,
\end{align}
which has dimension $45$, whereas the the second set of constraints in \eqref{QCL} takes values in this other one
\Yvcentermath1
\begin{align}\label{rep2}\left(\boldsymbol{3}, \tiny{\yng(2,2)}\,\right)\,,
\end{align}
which should not yet be thought of as only consisting of its irreducible (traceless) part and therefore it has dimension $63$. This leads us to $108$ as total amount of constraints, which can also be obtained by means of a computer. It turns out, though, that the number of independent constraints reduces to\footnote{This fact should be understood in the following way: the trace part of \eqref{rep2} is already implied by the remaining full set of constraints coming from both \eqref{rep1} and \eqref{rep2}.} $105$. We will come back to this point in the next section when investigating the superpotential formulation of our truncated theory.

\subsection*{Relation To Flux Compactifications}

So far, we have introduced the main features of the SO($3$) truncation of half-maximal supergravity in four dimensions. As we have seen in the previous section, the scalar manifold in the truncated theory reduces to
\be
\frac{\mathrm{SL}(2)}{\mathrm{SO}(2)}\,\times\,\frac{\mathrm{SO}(2,2)}{\mathrm{SO}(2)\,\times\,\mathrm{SO}(2)}\,\sim\,\left(\frac{\mathrm{SL}(2)}{\mathrm{SO}(2)}\right)^{3} \ ,
\ee
where each of the SL($2$) factors can be parameterised by a complex scalar field. The resulting supergravity models are commonly referred to in the literature as $STU$-models (see section~\ref{Geom_Flux_Comp}). They consist of three complex fields which are related to those entering the
$\,M_{\alpha \beta}\,$ matrix in (\ref{chi,phi}) and the $\,M_{MN}\,$ matrix in (\ref{M_MN}) -- through the metric $G$ and the $B$-field in (\ref{G&B}) -- by
\be
\label{complex_STU}
S \equiv \chi+i \, e^{-\phi}
\hspace{8mm},\hspace{8mm}
T \equiv \chi_1 + i \, e^{-\varphi_1}
\hspace{8mm} \textrm{and} \hspace{8mm}
U \equiv \chi_2 + i \, e^{-\varphi_2} \ .
\ee
Furthermore, the splitting $\,\textbf{4}\,\rightarrow\,\textbf{1}\,\oplus\,\textbf{3}\,$ of the fundamental representation of $\,\textrm{SU}(4) \sim \textrm{SO}(6)\,$ $R$-symmetry under the action of $\,\textrm{SO}(3)\,$ ensures an $\,\mathcal{N}=1\,$ structure of the supergravity describing the truncated theory. This implies that it has to be possible to formulate it in terms of a real K\"ahler potential $\,K(\Phi,\bar{\Phi})\,$ and a holomorphic superpotential $\,W(\Phi)\,$ (see section~\ref{section:N=1W}), where $\,\Phi=(S,T,U)\,$, by using the standard minimal supergravity formalism. According to it, the scalar potential can be worked out as
\be
\label{scalar_potential}
V = e^K \left(  \sum_{\Phi} K^{\Phi\bar \Phi} |D_\Phi W|^2 - 3|W|^2 \right)  \ ,
\ee
where $\,K^{\Phi \bar{\Phi}}\,$ denotes the inverse of the K\"ahler metric $\,K_{I\bar{J}} = \frac{\partial K}{\partial \Phi^{I} \partial \bar{\Phi}^{\bar{J}}}\,$, and $\,D_\Phi W = \frac{\partial W}{\partial \Phi} + \frac{\partial K}{\partial \Phi} W\,$ is the K\"ahler derivative.

\subsubsection*{The K\"ahler Potential}

Let us start by noticing that the kinetic Lagrangian in \eqref{L_kin} can be rewritten in terms of the complex fields in (\ref{complex_STU}) as
\be
\mathcal{L}_{\textrm{kin}}=K_{I\bar{J}}\,\partial\Phi^I\partial\bar{\Phi}^{\bar{J}}=\frac{\partial
S\partial \bar{S}}{\left(-i(S-\bar{S})\right)^2}+3 \, \frac{\partial
T\partial \bar{T}}{\left(-i(T-\bar{T})\right)^2}+3 \, \frac{\partial
U\partial \bar{U}}{\left(-i(U-\bar{U})\right)^2} \ ,
\ee
with $\,K_{I\bar{J}}\,$ being again the K\"ahler metric. The above kinetic terms are then reproduced from the K\"ahler potential 
\be
\label{Kahler_pot}
K = - \,\log\left( -i\,(S-\bar{S})\right)  - 3 \,\log\left(-i\,(T-\bar{T})\right)  -3 \,\log\left( -i\,(U-\bar{U})\right) \ ,
\ee
which matches the one obtained in string compactifications given in \eqref{Kaehler_STU} in the isotropic case and is valid to first order in the string and the sigma model perturbative expansions.

\subsubsection*{The Superpotential: Flux Backgrounds and Embedding Tensor}

Finding out the precise superpotential $\,W_{\textrm{SO}(3)}(\Phi)\,$ from which to reproduce the scalar potential in (\ref{V}) is certainly not an easy task. The reason for this is that both scalar potentials, namely the one computed from the superpotential and that of (\ref{V}), do not have to perfectly match each other but they have to coincide up to the QC in (\ref{QCL}).

As for the above  K\"ahler potential, we want the superpotential $\,W_{\textrm{SO}(3)}(\Phi)\,$ also to stem from (orientifolds of) some string compactifications from ten to four dimensions. Their compatibility with producing an SO($3$) truncation of half-maximal supergravity in four dimensions allows for a simple interpretation of the internal space of the compactification. It can be taken to be the factorised six-torus of figure~\ref{fig:Torus_Factor1} in appendix~\ref{App:fluxes} (see also section~\ref{Geom_Flux_Comp}) whose coordinate basis is denoted $\eta^m$ with $m=1,\ldots,\,6\,$, supplemented with a set of flux objects fitting the embedding tensor components $\,f_{\pm MNP}\,$ surviving the truncation.

The identification between the embedding tensor components (gauging parameters) in the supergravity side and the flux objects in the string compactification side crucially depends on the string theory under investigation. As an example, when considering $\,\cN=1\,$ type IIA orientifold compactifications including O$6$-planes and D$6$-branes, only a few embedding tensor components in the supergravity side are known to correspond to flux components on the string theory side. In contrast, all of them correspond to (generalised) fluxes in $\,\cN=1\,$ orientifold compactifications of type IIB string theory including O3/O7-planes and D3/D7-branes. In this type IIB scheme \cite{Aldazabal:2008zza,Dibitetto:2010rg}, as we just saw, the correspondence between embedding tensor components and fluxes entering the superpotential reads
\be
\begin{array}{cccccccc}
f_{+mnp} = \tilde{F'}_{mnp} & \hspace{1mm},\hspace{1mm} & {f_{+\,mn}}^{p} = {{Q'}_{mn}}^{p} &  \hspace{1mm},\hspace{1mm}  &  f_{+\,\phantom{mn}p}^{\phantom{\a}mn\phantom{p}} = Q_{\,\phantom{mn}p}^{\,mn\phantom{p}} &  \hspace{1mm},\hspace{1mm}  &  f_{+}^{\phantom{+}mnp} = \tilde{F}^{mnp} & ,\\[2mm]
f_{-mnp} = \tilde{H'}_{mnp} & \hspace{1mm},\hspace{1mm} & {f_{-\,mn}}^{p} = {{P'}_{mn}}^{p} &  \hspace{1mm},\hspace{1mm}  &  f_{-\,\phantom{mn}p}^{\phantom{\a}mn\phantom{p}} = P_{\,\phantom{mn}p}^{\,mn\phantom{p}} &  \hspace{1mm},\hspace{1mm}  &  f_{-}^{\phantom{-}mnp} = \tilde{H}^{mnp} & ,
\end{array}
\ee
where, for instance, $\tilde{F}^{mnp}\equiv \dfrac{1}{3!}\,\epsilon^{m \, n \, p \, m'n'p'} \, F_{m'n'p'}$. The correspondence between SO$(6,6)$ and SO$(2,2)$ embedding tensor components with known/conjectured flux objects in both type IIA and type IIB orientifold compactifications is presented in appendix~\ref{App:fluxes}.

Irrespective of the particular string theory realisation, we have explicitly checked that the scalar potential (\ref{V}) induced by the gaugings in the SO$(3)$ truncated theory is correctly reproduced, up to $\,\mathcal{N}=4\,$ QC, from the flux-induced superpotential given in \eqref{W_fluxes4}, using the standard results in minimal supergravity. 

The superpotential in (\ref{W_fluxes4}) was originally derived from a type II string theory approach in ref.~\cite{Aldazabal:2006up} by using duality arguments. Concretely, they worked out the $\,\cN=1\,$ duality invariant effective supergravity arising as the low energy limit of type II orientifold compactifications on the
$\,T^{6}/(\mathbb{Z}_{2} \times \mathbb{Z}_{2})\,$ toroidal orbifold. More recently, this has been put in the context of type IIB (with O3/O7-planes)/F-theory compactifications in ref.~\cite{Aldazabal:2008zza} and connected to generalised geometry in ref.~\cite{Aldazabal:2010ef}. Finally, some aspects of the vacua
structure of this supergravity have been explored in refs~\cite{Font:2008vd, Guarino:2008ik, deCarlos:2009qm} where only the unprimed fluxes inducing the polynomials in (\ref{Poly_unprim}) were considered.

A worthwhile final remark about the SO$(3)$ truncation of half-maximal supergravity in four dimensions is that the resulting scalar potential $\,V\,$ is left invariant by the action of a discrete $\,\mathbb{Z}_{2}=\left\lbrace 1 \,,\,\alpha_{1} \right\rbrace\,$ symmetry. This parity symmetry transforms simultaneously the moduli fields $\,\Phi=(S,T,U)\,$ and the different fluxes $\,f_{i}\,$ as
\be
\label{alpha1_sym}
\begin{array}{clccll}
 \alpha_{1} & : & \Phi   & \longrightarrow &  -\bar{\Phi} & , \\[2mm]
            &   &  f_{i} & \longrightarrow &  (-1)^{n_{1}+n_{2}+n_{3}} \, f_{i} & ,
\end{array}
\ee
where $\,f_{i}\,\, S^{n_{1}} T^{n_{2}} U^{n_{3}}\,$ denotes a generic term in the superpotential (\ref{W_fluxes4}). This transformation can be equivalently viewed as taking the superpotential from holomorphic to anti-holomorphic, \emph{i.e.}, $\,W(\Phi) \rightarrow$ $\,W(\bar{\Phi})$, without modifying the K\"ahler potential. This additional generator extends the SO$(2,2)$ part of the duality group to O$(2,2)$, while also acting with an element of determinant $-1$ on the SL$(2)$ indices.

\subsubsection*{Understanding the matching: are there unnecessary quadratic constraints?}

Let us go deeper into the matching between the $\,\mathcal{N}=1\,$ and $\cN = 4$ supergravity formulations of the theory. This equivalence happens to hold only after the $\,\mathcal{N}=4\,$ QC in (\ref{QCL}) are imposed on the $\,\mathcal{N}=1\,$ side as well. Some of those constraints happen
to kill some moduli dependences which are not allowed by $\,\mathcal{N}=4$, since they cannot be expressed in an $\,\textrm{SL}(2) \times \textrm{SO}(6,6)\,$ covariant way, whereas some others are only needed in order to recover the same coefficients in front of terms which are present in both of the theories. A further subtlety is that, in total, one only needs to impose $96$ out of the $105$ independent QC. This means that there are $9$ QC which do not seem to be needed in order for the matching to work. Going back to the representation theory analysis we started in \eqref{rep1} and \eqref{rep2}, one realises that \eqref{rep1} splits in the following
irrep's of SO($2,2$) in the case of the SO($3$) truncated theory
\Yvcentermath1 
\be \label{rep3}
\left(\,{\yng(2)\,\otimes\,\yng(2)}\,\right)_{a}={\yng(1,1)\,\oplus\,\yng(2)\,\oplus\,\yng(3,1)\ ,}
\ee
that is to say, a splitting of the $\boldsymbol{45}$ into $\boldsymbol{6}\,\,\oplus\,\,\boldsymbol{9}\,\,\oplus\,\,\boldsymbol{30}$. It turns out that all of the unneeded constraints combine together to give the $\boldsymbol{9}$ irreducible component in the right-hand side of \eqref{rep3}. The reason why these constraints are not needed still remains unclear but it is a peculiar feature of the SO($3$) truncation. This can be understood by going back to the full theory, where those constraints combine together with other ones into a bigger irrep of $\,\textrm{SL}(2) \times \textrm{SO}(6,6)\,$ and hence they have to be necessary as well as the other constraints in order to have a complete matching between the $\,\mathcal{N}=4\,$ and $\,\mathcal{N}=1\,$ scalar potentials.

Up to our knowledge, these results represent the first general demonstration\footnote{This point was also discussed in ref.~\cite{Aldazabal:2008zza} and we thank the authors for correspondence on their results.} of the explicit relation between the embedding tensor formulation of $\cN = 4$ supergravity and the superpotential formulation of $\cN =1$ supergravity in this particular truncation.

\section{Vacua of Geometric Type IIA Compactifications}
\label{sec:geom_IIA}

In this section we will analyse the complete vacua structure of the SO$(3)$ truncation of $\,\cN=4$ supergravity which arises as the low energy limit of certain type IIA orientifold compactifications including background fluxes, D6-branes and O6-planes. More concretely, it is obtained from type IIA orientifold compactifications on a $T^{6}/(\mathbb{Z}_{2} \times \mathbb{Z}_{2})$ isotropic orbifold in the presence of gauge R-R ($F_{0}$, $F_{2}$, $F_{4}$, $F_{6}$) and NS-NS $H_{3}$ fluxes, together with metric $\omega$ fluxes, D$6$-branes and O$6$-planes. In order to preserve half-maximal supersymmetry in four dimensions, the D$6$-branes have to be parallel to the O$6$-planes, \emph{i.e.} they wrap the $3$-cycle in the internal manifold which is invariant under the action of the orientifold involution\footnote{Sources invariant under the combined action of the orientifold involution and the orbifold group break from half-maximal to minimal supersymmetry in four dimensions.}.

According to the mapping between fluxes and SO$(3)$-invariant embedding tensor components listed in table~\ref{table:unprimed_fluxes}, this type IIA flux compactification gives rise to an $\,\cN=4\,$ gauged supergravity for which the possible gaugings are determined in terms of the electric and magnetic flux parameters
\be
\begin{array}{lclclclclc}
f_{+ \bar{a}\bar{b}\bar{c}} = -a_{0} & , & f_{+ \bar{a}\bar{b}\bar{k}}=a_{1}   &  ,  &  f_{+ \bar{a}\bar{j}\bar{k}}=-a_2  & , & f_{+ \bar{i}\bar{j}\bar{k}}=a_{3} & , \\[2mm]
f_{- \bar{a}\bar{b}\bar{c}}=-b_0 & , & f_{- \bar{a}\bar{b}\bar{k}}=b_{1} & , & f_{+ \bar{a}\bar{b}k} = c_{0} & , & f_{+ \bar{a}\bar{j} k}=f_{+ \bar{i}\bar{b} k}= c_{1} & , & f_{+ a\bar{b}\bar{c}}=\tilde{c}_{1} & .
\end{array}
\ee
It is worth noticing here that in the type IIA scheme: $\,(a_{0}, \,a_{1}, \,a_{2}, \,a_{3})\,$ are R-R fluxes, $\,(b_{0}, \,c_{0})\,$ are NS-NS $\,H_{3}$-fluxes and $\,(b_{1}, \,c_{1}, \,\tilde{c}_{1})\,$ are metric $\,\omega$-fluxes. As we just showed, this effective supergravity admits an $\cN=1$ formulation in terms of the K\"ahler potential in (\ref{Kahler_pot}) and the superpotential
\be
\label{W_IIA}
W_{\textrm{IIA}}=a_0 - 3 \, a_1 \, U + 3 \, a_2 \, U^2 - a_3 \, U^3 - b_0\,S + 3 \, b_1 \, S\,U + 3 \, c_0 \, T + (6 \, c_{1} - 3\, \tilde{c}_{1}) \, T\,U\ .
\ee
Observe how acting upon this supergravity with the non-compact part of the duality group, \emph{i.e.} rescalings and real shifts of the moduli fields, will not turn on new couplings in the superpotential (\ref{W_IIA}).

The QC in (\ref{QC}) coming from the consistency of the $\,\cN=4\,$ gauging give rise to the three flux relations
\be
\label{QC_IIA}
c_{1}\,(c_{1}-\tilde{c}_{1})=0 \hspace{7mm} , \hspace{7mm} b_{1}\,(c_{1}-\tilde{c}_{1})=0 \hspace{7mm} , \hspace{7mm} - a_{3} \, c_{0} - a_{2} \, (2\, c_{1}-\tilde{c}_{1})=0 \ .
\ee
The first and the second are respectively identified with the nilpotency ($d^{2}=\omega^{2}=0$) of the exterior derivative operator $d=\partial + \omega \, \wedge\,$ and the closure of the NS-NS flux background $\,dH_{3}=\omega \wedge H_{3}=0\,$. The third one is however related to the flux-induced tadpole
\be
\label{Tad6}
\int_{10\textrm{d}} (\omega \wedge F_{2} + H_{3} \wedge F_{0}) \wedge C_{7}
\hspace{5mm} \Rightarrow \hspace{5mm}
N_{6}=\omega \wedge F_{2} + H_{3} \wedge F_{0} \ ,
\ee
for the R-R gauge potential $\,C_{7}\,$ that couples to the D$6$-branes. In particular, it corresponds to the vanishing of the components along the internal directions orthogonal to the O$6$-planes,
\be
\label{N6_orth}
N^{\bot}_{6}= - a_{3} \, c_{0} - a_{2} \, (2\, c_{1}-\tilde{c}_{1}) = 0 \ .
\ee
In contrast, the component parallel to the O$6$-planes, denoted $\,N^{||}_{6}\,$, remains unrestricted since it can be cancelled by adding sources still preserving half-maximal supersymmetry
\be
N^{||}_{6}= 3 \, a_{2} \, b_{1} - a_{3} \, b_{0} \ .
\ee
Nevertheless, whenever $\,N^{||}_{6}=0\,$ for a consistent flux background, then the resulting gauged supergravity admits an embedding into an $\,\cN=8\,$ theory. As a result, the flux background does not induce a tadpole for the $\,C_{7}$ gauge potential, \emph{i.e.}, $N^{\bot}_{6}=N^{||}_{6}=0$, and an enhanced four-elements discrete $\,\mathbb{Z}_{2} \times \mathbb{Z}_{2}=\left\lbrace 1 \,,\, \alpha_{1} \,,\, \alpha_{2} \,,\, \alpha_{1}\alpha_{2} \right\rbrace\,$ symmetry group shows up when it comes to relate non-equivalent vacuum configurations.

This $\,\mathbb{Z}_{2} \times \mathbb{Z}_{2}\,$ discrete group is generated by the $\,\alpha_{1}$-transformation in (\ref{alpha1_sym}) and an extra parity transformation defined by
\be
\label{alpha2_sym}
\begin{array}{clccll}
 \alpha_{2} & : & U   & \longrightarrow &  -\bar{U} & , \\[2mm]
            &   &  f_{i} & \longrightarrow &  (-1)^{n_{3}+1} \, f_{i} & ,
\end{array}
\ee
where now $\,f_{i}\,\, S^{n_{1}} T^{n_{2}} U^{n_{3}}\,$ denotes a generic term in the superpotential of (\ref{W_IIA}). The action of the $\alpha_{2}$-transformation can equivalently be viewed as taking the original superpotential to a ``fake'' new one
\be
\label{fake_W}
W_{\textrm{IIA}}(S,T,U) \rightarrow -W_{\textrm{IIA}}(S,T,\bar{U}) \ .
\ee
As a consequence, the scalar potential gets also modified as $\,V \rightarrow V + \delta V\,$ where $\,\delta V\,$ takes the form
\be
\label{delta_V}
\delta V = \frac{1}{8\,(\textrm{Im}T)^{3}} \left[ \, 3 \, \left(\dfrac{\textrm{Im}T} {\textrm{Im}S}\right)\,N^{\bot}_{6} - N^{||}_{6} \, \right] \ .
\ee
Therefore, having $\,N^{\bot}_{6}= N^{||}_{6} = 0\,$ (equivalently an $\,\cN=8\,$ flux background) ensures $\,\delta V=0\,$ and hence a complete realisation of the $\,\mathbb{Z}_{2} \times \mathbb{Z}_{2}\,$ discrete group on the vacua distribution. The first relates a supersymmetric critical point to another supersymmetric one, while the second brings one to a pair of fake supersymmetric critical points \cite{fake}.

The aim of this section is to completely map out the vacua structure of these $\,\cN=4\,$ type IIA compactifications. In particular, we are computing the complete set of extrema of the flux-induced scalar potential as well as the number of supersymmetries which they preserve and their mass spectrum. In the appendix~\ref{App:N1_vacua}, we have also studied the effect of introducing O$6$/D$6$ sources breaking from half-maximal to minimal supersymmetry, namely $\,N^{\bot}_{6}\neq 0\,$, and their consequences from the moduli stabilisation perspective.

\subsection*{Full vacua analysis of the $\,\mathcal{N}=4\,$ theory}

Here we will present the complete vacua data of the $\,\mathcal{N}=4\,$ supergravity theory introduced above. By this we mean to specify:
\begin{enumerate}
\item The complete set of vacua forming the landscape of the theory and the connections among themselves.

\item The associated data for each of these solutions: vacuum energy, supersymmetries preserved, mass spectrum and stability under fluctuations of all the scalar fields in the $\,\cN=4\,$ theory.
\item The gauge group $\,G_{0}\,$ underlying the solutions.
\end{enumerate}

As it was explained in the previous section, algebraic geometry techniques are found to be powerful enough to find the entire set of extrema of the flux-induced scalar potential but, unfortunately, they will not give us any information about whether, and if so how, these extrema are linked to each other. To this respect, we will use the non-compact part $G_{n.c.}$ of the duality group in (\ref{Gnc}) together with the discrete group generated by the transformations in (\ref{alpha1_sym}) and (\ref{alpha2_sym}) as an organising principle to connect different vacuum solutions. These connections will shed light upon the often confusing landscape of $\,\cN=4\,$ flux vacua.

Our starting point is the ideal $\,I\,$ in (\ref{ideal_I}) consisting of the set of $\,\cN=4\,$ QC in (\ref{QC_IIA}) together with the six extremisation conditions of the scalar potential with respect to the real and imaginary parts of the $S$, $T$ and $U$ fields evaluated at the origin of the moduli space. After decomposing it into prime factors, as explained before, we are left with a set of simpler pieces which can be solved analytically. The outcome of this process is a splitting of the landscape of vacua into sixteen pieces of dim$=1$ and an extra piece of dim$=2$. Let us go deeper into the features of these critical points.

\subsubsection*{The sixteen critical points of dim$=1$}

The sixteen critical points of $\,\dim=1\,$ in the $\,\cN=4\,$ theory are presented in table~\ref{table:N=4_vacua}. More concretely, we list the associated flux backgrounds after having brought these moduli solutions to the origin of the moduli space, as it was explained in detail in the subsection about the analysis of critical points. The vacuum energy at the solutions turns out to be
\be
\label{V0_N=4}
V_{0}\left[ 1_{(s_1,s_2)} \right] = -\lambda^{2}
\hspace{3mm} , \hspace{3mm}
V_{0}\left[ 2_{(s_1,s_2)} \right] = V_{0}\left[ 4_{(s_1,s_2)} \right]= -\dfrac{32 \, \lambda^{2}}{27}
\hspace{3mm} , \hspace{3mm}
V_{0}\left[ 3_{(s_1,s_2)} \right] = -\dfrac{8 \, \lambda^{2}}{15}  \ .
\ee

\begin{table}[t!]
\renewcommand{\arraystretch}{1.80}
\begin{center}
\scalebox{0.75}[0.77]{
\begin{tabular}{ | c || c | c |c | c | c | c |c | c |}
\hline
\textrm{\textsc{id}} & $a_{0}$ & $a_{1}$ & $a_{2}$ & $a_{3}$ & $b_{0}$ & $b_{1}$ & $c_{0}$ & $c_{1}=\tilde{c}_{1}$ \\[1mm]
\hline \hline
$1_{(s_1,s_2)}$ & $s_{2} \,  \dfrac{3 \,\sqrt{10}}{2}\, \lambda $ & $s_{1} \,\dfrac{\sqrt{6}}{2} \, \lambda$ & $ - s_{2} \,\dfrac{\sqrt{10}}{6} \, \lambda$ & $s_{1} \, \dfrac{5\,\sqrt{6}}{6} \, \lambda$ & $-s_{1} \,s_{2} \, \dfrac{\sqrt{6}}{3} \, \lambda$ & $\dfrac{\sqrt{10}}{3}\,\lambda$ & $s_{1} \,s_{2} \, \dfrac{\sqrt{6}}{3}\,\lambda$ & $\sqrt{10} \, \lambda$   \\[1mm]
\hline \hline
$2_{(s_1,s_2)}$ & $s_{2} \,\dfrac{16 \, \sqrt{10}}{9} \,\lambda$ & $0$ & $0$ & $s_{1} \, \dfrac{16 \, \sqrt{2}}{9} \, \lambda$ & $0$ & $\dfrac{16 \, \sqrt{10}}{45} \, \lambda$ & $0$ & $\dfrac{16 \, \sqrt{10}}{15} \, \lambda$   \\[1mm]
\hline
$3_{(s_1,s_2)}$ & $s_{2} \,\dfrac{4\,\sqrt{10}}{5}\,\lambda$ & $-s_{1} \, \dfrac{4\,\sqrt{30}}{15}\,\lambda$ & $s_{2} \, \dfrac{4\,\sqrt{10}}{15}\,\lambda$ & $s_{1} \,\dfrac{4\,\sqrt{30}}{15}\,\lambda$ & $s_{1} \, s_{2} \,\dfrac{4\,\sqrt{30}}{15}\,\lambda$ & $\dfrac{4\,\sqrt{10}}{15}\,\lambda$ & $-s_{1} \, s_{2} \,\dfrac{4\,\sqrt{30}}{15}\,\lambda$ & $\dfrac{4\,\sqrt{10}}{5}\,\lambda$   \\[1mm]
\hline
$4_{(s_1,s_2)}$ & $s_{2} \,\dfrac{16 \, \sqrt{10}}{9} \,\lambda$ & $0$ & $0$ & $s_{1} \,\dfrac{16 \, \sqrt{2}}{9} \,\lambda$ & $0$ & $\dfrac{16 \, \sqrt{2}}{9} \,\lambda$ & $0$ & $\dfrac{16 \, \sqrt{2}}{9} \,\lambda$  \\
\hline
\end{tabular}
}
\end{center}
\caption{{\it List of the sixteen critical points of the $\mathcal{N}=4$ theory generated by R-R $\,(a_{0,1,2,3})\,$, NS-NS $\,(b_{0},c_{0})\,$ and metric $\,(b_{1},c_{1},\tilde{c}_{1})\,$ flux backgrounds in type IIA scenarios. They can be organised into four groups each of which consists of four equivalent solutions labelled by a pair $\,(s_{1},s_{2})\equiv\left\lbrace (+,+) , (+,-) , (-,+) , (-,-) \right\rbrace$. The quantity $\,\lambda\,$ is a free parameter setting the AdS energy scale $\,V_{0} \propto -\lambda^{2}\,$ at the solutions.}}
\label{table:N=4_vacua}
\end{table}

As we already discussed before, the number of supersymmetries preserved in these solutions can be computed from the gravitini mass matrix $\,A^{ij}_{1}\,$ in (\ref{kappa-eigenvalues}). After solving the eigenvalues equation of (\ref{susy_cond}), we find that all the solutions of the $\,\mathcal{N}=4\,$ theory are non-supersymmetric except those ones labelled by $\,1_{(+,+)}\,$ and $\,1_{(-,+)}\,$ which turn out to preserve $\mathcal{N}=1$ supersymmetry. Nevertheless, it is worth noticing here that they all actually enjoy an embedding in an $\,\mathcal{N}=8\,$ theory due to the lack of flux-induced tadpoles for the local sources\footnote{The condition $\,N_{6}^{||}=0\,$ is in fact implied by the $\mathcal{N}=4$ QC and two of the three axionic field equations provided $c_0\,a_1\neq 0$. This is the case for the solutions $1_{(s_1,s_2)}$ and $3_{(s_1,s_2)}$ in table~\ref{table:N=4_vacua}, whereas for the flux background in the remaining cases it is straightforward.}, \emph{i.e.},
\be
N_{6}^{\bot} = N_{6}^{||} =  0 \ .
\ee
This observation was previously made for the $\,\cN=1\,$ type IIA supersymmetric solution found in ref.~\cite{Dall'Agata:2009gv}. Now we are extending the statement about the existence of an $\,\cN=8\,$ lifting to the complete vacuum structure of the theory including both minimal supersymmetric and non-supersymmetric solutions. This fact has two immediate implications, the second actually been a direct consequence of the first:
\begin{itemize}
\item[$i)$] The discrete $\,\mathbb{Z}_{2}\,$ group generated by the $\,\alpha_{2}$-transformation in (\ref{alpha2_sym}) is ``accidentally'' realised as a symmetry of the flux-induced scalar potential $V(\Phi)$.  Then a complete discrete symmetry group $\,\mathbb{Z}_{2} \times \mathbb{Z}_{2}=\left\lbrace 1 \,,\, \alpha_{1} \,,\, \alpha_{2} \,,\, \alpha_{1}\alpha_{2} \right\rbrace\,$ appears in the landscape of the $\,\cN=4\,$ theory connecting solutions through the chain
\be
\label{disc_chain}
N_{(+,+)} \,\,\overset{\alpha_{1}}{\longrightarrow}\,\,
N_{(-,+)} \,\,\overset{\alpha_{2}}{\longrightarrow}\,\,
N_{(-,-)} \,\,\overset{\alpha_{1}}{\longrightarrow}\,\,
N_{(+,-)} \,\,\overset{\alpha_{2}}{\longrightarrow}\,\,
N_{(+,+)} \ ,
\ee
where $\,N=1,2,3,4\,$ stands for the four groups of solutions $\,N_{(s_{1},s_{2})}\,$ in table~\ref{table:N=4_vacua}. In fact, we have checked that combining these discrete transformations with the continuous non-compact part $\,G_{n.c.}\,$ in (\ref{Gnc}) of the duality group, the vacua structure of the theory turns out to be a net of extrema connected by elements of the enhanced group
\be
\label{Gvac}
G_{vac} = G_{n.c.} \times \mathbb{Z}_{2} \times \mathbb{Z}_{2} \ .
\ee
As it is shown in figure~\ref{fig:net}, all the sixteen critical points of $\,\dim=1\,$ in the $\,\cN=4\,$ theory are then connected to each other by an element of $G_{vac}$.

\begin{figure}[h!]
\vspace{5mm}
\begin{center}
\scalebox{0.7}[0.7]{
\includegraphics[keepaspectratio=true]{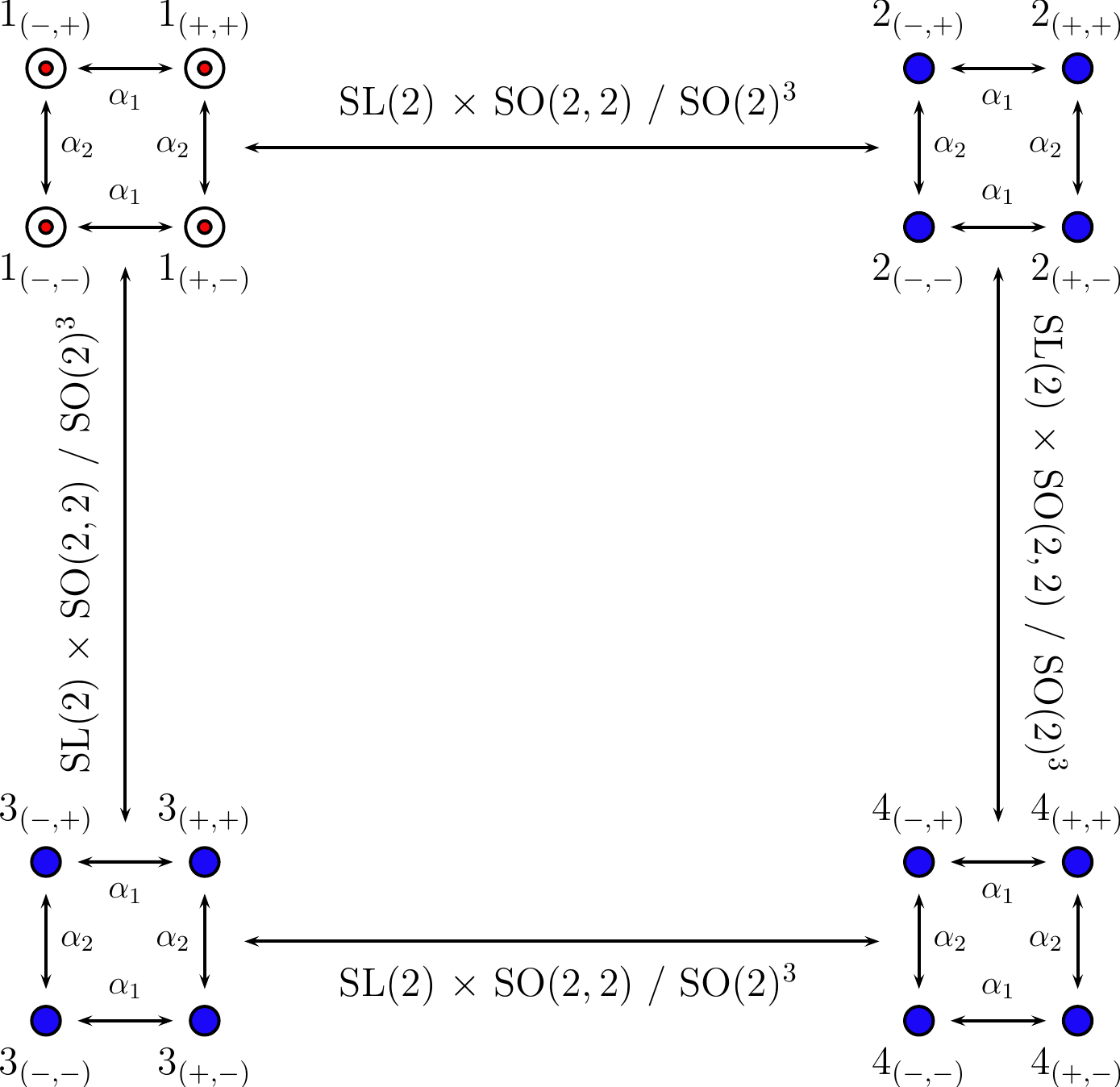}
}
\end{center}
\caption{\it{Net of connections between the $\,\dim=1\,$ sixteen critical points of the $\,\cN=4\,$ theory.}}
\label{fig:net}
\end{figure}

\item[$ii)$] Since the $\,\alpha_{2}$-transformation in (\ref{alpha2_sym}) is an accidental symmetry of the scalar potential but not of the superpotential, then the existence of non-supersymmetric and nevertheless stable solutions is guaranteed as long as there are supersymmetric ones. The reason is that these non-supersymmetric solutions would be ``fake'' supersymmetric in the sense that they do correspond to supersymmetric solutions of the ``fake'' superpotential in (\ref{fake_W}). Consequently, all the results concerning stability of supersymmetric solutions still apply to these non-supersymmetric ones since the scalar potential is left invariant. Supersymmetric and ``fake'' supersymmetric (non-supersymmetric) solutions of the theory are then connected by
\be
\label{susy/fake_chain}
\begin{array}{ccccccc}
\textrm{\footnotesize{SUSY}}&  & \textrm{\footnotesize{SUSY}}  &  & \textrm{\footnotesize{FAKE SUSY}}  &  &  \textrm{\footnotesize{FAKE SUSY}} \\
1_{(+,+)} &\overset{\alpha_{1}}{\longrightarrow}&
1_{(-,+)} &\overset{\alpha_{2}}{\longrightarrow}&
1_{(-,-)} &\overset{\alpha_{1}}{\longrightarrow}&
1_{(+,-)}
\end{array}
\,\,\,  \ .
\nonumber
\ee
We will see this explicitly by computing the full mass spectrum associated to these solutions and checking that they coincide.
\end{itemize}

The first step to check stability involves computing the masses only for the SO$(3)$-invariant fields, namely the $\,\textrm{SL}(2)/\textrm{SO}(2)\,$ axiodilaton $\,S\,$ and the two $\,\textrm{SO}(2,2)/\textrm{SO}(2)^{2}\,$ moduli fields $\,T\,$ and $\,U$. Nonetheless, stability of a solution under fluctuations of these $\,2+4=6\,$ real fields does not imply stability with respect to the rest of the $\,\mathcal{N}=4\,$ scalars which may render it unstable. The set of normalised masses of the SO$(3)$-invariant scalars at the sixteen $\,\dim=1\,$ extrema of the $\,\mathcal{N}=4\,$ theory are summarised in table~\ref{table:N=4_masses}. As we anticipated, they do not depend on the choice of a particular $\,(s_{1},s_{2})\,$ solution within a $\,N_{(s_{1},s_{2})\,}$ group.

\begin{table}[h!]
\renewcommand{\arraystretch}{1.80}
\begin{center}
\scalebox{0.9}[0.9]{
\begin{tabular}{ | c || c | c | c | c | c | c | c |}
\hline
\textrm{\textsc{id}} & $m_{1}^{2}$ & $m_{2}^{2}$ & $m_{3}^{2}$ & $m_{4}^{2}$ & $m_{5}^{2}$ & $m_{6}^{2}$ & BF \\
\hline \hline
$1_{(s_1,s_2)}$ & $0$ & $-\dfrac{2}{3}$ & $\dfrac{4 + \sqrt{6}}{3}$ & $\dfrac{4 - \sqrt{6}}{3}$ & $\dfrac{47 + \sqrt{159}}{9}$ & $\dfrac{47 - \sqrt{159}}{9}$ & $m^{2}=-\dfrac{2}{3} \rightarrow \textrm{stable}$ \\[1mm]
\hline\hline
$2_{(s_1,s_2)}$ & $0$ & $-\dfrac{4}{5}$ & $-\dfrac{2}{5}$ & $2$ & $\dfrac{64}{15}$ & $\dfrac{20}{3}$ & $m^{2}=-\dfrac{4}{5} \rightarrow \textrm{unstable}$ \\[1mm]
\hline
$3_{(s_1,s_2)}$ & $0$ & $0$ & $2$ & $2$ & $\dfrac{20}{3}$ & $\dfrac{20}{3}$ & $\min$ \\[1mm]
\hline
$4_{(s_1,s_2)}$ & $0$ & $0$ & $\dfrac{4}{3}$ & $2$ & $6$ & $\dfrac{20}{3}$ & $\min$ \\[1mm]
\hline
\end{tabular}
}
\end{center}
\caption{{\it Eigenvalues of the SO($3$)-truncated canonically normalised mass matrix at the AdS$_{4}$ extrema of the scalar potential in the $\mathcal{N}=4$ theory. For those being saddle points, the last column shows their stability according to the BF bound in (\ref{BF_bound}).}}
\label{table:N=4_masses}
\end{table}

Up to this point, the given information about the mass spectrum and stability of solutions is still incomplete. In order to determine whether these critical points are actually stable under fluctuations of all the scalar fields in the $\,\mathcal{N}=4\,$ theory, we have to compute the full mass spectrum. As already anticipated, we have made use of the mass formula provided in ref.~\cite{Borghese:2010ei} to address the issue of stability. The computation of the complete mass spectrum for the sixteen $\,\dim=1\,$ solutions of the $\,\mathcal{N}=4\,$ geometric type IIA compactifications gives the following results:

\begin{itemize}
\item The normalised scalar field masses and their multiplicities for the four solutions $\,1_{(s_{1},s_{2})}\,$ take the values of
\be
\begin{array}{lcrr}
\dfrac{1}{9} \left(47 \pm \sqrt{159}\right)\,\,(\times 1)
\hspace{8mm} , \hspace{8mm} \dfrac{1}{3} \left(4 \pm
\sqrt{6}\right)\,\,(\times 1) & , &
\dfrac{29}{9}\,\,(\times 3) & ,\\[4mm]
\dfrac{1}{18} \left( \, 89 + 5 \, \sqrt{145} \pm \sqrt{606 + 30 \, \sqrt{145}} \, \right)\,\,(\times 5)
& \hspace{5mm} , \hspace{5mm} &
0\,\,(\times 10) & , \\[4mm]
\dfrac{1}{18} \left(\, 89 - 5 \, \sqrt{145} \pm \sqrt{606 - 30 \, \sqrt{145}} \, \right)\,\,(\times 5)
& \hspace{5mm} ,\hspace{5mm} &
-\dfrac{2}{3}\,\,(\times 1) & .
\end{array}
\nonumber
\ee
The unique tachyonic scalar then implies $\,m^2= - \tfrac23\,$ so these AdS$_4$ solutions satisfy the BF bound in (\ref{BF_bound}) hence being totally stable. Notice that the dangerous tachyonic mode has a special mass value, corresponding to a massless supermultiplet and being identical to that of a conformally coupled scalar field in AdS$_4$ \cite{Townsend}. In terms of group theory, it corresponds to the discrete unitary irreducible representation for AdS$_4$, while all other masses with $m^2 \geq - \tfrac34$ comprise a continuous family of such irreps.

\item The normalised scalar field masses and their multiplicities for the four solutions $\,2_{(s_{1},s_{2})}\,$ take the values of
\be
\begin{array}{c}
\dfrac{1}{15} \left(77 \pm 5 \, \sqrt{145}\right)\,\,(\times
5) \hspace{4mm} , \hspace{4mm} \dfrac{2}{15} \left(31 \pm
\sqrt{145}\right)\,\,(\times 5) \hspace{4mm} , \hspace{4mm}
\dfrac{64}{15}\,\,(\times 1) \hspace{4mm} , \hspace{4mm}
\dfrac{20}{3}\,\,(\times 1) \ , \\[4mm]
\dfrac{46}{15}\,\,(\times 3) \hspace{5mm} , \hspace{5mm}
2\,\,(\times 1) \hspace{5mm} , \hspace{5mm}
0\,\,(\times 10) \hspace{5mm} , \hspace{5mm}
-\dfrac{2}{5}\,\,(\times 1) \hspace{5mm} , \hspace{5mm}
-\dfrac{4}{5}\,\,(\times 1) \ .
\end{array}
\nonumber
\ee
In this case the most tachyonic mode gives rise to $\,m^2=-4/5\,$ that is below the BF bound in (\ref{BF_bound}), so these AdS$_4$ solutions become unstable under fluctuations of this mode.

\item The normalised scalar field masses and their multiplicities for the four solutions $\,3_{(s_{1},s_{2})}\,$ take the values of
\be
\frac{1}{3} \left(19\pm\sqrt{145}\right)\,\,(\times
10) \hspace{4mm} , \hspace{4mm} \frac{20}{3}\,\,(\times 2)
\hspace{4mm} , \hspace{4mm} \frac{14}{3}\,\,(\times 3)
\hspace{4mm} , \hspace{4mm} 2\,\,(\times 2)
\hspace{4mm} , \hspace{4mm} 0\,\,(\times 11) \ , \nonumber
\ee
whereas those corresponding to the four solutions $\,4_{(s_{1},s_{2})}\,$ are given by
\be
\dfrac{20}{3}\,\,(\times 1) \hspace{4mm} , \hspace{4mm}
6\,\,(\times 6) \hspace{4mm} , \hspace{4mm}
\dfrac{8}{3}\,\,(\times 5) \hspace{4mm} , \hspace{4mm}
2\,\,(\times 4) \hspace{4mm} , \hspace{4mm}
\dfrac{4}{3}\,\,(\times 6) \hspace{4mm} , \hspace{4mm}
0\,\,(\times 16) \,\ .
\nonumber
\ee
One observes that all the normalised masses are non-negative so these AdS$_{4}$ solutions do actually correspond to stable extrema of the scalar potential.
\end{itemize}
Therefore, this shows that most of the AdS$_{4}$ moduli solutions of the $\,\mathcal{N}=4\,$ theories coming from geometric type IIA flux compactifications are non-supersymmetric and nevertheless stable even when considering all the $\,2+36=38\,$ scalar fields\footnote{It would be interesting to understand the (dis-)similarities with the non-supersymmetric vacua of \cite{Lust:2008zd,Koerber:2010rn}.}.

A point to be highlighted is that, in this type IIA case, the SO($3$) truncation turns out to capture the interesting dynamics of the scalars, in the sense that the lightest mode is always kept by the truncation. This is by no means guaranteed by the consistency of the truncation. Indeed, there are $\cN = 8$ examples of consistent truncations where the non-singlets lead to instabilities of critical points that are stable with respect to the singlet sector \cite{Warner:1006}. The situation for the critical points here differs from this in two respects. Firstly, the non-singlet masses always lie above the lightest mode in the singlet sector. Moreover, the non-singlet masses are in fact always non-negative. 

Another remarkable feature is that the supersymmetric solutions $\,1_{(+,+)}\,$ and $\,1_{(-,+)}\,$ are not the (stable) ones with highest potential energy. Indeed, the solutions $\,3_{(s_{1},s_{2})}\,$ are non-supersymmetric and still stable with a higher vacuum energy, as can be read from (\ref{V0_N=4}).  This again differs from the situation in the prototypical $\cN = 8$ supergravity with SO$(8)$ gauging, where the vacuum that preserves all supersymmetry has the highest potential energy of all known critical points \cite{Fischbacher:2009cj}.

Finally we want to identify the gauge group(s) $\,G_{0}\,$ underlying these solutions. The antisymmetry of the brackets in (\ref{SW_algebra_ZX}), when restricted to the fluxes compatible with type IIA geometric backgrounds, allows to write the magnetic generators in terms of the electric ones
\be
X_{-}{}^{a} = - \frac{(b_1\,c_0 + b_0\,c_1)}{c_1 \, \tilde{c}_{1}} \,\, Z_{+a} \,+\, \frac{b_{1}}{\tilde{c}_{1}} \,\, Z_{+i}
\hspace{5mm} , \hspace{5mm}
X_{-}{}^{i} = \frac{b_1}{c_1} \,\, Z_{+a}
\hspace{5mm} , \hspace{5mm}
Z_{-a}=Z_{-i} = 0 \ ,
\ee
with pairs $\,(a,i)=\left\lbrace (1,2),(3,4),(5,6) \right\rbrace\,$. Notice that $\,c_1 \, \tilde{c}_{1} \neq 0\,$ for all the solutions listed in table~\ref{table:N=4_vacua}. In terms of electric generators, the algebra $\,\mathfrak{g}_{0}\,$ of $\,G_{0}\,$ is expressed as a twelve-dimensional algebra which is now suitable to define a consistent gauging of the theory. The brackets involving isometry-isometry generators are given by
\be
[Z_{+a},Z_{+b}]=[Z_{+a},Z_{+j}]=[Z_{+i},Z_{+j}]=0 \ ,
\ee
and then span an abelian $\,\mathfrak{u}(1)^{6\,}$ subalgebra of $\,\mathfrak{g}_{0}$. Furthermore, the mixed non-vanishing isometry-gauge brackets read
\be
[Z_{+a},X_{+}{}^{b}] = \tilde{c}_{1}\,Z_{+c}
\hspace{5mm} , \hspace{5mm}
[Z_{+i},X_{+}{}^{b}] = c_{0}\,Z_{+c} \,+\, c_{1} \, Z_{+k}
\hspace{5mm} , \hspace{5mm}
[Z_{+i},X_{+}{}^{j}] = c_{1} \, Z_{+c} \ ,
\ee
so the isometry generators actually determine an abelian ideal within $\,\mathfrak{g}_{0}$. Accordingly to the Levi's decomposition theorem, the algebra $\,\mathfrak{g}_{0}\,$ can then be written as
\be
\mathfrak{g}_{0} = \mathfrak{g}_{\textrm{gauge}} \oplus \mathfrak{u}(1)^{6} \ ,
\ee
where $\,\mathfrak{g}_{\textrm{gauge}}\,$ has to be read off from the gauge-gauge brackets after quotienting $\,\mathfrak{g}_{0}\,$ by the abelian ideal. They take the form of
\be
[X_{+}{}^{a},X_{+}{}^{b}] = \tilde{c}_{1} \, X_{+}{}^{c} + c_{0}\,X_{+}{}^{k}
\hspace{5mm} , \hspace{5mm}
[X_{+}{}^{a},X_{+}{}^{j}] = c_{1}\,X_{+}{}^{k}
\hspace{5mm} , \hspace{5mm}
[X_{+}{}^{i},X_{+}{}^{j}] = 0 \ ,
\ee
so the gauge-gauge brackets are identified with $\,\mathfrak{g}_{\textrm{gauge}}=\mathfrak{iso}(3)$. As a result, the algebra $\,\mathfrak{g}_{0}\,$ turns out to be
\be
\label{g_IIA_algebra}
\mathfrak{g}_{0} \,\,=\,\,  \mathfrak{iso}(3) \oplus \mathfrak{u}(1)^{6} \,\,\sim\,\, \mathfrak{so}(3) \oplus \mathfrak{nil}_{9}(2) \ ,
\ee
where $\,\mathfrak{nil}_{9}(2)\,$ denotes a nilpotent $9$-dimensional ideal of order two (three steps) spanned by the generators $\,\left\lbrace X_{+}{}^{i} \,,\, Z_{+a} \,,\, Z_{+i} \right\rbrace\,$ and with lower central series
\be
\left\lbrace X_{+}{}^{i} \,,\, Z_{+a} \,,\, Z_{+i} \right\rbrace \supset \left\lbrace Z_{+a} \,,\, Z_{+i} \right\rbrace \supset 0 \ .
\ee
The main property to be highlighted is that there is an unique gauge group, \emph{i.e.},
\be
G_{0} = \textrm{ISO}(3) \ltimes \textrm{U}(1)^{6} \ ,
\ee
underlying all the solutions of the IIA geometric theory. This was already noted for the supersymmetric solution in ref.~\cite{Dall'Agata:2009gv}. As a final remark, none of the generators in the adjoint representation vanishes at these solutions, so the algebra $\,\mathfrak{g}_{0}\,$ in (\ref{g_IIA_algebra}) is actually embeddable within the $\,\mathfrak{so}(6,6)\,$ duality group.

The above gauge group has three compact and nine non-compact generators. The latter are spontaneously broken at all critical points. The corresponding vector bosons in such cases acquire a mass due to gauge symmetry breaking by absorbing a scalar degree of freedom. In the scalar mass spectra listed above, there will always be nine scalar fields that  do not correspond to propagating degrees of freedom. Being pure gauge, these do not appear in the scalar potential and hence have $m^2 = 0$.

In all critical points considered above, the number of scalar fields with $m^2 = 0$ exceeds nine. This implies that there will always be a number of propagating degrees of freedom whose value is not fixed by the quadratic terms in $V$. Of course there could be higher-order terms that do give rise to moduli stabilisation, or could lead to a negative potential energy. However, in contrast to the Minkowski case, such scalar fields do not represent a potential instability due to the additional contribution from the space-time curvature. Instead, in Anti-de Sitter one should be worried about fields whose quadratic mass term is at the BF bound, and if possible verify if their higher-order terms give rise to stability or rather to tachyons. Having no such mass values in our spectra, this issue plays no role here.

\subsubsection*{The critical point solution of $\dim=2$}

Besides the previous sixteen critical points, the landscape of the $\,\cN=4\,$ type IIA geometric theory still has a $\,\dim=2\,$ piece. In terms of the flux background, it is given by\footnote{We would like to stress that the superpotential induced by gauge fluxes in type IIB enjoys an extra compact SO($2)_{S}$ duality symmetry which can be used to rewrite the solution in \eqref{IIA_GKP} in terms of only one physical parameter.\label{footnote:GKP}}
\be
\label{IIA_GKP}
c_{0} = c_{1} = \tilde{c}_{1} = 0
\hspace{5mm},\hspace{5mm}
a_0 = a_1 = 0
\hspace{5mm},\hspace{5mm}
b_1 = a_2
\hspace{5mm},\hspace{5mm}
b_0 =-a_3 \ .
\ee
After three T-dualities along the $\,\eta^{a}\,$ directions, where $\,a=1,3,5$, this type IIA background is mapped to a type IIB one only involving certain gauge fluxes (see table~\ref{table:unprimed_fluxes}). We postpone the discussion of this solution to the next section where type IIB backgrounds including gauge fluxes, O$3$-planes and D$3$-branes are being explored in full generality.

\section{Vacua of (Non-)Geometric Type IIB Compactifications}
\label{non-geom_IIB}

In this final part we study other realisations of the SO$(3)$-truncation of half-maximal supergravity in four dimensions. This time it will be in the context of isotropic type IIB compactifications on $T^{6}/(\mathbb{Z}_{2} \times \mathbb{Z}_{2})$ including generalised background fluxes.

\subsection*{GKP Flux Compactifications: Stability and Gaugings}
\label{sec:extrema_N=4_IIB}

Let us start with the well known type IIB string compactifications including a background for the gauge fluxes $\,(H_{3},F_{3})\,$ and eventually O$3$-planes and/or D$3$-branes sources in order to cancel a flux-induced tadpole
\be
\label{Tad3}
\int_{10\textrm{d}} ( H_{3} \wedge F_{3} ) \wedge C_{4}
\hspace{5mm} \Rightarrow \hspace{5mm}
N_{3} = H_{3} \wedge F_{3} \ ,
\ee
for the R-R gauge potential $\,C_{4}$. These compactifications were presented in the seminal GKP paper of ref.~\cite{Giddings:2001yu} (see also section~\ref{Geom_Flux_Comp}) and deeply explored from the moduli stabilisation point of view in refs~\cite{Kachru:2002he,Frey:2002hf,DeWolfe:2004ns,DeWolfe:2005uu} among many others.

When compatible with an SO($3$) truncation of half-maximal supergravity, these compactifications correspond to having non-vanishing
$\,(a_{0}, \,a_{1}, \,a_{2}, \,a_{3})\,$ as well as $\,(b_{0},\,b_{1}, \,b_{2}, \,b_{3})\,$ flux components in table~\ref{table:unprimed_fluxes4}. The flux-induced superpotential for the resulting $STU$-models then reads
\be
\label{W_IIB_GKP}
W_{\textrm{GKP}}= a_0 - 3 \, a_1 \, U + 3 \, a_2 \, U^2 - a_3 \, U^3 + \left( \,b_0 - 3 \, b_1 \, U + 3 \, b_2 \, U^2 - b_3 \, U^3 \,\right) S \ ,
\ee
and the theory comes out with a non-scale structure \cite{Cremmer:1983bf}. It is worth noticing at this point that in these IIB models with only gauge fluxes there are no QC from (\ref{QCL}) to fulfill.

At the origin of the moduli space, the potential energy arranges into a sum of square terms hence being non-negative defined
\be
\label{Mink_cond}
V_{0}= \frac{1}{32} \left( \,\,(a_0 - b_3)^2 + 3 \,(a_1+b_2)^2 + 3 \, (a_2 - b_1)^2 + (a_3 + b_0)^2 \,\,\right) \ .
\ee
Using the stabilisation of the imaginary part of the modulus $T$, it can be shown that there is no solution to the extremum conditions without satisfying $\,V_{0}=0\,$, \emph{i.e.}, any solution will be a Minkowski extremum. Then the $\,H_{3}\,$ flux background is related to the $\,F_{3}\,$ one via\footnote{The same comment made in footnote~\ref{footnote:GKP} applies here for the solution in \eqref{FH_GKP}.}
\be
\label{FH_GKP}
b_3 = a_0 \hspace{5mm},\hspace{5mm} b_2 =- a_1 \hspace{5mm},\hspace{5mm} b_1 = a_2 \hspace{5mm},\hspace{5mm} b_0 =-a_3 \ ,
\ee
and the flux-induced tadpole in (\ref{Tad3}) simply reads
\be
N_{3}= a_{0}^{2} + 3 \, a_{1}^{2} + 3 \, a_{2}^{2} + a_{3}^{2} \ .
\ee
The $\,\kappa_{1}\,$ and $\,\kappa_{2}\,$ values entering the gravitini mass matrix $\,A^{ij}_{1}\,$ in (\ref{kappa-eigenvalues}), and then determining the amount of supersymmetry preserved at an extremum, are given by
\be \kappa_{1} = \frac{3}{4 \, \sqrt{2}} \, \sqrt{\left( a_{0}- 3\,
a_{2}\right)^{2} + \left( a_{3}- 3\, a_{1}\right)^{2}}
\hspace{5mm},\hspace{5mm} \kappa_{2} = \frac{3}{4 \, \sqrt{2}} \,
\sqrt{\left( a_{0} + a_{2}\right)^{2} + \left( a_{1} +
a_{3}\right)^{2}} \ . \ee
As a consequence, a generic GKP solution will be non-supersymmetric. However, let us comment about two interesting limits which give rise to solutions that preserve certain amount of supersymmetry:
\begin{itemize}

\item The first limit is that of taking $\,a_{0} = 3 \, a_{2}\,$ and $\,a_{3} = 3 \, a_{1}$. This limit results in $\,\kappa_{1} = 0\,$ and $\,\kappa_{2}=\frac{3 \sqrt{a_{1}^2 + a_{2}^2}}{\sqrt{2}}\,$ so that the solutions preserve $\mathcal{N}=1$ supersymmetry.

\item The second limit is that of taking $\,a_{0} = - a_{2}\,$ and $\,a_{3} = -a_{1}$. This limit results in $\,\kappa_{2} = 0\,$ and $\,\kappa_{1}=\frac{3 \sqrt{a_{1}^2 + a_{2}^2}}{\sqrt{2}}\,$ so that the solutions preserve $\,\mathcal{N}=3\,$ supersymmetry \cite{Frey:2002hf}.
\end{itemize}

Let us now present the mass spectrum of these $\,\mathcal{N}=4\,$ compactifications\footnote{The numerical values of the eigenvalues of the mass matrix were computed in ref.~\cite{Saltman:2004sn} for some de Sitter GKP examples corresponding to non-isotropic moduli VEVs.}. In terms of the quantities
\be
\begin{array}{cclc}
M&=&\dfrac{1}{16} \, \Big(\, 9 \, \left( a_1^2 + a_2^2\right) + 6 \, (a_0 \, a_2 + a_1 \, a_3) + 5 \, (a_0^2 + a_3^2) \, \Big)  & , \\[5mm]
N&=&\dfrac{1}{16} \, \Big( \, 5 \, \left( a_1^2 + a_2^2\right) - 2 \,
(a_0 \, a_2 + a_1 \, a_3) + (a_0^2 + a_3^2) \, \Big) & ,
\\[5mm]
Q&=& \dfrac{1}{16} \, \sqrt{ \Big( \,(a_0 - 3 \, a_2)^2 + ( a_3 - 3 \,
a_1)^2 \, \Big) \, \Big( \, (a_0 + a_2)^2 + ( a_1 + a_3)^2 \,
\Big)} & ,
\end{array}
\ee
the moduli (masses)$^{2}$ as well as their multiplicities are given by
\be
\begin{array}{c}
M \pm 3 \, Q \,\,(\times 1) \hspace{5mm} , \hspace{5mm} N \pm
Q\,\,(\times 6) \hspace{5mm} , \hspace{5mm} \dfrac{1}{8} \, \left(
\, (a_0 + a_2)^{2} + (a_1 + a_3)^{2} \, \right) \,\,(\times 3)
\hspace{5mm} , \hspace{5mm} 0\,\,(\times 21) \ . \nonumber
\end{array}
\ee
Only the third of the above masses is not recovered when considering only the scalars of the SO$(3)$ truncation. Clearly though, these solutions can never be stable because of the general presence of flat directions.

The last question we will address is to determine the gauging underlying this GKP backgrounds. The brackets in (\ref{SW_algebra_ZX}) get now simplified to
\be
\label{brackets_GKP}
\begin{array}{cccc}
\left[ {X_{+}}^{m} , {X_{+}}^{n} \right] =  {\tilde{F}}^{mnp} \,\,Z_{+ p}
&\hspace{5mm},\hspace{5mm}&
\left[  {X_{+}}^{m} , {X_{-}}^{n} \right] =  {\tilde{F}}^{mnp} \,\,Z_{- p} & , \\[2mm]
\left[  {X_{-}}^{m} , {X_{-}}^{n} \right] =  {\tilde{H}}^{mnp} \,\,Z_{- p}
&\hspace{5mm},\hspace{5mm}&
\left[  {X_{-}}^{m} , {X_{+}}^{n} \right] =  {\tilde{H}}^{mnp} \,\,Z_{+ p} & .
\end{array}
\ee
Even when there are no QC for the fluxes to obey, the antisymmetry of the brackets in (\ref{brackets_GKP}) when substituting (\ref{FH_GKP}) is guaranteed iff
\be
\label{antisym_cond}
Z_{+ a} = - Z_{- i}
\hspace{5mm} , \hspace{5mm}
(a_{0} + a_{2}) \, Z_{+ i} = (a_{1} + a_{3}) \, Z_{- i}
\hspace{5mm} , \hspace{5mm}
(a_{0} + a_{2}) \, Z_{- a} = (a_{1} +
a_{3}) \, Z_{- i} \ ,
\ee
again with pairs $\,(a,i)=\left\lbrace (1,2),(3,4),(5,6) \right\rbrace$. As a result, the isometry $\,Z_{\alpha m}\,$ generators span a central extension of a $\,\mathfrak{u}(1)^{12}\,$ algebra specified by the $\,X_{\alpha}{}^{m}\,$ generators in (\ref{brackets_GKP}). Consequently, $\,\textrm{R}_\textrm{Adj}\left[Z_{\alpha m}\right]=0\,$ and the antisymmetry conditions in (\ref{antisym_cond}) are trivially satisfied in this representation\footnote{In other words, the adjoint representation is no longer faithful.}. This is the representation of the gauging which has to be embeddable into the $\mathfrak{so}(6,6)$ duality algebra, so the gauging is the abelian group $\,G_{0}=\textrm{U}(1)^{12}$.

\subsection*{Non-Geometric Backgrounds: The $\textrm{SO}(3,3) \times \textrm{SO}(3,3)$ Splitting}

In this final section we move to study some gaugings which cannot be realised as geometric type II string compactifications. Specifically, we will focus on those based on the direct product splitting $\,\textrm{SO}(3,3) \times \textrm{SO}(3,3)\,$ discussed in refs~\cite{deRoo:2002jf, deRoo:2003rm,deRoo:2006ms} and further interpreted as non-geometric flux compactifications in refs~\cite{Roest:2009dq, Dibitetto:2010rg}.

This splitting implies the factorisation of the gauge group in terms of $G_1 \times G_2$, where furthermore $\,G_1\,$ and $\,G_2\,$ were chosen in ref.~\cite{deRoo:2003rm} to be electric and magnetic respectively. This provides the simplest solution to the the second set of QC in \eqref{QC} and moreover a non-trivial gauging at angles which is necessary in order to guarantee moduli stabilisation \cite{deRoo:1985jh}. In ref.~\cite{deRoo:2003rm} some de Sitter solutions have been found by investigating the case in which $G_1$ and $G_2$ are chosen to be some SO($p,q$), with $\,p+q=4$. Later on non-semi-simple gaugings of the form CSO($p,q,r$)$\,\times\,$CSO($p,q,r$) have been investigated in ref.~\cite{deRoo:2006ms}, but no de Sitter solutions were found.

Let us go deeper into the vacua structure of these CSO($p,q,r$)$\,\times\,$CSO($p,q,r$) gaugings. In order to do so, we will use the parameterisation of the embedding of each CSO factor inside {SO}($3,3$) in terms of the two real symmetric matrices $M_{\pm}$ and $\tilde{M}_{\pm}\,$ as explained in ref.~\cite{Roest:2009tt}. In the case of the SO($3$) truncation, these are given by
\be
M_{+} \equiv \textrm{diag}\left(-a'_{0} \,\,,\,\, \tilde{c}_{1}\,,\,\tilde{c}_{1}\,,\,\tilde{c}_{1}\right)
\hspace{5mm},\hspace{5mm}
\tilde{M}_{+} \equiv \textrm{diag}\left(-a_{0} \,\,,\,\, \tilde{c}'_{1}\,,\,\tilde{c}'_{1}\,,\,\tilde{c}'_{1}\right) \ ,
\ee
together with
\be
M_{-} \equiv \textrm{diag}\left(b'_{3} \,\,,\,\, \tilde{d}_{2}\,,\,\tilde{d}_{2}\,,\,\tilde{d}_{2}\right)
\hspace{5mm},\hspace{5mm}
\tilde{M}_{-} \equiv \textrm{diag}\left(b_{3} \,\,,\,\,\tilde{d}'_{2}\,,\,\tilde{d}'_{2}\,,\,\tilde{d}'_{2}\right) \,\ ,
\ee
where the relation between the entries of the above matrices and the embedding tensor components can be read off from tables~\ref{table:unprimed_fluxes4} and \ref{table:primed_fluxes4}. The flux-induced superpotential in (\ref{W_fluxes4}) then reduces to
\be
\label{W_SO(3,3)xSO(3,3)}
\begin{array}{cclcc}
W_{\textrm{SO}(3,3)^{2}} &=& a_0 + b_3 \, S \, U^3 - 3 \, \tilde{c}_{1} \, T \,U  - 3 \, \tilde{d}_{2} \, S  \,T \,U^2 & + & \\[3mm]
&+& a_0' \,T^3 \,U^3 + b_3' \, S \, T^3  - 3 \, \tilde{c}_{1}' \,T^2 \, U^2   + 3 \, \tilde{d}'_{2} \, S  \,T^2 \,U  &  .
\end{array}
\ee

The antisymmetry of the brackets in (\ref{SW_algebra_ZX}) now translates into
\be
Z_{+i} = X_{+}{}^{i} = Z_{-a} = X_{-}{}^{a} = 0 \ ,
\ee
and the resulting twelve-dimensional algebra $\,\mathfrak{g}_{0}\,$ is written as
\be
\begin{array}{cccl}
\label{SO(3,3)xSO(3,3)_brackets}
[Z_{+a},Z_{+b}] = \phantom{-}\tilde{c}'_{1} \, Z_{+c} - a'_{0} \, X_{+}{}^{c}    &     \hspace{5mm} , \hspace{5mm}      &
[Z_{-i},Z_{-j}] = \tilde{d}'_{2} \, Z_{-k} + b'_{3} \, X_{-}{}^{k} & , \\[2mm]
[Z_{+a},X_{+}{}^{b}] = \phantom{-}\tilde{c}_{1} \, Z_{+c} + \tilde{c}'_{1} \, X_{+}{}^{c} &     \hspace{5mm} , \hspace{5mm}      &
\,\,[Z_{-i},X_{-}{}^{j}] =  \tilde{d}_{2} \, Z_{-k} + \tilde{d}'_{2} \, X_{-}{}^{k} & , \\[2mm]
\,\,[X_{+}{}^{a},X_{+}{}^{b}] = -a_{0} \, Z_{+c} + \tilde{c}_{1} \, X_{+}{}^{c}                                 &     \hspace{5mm} , \hspace{5mm}      &
\,\,\,[X_{-}{}^{i},X_{-}{}^{j}] =  b_{3} \, Z_{-k} + \tilde{d}_{2} \, X_{-}{}^{k} & .
\end{array}
\ee
The first set of QC in (\ref{QC}) gets also simplified and forces the product $\,M_{+} \, \tilde{M}_{+}\,$ and $\,M_{-} \, \tilde{M}_{-}\,$ to be proportional to the identity matrix.

For the sake of simplicity we will consider the case of having only unprimed fluxes, \emph{i.e.} having a type IIB background including gauge $\,(F_{3},H_{3})\,$ and non-geometric $\,(Q,P)\,$ fluxes. Such backgrounds, although being non-geometric, still admit a locally geometric description and in accord with ref.~\cite{Dibitetto:2010rg}, they can never give rise to semi-simple gaugings. Their associated flux-induced superpotential takes the quite simple form of
\be
\label{W_SO(3,3)xSO(3,3)-unprim}
\begin{array}{ccc}
W_{\textrm{SO}(3,3)^{2}}^{\textrm{loc. geom.}} &=& a_0 + b_3 \, S \, U^3 - 3 \, \tilde{c}_{1} \,T \, U   - 3 \, \tilde{d}_{2} \, S  \,T \,U^2 \ .
\end{array}
\ee
These backgrounds already satisfy all of the QC as well as the extremality conditions for the axions at the origin of moduli space\footnote{This fact points out that the origin of moduli space is an especially interesting point even though it is not the most general solution since this flux background is not
duality invariant.}. In addition, their corresponding flux-induced tadpoles are given by
\be
\label{Tad37}
N_{3}=a_{0} \, b_{3} \hspace{10mm} , \hspace{10mm} N_{7}=\tilde{N}_{7}=N'_{7}=0 \ ,
\ee
where $N_{7}$, $\tilde{N}_{7}$ and $N'_{7}$ relate to the SL$(2)$-triplet of $7$-branes in a type IIB S-duality invariant realisation of the theory \cite{Bergshoeff:2006ic,Bergshoeff:2006jj}. In fact, the second condition in (\ref{Tad37}) is actually identified with $\,\cN=4\,$ QC since these $7$-branes would break from half-maximal to minimal supergravity.

\begin{table}[h!]
\renewcommand{\arraystretch}{1.80}
\begin{center}
\scalebox{0.9}[0.9]{
\begin{tabular}{ | c || c | c | c | c | c | c |}
\hline
\textrm{ID} & $a_{0}$ & $\tilde{c}_{1}$ & $b_{3}$ & $\tilde{d}_{2}$ & $V_0$  & BF  \\[1mm]
\hline \hline
$1$ & $-\lambda$ & $\lambda$ & $-\lambda$ & $-\lambda$ & $-\dfrac{3 \lambda^{2}}{8}$ & $m^2=-\dfrac{2}{3}\rightarrow$ stable \\[1mm]
\hline
$2$ & $\lambda$ & $-\lambda$ & $-\lambda$ & $-\lambda$ & $\dfrac{\lambda^{2}}{8}$ & unstable de Sitter\\[1mm]
\hline
$3_a$ & $5 \, \lambda$ & $3 \, \lambda$ & $-\lambda$ & $-\lambda$ & $-\dfrac{15 \lambda^{2}}{8}$ & $m^2=-\dfrac{26}{15}\rightarrow$ unstable \\[1mm]
\hline
$3_b$ & $-\lambda$ & $\lambda$ & $5 \, \lambda$ & $-3 \, \lambda$ & $-\dfrac{15 \lambda^{2}}{8}$ & $m^2=-\dfrac{26}{15}\rightarrow$ unstable \\[1mm]
\hline
\end{tabular}
}
\end{center}
\caption{{\it Set of extrema of the scalar potential (at the origin of the moduli space) for the $\,\textrm{SO}(3,3) \times \textrm{SO}(3,3)\,$ embeddable type IIB backgrounds admitting a locally geometric description. We also present their stability according to the BF bound in (\ref{BF_bound}).}}
\label{table:elecxmag}
\end{table}

Restricting our search of extrema to the origin of the moduli space, we find five critical points some of them with novel features compared to the ``geometric'' results obtained in the previous sections. Apart from the GKP-like solution appearing when switching off the non-geometric fluxes, i.e, $\,\tilde{c}_{1}=\tilde{d}_{2}=0\,$, the set of extrema of the scalar potential and their vacuum energy are summarised in table~\ref{table:elecxmag}. Notice that solutions $3_a$ and $3_b$ are related to each other by a simultaneous inversion of the $S$ and $U$ moduli fields, \emph{i.e.}, by an element of the compact subgroup $\,\textrm{SO}(2)^{3}\,$ of the duality group. The critical points labelled $1$ and $2$ are invariant under this transformation. This is similar to the $\mathbb{Z}_2 \times \mathbb{Z}_2$ structure in the geometric IIA case. However, in contrast to that situation, the other critical points in table~\ref{table:elecxmag} cannot be related by non-compact duality transformations. Therefore these are solutions to different theories.

The computation of the gravitini mass matrix $\,A^{ij}_{1}\,$ in (\ref{kappa-eigenvalues}) shows that the solution $1$ in table~\ref{table:elecxmag} preserves $\,\cN=4\,$ supersymmetry whereas all the others turn out to be non-supersymmetric. The normalised mass spectra for these solutions are as follows:
\begin{itemize}

\item The normalised masses and their multiplicities for the solution $1$ are given by
\be
\dfrac{4}{3} \,\,(\times \, 2) \hspace{5mm} , \hspace{5mm} 0 \,\,(\times \, 24) \hspace{5mm} , \hspace{5mm} -\dfrac{2}{3} \,\,(\times \, 12) \ .
\ee
The twelve tachyonic modes imply $\,m^2=-2/3\,$ and then satisfy the BF bound in (\ref{BF_bound}) ensuring the stability of this AdS$_4$ solution.

\item The normalised masses and their multiplicities for the solution $2$ are given by
\be
6 \,\,(\times \, 10) \hspace{5mm} , \hspace{5mm} 4 \,\,(\times \, 18) \hspace{5mm} , \hspace{5mm} -2 \,\,(\times \, 2) \hspace{5mm} , \hspace{5mm} 0 \,\,(\times \, 8) \ ,
\ee
so this de Sitter solution is automatically unstable since it contains two tachyons.

\item The normalised masses and their multiplicities for the solutions $3_{a,b}$ are given by
\be
\begin{array}{cccccccc}
-\dfrac{26}{15} \,\,(\times 5)  & \hspace{3mm},\hspace{3mm} & -\dfrac{4}{5}\,\,(\times 9) & \hspace{3mm},\hspace{3mm} & -\dfrac{2}{15}\,\,(\times 1) & \hspace{3mm},\hspace{3mm} &  \dfrac{1}{15} \left( \, 23 \pm \sqrt{1009} \, \right)\,\,(\times 1) & , \\[4mm]
\dfrac{2}{5}\,\,(\times 5) & , &   \dfrac{16}{15}\,\,(\times 1)  & , &  \dfrac{4}{3}\,\,(\times 9)   & , &  0\,\,(\times 6) & ,
\end{array}
\nonumber
\ee
so these AdS$_{4}$ solutions do not satisfy the BF bound in (\ref{BF_bound}) for fourteen tachyonic modes hence becoming unstable.

\end{itemize}
\noindent
We would like to point out that in these non-geometric flux vacua the lightest mode generically no longer belongs to the SO($3$) truncation.

Concerning the gauge group underlying these locally geometric type IIB backgrounds, it is directly identified with
\be
\textrm{G}_{0} = \textrm{ISO}(3) \times \textrm{ISO}(3) \ ,
\ee
when keeping only unprimed fluxes in the brackets of (\ref{SO(3,3)xSO(3,3)_brackets}). The three different theories correspond to inequivalent embeddings of this gauge group in the global symmetry group. All critical points break the non-compact generators of this gauge group, and hence six of the massless scalars in the mass spectra listed above correspond to non-physical scalars.

As a final remark, we want to highlight that table \ref{table:elecxmag}, even though not being exhaustive, contains interesting solutions such as an example of $\mathcal{N}=4$ supersymmetric AdS vacuum and an example of dS solution obtained from a non-semi-simple gauging. The latter is the first example with such a gauge group; all previously constructed dS solutions are based on semi-simple groups \cite{deRoo:2002jf, deRoo:2003rm}.

\chapter{\mbox{Exceptional Flux Compactifications}}
\markboth{Exceptional Flux Compactifications}{Exceptional Flux Compactifications}
\label{Maximal}
In the previous chapter we have studied the landscape of geometric $\cN=4$ compactifications and we saw that non-geometric fluxes seem to provide a crucial ingredient for dS extrema. In section~\ref{sec:non-geom_fluxes} we have seen that their existence was first conjectured in order for the low energy effective theory to be duality covariant. This duality is correctly encoded in the global symmetry of the underlying gauged supergravity in four dimensions \cite{Samtleben:2008pe}. In this sense, T-duality singles out the important role of half-maximal supergravities (see realtions to DFT in chapter~\ref{DFT}), whereas, in order to supplement it with non-perturbative dualities to generate the full U-duality group, one has to consider maximal supergravity.

The combination of some recent developments makes it interesting to further investigate the structure of maximal gauged supergravities in order to better understand which role U-dualities play in the context of flux compactifications. The embedding of half-maximal into maximal supergravity \cite{Aldazabal:2011yz, Dibitetto:2011eu} allows one to study flux backgrounds that preserve maximal supersymmetry. An interesting fact is that the completion of half-maximal supergravity deformations to maximal is given by objects which behave as spinors under T-duality. 

Our goal in this work will be to elaborate on the results of ref.~\cite{Dibitetto:2011gm} and explicitly show how these IIA geometric flux backgrounds and any other type II background can be embedded in maximal supergravity. To this end we will therefore need to relate different formulations of $\,\cN=8\,$ gauged supergravity. The embedding tensor formalism provides an $\,\textrm{E}_{7(7)}\,$ covariant formulation of maximal gauged supergravity in $D=4$ (see table~\ref{table:max}). However, in order to make contact with flux compactifications, we need a rewriting of this theory in terms of irrep's of the S- and T-duality groups, \emph{i.e.} $\,\textrm{SL}(2)\,\times\,\textrm{SO}(6,6)$, following the philosophy of ref.~\cite{Aldazabal:2010ef}. Finally, in order to study the physical properties of scalars, such as equations of motions and the mass matrix, $\,\textrm{SU}(8)\,$is the correct group rearranging all the 70 scalar physical degrees of freedom into an irrep. This can be summarised as

\begin{center}
\scalebox{0.9}[0.9]{\xymatrix{
*+[F-,]{\begin{array}{c}\boldsymbol{\cN=8\textbf{ SUGRA}}\\
\textrm{E}_{7(7)} \end{array} } & \leftrightarrow &
*+[F-,]{\begin{array}{c}\textbf{Fluxes}\\
\textrm{SL}(2)\,\times\,\textrm{SO}(6,6) \end{array} } &
\leftrightarrow &
*+[F-,]{\begin{array}{c}\textbf{Mass spectra}\\
\textrm{SU}(8) \end{array} }  }}
\end{center}
\vspace{2mm}

\noindent Employing this mapping, we will derive the mass spectrum and the gauge group of such ``exceptional'' cases of flux backgrounds without branes.

In this chapter we will first briefly review gauged $\cN=8$ supergravities in the so-called SU($8$) formulation. Secondly, we will see how to embed $\cN=4$ backgronds into $\cN=8$. This will turn out to be natural in what we will call the $\textrm{SL}(2)\,\times\,\textrm{SO}(6,6)$ formulation. Finally, we will show how to connect these formulation in order to compute all the physical quantities for $\cN=4$ backgrounds admitting an uplift to $\cN=8$, just like in the case of the geometric type IIA vacua presented in table~\ref{table:N=4_vacua}. Most of the results of these chapter where obtained and first presented in refs~\cite{Dibitetto:2011eu, Dibitetto:2011qs, Dibitetto:2012ia}. Some additional technical material relevant here is collected in appendix~\ref{appendix:Exceptional}.

\section{Gauged $\cN=8\,$, $D=4$ Supergravities}
\label{sec:maximal_supergravity}

Maximal supergravity appears when reducing type II ten-dimensional supergravities on a torus down to four dimensions. The embedding tensor formalism in maximal supergravity \cite{deWit:2007mt} describes the gauging procedure, \emph{i.e.} promoting to local a part of the $\textrm{E}_{7(7)}$ global symmetry of the 4D theory. After applying a gauging, a non-Abelian gauge symmetry is realised in a way compatible with still keeping $\,\mathcal{N}=8\,$ supersymmetry in four dimensions. Moreover, a non-trivial potential $\,V\,$ for the $70$ (physical) scalar fields in the lower-dimensional theory (a.k.a. moduli fields) is also generated, hence opening the possibility for them to get stabilised, \emph{i.e.} to acquire a mass, due to the gauging. The aim of this chapter is to explore the interplay between gaugings and moduli stabilisation in the context of maximal supergravity.

\subsection*{Embedding tensor and the $\,\textrm{E}_{7(7)}\,$ formulation}

A gauging is totally encoded inside the embedding tensor ${\Theta_{\mathbb{M}}}^{\mathpzc{A}}$, where $\mathbb{M}=1,...,56$ and ${\mathpzc{A}=1,...,133}$ respectively denote indices in the fundamental $\textbf{56}$ and adjoint $\textbf{133}$ representations of $\textrm{E}_{7(7)}$. The tensor ${\Theta_{\mathbb{M}}}^{\mathpzc{A}}$ lives in the $\,\textbf{56} \, \otimes \, \textbf{133} = \textbf{56} \,\oplus\, \textbf{912} \,\oplus\, \textbf{6480}\,$ irrep's of $\textrm{E}_{7(7)}$ and specifies which subset of the $\textrm{E}_{7(7)}$ generators $\left\lbrace t_{\mathpzc{A}=1,...,133} \right\rbrace$ become gauge symmetries after the gauging procedure and hence have an associated gauge boson $V_{\mathbb{M}}$ in four dimensions. As in standard gauge theories, the ordinary derivative is replaced by a covariant one, $\nabla \rightarrow \nabla -g\, V^{\mathbb{M}} \,{\Theta_{\mathbb{M}}}^{\mathpzc{A}} \, t_{\mathpzc{A}} $, and a non-Abelian gauge algebra
\be
\label{gauge_algebra}
\left[ X_{\mathbb{M}} , X_{\mathbb{N}} \right] = - {X_{\mathbb{M} \mathbb{N}}}^{\mathbb{P}} \, X_{\mathbb{P}} \hspace{15mm} \textrm{with} \hspace{15mm} {X_{\mathbb{M} \mathbb{N}}}^{\mathbb{P}} = {\Theta_{\mathbb{M}}}^{\mathpzc{A}} \, {[t_{\mathpzc{A}}]_{\mathbb{N}}}^{\mathbb{P}} \ ,
\ee
is spanned by the generators $X_{\mathbb{M}}$. Since $\textrm{E}_{7(7)} \subset \textrm{Sp}(56,\mathbb{R})$, and even though $\textrm{E}_{7(7)}$ does not have any invariant metric, one can still use the $\textrm{Sp}(56,\mathbb{R})$ invariant matrix $\Omega_{\mathbb{M}\mathbb{N}}$ (skew-symmetric) in order to raise and lower $\textrm{E}_{7(7)}$ fundamental indices. In what follows, we adopt the SouthWest-NorthEast (SW-NE) convention, \emph{e.g.} $X^{\mathbb{M}}=X_{\mathbb{N}} \, \Omega^{\mathbb{N} \mathbb{M}}$, together with $\,\Omega_{\mathbb{M}\mathbb{P}}\,\Omega^{\mathbb{N}\mathbb{P}}=\delta_{\mathbb{M}}^{\mathbb{N}}$.

Maximal supersymmetry requires the tensor $X_{\mathbb{M} \mathbb{N} \mathbb{P}}=X_{\mathbb{M} (\mathbb{N} \mathbb{P})}=-{X_{\mathbb{M} \mathbb{N}}}^{\mathbb{Q}} \, \Omega_{\mathbb{Q} \mathbb{P}}$ to live in the $\textbf{912}$ irrep of $\textrm{E}_{7(7)}$. This translates into the following set of LC
\be
\label{linear_const}
X_{(\mathbb{M} \mathbb{N} \mathbb{P})}=0
\hspace{10mm},\hspace{10mm}
{X_{\mathbb{P} \mathbb{M}}}^{\mathbb{P}}=0 \ .
\ee
On the other hand, the tensor $X_{\mathbb{M} \mathbb{N} \mathbb{P}}$ must also satisfy a set of QC coming from the consistency of the gauge algebra in (\ref{gauge_algebra}). These QC sit in the $(\textbf{133} \,\otimes\, \textbf{133})_{a} = \textbf{133} \,\oplus\, \textbf{8645}$ irrep's of $\textrm{E}_{7(7)}$ and are given by
\be
\label{quadratic_const}
\Omega^{\mathbb{R} \mathbb{S}} \, X_{\mathbb{R} \mathbb{M} \mathbb{N}} \, X_{\mathbb{S} \mathbb{P} \mathbb{Q}}=0 \ .
\ee
The above set of linear (\ref{linear_const}) and quadratic (\ref{quadratic_const}) constraints guarantee the consistency of the maximal gauged supergravity in four dimensions \cite{deWit:2007mt}.

Switching on a gauging has strong implications for the scalar sector of the four-dimensional theory. It consists of $133$ scalars out of which only $70$ are physical degrees of freedom -- the remaining $63$ can be removed from the theory after gauge fixing -- and parameterise an $\,\,\textrm{E}_{7(7)}/\textrm{SU}(8)$ coset element $\mathcal{M}_{\mathbb{M}\mathbb{N}}=\mathcal{M}_{(\mathbb{M}\mathbb{N})}$. Because of the gauging, a non-trivial scalar potential appears
\be
\begin{array}{ccc}
\label{V_N=8}
V &=& \dfrac{g^{2}}{672} \, X_{\bbM \bbN \bbP}  \, X_{\bbQ \bbR \bbS} \left( \mathcal{M}^{\bbM \bbQ} \, \mathcal{M}^{\bbN \bbR} \, \mathcal{M}^{\bbP \bbS}   +   7 \, \mathcal{M}^{\bbM \bbQ} \, \Omega^{\bbN \bbR} \, \Omega^{\bbP \bbS}    \right) \ ,
\end{array}
\ee
which is invariant under the linear action of $\textrm{E}_{7(7)}$ transformations. This scalar potential might contain a rich structure of critical points where to stabilise all the moduli fields in the four-dimensional theory.

\subsection*{Fermionic Mass Terms and the $\,\textrm{SU}(8)\,$ Formulation}

The Lagrangian of maximal supergravity in four dimensions can be unambiguously written in terms of $\textrm{SU}(8)$ tensors, since $\textrm{SU}(8)$ is one of the maximal subgroups of $\textrm{E}_{7(7)}$. More concretely, it is its maximal compact subgroup and is identified with the $R$-symmetry group under which the eight gravitini of the theory get rotated amongst themselves.

\subsubsection*{Bosonic Field Content}

Under its $\textrm{SU}(8)$ maximal subgroup, we have the following branching for some relevant $\textrm{E}_{7(7)}$ representations
\be
\begin{array}{cclcl}
\textrm{E}_{7(7)} & \supset &  \textrm{SU}(8) & \hspace{5mm} &   \hspace{40mm} \textrm{\textsc{fields}}\\[2mm]
\textbf{56} & \rightarrow & \textbf{28} \oplus \overline{\textbf{28}} & &  \textrm{vectors: } \,\,\,\,\,\,V_{\mathcal{IJ}} \oplus V^{\mathcal{IJ}}\\[2mm]
\textbf{133} & \rightarrow & \textbf{63} \oplus \textbf{70} & & \textrm{scalars: } \,\,\,\,\,\,{\phi_{\mathcal{I}}}^{\mathcal{J}} \,\,(\textrm{\footnotesize{unphysical}})\oplus \phi_{\mathcal{I} \mathcal{J} \mathcal{K} \mathcal{L}}  \,\,(\textrm{\footnotesize{physical}}) \\[2mm]
\textbf{912} & \rightarrow & \textbf{36} \oplus \textbf{420} \oplus \overline{\textbf{36}} \oplus \overline{\textbf{420}} &  & \textrm{emb tens: } \mathcal{A}^{\mathcal{IJ}} \oplus {\mathcal{A}_{\mathcal{I}}}^{\mathcal{JKL}} \oplus \mathcal{A}_{\mathcal{IJ}} \oplus {\mathcal{A}^{\mathcal{I}}}_{\mathcal{JKL}}
\end{array}
\nonumber
\ee
related to the vectors, scalars and embedding tensor\footnote{Strictly speaking, the $\mathcal{A}^{\mathcal{IJ}}$ and ${\mathcal{A}_{\mathcal{I}}}^{\mathcal{JKL}}$ fermionic mass terms as well as their complex conjugates correspond to the irrducible components of the $T$-tensor (see section~\ref{section:Theta}), obtained by dreesing up the embedding tensor with scalar vielbeins.} in the four-dimensional theory. When expressed in terms of $\textrm{SU}(8)$ fundamental indices $\mathcal{I}=1,...,8$, the above fields have the following symmetry properties according to the irrep's they are associated to:
\begin{itemize}

\item[$i)$] $V_{\mathcal{IJ}}=V_{\mathcal{[IJ]}}\,$ and $\,V^{\mathcal{IJ}}=(V_{\mathcal{IJ}})^{*}\,$ for the complex vector fields.

\item[$ii)$] ${\phi_{\mathcal{I}}}^{\mathcal{I}}=0\,$
and $\,\phi_{\mathcal{I} \mathcal{J} \mathcal{K} \mathcal{L}} = \phi_{[\mathcal{I} \mathcal{J} \mathcal{K} \mathcal{L}]}\,$ for the scalar fields which are further restricted by the pseudo-reality condition
\be
\phi_{\mathcal{I} \mathcal{J} \mathcal{K} \mathcal{L}} = \frac{1}{24} \, \epsilon_{\mathcal{IJKL MNPQ}} \,  \phi^{\mathcal{MNPQ}}  \hspace{10mm} \textrm{with} \hspace{10mm} \phi^{\mathcal{MNPQ}} = (\phi_{\mathcal{MNPQ}})^{*} \ .
\ee
It is worth noticing that the physical scalars $\,\phi_{\mathcal{I} \mathcal{J} \mathcal{K} \mathcal{L}}\,$ in the theory fit an irrep, namely the $\textbf{70}$ of $\textrm{SU}(8)$. This will no longer be the case when using another formulation of the theory, as we will see in the next section.

\item[$iii)$] $\mathcal{A}^{\mathcal{IJ}}=\mathcal{A}^{(\mathcal{IJ})}$, $\,{\mathcal{A}_{\mathcal{I}}}^{\mathcal{J} \mathcal{K} \mathcal{L}} = {\mathcal{A}_{\mathcal{I}}}^{[\mathcal{J} \mathcal{K} \mathcal{L}]}\,$ and $\,{\mathcal{A}_{\mathcal{Q}}}^{\mathcal{Q} \mathcal{I} \mathcal{J}} =0\,$ and equivalently for their complex conjugate counterparts $\mathcal{A}_{\mathcal{IJ}}=(\mathcal{A}^{\mathcal{IJ}})^{*}\,$ and $\,{\mathcal{A}^{\mathcal{I}}}_{\mathcal{J} \mathcal{K} \mathcal{L}} =({\mathcal{A}_{\mathcal{I}}}^{\mathcal{J} \mathcal{K} \mathcal{L}} )^{*}\,$.

\end{itemize}

In the $\,\textrm{SU}(8)\,$ formulation, the $\,\textrm{Sp}(56,\mathbb{R})$ invariant (skew-symmetric) matrix $\,\Omega_{\mathbb{M} \mathbb{N}}\,$ takes the form
\be
\Omega_{\mathbb{M} \mathbb{N}} = - i
\left(
\begin{array}{c|c}
0 & \delta_{\mathcal{IJ}}^{\mathcal{KL}} \\[1mm]
\hline
\\[-4mm]
-\delta^{\mathcal{IJ}}_{\mathcal{KL}} & 0
\end{array}
\right) \ .
\ee

\subsubsection*{Fermionic Mass Terms and Scalar Potential}

The $\mathcal{A}^{\mathcal{IJ}}$ and ${\mathcal{A}_{\mathcal{I}}}^{\mathcal{JKL}}$ tensors play a central role in the $\textrm{SU}(8)$ formulation of maximal supergravity. They determine the fermionic mass terms for the gravitini $\psi^{\,\,\mathcal{I}}_{\mu}$ and the dilatini $\chi_{\mathcal{IJK}}$ in the four-dimensional Lagrangian \cite{deWit:2007mt} (where in this formula $\mu, \nu$ are understood as space-time indices)
\be
\label{Fermi_Lagrangian}
e^{-1} \, g^{-1} \, \mathcal{L}_{\textrm{fermi}} = \frac{\sqrt{2}}{2} \, \mathcal{A}_{\mathcal{I} \mathcal{J}} \,\overline{\psi}^{\,\,\mathcal{I}}_{\mu} \, \gamma^{\mu \nu} \,\psi^{\,\,\mathcal{J}}_{\nu} + \frac{1}{6} \, {\mathcal{A}_{\mathcal{I}}}^{\mathcal{JKL}} \,\overline{\psi}^{\,\,\mathcal{I}}_{\mu} \, \gamma^{\mu} \, \chi_{\mathcal{JKL}} + \mathcal{A}^{\mathcal{IJK},\mathcal{LMN}} \, \overline{\chi}_{\mathcal{IJK}} \, \chi_{\mathcal{LMN}} + \textrm{h.c.} \ ,
\ee
where $\,\mathcal{A}^{\mathcal{IJK},\mathcal{LMN}} \equiv \frac{\sqrt{2}}{144} \, \epsilon^{\mathcal{IJKPQR[LM}} \, {\mathcal{A}^{\mathcal{N}]}}_{\mathcal{PQR}}\,$. The number of supersymmetries preserved by an AdS ($V_{0}<0$) or Minkowski ($V_{0}=0$) solution of the theory is related to the number of spinors satisfying the Killing equations
\be
\label{Killing_equations}
g \, \mathcal{A}_{\mathcal{IJ}} \, \epsilon^{\mathcal{J}} \, = \, \sqrt{-\frac{1}{6}\,V_{0}} \,\, \epsilon_{\mathcal{I}} \ .
\ee

The scalar potential in (\ref{V_N=8}) can also be rewritten in terms of the fermionic mass terms as
\be
\label{V_SU8}
g^{-2} \,V = -\frac{3}{4} \, |\mathcal{A}_1|^{2} + \frac{1}{24} \, |\mathcal{A}_2|^{2} \ ,
\ee
where $\,|\mathcal{A}_1|^{2}=\mathcal{A}_{\mathcal{IJ}} \, \, \mathcal{A}^{\mathcal{IJ}}\,$ and $\,|\mathcal{A}_2|^{2}={\mathcal{A}_{\mathcal{I}}}^{\mathcal{JKL}} \, {\mathcal{A}^{\mathcal{I}}}_{\mathcal{JKL}}\,$. This potential will possess a structure of critical points satisfying
\be
\left. \frac{\partial V}{\partial \phi_{\mathcal{IJKL}}} \right|_{\left< \phi_{\mathcal{IJKL}} \right>}= 0 \ ,
\ee
where $\left< \phi_{\mathcal{IJKL}} \right>$ denotes the VEV for the $70$ physical scalar fields. Provided $\,\left< V_{\mathcal{IJ}}\right>=0\,$ for the vector fields, maximally symmetric solutions of the theory are obtained by solving the equations of motion \cite{Diffon:2011wt} of the physical scalars
\be
\label{scalars_eom}
\begin{array}{ccc}
\mathcal{C}_{\mathcal{IJKL}} \,+\, \dfrac{1}{24} \, \epsilon_{\mathcal{IJKLMNPQ}} \, \mathcal{C}^{\mathcal{MNPQ}} &=& 0 \ ,
\end{array}
\ee
where $\,\mathcal{C}_{\mathcal{IJKL}}={\mathcal{A}^{\mathcal{M}}}_{[\mathcal{IJK}}\,\mathcal{A}_{\mathcal{L}]\mathcal{M}} \, +\, \frac{3}{4} \, {\mathcal{A}^{\mathcal{M}}}_{\mathcal{N}[\mathcal{IJ}} \, {\mathcal{A}^{\mathcal{N}}}_{\mathcal{KL}]\mathcal{M}}\,$. At these solutions, the mass matrix for the physical scalars \cite{Diffon:2011wt,Borghese:2011en} reads
\be
\label{Mass-matrix}
\begin{array}{ccl}
g^{-2} \, {\left(\textrm{mass}^{2}\right)_{\mathcal{IJKL}}}^{\mathcal{MNPQ}}  & =  &  \delta_{\mathcal{IJKL}}^{\mathcal{MNPQ}} \, \left( \frac{5}{24} \, \mathcal{A}^{\mathcal{R}}{}_{\mathcal{STU}} \, \mathcal{A}_{\mathcal{R}}{}^{\mathcal{STU}} - \frac{1}{2} \, \mathcal{A}_{\mathcal{RS}} \, \mathcal{A}^{\mathcal{RS}} \right) \\[2mm]
& + & 6 \, \delta_{[\mathcal{IJ}}^{[\mathcal{MN}} \, \left( \mathcal{A}_{\mathcal{K}}{}^{\mathcal{RS} |\mathcal{P}} \, \mathcal{A}^{\mathcal{Q}]}{}_{\mathcal{L}]\mathcal{RS}} - \frac{1}{4} \, \mathcal{A}_{\mathcal{R}}{}^{\mathcal{S} |\mathcal{PQ}]} \, \mathcal{A}^{\mathcal{R}}{}_{\mathcal{S}|\mathcal{KL}]} \right) \\[2mm]
&-& \frac{2}{3} \, A_{[\mathcal{I}}{}^{[\mathcal{MNP}} \, \mathcal{A}^{\mathcal{Q}]}{}_{\mathcal{JKL}]} \ .
\end{array}
\ee
Defining the normalised mass as $\,\left(\textrm{mass}^{2}\right)_{\textrm{norm}}= \frac{1}{|V_{0}|}\,\left(\textrm{mass}^{2}\right)\,$, then the BF bound for the stability of an AdS solution is again given by \eqref{BF_bound} for the lowest eigenvalue of the normalised mass matrix at the AdS extremum with energy $\,V_{0}<0\,$. We will make extensive use of (\ref{Mass-matrix}) and (\ref{BF_bound}) in the last part of the chapter when discussing stability of solutions in specific maximal supergravity models arising from flux compactifications of type II strings.

\subsubsection*{Quadratic Constraints}

The set of QC in (\ref{quadratic_const}) can also be expressed in terms of the $\mathcal{A}^{\mathcal{IJ}}$ and ${\mathcal{A}_{\mathcal{I}}}^{\mathcal{JKL}}$ tensors. Using the branching relations
\bea
\textrm{E}_{7(7)} & \supset &  \textrm{SU}(8) \\[2mm]
\label{irrep_133}
\textbf{133} & \longrightarrow  & \textbf{63}\,\oplus\,\textbf{70} \ , \\[2mm]
\label{irrep_8640}
\textbf{8645} & \longrightarrow  & \textbf{63}\,\oplus \, \textbf{378} \,\oplus \,\overline{\textbf{378}} \,\oplus\,\textbf{945}\,\oplus\,\overline{\textbf{945}} \,\oplus \,\textbf{2352}\, \oplus \,\textbf{3584} \ ,
\eea
as an organising principle, one gets the following QC \cite{deWit:2007mt} \newline

\resizebox{.99\hsize}{!}{$
\label{QC_SU8}
\hspace{-2mm}
\begin{array}{crcl}
\hspace{-5mm} & 9 \, \mathcal{A}_{\mathcal{R}}{}^{\mathcal{STM}} \mathcal{A}^{\mathcal{R}}{}_{\mathcal{STI}} - \mathcal{A}_{\mathcal{I}}{}^{\mathcal{RST}} \mathcal{A}^{\mathcal{M}}{}_{\mathcal{RST}} - \delta_{\mathcal{I}}^{\mathcal{M}} \, |\mathcal{A}_2|^{2} & = & 0 \ , \\[3mm]
\hspace{-5mm} &3 \, \mathcal{A}_{\mathcal{R}}{}^{\mathcal{STM}} \mathcal{A}^{\mathcal{R}}{}_{\mathcal{STI}} - \mathcal{A}_{\mathcal{I}}{}^{\mathcal{RST}} \mathcal{A}^{\mathcal{M}}{}_{\mathcal{RST}} + 12 \, \mathcal{A}_{\mathcal{IR}} \mathcal{A}^{\mathcal{MR}} - \frac{1}{4} \, \delta_{\mathcal{I}}^{\mathcal{M}} \, |\mathcal{A}_2|^{2} - \frac{3}{2} \, \delta_{\mathcal{I}}^{\mathcal{M}} \, |\mathcal{A}_1|^{2} & = & 0 \ , \\[3mm]
\hspace{-5mm} &\mathcal{A}^{\mathcal{I}}{}_{\mathcal{JV}[\mathcal{M}} \, \mathcal{A}^{\mathcal{V}}{}_{\mathcal{NPQ}]} + \mathcal{A}_{\mathcal{JV}} \, \delta^{\mathcal{I}}_{[\mathcal{M}} \, \mathcal{A}^{\mathcal{V}}{}_{\mathcal{NPQ}]} - \mathcal{A}_{\mathcal{J}[\mathcal{M}} \, \mathcal{A}^{\mathcal{I}}{}_{\mathcal{NPQ}]}  &  & \\[1mm]
\hspace{-5mm} &+ \, \frac{1}{24} \, \epsilon_{\mathcal{MNPQRSTU}} \, \left( \mathcal{A}_{\mathcal{J}}{}^{\mathcal{IVR}} \, \mathcal{A}_{\mathcal{V}}{}^{\mathcal{STU}} + \mathcal{A}^{\mathcal{IV}} \, \delta_{\mathcal{J}}^{\mathcal{R}} \, \mathcal{A}_{\mathcal{V}}{}^{\mathcal{STU}} - \mathcal{A}^{\mathcal{IR}} \, \mathcal{A}_{\mathcal{J}}{}^{\mathcal{STU}} \right) & = & 0 \ , \\[3mm]
\hspace{-5mm} & - \frac{1}{8} \, \delta_{\mathcal{I}}^{\mathcal{N}} \left( \mathcal{A}_{\mathcal{R}}{}^{\mathcal{STM}} \mathcal{A}^{\mathcal{R}}{}_{\mathcal{STJ}} - \mathcal{A}_{\mathcal{J}}{}^{\mathcal{RST}} \mathcal{A}^{\mathcal{M}}{}_{\mathcal{RST}} \right) + \frac{1}{8} \, \delta_{\mathcal{J}}^{\mathcal{M}} \left( \mathcal{A}_{\mathcal{R}}{}^{\mathcal{STN}} \mathcal{A}^{\mathcal{R}}{}_{\mathcal{STI}} - \mathcal{A}_{\mathcal{I}}{}^{\mathcal{RST}} \mathcal{A}^{\mathcal{N}}{}_{\mathcal{RST}} \right) &  & \\[1mm]
\hspace{-5mm} &+\, \mathcal{A}_{\mathcal{I}}{}^{\mathcal{RSM}} \mathcal{A}^{\mathcal{N}}{}_{\mathcal{JRS}} - \mathcal{A}_{\mathcal{J}}{}^{\mathcal{RSN}} \mathcal{A}^{\mathcal{M}}{}_{\mathcal{IRS}} + 4 \, \mathcal{A}^{(\mathcal{M}}{}_{\mathcal{IJR}} \mathcal{A}^{\mathcal{N})\mathcal{R}} - 4 \, \mathcal{A}_{(\mathcal{I}}{}^{\mathcal{MNR}} \mathcal{A}_{\mathcal{J})\mathcal{R}}  & = & 0 \ , \\[3mm]
\hspace{-5mm} & - 9 \, \mathcal{A}_{[\mathcal{I}}{}^{\mathcal{R}[\mathcal{MN}} \mathcal{A}^{\mathcal{P}]}{}_{\mathcal{JK}]\mathcal{R}} - 9 \, \delta_{[\mathcal{I}}^{[\mathcal{M}} \, \mathcal{A}_{\mathcal{J}}{}^{\mathcal{RS}|\mathcal{N}} \mathcal{A}^{\mathcal{P}]}{}_{\mathcal{K}]\mathcal{RS}}  - 9 \, \delta_{[\mathcal{IJ}}^{[\mathcal{MN}} \, \mathcal{A}_{\mathcal{R}}{}^{\mathcal{P}]\mathcal{ST}} \mathcal{A}^{\mathcal{R}}{}_{\mathcal{K}]\mathcal{ST}} &  & \\[1mm]
\hspace{-5mm} &+ \, \delta_{\mathcal{IJK}}^{\mathcal{MNP}} \, |\mathcal{A}_{2}|^{2}  + \mathcal{A}_{\mathcal{R}}{}^{\mathcal{MNP}} \mathcal{A}^{\mathcal{R}}{}_{\mathcal{IJK}}  & = & 0 \ ,
\end{array}
$}

\noindent living in the $\,{\bf 63}\,$ (the first two), $\,{\bf 70} \oplus {\bf 378} \oplus {\bf 3584}\,$, $\,{\bf 945} \oplus {\bf \overline{945}}\,$ and $\,{\bf 2352}\,$ irrep's of $\textrm{SU}(8)$, respectively. The above set of QC in (\ref{QC_SU8}) automatically guarantees the consistency of the maximal gauged supergravity.
\\[-4mm]

The $\,\textrm{SO}(8)\,$ gauging \cite{Hull:1984ea} as well as the $\,\textrm{CSO}(p,q,r)\,$ with $p+q+r=8$ contractions thereof \cite{Roest:2009tt,DallAgata:2011aa} are either simple gauge groups or straightforward contractions thereof. For this reason it has been relatively easy to explore these from a pure supergravity viewpoint, irrespective of their realisation in string theory. However, as the relation between gauged supergravities and flux compactifications of string theory became better understood \cite{Samtleben:2008pe}, more complicated non-semisimple gauge groups other than the CSO gaugings have gained interest both in maximal \cite{deWit:2003hq} and half-maximal \cite{Dall'Agata:2009gv,Roest:2009dq,Dibitetto:2010rg,Dibitetto:2011gm} supergravity. The reason is that, as we will show later, the CSO gaugings turn out to correspond to non-geometric flux backgrounds for which an origin in string theory remains unknown\footnote{Nevertheless, some of them can still be obtained from M-theory reductions, as the $\textrm{SO}(8)$ gauging that appears after reducing eleven-dimensional supergravity on a $S^{7}$ sphere.}, whereas gaugings corresponding to flux backgrounds with a higher-dimensional origin are in general not CSO. The latter are called geometric backgrounds and include fluxes associated to the NS-NS and R-R gauge fields present in the spectrum of the string \cite{deWit:2003hq} together with a metric flux associated to a spin connection in the internal space. However, the $\textrm{SU}(8)$ formulation of maximal supergravity is not the most intuitive when it comes to describe gauged supergravities arising from flux compactifications of string theory. Instead, an alternative formulation in terms of $\textrm{SL}(2) \times \textrm{SO}(6,6)$ tensors becomes more adequate as we discuss in the next section.

\section{Embedding $\,\cN=4\,$ Inside $\,\cN=8\,$}
\label{sec:fluxes_A1D6}

Compactifications of string theory in the presence of background fluxes have become a very active research line when it comes to address the problem of moduli stabilisation. Non-geometric fluxes and their further extension to \textit{generalised fluxes} or \textit{dual fluxes} were originally introduced in order to recover invariance of four-dimensional supergravity under the action of duality transformations: more concretely, under non-perturbative S-duality and target space T-duality. The combined action of S-duality and T-duality relates ``apparently'' different four-dimensional backgrounds amongst themselves via an $\,\textrm{SL}(2) \,\times\, \textrm{SO}(6,6)\,$ transformation. This group of transformations corresponds with the global symmetry group of half-maximal $\,\mathcal{N}=4\,$ supergravity in four dimensions \cite{Schon:2006kz}. The relation between half-maximal supergravity and string compactifications with fluxes has been explored in refs~\cite{Aldazabal:2008zza,Dibitetto:2010rg,Dibitetto:2011gm}. As a speculative remark -- and up to quantum requirements such as the discrete nature of the gaugings when understood as fluxes --, by covering the different $\,\textrm{SL}(2) \times \textrm{SO}(6,6)\,$ orbits of half-maximal supergravities, one might have access to intrinsically \textit{stringy} effects involving winding modes and/or dyonic backgrounds, even though it is formulated as a supersymmetric field theory of point-like particles.

Both S-duality and T-duality belong to a larger U-duality group, the $\,\textrm{E}_{7(7)}\,$ global symmetry group of maximal supergravity in four dimensions. Consequently, in order to go from half-maximal to maximal supergravity \cite{Dibitetto:2011eu}, one has to enlarge the field content of the theory, \emph{i.e.} vectors, scalars and embedding tensor components, to complete irrep's of $\,\textrm{E}_{7(7)}$. It is at this point where an alternative formulation of maximal supergravity in terms of $\,\textrm{SL}(2) \,\times\, \textrm{SO}(6,6)\,$ tensors becomes mandatory in order to understand the relation between flux compactifications of string theory and maximal supergravity.

\subsubsection*{Bosonic field content}

Complementary to the $\textrm{SU}(8)$ formulation of the previous section, maximal supergravity can also be unambiguously expressed in terms of $\,\textrm{SL}(2) \,\times\, \textrm{SO}(6,6)\,$ tensors since that is a maximal subgroup of $\,\textrm{E}_{7(7)}$ as well. Under $\,\textrm{SL}(2) \,\times\, \textrm{SO}(6,6)\,$, we now have the following branching
\be
\begin{array}{cclcl}
\textrm{E}_{7(7)} & \supset &  \textrm{SL}(2) \times \textrm{SO}(6,6) & \hspace{5mm} &   \hspace{20mm} \textrm{\textsc{fields}}\\[2mm]
\textbf{56} & \rightarrow & \hspace{-1mm}(\textbf{2},\textbf{12}) \oplus (\textbf{1},\textbf{32}) &  & \hspace{-8mm} \textrm{vectors: } \,\,\,\,\,\,V_{\alpha M} \oplus V_{\mu}\\[2mm]
\textbf{133} & \rightarrow & \hspace{-1mm}(\textbf{1},\textbf{66}) \oplus (\textbf{3},\textbf{1}) \oplus (\textbf{2},\textbf{32}^{\prime}) & & \hspace{-8mm} \textrm{scalars: } \,\,\,\,\,\,\phi_{MN} \oplus \phi_{\alpha \beta} \oplus \phi_{\alpha \dot{\mu}} \\[2mm]
\textbf{912} & \rightarrow & \hspace{-1mm} (\textbf{2},\textbf{220}) \oplus (\textbf{2},\textbf{12}) \oplus (\textbf{1},\textbf{352}^{\prime}) \oplus(\textbf{3},\textbf{32}) &  & \hspace{-10mm} \textrm{emb ten: } f_{\alpha MNP} \oplus \xi_{\alpha M} \oplus F_{M \dot{\mu}} \oplus \Xi_{\a \b \m}
\end{array}
\nonumber
\ee
for the $\textrm{E}_{7(7)}$ representations associated to vectors, scalars and embedding tensor respectively. We follow the conventions in ref.~\cite{Dibitetto:2011eu} for the indices: $\,\alpha=+,-\,$ is a fundamental $\,\textrm{SL}(2)\,$ index, $\,M=1,...,12\,$ is a fundamental $\,\textrm{SO}(6,6)\,$ index and $\,\mu \,\,(\dot{\mu})=1,...,32\,$ denotes a left (right) Majorana-Weyl spinor transforming in the $\textbf{32} \,\, (\textbf{32}^{\prime})$ of $\,\textrm{SO}(6,6)$. In order to fit the irrep's, the above set of fields come out with the following symmetry properties:

\begin{itemize}

\item[$i)$] The real vectors $\,V_{\alpha M}\,$ and $\,V_{\mu}\,$ are unrestricted.

\item[$ii)$] The scalars satisfy $\,\phi_{MN}=\phi_{[MN]}\,$ and $\,\phi_{\a \b}=\phi_{(\a \b)}\,$ whereas $\,\phi_{\a \dot{\mu}}\,$ remain unrestricted. However, in contrast to the $\,\textrm{SU}(8)\,$ formulation, the $\,70\,$ physical scalars no longer fit an irrep of $\,\textrm{SL}(2) \times \textrm{SO}(6,6)\,$. Instead, $38$ of them parameterise an element of the coset space $\,\frac{\textrm{SL}(2)}{\textrm{SO}(2)} \times \frac{\textrm{SO}(6,6)}{\textrm{SO}(6) \times \textrm{SO}(6)}\,$, whereas the remaining 32 extend it to an $\,\frac{\textrm{E}_{7(7)}}{\textrm{SU}(8)}\,$ coset element. At the origin of the moduli space, \emph{i.e.} $\,{\phi_{MN}=\phi_{\a \b}=\phi_{\a \dot{\mu}}=0}\,$, we are left with a symmetric $\,\textrm{E}_{7(7)}\,$ scalar matrix of the form
\be
\label{M_origin}
\mathcal{M}_{\mathbb{M} \mathbb{N}} \Big|_{\textrm{origin}} =
\left(
\begin{array}{c|c}
\d_{\a \b} \, \d_{MN} & 0 \\[1mm]
\hline
\\[-4mm]
0 & B_{\m \n}
\end{array}
\right)
 \ ,
\ee
where the symmetric and unitary matrix $\,B_{\mu \nu}\,$ is the conjugation matrix introduced in appendix~\ref{App:spinors} to define a reality condition upon gamma matrices of $\,\textrm{SO}(6,6)\,$.

It is worth mentioning here that the explicit form of $\,B_{\mu \nu}\,$ crucially depends on the choice of the gamma matrices representation. For instance, if taking the real representation presented in appendix~\ref{App:spinors}, then $\,B_{\m \n}=\mathds{1}_{32}\,$. On the other hand, if taking the complex representation that makes $\,\textrm{SU}(4) \times \textrm{SU}(4)\,$ covariance explicit (see also the appendix~\ref{App:spinors}), one finds that
$\,B_{\m \n}=
{\footnotesize{
\left(
\begin{array}{cc}
0 & \mathds{1}_{16} \\
\mathds{1}_{16} & 0
\end{array}
\right) }}
\,$. Then, different choices of gamma matrices representation do change the notion of what is called the origin of the moduli space according to its definition in (\ref{M_origin}), even though they are related via a unitary $\,\textrm{U}(32)\,$ transformation.

In order to avoid confusion, we will adopt the convention of $\,B_{\m \n}=\mathds{1}_{32}\,$ when referring to the origin of the moduli space, hence being compatible with the natural choice of
\be
\label{M_origin_choice}
\mathcal{M}_{\mathbb{MN}}\Big|_{\textrm{origin}}=\mathds{1}_{56} \ ,
\ee
as the origin of field space.

\item[$iii)$] The different pieces of the embedding tensor $\,f_{\a MNP},\,\xi_{\a M},\,F_{M \dot{\mu}}$ and $\,\Xi_{\a \b \mu}$ satisfy $\,f_{\a MNP}=f_{\a [MNP]}\,$ together with $\,\slashed{F}^{\mu} \equiv F_{M \dot{\nu}} \, [\gamma^{M}]^{\mu \dot{\nu}}=0\,$ and $\,\Xi_{\a \b \mu}=\Xi_{(\a \b) \mu}\,$.

\end{itemize}

In the $\,\textrm{SL}(2) \times \textrm{SO}(6,6)\,$ formulation, the $\,\textrm{Sp}(56,\mathbb{R})\,$ skew-symmetric invariant matrix $\,\Omega_{\mathbb{M} \mathbb{N}}\,$ becomes block-diagonal and reads
\be
\label{Omega_matrix}
\Omega_{\mathbb{M} \mathbb{N}} =
\left(
\begin{array}{c|c}
\Omega_{\a M \b N} & 0 \\[1mm]
\hline
\\[-4mm]
0 & \Omega_{\m \n}
\end{array}
\right)
=
\left(
\begin{array}{c|c}
\epsilon_{\a \b}\, \eta_{MN} & 0 \\[1mm]
\hline
\\[-4mm]
0 & \mathcal{C}_{\m \n}
\end{array}
\right) \ ,
\ee
where $\,\epsilon_{\a \b}\,$ is the Levi-Civita $\,\textrm{SL}(2)$-invariant tensor (normalised as $\,\epsilon_{+-}=1\,$) and where $\,\eta_{MN}\,$ and $\,\mathcal{C}_{\m \n}\,$ are the metric and the charge conjugation matrix of $\,\textrm{SO}(6,6)$, respectively. We have summarised our conventions for spinorial representations, gamma matrices, etc. of $\,\textrm{SO}(6,6)\,$ in the appendix~\ref{App:spinors}.

\subsubsection*{Fluxes and The Embedding Tensor}

The decomposition of the $\,\textbf{56}\,$ of $\,\textrm{E}_{7(7)}\,$ under $\,\textrm{SL}(2) \times \textrm{SO}(6,6)\,$ translates into the index splitting $\,\mathbb{M}=\alpha M \oplus \mu\,$. When expressed in terms of the different pieces of the embedding tensor, the tensor $\,X_{\mathbb{MNP}}\,$ entering the gauge brackets in (\ref{gauge_algebra}) can then be split into components involving an even number of fermionic indices
\be
\label{Xbosonic}
\begin{array}{cclc}
X_{\a M \b N \g P} & = & - \, \epsilon_{\b \g} \, f_{\a MNP} - \, \epsilon_{\b \g} \, \eta_{M [N}\, \, \xi_{ \a P]} - \, \epsilon_{\a (\b} \, \xi_{\g) M} \, \eta_{NP}& , \\[4mm]
X_{\a M \m \n} & = & -\dfrac{1}{4}  \, f_{\a MNP} \, \left[ \g^{NP} \right]_{\m \n}  -  \dfrac{1}{4} \, \xi_{\a N} \, \left[ {\g_{M}}^{N} \right]_{\m \n}         & , \\[4mm]
X_{\m \a M \n} = X_{\m \n \a M} & = & \dfrac{1}{8} \, f_{\a MNP} \, \left[ \g^{NP} \right]_{\m \n} - \dfrac{1}{24} \, f_{\a NPQ} \, \left[ {\g_{M}}^{NPQ} \right]_{\m \n} \\[3mm]
& + & \dfrac{1}{8} \, \xi_{\a N} \, \left[ {\g_{M}}^{N} \right]_{\m \n} - \dfrac{1}{8} \, \xi_{\a M} \, \cC_{\m \n}        & , \\[2mm]
\end{array}
\ee
which turn out to be sourced by $\,f_{\a MNP}\,$ and $\,\xi_{\a M}$, together with those involving an odd number of them
\be
\label{Xfermionic}
\begin{array}{cclc}
X_{\mu \nu \rho}  & = & - \dfrac{1}{2} \, {F_{M}}_{\dot{\n}} \, {{[\gamma_{N}]}_{\mu}}^{\dot{\n}} \, \left[ \g^{MN} \right]_{\n \r} & , \\[4mm]
X_{\mu \a M \b N}  & = & -2 \, \epsilon_{\a \b} \, F_{[M \dot{\n}} \, {\left[ \g_{N]} \right]_{\m}}^{\dot{\n}} \, - \, 2 \,  \eta_{MN} \, \Xi_{\a \b \m}   & , \\[4mm]
X_{\a M \m \b N} = X_{\a M \b N \m} & = & \epsilon_{\a \b} \, {[\gamma_{N}]_{\mu}}^{\dot{\n}} \, F_{M\dot{\n}} \, +  \, \Xi_{\a \b \n} \, {\left[ \g_{MN} \right]^{\n}}_{\m} \, +  \, \Xi_{\a \b \m} \, \eta_{MN}  & ,
\end{array}
\ee
which are sourced by $\,F_{M \dot{\mu}}\,$ and $\,\Xi_{\a \b \mu}\,$. The set of components in (\ref{Xbosonic}) specifies how half-maximal supergravity is embedded inside maximal \cite{Dibitetto:2011eu}, whereas the remaining components in (\ref{Xfermionic}) represent the completion from half-maximal to maximal supergravity. A derivation of the expression in (\ref{Xbosonic}) and (\ref{Xfermionic}) can be found in the appendix~\ref{App:spinors}.

The brackets of the gauge algebra in (\ref{gauge_algebra}) involving the $\,X_{\mathbb{M}}\,=\, X_{\a M}\,\oplus\,X_{\mu}\,$ generators in maximal supergravity then takes the form
\be
\label{algebra_maximal}
\begin{array}{ccclclc}
\left[ X_{\a M} , X_{\b N} \right] & = &  -  & {X_{\a M \b N}}^{\g P} \, X_{\g P} &-& {X_{\a M \b N}}^{\r} \, X_{\r}  & , \\[2mm]
\left[ X_{\a M} , X_{\m} \right] & = &  - & {X_{\a M \m}}^{\g P} \, X_{\g P} &-& {X_{\a M \m}}^{\r} \, X_{\r}  & , \\[2mm]
\left[ X_{\m} , X_{\n} \right] & = &  - & {X_{\m \n}}^{\g P} \, X_{\g P} &-& {X_{\m \n}}^{\r} \, X_{\r}  & ,
\end{array}
\ee
where, if looking at the part only involving the $\,X_{\a M}\,$ generators, namely
\be
\label{algebra_half-maximal}
\begin{array}{ccc}
\left[ X_{\a M} , X_{\b N} \right]  &=&   - \,\, {X_{\a M \b N}}^{\g P} \, X_{\g P} \ ,
\end{array}
\ee
we rediscover the gauge algebra of half-maximal supergravity \cite{Schon:2006kz} where the $\,X_{\m}\,$ generators have been projected out of the theory.

Moving to explicit string constructions, the $\,f_{\a MNP}\,$ and $\,\xi_{\a M}$ embedding tensor pieces have been related to different background fluxes, \emph{e.g.} to gauge, geometric and non-geometric fluxes, in compactifications of type II and Heterotic strings producing half-maximal supergravities \cite{Aldazabal:2008zza,Dall'Agata:2009gv,Dibitetto:2010rg}. As we have seen, these fluxes restore the invariance of the four-dimensional supergravity under T- and S-duality, \emph{i.e.} under $\,\textrm{SL}(2) \times \textrm{SO}(6,6)\,$ transformations. The $\,F_{M \dot{\mu}}\,$ and $\,\Xi_{\a \b \mu}\,$ embedding tensor pieces are related to additional background fluxes which restore the invariance of the theory under U-duality, \emph{i.e.} under $\,\textrm{E}_{7(7)}\,$ transformations \cite{Aldazabal:2010ef}. Nevertheless, the identification between embedding tensor components and fluxes strongly depends on the string theory under consideration. For instance, a component of the embedding tensor corresponding to a metric flux $\,\omega\,$ in a type IIA construction might correspond to a non-geometric $Q$ flux in a type IIB one and vice versa (see tables \ref{table:unprimed_fluxes4} and \ref{table:primed_fluxes4} in appendix~\ref{App:fluxes}). We will take this fact into account in the last section when analysing specific type II flux models.

\subsubsection*{Quadratic Constraints}

Plugging the expression for the components of the tensor $\,X_{\mathbb{MNP}}\,$ in (\ref{Xbosonic}) and (\ref{Xfermionic}) into the QC in (\ref{quadratic_const}) one finds a set of quadratic relations for the embedding tensor pieces $\,f_{\a MNP}\,$, $\,\xi_{\a M}\,$, $\,F_{M \dot{\mu}}\,$ and $\,\Xi_{\a \b \mu}\,$. In doing so we use the $\,\textrm{Sp}(56,\mathbb{R})\,$  invariant matrix $\,\Omega_{\mathbb{M} \mathbb{N}}\,$ in (\ref{Omega_matrix}). As for the $\,\textrm{SU}(8)\,$ formulation in the previous section, let us use the $\,\textrm{E}_{7(7)}\supset \textrm{SL}(2) \times \textrm{SO}(6,6)\,$ branching relations
\bea
\label{irrep_133}
\textbf{133} & \longrightarrow  & (\textbf{1},\textbf{66})\,\oplus\,(\textbf{3},\textbf{1})\,\oplus\,(\textbf{2},\textbf{32}^{\prime}) \ , \\[2mm]
\label{irrep_8645}
\textbf{8645}&\longrightarrow & (\textbf{1},\textbf{66})\,\oplus\,(\textbf{1},\textbf{2079})\,\oplus\,(\textbf{3},\textbf{66})\,\oplus\,(\textbf{3},\textbf{495})\,\oplus\,(\textbf{3},\textbf{1})\,\oplus\,(\textbf{1},\textbf{462}^{\prime}) \,\oplus \nonumber \\[1mm]
  & &  \oplus \,\,(\textbf{2},\textbf{32}^{\prime})\,\oplus\,(\textbf{2},\textbf{352})\,\oplus\,(\textbf{2},\textbf{1728}^{\prime}) \,\oplus\,(\textbf{4},\textbf{32}^{\prime}) \ ,
\eea
as an organising principle for the QC. After a straightforward but tedious computation, one finds the following set of QC:
\bea
\label{QC1}
&\hspace{-15mm} i)  & \xi_{\a M}\,\xi_{\b}^{\phantom{a}M} \,+\,4\, \epsilon^{\g \d}\, \mathcal{C}^{\mu \nu} \,  \Xi_{\alpha \gamma \mu} \, \Xi_{\beta \delta \nu} = 0 \ , \\[2mm]
\label{QC2}
&\hspace{-15mm} ii) & \xi_{(\a}^{\phantom{a}P}\,f_{\b)PMN} \, - \, 4\,\Xi_{\a \b \m} \, F_{[M \dot{\n}} \, [\gamma_{N]}]^{\mu \dot{\n}} = 0 \ , \\[2mm]
\label{QC3}
&\hspace{-15mm} iii)& 3\,f_{\a R[MN}\,f_{\b PQ]}^{\phantom{abcde}R}\,+\,2\,\xi_{(\a [M}\,f_{\b)NPQ]}   \\ \nonumber
& & - \, 4\,\Xi_{\a \b \m} \, F_{[M \dot{\n}} \, [\gamma_{NPQ]}]^{\mu \dot{\n}} \,-\, \epsilon^{\g \d}\, \Xi_{\alpha \gamma \mu} \, \Xi_{\beta \delta \nu} \, [\gamma_{MNPQ}]^{\mu \nu} = 0 \ , \\[2mm]
\label{QC4}
&\hspace{-15mm} iv) & \epsilon^{\a \b}\left(\xi_{\a}^{\phantom{a}P}\,f_{\b PMN}\,+\,\xi_{\a M}\,\xi_{\b N}\right)  \\ \nonumber
& & - \, 4\, F_{M\dot{\m}} \, {F_{N}}^{\dot{\m}}  \, - \, F_{P \dot{\mu}} \, {F^{P}}_{\dot{\nu}} \, [\gamma_{MN}]^{\dot{\mu} \dot{\nu}} \, + \, 2\, \eps^{\a \g} \, \eps^{\b \d} \, \Xi_{\a \b \m}  \, \Xi_{\g \d \n} \, [\g_{MN}]^{\m \n} = 0  \ , \\[2mm]
\label{QC5}
&\hspace{-15mm} v) & \epsilon^{\a \b}\left(f_{\a MNR}\,f_{\b PQ}^{\phantom{abcde}R}\,-\,\xi_{\a}^{\phantom{a}R}\,f_{\b R[M[P}\,\eta_{Q]N]}\,-\,\xi_{\a [M}\,f_{\b N]PQ}\,+\,\xi_{\a [P}\,f_{\b Q]MN}\right)  \\ \nonumber
& & + \, 4\, F_{[M\dot{\m}}\,[\g_{N][P}]^{\dot{\m}\dot{\n}}\,F_{Q] \dot{\n}} \, - \, F_{R \dot{\m}} \,  {F^{R}}_{\dot{\n}} \,[\bar{\g}_{[M}\,\eta_{N][P}\,\g_{Q]}]^{\dot{\m} \dot{\n}} \, + \, 2\,\Xi_{\a\b\m}\,{\Xi^{\a\b}}_{\n}\,[\g_{[M}\,\eta_{N][P}\,\bar{\g}_{Q]}]^{\m\n}  = 0 \ , \\[2mm]
\label{QC6}
&\hspace{-15mm} vi)  & f_{\a MNP} \, {f_{\b}}^{MNP} \,+\,30 \, \epsilon^{\g \d}\, \mathcal{C}^{\mu \nu} \,  \Xi_{\alpha \gamma \mu} \, \Xi_{\beta \delta \nu} = 0 \ , \\[2mm]
\label{QC7} &\hspace{-15mm} vii)  & \left. \epsilon^{\a \b}\,\,
f_{\a [MNP} \, f_{\b QRS]} \,\, \right|_{\textrm{SD}} \, - \,
\frac{1}{40}  \, F_{T \dot{\m}} \, {F^{T}}_{\dot{\n}} \,
[\gamma_{MNPQRS}]^{\dot{\m}\dot{\n}}  = 0  \ , \eea
associated to the irrep's
\be
\begin{array}{cccc}
i)\,\,(\textbf{3},\textbf{1}) & \hspace{10mm}   ii)\,\,(\textbf{3},\textbf{66}) & \hspace{10mm} iii)\,\,(\textbf{3},\textbf{495}) & \hspace{10mm}  iv)\,\,(\textbf{1},\textbf{66})
\end{array}
\ee
\vspace{-5mm}
\be
\begin{array}{ccc}
v)\,\,(\textbf{1},\textbf{66}) \, \oplus \, (\textbf{1},\textbf{2079}) &\hspace{20mm}  vi)\,\,(\textbf{3},\textbf{1})  & \hspace{20mm} vii)\,(\textbf{1},\textbf{462}^{\prime}) \ ,
\end{array}
\ee
together with three additional ones
\bea
\label{QC8}
&\hspace{-15mm} viii)  & f_{(\a MNP}\,\Xi_{\b\g)\n}\,[\g^{MNP}]^{\n\dot{\m}}\,-\,15\,\xi_{(\a M}\,\Xi_{\b\g)\n}\,[\g^{M}]^{\n\dot{\m}} = 0 \ , \\[2mm]
\label{QC9}
&\hspace{-15mm} ix) & - 3\, \xi_{\a M} \,F^{M \dot{\m}}  \, + \, \epsilon^{\beta \gamma} \, \left( \frac{1}{2} \, \xi_{\b M} \, [\gamma^{M}]^{\nu \dot{\m}} \, + \, \frac{1}{6} \, f_{\b MNP} \, [\gamma^{MNP}]^{\nu \dot{\m}}  \right) \, \Xi_{\a \g \n} = 0 \ , \\[2mm]
\label{QC10} &\hspace{-15mm} x)& {f_{\a MN}}^{P} \,
{F_{P}}^{\dot{\m}} \, + \, \frac{1}{4} \, f_{\a PQ [M} \, F_{N]
\dot{\n}} \, [\gamma^{PQ}]^{\dot{\n} \dot{\m}} \,-\, \frac{1}{12} \,
f_{\a PQR} \, F_{[M \dot{\n}} \, [{\gamma_{N]}}^{PQR}]^{\dot{\n}
\dot{\m}} \\ \nonumber & & + \, \frac{1}{4} \, \xi_{\a P} \, F_{[M
\dot{\n}} \, [{\gamma_{N]}}^{P}]^{\dot{\n} \dot{\m}} \, + \, \epsilon^{\beta \gamma} \, \left( f_{\b MNP}
\, [\gamma^{P}]^{\nu \dot{\m}} \, - \, \xi_{\b [M} \,
[\gamma_{N]}]^{\nu \dot{\m}} \right) \, \Xi_{\a \g \n} \\ \nonumber
& & - \, \frac{5}{4} \, \xi_{\a [M} \, {F_{N]}}^{\dot{\m}} \, = 0
\eea
associated to
\be
\begin{array}{ccc}
 viii)\,\,(\textbf{4},\textbf{32}^{\prime}) & \hspace{10mm} ix)\,\,(\textbf{2},\textbf{32}^{\prime}) & \hspace{10mm} x)\,\,(\textbf{2},\textbf{32}^{\prime}) \, \oplus \, (\textbf{2},\textbf{352}) \, \oplus \, (\textbf{2},\textbf{1728}^{\prime}) \ .
\end{array}
\ee
Let us comment a bit more about the above set of QC. If we refer to the embedding tensor pieces $\,f_{\a MNP}\,$ and $\,\xi_{\a M}\,$  as ``bosonic'' and to $\,F_{M\dot{\m}}\,$ and $\,\Xi_{\a \b \m}\,$ as ``fermionic'', then the first seven conditions can be understood as $\,\textrm{(bos $\times$ bos) $+$ (fermi $\times$ fermi}) = 0\,$ QC whereas the last three are of the form $\,\textrm{(bos $\times$ fermi) } = 0\,$. As a check of consistency, the first seven conditions reduce to those of the form $\,\textrm{(bos $\times$ bos) } = 0\,$ in ref.~\cite{Dibitetto:2011eu} by setting $\,F_{M\dot{\m}} \,=\, \Xi_{\a \b \m} = 0\,$, namely, by switching off fluxes associated to $\,\textrm{SO}(6,6)$-fermi irrep's of the embedding tensor. In particular, in such a case, the first five conditions characterise a consistent $\cN=4$ gauging \cite{Schon:2006kz} and the remaining two extra conditions select those $\cN=4$ gaugings given by $f_{\a MNP}$ and $\xi_{\a M}$ which admit an uplift to the maximal theory \cite{Dibitetto:2011eu}. In this case, the last three conditions are trivially satisfied.

As mentioned before, looking for a higher-dimensional origin of dual fluxes is becoming a very exciting line of research. As far as fluxes related to the $\,f_{\a MNP}\,$ components of the embedding tensor are concerned, only purely electric $\,\textrm{SO}(6,6)\,$ gaugings have been at first formally addressed by DFT \cite{Aldazabal:2011nj, Geissbuhler:2011mx, Grana:2012rr}. However, the explicit twelve-dimensional twist matrices producing such gaugings have only been built in some particular cases \cite{Hull:2007jy, Dall'Agata:2007sr, Hull:2009sg}. In chapter~\ref{DFT} based on ref.~\cite{Dibitetto:2012rk} the construction has been presented for $D\ge 7$. In order to firstly extend to $\,\textrm{SL}(2) \times \textrm{SO}(6,6)\,$ gaugings including fluxes related to $\,\xi_{\a M}\,$ and secondly to $\,\textrm{E}_{7(7)}\,$ gaugings involving also fluxes related to the $\,F_{M\dot{\m}}\,$ and $\,\Xi_{\a \b \m}\,$ components (such as R-R gauge fluxes amongst others), a generalisation to a $56$-dimensional ``twisted megatorus'' reduction has been proposed \cite{Dall'Agata:2007sr}. The restrictions upon the $56$-dimensional twist matrices on this megatorus have not been worked out yet, but, when expressed in terms of fluxes, they should at least imply those in (\ref{QC1})-(\ref{QC7}) and (\ref{QC8})-(\ref{QC10}) whenever the twist is compatible with maximal supersymmetry in four dimensions.

\section{Connecting SU($8$) and SL($2)\,\times\,$SO($6,6$)}
\label{sec:connecting_A7&A1D6}

In order to relate the  $\,\textrm{SL}(2) \,\times \, \textrm{SO}(6,6)\,$ and the $\,\textrm{SU}(8)\,$ formulations of maximal supergravity, it is mandatory to derive the expression of the $\,X_{\mathbb{MNP}}\,$ tensor entering the brackets in (\ref{algebra_maximal}) as a function of the fermionic mass terms in (\ref{Fermi_Lagrangian}).\, This can be done in a two-step procedure as follows:

\begin{itemize}

\item[$1)$] By using the tensors $\mathcal{A}^{\mathcal{IJ}}$ and ${\mathcal{A}_{\mathcal{I}}}^{\mathcal{JKL}}$, we can build the so-called  $T$-tensor \cite{deWit:2007mt,Diffon:2011wt}. The components of this $T$-tensor take the form
\be
\hspace{-12mm}
\begin{array}{ccll}
\label{T(A)}
T_{\mathcal{IJ KL MN}} & = & \frac{1}{24} \, \epsilon_{\mathcal{KLMNRSTU}}\, \delta_{[\mathcal{I}}^{\mathcal{R}} \, {\mathcal{A}_{\mathcal{J}]}}^{\mathcal{STU}} & , \\[4mm]
{T_{\mathcal{IJ KL}}}^{\mathcal{MN}} & = & \phantom{-} \frac{1}{2} \,\delta_{[\mathcal{K}}^{[\mathcal{M}} \, {\mathcal{A}^{\mathcal{N}]}}_{\mathcal{L}]\mathcal{IJ}}  + \delta_{[\mathcal{I}[\mathcal{K}}^{\mathcal{MN}} \, \mathcal{A}_{\mathcal{L}]\mathcal{J}]} & , \\[4mm]
T_{\mathcal{IJ \phantom{KL} MN}}^{\mathcal{\phantom{IJ} KL}} & = & - \frac{1}{2} \,\delta_{[\mathcal{M}}^{[\mathcal{K}} \, {\mathcal{A}^{\mathcal{L}]}}_{\mathcal{N}]\mathcal{IJ}}  - \delta_{[\mathcal{I}[\mathcal{M}}^{\mathcal{KL}} \, \mathcal{A}_{\mathcal{N}]\mathcal{J}]} & ,\\[4mm]
{T_{\mathcal{IJ}}}^{\mathcal{KL MN}} & = & \delta_{[\mathcal{I}}^{[\mathcal{K}} \, {\mathcal{A}_{\mathcal{J}]}}^{\mathcal{LMN}]} & , \\[2mm]
\end{array}
\ee
together with their complex conjugates. We can arrange them into a $\,T_{\underline{\mathbb{M}} \underline{\mathbb{N}}}{}^{\underline{\mathbb{P}}}\,$ tensor by using the decomposition $\,\underline{\mathbb{M}}=[\mathcal{IJ}] \oplus [\mathcal{\bar{I}\bar{J}}] \equiv \left\lbrace _{\mathcal{IJ}} \, , \, ^{\mathcal{IJ}} \right\rbrace\,$ of the $\,\bold{56}\,$ of $\,\textrm{E}_{7(7)}\,$ under $\,\textrm{SU}(8)\,$.

\item[$2)$] The constant $\,X_{\mathbb{MNP}}\,$ tensor in the $\,\textrm{SU}(8)\,$ formulation, let us denote it $\,\widetilde{X}_{\mathbb{MNP}}\,$ to avoid confusion with that in the $\,\textrm{SL}(2) \times \textrm{SO}(6,6)\,$ formulation, can then be obtained by removing the dependence of the $\,T_{\underline{\mathbb{M}} \underline{\mathbb{N}} \underline{\mathbb{P}}}\,$ tensor on the scalar fields (see footnote~1)
\be
\label{tilde_X(T)}
\widetilde{X}_{\mathbb{MNP}} = 2 \, \widetilde{\mathcal{V}}_{\mathbb{M}}^{\phantom{\mathbb{M}}\underline{\mathbb{Q}}} \, \widetilde{\mathcal{V}}_{\mathbb{N}}^{\phantom{\mathbb{N}}\underline{\mathbb{R}}} \, \widetilde{\mathcal{V}}_{\mathbb{P}}^{\phantom{\mathbb{P}}\underline{\mathbb{S}}}\,\,\, T_{\underline{\mathbb{Q}\mathbb{R}\mathbb{S}}} \ ,
\ee
where $\,\widetilde{\mathcal{V}}_{\mathbb{M}}^{\phantom{\mathbb{M}}\underline{\mathbb{Q}}}\,$ is the $\,\textrm{E}_{7(7)}/\textrm{SU}(8)\,$ vielbein in the $\,\textrm{SU}(8)\,$ formulation \cite{deWit:2007mt}. After removing the scalar dependence, the $\,\widetilde{X}_{\mathbb{MNP}}\,$ and $\,X_{\mathbb{MNP}}\,$ constant tensors in the $\textrm{SU}(8)$ and $\,{\textrm{SL}(2) \times \textrm{SO}(6,6)}\,$ formulations are related via a constant change of basis
\be
\label{X(tilde_X)}
X_{\mathbb{MNP}} = \mathring{\mathcal{V}}_{\mathbb{M}}^{\phantom{\mathbb{M}}\mathbb{Q}} \, \mathring{\mathcal{V}}_{\mathbb{N}}^{\phantom{\mathbb{N}}\mathbb{R}} \, \mathring{\mathcal{V}}_{\mathbb{P}}^{\phantom{\mathbb{P}}\mathbb{S}}\,\,\, \widetilde{X}_{\mathbb{QRS}} \ .
\ee
Composing (\ref{tilde_X(T)}) and (\ref{X(tilde_X)}), the resulting vielbein\footnote{For more details on the vielbein $\,\mathcal{V}^{\phantom{\mathbb{M}}\underline{\mathbb{N}}}_{\mathbb{M}}\,$, see appendix~\ref{App:vielbein}.} $\,\mathcal{V}_{\mathbb{M}}^{\phantom{\mathbb{M}}\underline{\mathbb{N}}} = \mathring{\mathcal{V}}_{\mathbb{M}}^{\phantom{\mathbb{M}}\mathbb{P}} \, \widetilde{\mathcal{V}}_{\mathbb{P}}^{\phantom{\mathbb{P}}\underline{\mathbb{N}}}\,$ directly connects the tensors $\,X_{\mathbb{MNP}}\,$ and $\,T_{\underline{\mathbb{MNP}}}\,$
\be
\label{X(T)}
X_{\mathbb{MNP}} = 2 \, \mathcal{V}_{\mathbb{M}}^{\phantom{\mathbb{M}}\underline{\mathbb{Q}}} \, \mathcal{V}_{\mathbb{N}}^{\phantom{\mathbb{N}}\underline{\mathbb{R}}} \, \mathcal{V}_{\mathbb{P}}^{\phantom{\mathbb{P}}\underline{\mathbb{S}}}\,\,\, T_{\underline{\mathbb{QRS}}} \ .
\ee
\end{itemize}

Schematically, the connection between the two formulations of maximal supergravity works in the following way
\be
\label{mapping_summary}
\begin{array}{ccccc}
 \underbrace{\left( \mathcal{A}^{\mathcal{IJ}} \,\,,\,\, {\mathcal{A}_{\mathcal{I}}}^{\mathcal{JKL}} \right)}_{\textrm{fermi. masses}} \hspace{5mm} & \vspace{-5mm} \Longrightarrow  &  T_{\underline{\mathbb{MNP}}}  & \Longrightarrow  & \hspace{5mm} \underbrace{\hspace{5mm}X_{\mathbb{MNP}}\hspace{5mm}}_{\textrm{flux background}} \ .\\[2mm]
  & (\ref{T(A)}) &  & (\ref{X(T)})
\end{array}
\ee
By inverting the above chain\footnote{The inversion of the relations in (\ref{T(A)}) gives $\,\mathcal{A}^{\mathcal{IJ}}=\frac{4}{21} \, {T^{\mathcal{IKJL}}}_{\mathcal{KL}}\,$ and $\,{\mathcal{A}_{\mathcal{I}}}^{\mathcal{JKL}}=2 \, {T_{\mathcal{MI}}}^{\mathcal{MJKL}}$, showing that there is some redundancy in the $T$-tensor components.} we are able to relate a flux background given in terms of $\,f_{\a MNP}\,$, $\,\xi_{\a M}\,$, $\,F_{M \dot{\mu}}\,$ and $\,\Xi_{\a \b \mu}\,$ to certain fermionic mass terms $\mathcal{A}^{\mathcal{IJ}}$ and ${\mathcal{A}_{\mathcal{I}}}^{\mathcal{JKL}}\,$. This amounts to know the relations
\be
\label{A's_fluxes}
\begin{array}{ccr}
\mathcal{A}^{\mathcal{IJ}} &=& \mathcal{A}^{\mathcal{IJ}} \, \left( \, f_{\a MNP} \,,\, \xi_{\a M} \,,\, F_{M \dot{\mu}} \,,\, \Xi_{\a \b \mu} \,\, ; \,\, \mathcal{V}_{\phantom{\mathbb{M}}\underline{\mathbb{N}}}^{\mathbb{M}} \, \right) \phantom{\ .} \\[2mm]
{\mathcal{A}_{\mathcal{I}}}^{\mathcal{JKL}} &=& {\mathcal{A}_{\mathcal{I}}}^{\mathcal{JKL}} \, \left( \, f_{\a MNP} \,,\, \xi_{\a M} \,,\, F_{M \dot{\mu}} \,,\, \Xi_{\a \b \mu} \,\, ; \,\, \mathcal{V}_{\phantom{\mathbb{M}}\underline{\mathbb{N}}}^{\mathbb{M}} \, \right) \ ,
\end{array}
\ee
where $\,\mathcal{V}_{\phantom{\mathbb{M}}\underline{\mathbb{N}}}^{\mathbb{M}}=(\mathcal{V}^{-1})_{\underline{\mathbb{N}}}^{\phantom{\mathbb{N}}\mathbb{M}}\,$ is the inverse vielbein. Having the relations (\ref{A's_fluxes}), we can make use of (\ref{V_SU8}), (\ref{scalars_eom}) and (\ref{Mass-matrix}) in order to compute the scalar potential, the E.O.M's and the masses of the $70$ physical scalars for a specific flux background.

\subsection*{The $\,F_{M\dot{\m}} \,=\, \Xi_{\a \b \m} = 0\,$ Case}

Let us derive the relations (\ref{A's_fluxes}) between fermionic mass terms and embedding tensor components when $\,F_{M\dot{\m}} \,=\, \Xi_{\a \b \m} \,=\, 0\,$. On the string theory side, this means that fluxes related to fermionic components of the embedding tensor are set to zero, so that
\be
\label{A's_fluxes_Bosonic}
\begin{array}{ccr}
\mathcal{A}^{\mathcal{IJ}} &=& \mathcal{A}^{\mathcal{IJ}} \, \left( \, f_{\a MNP} \,,\, \xi_{\a M}  \,\, ; \,\, \mathcal{V}_{\phantom{\mathbb{M}}\underline{\mathbb{N}}}^{\mathbb{M}} \, \right) \phantom{\ .} \\[2mm]
{\mathcal{A}_{\mathcal{I}}}^{\mathcal{JKL}} &=& {\mathcal{A}_{\mathcal{I}}}^{\mathcal{JKL}} \, \left( \, f_{\a MNP} \,,\, \xi_{\a M}  \,\, ; \,\, \mathcal{V}_{\phantom{\mathbb{M}}\underline{\mathbb{N}}}^{\mathbb{M}} \, \right) \ .
\end{array}
\ee

Before presenting the explicit form of the relations in (\ref{A's_fluxes_Bosonic}), we want to point out an issue that appears during the computation, the way to overcome it and the corresponding price to pay:
\begin{itemize}

\item[$i)$] In the $\,\textrm{SL}(2) \times \textrm{SO}(6,6)\,$ formulation of maximal supergravity, the scalar fields split into ``bosonic'' $\,\{\phi_{\a\b} \,,\, \phi_{MN}\}\,$ and ``fermionic'' $\,\phi_{\a\dot{\m}}\,$ ones. While the former enter the vielbein $\,{\mathcal{V}_{\mathbb{M}}}^{\underline{\mathbb{N}}}\,$ in a simple way, the latter do it in a very complicated way. In the derivation of the relations (\ref{A's_fluxes_Bosonic}), we will set $\,\phi_{\a\dot{\m}}=0\,$ which means that all the ``fermionic'' scalars are fixed to their values at the origin of the moduli space. Therefore, the relation between fluxes and fermionic mass terms that we present here is only valid in the submanifold of the moduli space where $\,\phi_{\a\dot{\m}}=0\,$.

\item[$ii)$] Being tight to the submanifold with $\,\phi_{\a\dot{\m}}=0\,$ is perfectly consistent with embedding $\,\cN=4\,$ flux compactifications (and truncations thereof) inside $\,\cN=8\,$ supergravity, since ``fermionic'' scalars are projected out (set to zero) when truncating from maximal to half-maximal supergravity in four dimension \cite{Dibitetto:2011eu}. A special point in this submanifold is the origin of the moduli space defined in (\ref{M_origin}), where both ``bosonic'' and ``fermionic'' scalars are set to zero.

\item[$iii)$] One of the main consequences of taking $\,F_{M\dot{\m}} \,=\, \Xi_{\a \b \m} \,=\, 0\,$ as well as $\,\phi_{\a\dot{\m}}=0\,$ is that the method introduced in ref.~\cite{Dibitetto:2011gm} (and further exploited in ref.~\cite{DallAgata:2011aa}) for charting critical points of the scalar potential becomes more subtle. This method relies on the fact that the manifold spanned by the scalars, \emph{i.e.} $\,\textrm{E}_{7(7)}/\textrm{SU}(8)\,$ in the case of maximal supergravity, is homogeneous so any critical point can be brought back to the origin of the moduli space by applying an $\,\textrm{E}_{7(7)}\,$ transformation. However, neither $\,{F_{M\dot{\m}} \,=\, \Xi_{\a \b \m} \,=\, 0}\,$ nor $\,\phi_{\a\dot{\m}}=0\,$ are U-duality covariant conditions: $\,\textrm{E}_{7(7)}\,$ transformations will mix ``fermionic'' and ``bosonic'' embedding tensor components and scalars, hence rendering the relations in (\ref{A's_fluxes_Bosonic}) no longer valid. We will be back to this point in the last section when discussing specific flux backgrounds yielding maximal supergravities.

\end{itemize}

Taking $\,F_{M\dot{\m}} \,=\, \Xi_{\a \b \m} \,=\, 0\,$ together with $\,\phi_{\a\dot{\m}}=0\,$ has strong implications for the mapping between fluxes and fermionic mass terms. By virtue of the decompositions (\ref{A1D6_decomp}) and (\ref{A7_decomp}), only those components (as well as their c.c.) of the form $\,\{ {\mathcal{V}_{\a M}}^{ij} \,,\,  {\mathcal{V}_{\a M}}^{\hi \hj} \,,\,  {\mathcal{V}_{i \hj}}^{k \hl}  \}\,$ inside the vielbein $\,\mathcal{V}^{\phantom{\mathbb{M}}\underline{\mathbb{N}}}_{\mathbb{M}}\,$ are non-vanishing. They are expressed in terms of an $\,\textrm{SL}(2)\,$ complexified vielbein $\,\mathcal{V}_{\a}\,$ and an $\,\textrm{SO}(6,6)\,$ vielbein $\,\mathcal{V}_{M}=\{ {\mathcal{V}_{M}}^{ij} , {\mathcal{V}_{M}}^{\hi \hj} \}\,$ where $\,i=1,...,4\,$ and $\,\hi=1,...,4\,$ respectively denote $\,\textrm{SU}(4)_{\textrm{time-like}}\,$ and $\,\textrm{SU}(4)_{\textrm{space-like}}\,$ fundamental indices (see appendix~\ref{App:vielbein}).

Considering this reduced set of vielbein components, we can build the explicit mapping between fermionic mass terms and fluxes by following the prescription in (\ref{mapping_summary}). It will be useful to define the tensors
\be
\label{fermi_mass_N=4_new}
\begin{array}{ccl}
A_{1}^{ij} & = & \epsilon^{\alpha \beta} \, (\mathcal{V}_{\alpha})^{*} \, {\mathcal{V}^{M}}_{kl} \, \mathcal{V}^{N ik} \, \mathcal{V}^{P jl} \, f_{\beta MNP} \\[2mm]
A_{2}^{ij} & = & \epsilon^{\alpha \beta} \,\,\, \mathcal{V}_{\alpha} \,\,\,\,\,\, {\mathcal{V}^{M}}_{kl} \, \mathcal{V}^{N ik} \, \mathcal{V}^{P jl} \, f_{\beta MNP} \,+ \frac{3}{2} \, \epsilon^{\alpha \beta} \, \mathcal{V}_{\alpha} \, {\mathcal{V}}^{M ij} \, {\xi_{\beta M}} \\[2mm]
{A_{2\,\hi \hj i}}^{j} & = & \epsilon^{\alpha \beta} \,\,\, \mathcal{V}_{\alpha} \,\,\,\,\,\, {\mathcal{V}^{M}}_{\hi \hj} \, {\mathcal{V}^{N}}_{ik} \, \mathcal{V}^{P jk} \, f_{\beta MNP} - \frac{1}{4} \,\delta_{i}^{j} \, \epsilon^{\alpha \beta} \, \mathcal{V}_{\alpha} \, {\mathcal{V}^{M}}_{\hi \hj} \, {\xi_{\beta M}}
\end{array}
\ee
which reproduce the fermionic mass terms in $\,\cN=4\,$ supergravity \cite{Schon:2006kz} (see expressions given in \eqref{fermi_mass_N=4}, where the SO($6)_{\textrm{space-like}}$ indices have been complexified by means of the space-like and SD 't Hooft symbols $\left[G_{a}\right]^{\hi\hj}$ given in \eqref{'tHooft_spacelike}), together with their counterparts
\be
\label{fermi_mass_N=4_extension}
\begin{array}{ccl}
A_{1}^{\hi \hj} & = & \epsilon^{\alpha \beta} \,\,\, \mathcal{V}_{\alpha} \,\,\,\,\,\, {\mathcal{V}^{M}}_{\hk \hl} \, \mathcal{V}^{N \hi \hk} \, \mathcal{V}^{P \hj \hl} \, f_{\beta MNP} \\[2mm]
A_{2}^{\hi \hj} & = & \epsilon^{\alpha \beta} \, (\mathcal{V}_{\alpha})^{*} \, {\mathcal{V}^{M}}_{\hk \hl} \, \mathcal{V}^{N \hi \hk} \, \mathcal{V}^{P \hj \hl} \, f_{\beta MNP} \,- \frac{3}{2} \, \epsilon^{\alpha \beta} \, (\mathcal{V}_{\alpha})^{*} \, {\mathcal{V}}^{M \hi \hj} \, {\xi_{\beta M}} \\[2mm]
{A_{2\,i j \hi}}^{\hj} & = & \epsilon^{\alpha \beta} \, (\mathcal{V}_{\alpha})^{*} \, {\mathcal{V}^{M}}_{i j} \, {\mathcal{V}^{N}}_{\hi \hk} \, \mathcal{V}^{P \hj \hk} \, f_{\beta MNP} + \frac{1}{4} \,\delta_{\hi}^{\hj} \, \epsilon^{\alpha \beta} \, (\mathcal{V}_{\alpha})^{*} \, {\mathcal{V}^{M}}_{i j} \, {\xi_{\beta M}}
\end{array}
\ee
which complete the $\,\mathcal{N}=8\,$ theory. In terms of these, the relation between fluxes and fermionic mass terms is given by
\be
g \, \mathcal{A}^{\mathcal{IJ}} = \frac{1}{3\,\sqrt{2}} \left(
\begin{array}{c|c}
A_{1}^{ij} & 0 \\[1mm]
\hline
\\[-4mm]
0 & i \, A_{1}^{\hi \hj}
\end{array}
\right)
\ee
for the components inside $\,{\mathcal{A}}^{\mathcal{IJ}}\,$ and
\be
\begin{array}{cclcccl}
g \,  {\mathcal{A}_{i}}^{jkl} & = &  - \dfrac{1}{3\,\sqrt{2}} \, \epsilon^{jklm} \, A_{2 \, m i} & \hspace{5mm} , \hspace{5mm} & g \,  {\mathcal{A}_{\hi}}^{\hj \hk \hl} & = &  \dfrac{i}{3\,\sqrt{2}} \, \epsilon^{\hj \hk \hl \hm} \, A_{2 \, \hm \hi} \\[4mm]
g \,  {\mathcal{A}_{i}}^{j \hk \hl} & = &  \dfrac{i}{2\,\sqrt{2}} \, \epsilon^{\hk \hl \hi \hj} \, {A_{2 \, \hi \hj i}}^{j} & \hspace{5mm} , \hspace{5mm} & g \,  {\mathcal{A}_{\hi}}^{\hj k l} & = &  -\dfrac{1}{2\,\sqrt{2}} \, \epsilon^{k l i j} \, {A_{2 \, i j \hi}}^{\hj}
\end{array}
\ee
for those inside $\,{\mathcal{A}_{\mathcal{I}}}^{\mathcal{JKL}}$. Further components involving an odd number of $\,\hat{i}\,$ indices, \emph{e.g.} $\,\mathcal{A}^{i\hat{j}}\,$ or $\,{\mathcal{A}_{ijk}}^{\hat{l}}\,$, are sourced by fermionic fields and fluxes, thus vanishing in our setup.

In the next section we present a series of consistent truncations of maximal supergravity yielding simpler theories with a smaller set of fields and embedding tensor components. Later on, we will investigate the lifting of (solutions of) these truncations to maximal supergravity  making use of the explicit correspondence between flux backgrounds and fermionic mass terms derived here.

\subsection*{A Web of Group-theoretical Truncations}

Starting from gauged maximal supergravity in four dimensions, we present a net of group-theoretical truncations (see figure~\ref{fig:truncations}) which connects various supergravity theories preserving different amounts of supersymmetry with different field contents and sets of deformation parameters (embedding tensor components). By group-theoretical truncation, we mean the following procedure: a certain subgroup $H$ of the global symmetry group $G$ of the theory under consideration is chosen, the branching of $G$ irrep's in which fields and deformations live are computed and only the fields and deformations which are singlets with respect to $H$ are kept.

\begin{figure}[h]
\hspace{-4mm}
\renewcommand{\arraystretch}{1.5}
\scalebox{0.67}[0.70]{\xymatrix{*+[F-,]{\begin{array}{cc} \cN=8 : &
\dfrac{\textrm{E}_{7(7)}}{\textrm{SU}(8)}\\\phantom{S} &
\phantom{(\textbf{1},\textbf{1},\textbf{3})\oplus
(\textbf{1},\textbf{3},\textbf{1})\oplus(\textbf{3},\textbf{1},\textbf{1})}
\\[-8mm] \hline \textrm{vectors} & \textbf{56} \vspace{2mm}\\ \hdashline
 \textrm{scalars} & \textbf{133} \vspace{2mm}\\
\hdashline \textrm{emb ten} & \textbf{912}\\
 & \hspace{-16mm}\text{--1--}\end{array}} &
*+[F-,]{\begin{array}{cc}\cN=2 : &
\left(\dfrac{\textrm{SL}(2)}{\textrm{SO}(2)}\right)_{T}\times
\dfrac{\textrm{G}_{2(2)}}{\textrm{SO}(4)}\\ \phantom{S} &
\phantom{(\textbf{1},\textbf{1},\textbf{3})\oplus
(\textbf{1},\textbf{3},\textbf{1})\oplus(\textbf{3},\textbf{1},\textbf{1})}\\[-8mm]
\hline
\textrm{vectors} & (\textbf{4},\textbf{1}) \vspace{2mm}\\\hdashline \textrm{scalars} &
(\textbf{3},\textbf{1})\oplus
(\textbf{1},\textbf{14})\vspace{2mm}\\\hdashline \textrm{emb ten} &
(\textbf{2},\textbf{1})\oplus(\textbf{4},\textbf{14})\oplus(\textbf{2},\textbf{7})\\
 & \hspace{-16mm}\text{--2--}\end{array}}
&*+[F-,]{\begin{array}{cc}\cN=2 : &
\left(\dfrac{\textrm{SL}(2)}{\textrm{SO}(2)}\right)_{T}\times
\dfrac{\textrm{SU}(2,1)}{\textrm{SU}(2)\times\textrm{U}(1)_{U}}\\\phantom{S}
& \phantom{(\textbf{1},\textbf{1},\textbf{3})\oplus
(\textbf{1},\textbf{3},\textbf{1})\oplus(\textbf{3},\textbf{1},\textbf{1})}
\\[-8mm]
\hline \textrm{vectors} & (\textbf{4},\textbf{1}) \vspace{2mm}\\\hdashline \textrm{scalars} &
(\textbf{3},\textbf{1})\oplus
(\textbf{1},\textbf{8})\vspace{2mm}\\\hdashline \textrm{emb ten} &
(\textbf{2},\textbf{1})\oplus(\textbf{4},\textbf{8})\oplus(\textbf{2},\textbf{1})\\
 & \hspace{-16mm}\text{--3--}\end{array}}\\
*+[F-,]{\begin{array}{cc}\cN=4 : &
\left(\dfrac{\textrm{SL}(2)}{\textrm{SO}(2)}\right)_{S}\times\dfrac{\textrm{SO}(6,6)}{\textrm{SO}(6)\times\textrm{SO}(6)}\\\phantom{S}
& \phantom{(\textbf{1},\textbf{1},\textbf{3})\oplus
(\textbf{1},\textbf{3},\textbf{1})+(\textbf{3},\textbf{1},\textbf{1})}
\\[-8mm]
\hline \textrm{vectors} & (\textbf{2},\textbf{12}) \vspace{2mm}\\\hdashline \textrm{scalars} &
(\textbf{3},\textbf{1})\oplus
(\textbf{1},\textbf{66})\vspace{2mm}\\\hdashline \textrm{emb ten} &
(\textbf{2},\textbf{12})\oplus(\textbf{2},\textbf{220})\\
 & \hspace{-16mm}\text{--4--}\end{array}}
& *+[F-,]{\begin{array}{cc}\cN=1 : &
\displaystyle\prod_{\Phi=S,T,U} \left(\dfrac{\textrm{SL}(2)}{\textrm{SO}(2)}\right)_{\Phi} \\[3mm]
\hline \textrm{vectors} &
\text{---} \vspace{2mm} \\\hdashline \textrm{scalars} &
(\textbf{3},\textbf{1},\textbf{1})\oplus(\textbf{1},\textbf{3},\textbf{1})\oplus(\textbf{1},\textbf{1},\textbf{3})\vspace{2mm}\\\hdashline
\textrm{emb ten} &
(\textbf{2},\textbf{2},\textbf{2})\oplus(\textbf{2},\textbf{4},\textbf{4})\\
 & \hspace{-16mm}\text{--5--}\end{array}}
&*+[F-,]{\begin{array}{cc}\cN=1 : &
\displaystyle\prod_{\Phi=S,T}\left(\dfrac{\textrm{SL}(2)}{\textrm{SO}(2)}\right)_{\Phi}\\ \phantom{S}
& \phantom{(\textbf{3},\textbf{1},\textbf{1})\oplus(\textbf{1},\textbf{3},\textbf{1})+(\textbf{1},\textbf{1},\textbf{3})}\\[-8mm]
\hline
\textrm{vectors} & \text{---} \vspace{2mm}\\\hdashline \textrm{scalars} &
(\textbf{3},\textbf{1}) \oplus (\textbf{1},\textbf{3}) \vspace{2mm}\\\hdashline \textrm{emb ten}
& (\textbf{2},\textbf{4}) \oplus(\textbf{2},\textbf{4}) \\
 & \hspace{-16mm}\text{--6--}\end{array}}
}}\caption{{\it Starting from gauged maximal supergravity (box --1-- in
the above diagram), one can move step by step downwards or towards
the right by performing group-theoretical truncations which are
described below in detail. The labels $S$, $T$ and $U$ are introduced in order to keep track of the different group factors along the truncations.}}
\label{fig:truncations}
\end{figure}

\begin{itemize}

\item \textbf{Step from --1-- to --2--:} a truncation with respect to an $\,H=\textrm{SO}(3)\,$ subgroup of the $\,G=\textrm{E}_{7(7)}\,$ global symmetry of maximal supergravity is performed by making use of the following chain of maximal subgroups
\be
\textrm{E}_{7(7)} \,\,\supset \,\,\textrm{SL}(2)_{T} \,\times \,\textrm{F}_{4(4)} \,\,\supset \,\, \textrm{SL}(2)_{T} \,\times \,\textrm{G}_{2(2)} \,\times \, \textrm{SO}(3) \ .
\ee
By looking at the decomposition of the fundamental representation of the $\textrm{SU}(8)$ $R$-symmetry group in --1-- under the $\,\textrm{SO}(3)\,$ diagonal subgroup\footnote{The diagonal subgroup $\,\textrm{SO}(3)_{\textrm{diag}}\,$ in the chain (\ref{chain:SO3_trunc}) is obtained by identifying the two $\,\textrm{SO}(3)\,$ factors, namely, the fundamental representation of the first with the fundamental of the second.}
\be
\label{chain:SO3_trunc}
\textrm{SU}(8) \,\,\supset \,\,\textrm{SU}(4) \,\times \,\textrm{SU}(4) \,\,\supset \,\, \textrm{SU}(3) \,\times \,\textrm{SU}(3) \,\,\supset \,\, \textrm{SO}(3) \,\times \,\textrm{SO}(3) \,\,\supset \,\, \textrm{SO}(3)_{\textrm{diag}} \ ,
\ee
one finds $\,\textbf{8} = \textbf{3} \oplus \textbf{3} \oplus \textbf{1} \oplus \textbf{1}\,$, hence containing two singlets. This implies that the theory preserves $\cN=2$ supersymmetry with
\be
\cM_{\textrm{SK}}= \left(\frac{\textrm{SL}(2)}{\textrm{SO}(2)}\right)_{T} \qquad\textrm{and}\qquad \cM_{\textrm{QK}}=\frac{\textrm{G}_{2(2)}}{\textrm{SO}(4)} \ ,
\ee
being the special K\"ahler (SK) and the quaternionic K\"ahler (QK) manifolds associated to one vector multiplet and two hypermultiplets respectively. From the diagram in figure~\ref{fig:truncations} one reads that the vector fields in maximal supergravity, transforming in the $\,\textbf{56}\,$ of $\,\textrm{E}_{7(7)}$, are branched into the sum of several irrep's of $\,{\textrm{SL}(2)_{T}\times\textrm{G}_{2(2)} \times \textrm{SO}(3)}$, of which, though, only the ones transforming in the $\,(\textbf{4},\textbf{1})\,$ of $\,\textrm{SL}(2)_{T}\times\textrm{G}_{2(2)}\,$ are $\,\textrm{SO}(3)$ singlets hence surviving the truncation. The resulting theory then comprises four vectors out of which only two (the graviphoton plus an extra vector coming from the vector multiplet) are linearly independent due to the $\,\textrm{Sp}(4,\mathbb{R})\,$ electric-magnetic duality. In addition, the theory contains $2 + 8$ physical scalars spanning $\cM_{\textrm{SK}}$ and $\cM_{\textrm{QK}}$ respectively, together with $72$ deformation parameters associated to the embedding tensor components surviving the truncation.
\\[-6mm]

\noindent Due to the presence of vectors, this theory might have interesting applications in holographic superconductivity as well as in Cosmology as far as the existence of dS solutions via D-terms uplifting is concerned. We hope to come back to these two issues in the near future.

\item \textbf{Step from --2-- to --3--:} the truncation is now with respect to an $\,H=\mathbb{Z}_3\,$ discrete subgroup of the $\,G=\textrm{SL}(2)_{T}\times\textrm{G}_{2(2)}\,$ global symmetry of the previous $\,\cN=2\,$ theory via the chain
\be
\label{chainN2Z3}
\textrm{SL}(2)_{T}\times \textrm{G}_{2(2)} \,\,\supset \,\,\textrm{SL}(2)_{T}\times\textrm{SU}(2,1) \,\,\supset \,\, \textrm{SL}(2)_{T} \times \textrm{SU}(2) \times \textrm{U}(1)_{U} \ .
\ee
More concretely we mod-out the different fields in the theory by a $\,\mathbb{Z}_{3}\,$ element of the form $e^{i \frac{2 \pi}{3} \, q}$, where $\,q\,$ $\textrm{mod}(3)$ denotes the charge of the fields with respect to the $\,\textrm{U}(1)_{U}\,$ factor in (\ref{chainN2Z3}). The field content inside the box --3-- follows from the $\,\textrm{G}_{2(2)}\,$ irrep decompositions
\be
\begin{array}{ccccl}
\hspace{-1mm}\textrm{G}_{2(2)} & \supset & \textrm{SU}(2,1) & \supset & \textrm{SU}(2) \times \textrm{U}(1)_{U} \\[2mm]
\textbf{1} & \rightarrow & \textbf{1} & \rightarrow & \textbf{1}_{(0)} \\[2mm]
\textbf{7} & \rightarrow & \textbf{1} \oplus \textbf{3} \oplus \overline{\textbf{3}} & \rightarrow & \textbf{1}_{(0)} \oplus (\, \textbf{1}_{(-2)} \oplus \textbf{2}_{(1)} \,) \oplus (\, \textbf{1}_{(2)} \oplus \textbf{2}_{(-1)}  \, ) \\[2mm]
\textbf{14} & \rightarrow & \textbf{8} \oplus \textbf{3} \oplus \overline{\textbf{3}} & \rightarrow &\hspace{-1mm} ( \,\textbf{1}_{(0)} \oplus \textbf{2}_{(0)} \oplus \textbf{2}_{(0)} \oplus \textbf{3}_{(0)} \, ) \oplus ( \, \textbf{1}_{(-2)} \oplus \textbf{2}_{(1)} \, ) \oplus (\, \textbf{1}_{(2)} \oplus \textbf{2}_{(-1)} \,)
\end{array}
\nonumber
\ee
where the subindex in $\,\textbf{n}_{(q)}\,$ refers to the $\,\textrm{U}(1)_{U}\,$ charge $\,q\,$ of the $\,\textrm{SU}(2)\,$ irrep $\,\textbf{n}\,$. The truncated theory has an $\,\textrm{SL}(2)_{T}\times\textrm{SU}(2,1)\,$ global symmetry and still keeps $\,\cN=2\,$ supersymmetry. This fact can be seen by obtaining the theory directly from an $\,\textrm{SU}(3)\,$ truncation of maximal supergravity without any intermediate step, as we see next.

\item \textbf{Step from --1-- to --3--:} truncating maximal supergravity with respect to a compact $\,H=\textrm{SU}(3)\,$ subgroup of its $\,G=\textrm{E}_{7(7)}\,$ global symmetry via the chain
\be
\textrm{E}_{7(7)} \,\,\supset \,\,\textrm{SL}(2)_{T} \,\times \,\textrm{F}_{4(4)} \,\,\supset \,\, \textrm{SL}(2)_{T} \,\times \,\textrm{SU}(2,1) \,\times \, \textrm{SU}(3) \ ,
\ee
gives rise to the theory inside the box --3-- of figure~\ref{fig:truncations}. The truncation preserves $\,{\cN=2}\,$ supersymmetry as results from the decomposition $\,\textbf{8} = \textbf{3} \,\oplus\, \textbf{3} \,\oplus\, \textbf{1} \,\oplus\, \textbf{1}\,$ of the fundamental representation of the $\textrm{SU}(8)$ $R$-symmetry group of the maximal theory under the $\,\textrm{SU}(3)\,$ diagonal subgroup\footnote{This time the diagonal subgroup $\,\textrm{SU}(3)_{\textrm{diag}}\,$ in the chain (\ref{chain:SU3_trunc}) is obtained by anti-identifying the two $\,\textrm{SU}(3)\,$ factors, namely, the fundamental representation of the first with the anti-fundamental of the second.}
\be
\label{chain:SU3_trunc}
\textrm{SU}(8) \,\,\supset \,\,\textrm{SU}(4) \,\times \,\textrm{SU}(4) \,\,\supset \,\, \textrm{SU}(3) \,\times \,\textrm{SU}(3) \,\,\supset  \,\, \textrm{SU}(3)_{\textrm{diag}} \ .
\ee
The $\,\cN=2\,$ truncated theory involves the special K\"ahler and the quaternionic K\"ahler manifolds
\be
\cM_{\textrm{SK}}= \left(\frac{\textrm{SL}(2)}{\textrm{SO}(2)}\right)_{T} \qquad\textrm{and}\qquad \cM_{\textrm{QK}}=\frac{\textrm{SU}(2,1)}{\textrm{SU}(2) \times \textrm{U}(1)_{U}} \ ,
\ee
associated to one vector multiplet and one hypermultiplet respectively. This theory can therefore be seen as a truncation of that in box  --2-- where one of the hypermultiplets is projected out after modding out by the $\,\mathbb{Z}_{3}\,$ discrete subgroup previously introduced. The same truncation was explored in ref.~\cite{Warner:1006} and further investigated in refs~\cite{Cassani:2011sv, Halmagyi:2011xh} as gravity dual of non-relativistic field theories.

\item \textbf{Step from --1-- to --4--:} this is the truncation connecting maximal supergravity and half-maximal supergravity coupled to six vector multiplets. It can be seen as a truncation with respect to an $\,H=\mathbb{Z}_2\,$ discrete subgroup of the $\,G=\textrm{E}_{7(7)}\,$ global symmetry of the maximal theory. The resulting theory keeps $\,\cN=4\,$ supersymmetry due to the branching of the $R$-symmetry group of the maximal theory
\be
\textrm{SU}(8) \,\, \supset \,\, \textrm{SU}(4) \times \textrm{SU}(4) \, \sim \, \textrm{SO}(6) \times \textrm{SO}(6) \ ,
\ee
where one of the $\,\textrm{SU}(4)\,$ factors, let us say the first, is parity even under the $\,\mathbb{Z}_2\,$ and the other is parity odd. The fundamental representation of the $R$-symmetry group of maximal supergravity then decomposes as $\,\textbf{8} = (\textbf{4} , \textbf{1}) \oplus (\textbf{1} , \textbf{4}) = \textbf{4}_{\textrm{even}} \oplus \textbf{4}_{\textrm{odd}}\,$ hence keeping only half of the supersymmetries, namely, those related to $\,\textbf{4}_{\textrm{even}}\,$. The action of the $\,\mathbb{Z}_2\,$ on the fundamental representation of $\textrm{E}_{7(7)}$ becomes more transparent by looking at the branching
\be
\begin{array}{ccl}
\textrm{E}_{7(7)} & \supset & \textrm{SL}(2)_{S} \times \textrm{SO}(6,6) \\[2mm]
\textbf{56} & \rightarrow & (\textbf{2},\textbf{12}) \oplus (\textbf{1},\textbf{32})
\end{array}
\nonumber
\ee
Under the $\,\mathbb{Z}_2\,$, the different $\textrm{E}_{7(7)}$ irrep's are modded-out according to the $\textrm{SO}(6,6)$ irrep's appearing in their branchings. In particular, $\textrm{SO}(6,6)$ bosonic irrep's, \emph{e.g.} the $\bf{1}$, $\bf{12}$, $\bf{66}$, etc., are parity even and survive the truncation. In contrast, $\textrm{SO}(6,6)$ fermionic irrep's involving an odd number of Majorana-Weyl indices, \emph{e.g.} the $\bf{32}$ and $\textbf{32}^{\prime}$, are parity odd and are projected out whereas those involving an even number of them are parity even hence surviving the truncation. As a result, the vectors $\,V_{\mu}\,$, the scalars $\,\phi_{\alpha \dot{\mu}}\,$ and the embedding tensor pieces $\,F_{M \dot{\mu}}\,$ and $\,\Xi_{\alpha \beta \mu}\,$ in the bosonic field content of maximal supergravity are truncated away when going to half-maximal. The remaining fields then describe an $\,\cN=4\,$ supergravity coupled to six vector multiplets with an associated
\be
\cM_{\textrm{scalar}} = \left(\frac{\textrm{SL}(2)}{\textrm{SO}(2)}\right)_{S} \times \frac{\textrm{SO}(6,6)}{\textrm{SO}(6) \times \textrm{SO}(6)} \ ,
\ee
coset space spanned by the $\,2 + 36\,$ physical scalars in the theory belonging to the gravity multiplet and the six vector multiplets respectively. Further details about this truncation can be found in ref.~\cite{Dibitetto:2011eu}.

\item \textbf{Step from --4-- to --5--:} this step corresponds to a truncation with respect to an $\,H=\textrm{SO}(3)\,$ subgroup of the $\,G=\textrm{SL}(2)_{S} \times \textrm{SO}(6,6)\,$ global symmetry of half-maximal supergravity coupled to six vector multiplets following the chain
\be
\textrm{SL}(2)_{S} \times \textrm{SO}(6,6) \,\,\supset \,\,\textrm{SL}(2)_{S} \times \textrm{SO}(2,2) \times \textrm{SO}(3) \,\sim\,  \prod_{\Phi=S,T,U} \textrm{SL}(2)_{\Phi} \,\times\, \textrm{SO}(3) \ .
\ee
The truncation (see the last part of section~\ref{FTheta_Dictionary}) breaks half-maximal to minimal $\,\cN=1\,$ supergravity due to the decomposition $\,\textbf{4} = \textbf{1} \oplus \textbf{3}\,$ of the fundamental representation of the $\,\textrm{SU}(4)\,$ $R$-symmetry group in $\cN=4$ supergravity under the $\,\textrm{SO}(3)\,$ subgroup
\be
\textrm{SU}(4) \,\,\supset \,\, \textrm{SU}(3) \,\,\supset \,\, \textrm{SO}(3) \ .
\ee
The resulting theory does not contain vectors since there are no $\,\textrm{SO}(3)$-singlets in the decomposition $\,\textbf{12}=(\textbf{4},\textbf{3})\,$ of the fundamental of $\,\textrm{SO}(6,6)\,$ under $\,\textrm{SO}(2,2) \times \textrm{SO}(3)$. The physical scalar fields span the coset space
\be
\cM_{\textrm{scalar}}= \prod_{\Phi=S,T,U}\left( \frac{\textrm{SL}(2)}{\textrm{SO}(2)} \right)_{\Phi} \ ,
\ee
involving three $\,\textrm{SL}(2)/\textrm{SO}(2)\,$ factors each of which can be parameterised by a complex scalar $\,\Phi=(S , T , U)$. In addition, the embedding tensor of the theory contains $\,40\,$ independent components fitting two irrep's of the $\,\textrm{SL}(2)_{S} \times \textrm{SL}(2)_{T} \times \textrm{SL}(2)_{U}\,$ global symmetry group, as shown inside box --5-- in figure~\ref{fig:truncations}. We will come back to this truncation in the next section when studying type II string models.
\\[-6mm]

\noindent This $\,\cN=1\,$ supergravity theory has been extensively investigated because of its direct connection to string theory via type II orientifold compactifications with fluxes \cite{Kachru:2002he, Derendinger:2004jn, Villadoro:2005cu, Camara:2005dc, Derendinger:2005ph, Aldazabal:2006up, Aldazabal:2007sn}. The resulting supergravity models are referred to as $STU$-models and different background fluxes in the string theory side correspond with different embedding tensor configurations in the supergravity side. However, not all the embedding tensor configurations in the supergravity side have a higher-dimensional interpretation since most of them correspond to non-geometric flux backgrounds for which an origin in string theory, if possible, remains to be found.

\noindent It is worth noticing here that this theory can be lifted to that in box --2-- by completing it with the fermionic irrep's removed by the $\,\mathbb{Z}_{2}\,$ truncation taking from the box --1-- to the box --4-- in figure~\ref{fig:truncations}.

\item \textbf{Step from --5-- to --6--:} this truncation is with respect to an $\,H=\mathbb{Z}_3\,$ discrete subgroup of the $\,G=\textrm{SL}(2)_{S} \times \textrm{SL}(2)_{T} \times \textrm{SL}(2)_{U}\,$ global symmetry in --5-- via the chain
\be
\label{chainN2}
\textrm{SL}(2)_{S} \times \textrm{SL}(2)_{T} \times \textrm{SL}(2)_{U} \,\,\supset \,\, \textrm{SL}(2)_{S} \times \textrm{SL}(2)_{T} \times \textrm{U}(1)_{U} \ .
\ee
As happened when truncating from --2-- to --3-- before, we mod-out again the different fields in the theory by a $\,\mathbb{Z}_{3}\,$ element of the form $e^{i \frac{2 \pi}{3} \, q}$, with $\,q\,$ $\textrm{mod}(3)$ being this time the charge of the fields with respect to the $\,\textrm{U}(1)_{U}\,$ factor in (\ref{chainN2}). Now, the relevant branchings in order to derive the field content inside the box --6-- are
\be
\begin{array}{ccl}
\textrm{SL}(2)_{U} & \supset & \textrm{U}(1)_{U} \\[2mm]
\textbf{1} & \rightarrow & \textbf{1}_{(0)} \\[2mm]
\textbf{2} & \rightarrow & \textbf{1}_{(-1)} \oplus \textbf{1}_{(1)} \\[2mm]
\textbf{3} & \rightarrow & \textbf{1}_{(-2)} \oplus \textbf{1}_{(0)} \oplus \textbf{1}_{(2)}\\[2mm]
\textbf{4} & \rightarrow & \textbf{1}_{(0)} \oplus \textbf{1}_{(-1)} \oplus \textbf{1}_{(1)} \oplus \textbf{1}_{(0)}
\end{array}
\nonumber
\ee
where, as before, the subindex in $\,\textbf{1}_{(q)}\,$ refers to the $\,\textrm{U}(1)_{U}\,$ charge $\,q\,$ of the state. The truncated theory still has $\,\cN=1\,$ supersymmetry since the gravitino in the parent theory was already a singlet with respect to both $\,\textrm{U}(1)_{T}\,$ and $\,\textrm{U}(1)_{U}\,$.

\noindent The scalars in the truncated theory span the scalar manifold
\be
\label{scalar_manifold_6}
\cM_{\textrm{scalar}}= \prod_{\Phi=S,T}\left( \frac{\textrm{SL}(2)}{\textrm{SO}(2)} \right)_{\Phi}  \ .
\ee
It can be parameterised by two complex scalars $\,S\,$ and $\,T\,$ associated to the $\,\textrm{SL}(2)/\textrm{SO}(2)\,$ factors. As summarised inside the  box --6-- in figure~\ref{fig:truncations}, the embedding tensor consists of two pieces sitting in the same irrep of the global symmetry group of the theory.

\end{itemize}

In the next section we concentrate back on the $\,\cN=1\,$ theory inside box --5-- which can be seen as a truncation of the $\,\cN=4\,$ theory inside box --4-- . We will investigate the lifting of some vacuum solutions found in chapter~\ref{Half_Max} to $\,\cN=8\,$ supergravity (box --1--) making use of the relations (\ref{A's_fluxes_Bosonic}) between fermionic masses and flux backgrounds when $\,F_{M\dot{\m}}=\Xi_{\a \b \m}=0\,$, \emph{i.e.} when spinorial fluxes do vanish. It would be interesting to explore the phenomenology of fluxes related to embedding tensor components fitting these fermionic irrep's and still having a higher-dimensional origin in string theory as gauge fluxes or metric fluxes.

\section{Exceptional Flux Backgrounds}
\label{sec:typeII_examples}

When compactifying type II ten-dimensional supergravities down to four dimensions, background fluxes threading the internal space can be switched on during the compactification procedure giving rise to gauged maximal supergravity models. As introduced in section~\ref{sec:fluxes_A1D6}, flux backgrounds on the string side correspond to deformation parameters related to the $\,f_{\a MNP}\,$, $\,\xi_{\a M}\,$, $\,F_{M \dot{\mu}}\,$ and $\,\Xi_{\a \b \mu}\,$ pieces of the embedding tensor on the supergravity side. For the sake of simplicity, we will restrict our study to the case
\be
\label{F&Xi=0}
F_{M\dot{\m}} \,\,=\,\, \Xi_{\a \b \m} \,\,=\,\, 0 \ ,
\ee
that is, to string backgrounds not including fluxes associated to $\,\textrm{SO}(6,6)\,$ fermionic irrep's of the embedding tensor. However, even though the remaining $\,f_{\a MNP}\,$ and $\,\xi_{\a M}\,$ pieces reproduce those of half-maximal supergravity, the set of QC they are restricted by will be that of maximal supergravity derived in section~\ref{sec:fluxes_A1D6}. Setting to zero spinorial fluxes as in (\ref{F&Xi=0}) does not amount to modding out maximal supergravity by a $\,\mathbb{Z}_2\,$ symmetry. While the former does not affect other fields in the theory (as scalars and vectors), the latter projects out some of them in order to truncate from maximal to half-maximal supergravity. On the string theory side, modding out by this $\,\mathbb{Z}_2\,$ symmetry is commonly referred to as applying an orientifold projection.

\subsection*{String Theory Embedding \,vs\, Moduli Stabilisation}

Thus far, we have discussed in detail the correspondence between maximal gauged supergravities and type II flux compactifications. However, one might also be interested in the interplay between gaugings, fluxes and moduli stabilisation: in short, fluxes were introduced in order to achieve moduli stabilisation. Sketchily, the picture in this respect seems to be the following

\begin{center}
\scalebox{0.80}[0.88]{\xymatrix{
*+[F-,]{\begin{array}{c}
\textbf{\textrm{semisimple gaugings}}\\[1mm]
\hline \\[-3mm]
\textrm{moduli stabilisation } $\ding{51}$ \\[2mm]
\textrm{string embedding } $\ding{55}$ \\[2mm]
\end{array} }
& \leftrightarrow &
*+[F-,]{\begin{array}{c}
\textbf{\textrm{intermediate gaugings}}\\[1mm]
\hline \\[-3mm]
\textrm{moduli stabilisation } ? \\[2mm]
\textrm{string embedding } ? \\[2mm]
\end{array} }
& \leftrightarrow &
*+[F-,]{\begin{array}{c}
\textbf{\textrm{nilpotent gaugings}}\\[1mm]
\hline \\[-3mm]
\textrm{moduli stabilisation } $\ding{55}$ \\[2mm]
\textrm{string embedding } $\ding{51}$ \\[2mm]
\end{array} }
}}
\end{center}
\vspace{2mm}

\noindent Semisimple gaugings are likely to produce critical points and moduli stabilisation \cite{Hull:1988jw, DallAgata:2011aa, Fischbacher:2009cj, Fischbacher:2011jx}, but we will show that their embedding as type II flux compactifications involves highly non-geometric backgrounds. On the other hand, nilpotent gaugings can be obtained from type II compactifications including gauge fluxes \cite{deWit:2003hq}, but they seem not to be enough to get moduli stabilisation. Intermediate gaugings containing a semisimple part and an Abelian part have recently been found in ref.~\cite{DallAgata:2011aa}, although their embedding into string theory/M-theory has not been explored yet.

Here we will present a novel intermediate gauging consisting of a semisimple and a nilpotent part which allows for moduli stabilisation and can be embedded into string theory as a type IIA flux compactification including gauge and metric fluxes.

\subsection*{Setting up the flux models}

Our starting point is the $\,\mathcal{N}=1\,$ supergravity theory inside box --5-- in figure~\ref{fig:truncations}. As explained at the end of  section~\ref{sec:connecting_A7&A1D6}, this theory can be obtained by truncating the $\,\mathcal{N}=4\,$ supergravity in box --4-- with respect to an $\,\textrm{SO}(3)\,$ subgroup which, in turn, can be obtained by a $\,\mathbb{Z}_{2}\,$ truncation of $\,\mathcal{N}=8\,$ supergravity in box --1--. As summarised in appendix~{\ref{App:fluxes}}, all the deformation parameters of this theory belong to the $\,f_{\alpha MNP}\,$ piece of the embedding tensor which comes out with forty independent components \cite{Dibitetto:2011gm}. These can be arranged into a tensor $\,{\Lambda_{\a ABC}=\Lambda_{\a (ABC)}}\,$, where $\,\a=+,-\,$ denotes an $\,\textrm{SL}(2)_{S}$ electric-magnetic index and $\,{A=1,...,4}\,$ denotes an $\,\textrm{SO}(2,2) \sim \textrm{SL}(2)_{T} \times \textrm{SL}(2)_{U}\,$ fundamental index of the global symmetry group. The theory comprises three complex scalars $\,S\,$, $\,T\,$ and $\,U\,$ parameterising the complex K\"ahler manifold
\be
\cM_{\textrm{scalar}}= \left( \frac{\textrm{SL}(2)}{\textrm{SO}(2)} \right)_{S}  \times  \left( \frac{\textrm{SL}(2)}{\textrm{SO}(2)} \right)_{T}   \times  \left( \frac{\textrm{SL}(2)}{\textrm{SO}(2)} \right)_{U}  \ ,
\ee
and no vector fields since they are projected out in the truncation. This supergravity theory can be obtained from type II orientifolds of $\,\mathbb{Z}_{2} \times \mathbb{Z}_{2}\,$ orbifold compactifications in the presence of generalised flux backgrounds, and the scalar potential can be derived from the $\,\mathcal{N}=1$ flux-induced superpotential in (\ref{W_fluxes4}).

Now we want to lift this $\,\cN=1\,$ theory firstly to $\,\cN=4\,$ by removing the $\,\textrm{SO}(3)$ truncation and secondly to $\,\cN=8\,$ by also removing the $\,\mathbb{Z}_2\,$ orientifold projection. This amounts to re-introduce the $\,28\,$ physical vectors in maximal supergravity and to complete the number of scalars from $\,6\,$ to $\,70\,$ without changing the set of embedding tensor components, in other words, without modifying the flux backgrounds. Nevertheless, in order for a flux background to be liftable to maximal supergravity, the set of QC found in section~\ref{sec:fluxes_A1D6} must be imposed. These guarantees the absence of supersymmetry-breaking branes and all their U-dual local sources \cite{Bergshoeff:2010xc, Bergshoeff:2012ex}.

\begin{figure}[t!]
\begin{center}
\scalebox{0.4}[0.4]{
\includegraphics[keepaspectratio=true]{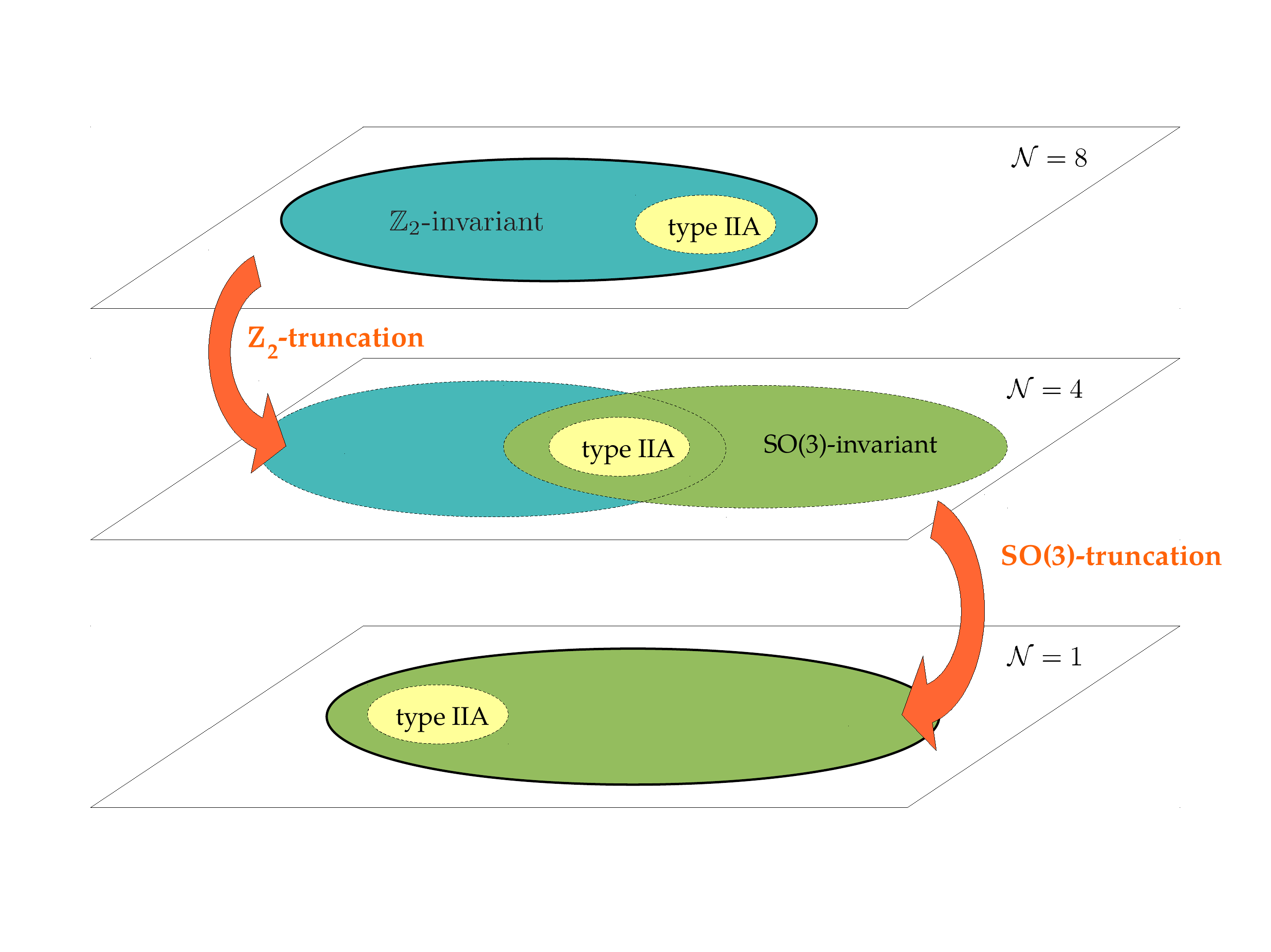}}
\end{center}
\vspace{-15mm}
\caption{{\it Diagram of the two-step lifting of $\,\cN=1\,$ flux backgrounds firstly to $\,\cN=4\,$ by removing the $\,\textrm{SO}(3)\,$ truncation and secondly to $\,\cN=8\,$ by removing the $\,\mathbb{Z}_2\,$ orientifold projection. As depicted in the figure, only a subset of $\,\cN=4\,$ theories can be truncated to $\,\cN=1\,$ theories via an $\,\textrm{SO}(3)\,$ truncation. On the other hand, only a subset of $\,{\cN=4}$ theories can be obtained from $\,\cN=8\,$ supergravity via a $\,\mathbb{Z}_{2}\,$ orientifold projection. The relevant fact is that the intersection between these two subsets of $\,\cN=4\,$ theories happens not to be empty and, furthermore, contains some theories for which a realisation in terms of type IIA string theory is known.}}
\label{fig:field/flux_pic}
\end{figure}

In the rest of the section we will concentrate on two specific type II fluxes models which are relevant from a string theory point of view:
\begin{itemize}
\item Type IIB non-geometric flux backgrounds with an $\,\textrm{SO}(3,3) \,\times\, \textrm{SO}(3,3)\,$ splitting in $\,{\cN=4}\,$ supergravity and lifting to $\,\textrm{SO}(8)\,$, $\,\textrm{SO}(4,4)\,$, $\,\textrm{SO}(3,5)\,$ and $\,\textrm{CSO}(2,0,6)\,$ gaugings in $\,\cN=8\,$ supergravity.

\item Type IIA geometric flux backgrounds lifting to $\,\textrm{ISO}(3) \,\ltimes\, \textrm{U}(1)^{6}\,$ gaugings in $\,{\cN=4}\,$ supergravity and further lifting to $\,\textrm{SO}(4) \,\ltimes\, \textrm{Nil}_{22}\,$ gaugings in $\,\cN=8\,$ supergravity.

\end{itemize}

We will reduce our search of critical points to the origin of the moduli space and discuss the issues of stability and supersymmetry at those solutions. However, due to the restriction (\ref{F&Xi=0}), our classification of critical points will no longer be exhaustive since spinorial fluxes might produce new solutions we do not have access to by only looking at the origin of the moduli space \cite{Dibitetto:2011gm,DallAgata:2011aa}.

\subsection*{$\textrm{CSO}(p,q,r)\,$ gaugings from type IIB with non-geometric fluxes}

Let us concentrate on a set of type IIB flux backgrounds for which one has the direct product splitting $\,\textrm{SL}(2)_{S} \times \textrm{SO}(6,6) \supset \textrm{SO}(3,3)_{+} \times \textrm{SO}(3,3)_{-}\,$, where the labels $\,+\,$ and $\,-\,$ stand for $\,\textrm{SL}(2)_{S}\,$ electric and magnetic pieces, respectively. These backgrounds can be obtained from type IIB flux compactifications including the following set of generalised fluxes: R-R and NS-NS gauge fluxes $\,(F_{3},H_{3})\,$, non-geometric fluxes $\,(Q,P)\,$ and their primed counterparts which have been less studied in the literature.

The above set of fluxes gives rise to maximal gauged supergravities based on $\,\textrm{CSO}(p,q,r)\,$ gauge groups with $\,p+q+r=8\,$. By applying the $\,\mathbb{Z}_{2}\,$ orientifold projection truncating from maximal to half-maximal supergravity, the $\,\textrm{CSO}(p,q,r)\,$ gauge groups get broken to the direct product of two smaller $\,\textrm{CSO}_{\pm}\,$ groups as
\be
\label{CSO_chain}
\begin{array} {ccc}
\mathcal{N}=8 &   & \mathcal{N}=4 \\[2mm]
\textrm{CSO}(p,q,r) & \longrightarrow & \textrm{CSO}_{+}(p_{+},q_{+},r_{+}) \, \times \, \textrm{CSO}_{-}(p_{-},q_{-},r_{-}) \ ,
\end{array}
\ee
with $\,p_{\pm} + q_{\pm} + r_{\pm} =4\,$. As explained in ref.~\cite{Roest:2009tt}, each of the $\,\textrm{CSO}_{\pm}\,$ factors in the r.h.s of (\ref{CSO_chain}) can be parameterised in terms of two real symmetric $\,4 \times 4\,$ matrices $M_{\pm}$ and $\tilde{M}_{\pm}\,$ which determine their embedding into an $\,\textrm{SO}(3,3)_{\pm}\,$ group, respectively. In terms of generalised flux components, these matrices read
\be
M_{+} =
\left(
\begin{array}{cc}
-a'_{0} & 0 \\
0 & \tilde{c}_{1} \times \mathds{1}_{3}
\end{array}
\right)_{(\,F'_{3}\,,\,Q\,)}
\hspace{6mm} \textrm{ and } \hspace{6mm}
\tilde{M}_{+} =
\left(
\begin{array}{cc}
-a_{0} & 0 \\
0 & \tilde{c}'_{1} \times \mathds{1}_{3}
\end{array}
\right)_{(\,F_{3}\,,\,Q'\,)}
\ee
together with
\be
M_{-} =
\left(
\begin{array}{cc}
b'_{3} & 0 \\
0 & \tilde{d}_{2} \times \mathds{1}_{3}
\end{array}
\right)_{(\,H'_{3}\,,\,P\,)}
\hspace{10mm} \textrm{ and } \hspace{10mm}
\tilde{M}_{-} =
\left(
\begin{array}{cc}
b_{3} & 0 \\
0 & \tilde{d}'_{2} \times \mathds{1}_{3}
\end{array}
\right)_{(\,H_{3}\,,\,P'\,)}
\ee
where the concrete identification between flux entries in $M_{\pm}$ and $\tilde{M}_{\pm}\,$ and embedding tensor components $\,f_{\a MNP}\,$ can be read off from tables~\ref{table:unprimed_fluxes4} and \ref{table:primed_fluxes4} in appendix~\ref{App:fluxes}. By substituting into the set of QC derived in section~\ref{sec:fluxes_A1D6}, one finds three families of solutions:
\begin{itemize}
\item[$i)$] Flux matrices corresponding to a $\,(Q,F_{3})\,$-flux background
\be
M_{+} =
\left(
\begin{array}{cc}
0 & 0 \\
0 & \lambda_{1} \times \mathds{1}_{3}
\end{array}
\right)
\hspace{2mm} \textrm{ , } \hspace{2mm}
\tilde{M}_{+} =
\left(
\begin{array}{cc}
\lambda_{2} & 0 \\
0 & 0  \times \mathds{1}_{3}
\end{array}
\right)
\hspace{5mm} \textrm{ and } \hspace{5mm}
M_{-} = \tilde{M}_{-} =0
\ee

\item[$ii)$] Flux matrices corresponding to a $\,(P,H_{3})\,$-flux background
\be
M_{-} =
\left(
\begin{array}{cc}
0 & 0 \\
0 & \lambda_{1} \times \mathds{1}_{3}
\end{array}
\right)
\hspace{2mm} \textrm{ , } \hspace{2mm}
\tilde{M}_{-} =
\left(
\begin{array}{cc}
\lambda_{2} & 0 \\
0 & 0 \times \mathds{1}_{3}
\end{array}
\right)
\hspace{5mm} \textrm{ and } \hspace{5mm}
M_{+} = \tilde{M}_{+} =0
\ee

\item[$iii)$] Flux matrices corresponding to a $\,(F'_{3}\,,\,Q\,,\,H_{3}\,,\,P')\,$-flux background
\be
M_{+} = \textrm{unrestricted}
\hspace{5mm} \textrm{ , } \hspace{5mm}
\tilde{M}_{-} = \textrm{unrestricted}
\hspace{5mm} \textrm{ and } \hspace{5mm}
\tilde{M}_{+} = M_{-} = 0
\ee

\end{itemize}
together with three additional ones obtained by swapping $\,M_{\pm} \leftrightarrow \tilde{M}_{\pm}\,$. This amounts to interchange primed and unprimed fluxes, \emph{i.e.} to apply six T-dualities along the internal space directions, so the resulting theories are physically equivalent.

\begin{table}[h!]
\renewcommand{\arraystretch}{1.50}
\begin{center}
\hspace{-6.2mm}
\scalebox{0.67}[0.7]{
\begin{tabular}{ | c | c | c | c | c  | c |}
\hline
\textrm{ID} & $\frac{1}{\lambda} \, M_{+}\,\,\,$ and $\,\,\,\frac{1}{\lambda} \, \tilde{M}_{-}$ & $\,\cN=8\,$ gauging & $\,\cN=4\,$ gauging &  $\frac{1}{\lambda^{2}} \, V_0$  & Mass spectrum  \\[1mm]
\hline \hline
$1$ & $\begin{array}{l} M_{+}=(1,1,1,1) \\[-3mm] \tilde{M}_{-}=(1,1,1,1) \end{array} $ & \multirow{3}{*}{$\textrm{SO}(8)$} & \multirow{3}{*}{$\textrm{SO}(4)^2$}  & $-\dfrac{3}{2}$ & $(70 \, \times)\, -\frac{2}{3}$  \\[1mm]
\cline{1-2}\cline{5-6} $2$ & $\begin{array}{l} M_{+}=(5 ,1,1,1) \\[-3mm] \tilde{M}_{-}=(1,1,1,1) \end{array} $ &  &   & $-\dfrac{5}{2}$ & $2$ , $(27 \, \times)\, -\frac{4}{5}$ , $(35 \, \times)\, -\frac{2}{5}$ , $(7 \, \times) \, 0$ \\[1mm]
\hline
\hline
$3$ & $\begin{array}{l} M_{+}=(1,1,1,1) \\[-3mm] \tilde{M}_{-}=(1,-3,-3,-3) \end{array} $ & $\textrm{SO}(5,3)$ & $\textrm{SO}(4) \times \textrm{SO}(1,3)$  & $\dfrac{3}{2}$ & $-2$ , $(5 \, \times)\, 4$ , $(30 \, \times)\, 2$ , $(14 \, \times) \, \frac{4}{3}$ , $(5 \, \times) \, -\frac{2}{3}$ , $(15 \, \times) \, 0$ \\[1mm]
\hline
\hline
$4$ & $\begin{array}{l} M_{+}=(1,-1,-1,-1) \\[-3mm] \tilde{M}_{-}=(-1,1,1,1) \end{array} $ & \multirow{3}{*}{$\textrm{SO}(4,4)$} & $\textrm{SO}(1,3) \times \textrm{SO}(3,1)$  & \multirow{3}{*}{$\dfrac{1}{2}$} & \multirow{3}{*}{$(2 \, \times)\, -2$ , $(36 \, \times)\, 2$ , $(16 \, \times) \, 1$ , $(16\, \times) \, 0$} \\[1mm]
\cline{1-2}\cline{4-4} $5$ & $\begin{array}{l} M_{+}=(1,1,1,1) \\[-3mm] \tilde{M}_{-}=(-1,-1,-1,-1) \end{array} $ &  & $\textrm{SO}(4,0)\times \textrm{SO}(0,4)$  &  &  \\[1mm]
\hline
\hline
$6$ & $\begin{array}{l} M_{+}=(1,0,0,0) \\[-3mm] \tilde{M}_{-}=(1,0,0,0) \end{array} $ & $\textrm{CSO}(2,0,6)$ & $\textrm{CSO}(1,0,3)^2$  & $0$ & $(20 \, \times)\, \frac{\lambda^{2}}{8}$ , $(2 \, \times)\, \frac{\lambda^{2}}{2}$ , $(48\, \times) \, 0$ \\[1mm]
\hline
\end{tabular}
}
\end{center}
\caption{{\it Set of critical points of the scalar potential for generalised type IIB flux backgrounds compatible with an $\,\textrm{SL}(2)_{S} \times \textrm{SO}(6,6) \supset \textrm{SO}(3,3)_{+} \times \textrm{SO}(3,3)_{-}\,$ splitting. The first two correspond to AdS solutions, the next three to dS solutions and the last one is a Minkowski solution. By looking for solutions of the Killing equations (\ref{Killing_equations}), we find that all the solutions break all the supersymmetries except the first one which preserves $\,\cN=8\,$ supersymmetry.}}
\label{table:typeIIB_vacua}
\end{table}

The next step is to build the fermionic mass terms $\,\mathcal{A}^{\mathcal{IJ}}\,$ and $\,{\mathcal{A}_{\mathcal{I}}}^{\mathcal{JKL}}\,$ in maximal supergravity as a function of the matrices $M_{\pm}$ and $\tilde{M}_{\pm}\,$ by using the relations (\ref{A's_fluxes_Bosonic}). Plugging them into the set of E.O.M's for the scalars (\ref{scalars_eom}), it turns out that gaugings belonging to the first and the second family of solutions to the QC do not generate critical points at the origin of the moduli space. This is related to the fact that they are purely electric and magnetic gaugings in half-maximal supergravity, so moduli stabilisation is not possible \cite{deRoo:1985jh}. In contrast, six inequivalent patterns\footnote{Additional solutions apart from those shown in table~\ref{table:typeIIB_vacua} can be obtained by exchanging the flux matrices $\,M_{+} \leftrightarrow \tilde{M}_{-}\,$. However, they go back to those in the table via the composition of an S-duality and three T-duality transformations, hence being physically equivalent.} of flux matrices belonging to the third family of solutions, \emph{i.e.} $\,M_{+} , \tilde{M}_{-} \neq 0\,$ and $\,\tilde{M}_{+} = M_{-} = 0\,$, are compatible with moduli stabilisation at the origin of the moduli space (\ref{M_origin_choice}). They correspond to dyonic gaugings in half-maximal supergravity even though their parent CSO gaugings in the maximal theory turn out to be purely electric. These are determined by the signature of the block-diagonal flux matrix
\be
M_{\textrm{electric}} =
\left(
\begin{array}{c|c}
M_{+} & 0 \\[1mm]
\hline
\\[-4mm]
0 & \tilde{M}_{-}
\end{array}
\right) \ .
\ee
We have computed the value of the energy\footnote{We are setting $\,g=\frac{1}{2}\,$ in analogy to ref.~\cite{Dibitetto:2011gm}.} $\, V_{0}\,$, the normalised mass spectrum (for those solutions with $\,V_{0} \neq 0\,$) and the amount of residual supersymmetry at the solutions together with the corresponding gauge group in maximal and half-maximal supergravity according to the chain (\ref{CSO_chain}). The results are summarised in table~\ref{table:typeIIB_vacua}, matching perfectly those in ref.~\cite{DallAgata:2011aa}.

\subsection*{Type IIA with gauge and metric fluxes}

Now we investigate specific flux backgrounds having a higher-dimensional interpretation in terms of type IIA string compactifications including geometric fluxes: these are R-R $\,F_{0,2,4,6}\,$ and NS-NS $\,H_{3}\,$ gauge fluxes together with a metric flux $\,\omega\,$ associated to the spin connection of the internal space.

By using again the fluxes/embedding tensor correspondence of table~\ref{table:unprimed_fluxes4} and the relations~(\ref{A's_fluxes_Bosonic}), we can build the fermionic mass terms $\,\mathcal{A}^{\mathcal{IJ}}\,$ and $\,{\mathcal{A}_{\mathcal{I}}}^{\mathcal{JKL}}\,$ in maximal supergravity associated to these type IIA backgrounds. Imposing the
set of $\,\cN=8\,$ QC (\ref{QC_SU8}) and the E.O.M's for the scalar fields (\ref{scalars_eom}), we obtain exactly the same sixteen AdS critical points at the origin of the moduli space (\ref{M_origin_choice}) which we collected in table~\ref{table:N=4_vacua}. As anticipated in ref.~\cite{Dibitetto:2011qs}, they can be seen as the uplifting to
maximal supergravity of half-maximal supergravity solutions compatible with the total absence of sources\footnote{The presence of sources as O$6$-planes and D$6$-branes in these type IIA scenarios modifies the set of $\,\cN=8\,$ QC.}.

With the fermionic mass terms $\,\mathcal{A}^{\mathcal{IJ}}\,$ and $\,{\mathcal{A}_{\mathcal{I}}}^{\mathcal{JKL}}\,$ at our disposal, we can now compute the different values of the cosmological constant (\ref{V_SU8}) at the above set type IIA solutions. These are given by
\be
\label{V0_IIA}
V_{0}\left[ 1_{(s_1,s_2)} \right] = -\lambda^{2}
\hspace{3mm} , \hspace{3mm}
V_{0}\left[ 2_{(s_1,s_2)} \right] = V_{0}\left[ 4_{(s_1,s_2)} \right]= -\dfrac{32 \, \lambda^{2}}{27}
\hspace{3mm} , \hspace{3mm}
V_{0}\left[ 3_{(s_1,s_2)} \right] = -\dfrac{8 \, \lambda^{2}}{15}  \ .
\ee
In addition, we can also obtain the complete mass spectrum for the $70$ physical scalars by using the mass formula (\ref{Mass-matrix}) and check stability as well as the amount of supersymmetry preserved. The mass spectrum at the critical points in table~\ref{table:N=4_vacua} turns out to be the following:
\begin{itemize}
\item At the solution $\,1_{(s_{1},s_{2})}\,$, the normalised scalar field masses and their multiplicities are given by
\be
\begin{array}{lcrr}
\dfrac{1}{9} \left(47 \pm \sqrt{159}\right)\,\,(\times 1)
\hspace{8mm} , \hspace{8mm} \dfrac{1}{3} \left(4 \pm
\sqrt{6}\right)\,\,(\times 1) & , &
\dfrac{29}{9}\,\,(\times 3) & ,\\[4mm]
\dfrac{1}{18} \left( \, 89 + 5 \, \sqrt{145} \pm \sqrt{606 + 30 \, \sqrt{145}} \, \right)\,\,(\times 5)
& \hspace{5mm} , \hspace{5mm} &
0\,\,(\times 10) & , \\[4mm]
\dfrac{1}{18} \left(\, 89 - 5 \, \sqrt{145} \pm \sqrt{606 - 30 \, \sqrt{145}} \, \right)\,\,(\times 5)
& \hspace{5mm} ,\hspace{5mm} &
-\dfrac{2}{3}\,\,(\times 1) & ,
\end{array}
\nonumber
\ee
for the $\,38\,$ scalars surviving the truncation from maximal to half-maximal supergravity, together with
\be
\dfrac{1}{3} \left(4 \pm \sqrt{6}\right)\,\,(\times 3)
\hspace{5mm} , \hspace{5mm}
6\,\,(\times 3)
\hspace{5mm} , \hspace{5mm}
\dfrac{13}{3} \,\,(\times 5)
\hspace{5mm} , \hspace{5mm}
-\dfrac{2}{3}\,\,(\times 1)
\hspace{5mm} , \hspace{5mm}
0 \,\,(\times 17) \ , \nonumber
\ee
for the additional $\,32\,$ scalars in the maximal theory. There are two tachyons in the spectrum both with the same normalised mass $\,m^2= - \frac{2}{3}\,$, so this AdS solution is completely stable since it satisfies the BF bound in (\ref{BF_bound}).

\item At the solution $\,2_{(s_{1},s_{2})}\,$, the values of the normalised scalar masses and their multiplicities read
\be
\begin{array}{c}
\dfrac{1}{15} \left(77 \pm 5 \, \sqrt{145}\right)\,\,(\times
5) \hspace{4mm} , \hspace{4mm} \dfrac{2}{15} \left(31 \pm
\sqrt{145}\right)\,\,(\times 5) \hspace{4mm} , \hspace{4mm}
\dfrac{64}{15}\,\,(\times 1) \hspace{4mm} , \hspace{4mm}
\dfrac{20}{3}\,\,(\times 1) \ , \\[4mm]
\dfrac{46}{15}\,\,(\times 3) \hspace{5mm} , \hspace{5mm}
2\,\,(\times 1) \hspace{5mm} , \hspace{5mm}
0\,\,(\times 10) \hspace{5mm} , \hspace{5mm}
-\dfrac{2}{5}\,\,(\times 1) \hspace{5mm} , \hspace{5mm}
-\dfrac{4}{5}\,\,(\times 1) \ ,
\end{array}
\nonumber
\ee
for the scalars surviving the truncation to half-maximal supergravity, and also
\be
6\,\,(\times 3)
\hspace{4mm} , \hspace{4mm}
4\,\,(\times 5)
\hspace{4mm} , \hspace{4mm}
2\,\,(\times 3)
\hspace{4mm} , \hspace{4mm}
-\frac{4}{5}\,\,(\times 1)
\hspace{4mm} , \hspace{4mm}
0\,\,(\times 20) \ ,
\nonumber
\ee
for the scalars been projected out by the $\mathbb{Z}_{2}$ orientifold projection. There are three tachyons in the spectrum, two of them with $\,m^2=-4/5\,$. This value is below the BF bound in (\ref{BF_bound}), rendering this AdS solution unstable.

\item At the solution $\,3_{(s_{1},s_{2})}\,$, the normalised scalar field masses and their multiplicities take the values of
\be
\frac{1}{3} \left(19\pm\sqrt{145}\right)\,\,(\times
10) \hspace{4mm} , \hspace{4mm} \frac{20}{3}\,\,(\times 2)
\hspace{4mm} , \hspace{4mm} \frac{14}{3}\,\,(\times 3)
\hspace{4mm} , \hspace{4mm} 2\,\,(\times 2)
\hspace{4mm} , \hspace{4mm} 0\,\,(\times 11) \ , \nonumber
\ee
for those scalars still present in half-maximal supergravity, and
\be
2\,\,(\times 6)
\hspace{4mm} , \hspace{4mm}
6\,\,(\times 5)
\hspace{4mm} , \hspace{4mm}
8\,\,(\times 3)
\hspace{4mm} , \hspace{4mm}
0\,\,(\times 18) \ ,
\nonumber
\ee
for those ones completing to maximal supergravity. It is worth noticing that all the masses are non-negative at this critical point, hence corresponding to an AdS stable extremum of the theory.

\item At the solution $\,4_{(s_{1},s_{2})}\,$, the set of normalised scalar masses and their multiplicities are
\be
\dfrac{20}{3}\,\,(\times 1) \hspace{4mm} , \hspace{4mm}
6\,\,(\times 6) \hspace{4mm} , \hspace{4mm}
\dfrac{8}{3}\,\,(\times 5) \hspace{4mm} , \hspace{4mm}
2\,\,(\times 4) \hspace{4mm} , \hspace{4mm}
\dfrac{4}{3}\,\,(\times 6) \hspace{4mm} , \hspace{4mm}
0\,\,(\times 16) \,\ .
\nonumber
\ee
again for the $38$ scalars present in half-maximal supergravity, as well as
\be
\dfrac{8}{3}\,\,(\times 5) \hspace{4mm} , \hspace{4mm}
6\,\,(\times 3) \hspace{4mm} , \hspace{4mm}
-\dfrac{4}{3}\,\,(\times 1) \hspace{4mm} , \hspace{4mm}
2\,\,(\times 3) \hspace{4mm} , \hspace{4mm}
\dfrac{4}{3}\,\,(\times 3) \hspace{4mm} , \hspace{4mm}
0\,\,(\times 17) \,\ .
\nonumber
\ee
for the $32$ extra ones in the maximal theory. Amongst the latter, there is a tachyon with a normalised mass $\,m^2=-4/3\,$ lying below the BF bound (\ref{BF_bound}). Therefore, this solution, while stable with respect to the scalars in half-maximal supergravity, becomes unstable when lifted to the maximal theory.
\end{itemize}

In order to determine the amount of residual supersymmetry preserved by the above set of critical points, we have to look for solutions to the Killing equations (\ref{Killing_equations}). The final outcome is that supersymmetry becomes completely broken in all the solutions except in $\,1_{(s_{1},s_{2})}\,$ which preserves $\,\cN=1\,$ supersymmetry. Let us go deeper into the way in which minimal supersymmetry is preserved by the $\,1_{(s_{1},s_{2})}\,$ solution. Recalling the decomposition of the $\,\textrm{SU}(8)\,$ $R$-symmetry group of maximal supergravity first under the $\,\mathbb{Z}_{2}\,$ orientifold projection truncating to half-maximal supergravity and then under the $\,\textrm{SO}(3)\,$ truncation yielding minimal supergravity
\be
\begin{array}{ccccc}
 \textrm{SU}(8) & \overset{\mathbb{Z}_{2}}{\supset} & \textrm{SU}(4)_{\textrm{even}} \times \textrm{SU}(4)_{\textrm{odd}}  & \supset  & \textrm{SO}(3)_{\textrm{even}}  \times \textrm{SO}(3)_{\textrm{odd}} \\[2mm]
\textbf{8} & \rightarrow & (\textbf{4},\textbf{1})_{\textrm{even}} \oplus (\textbf{1},\textbf{4})_{\textrm{odd}}  & \rightarrow & (\textbf{1},\textbf{1})_{\textrm{even}} \oplus (\textbf{3},\textbf{1})_{\textrm{even}} \oplus (\textbf{1},\textbf{1})_{\textrm{odd}} \oplus (\textbf{1},\textbf{3})_{\textrm{odd}}
\end{array}
\nonumber
\ee
one observes that there are two invariant (covariantly constant) spinors associated to the $\,(\textbf{1},\textbf{1})_{\textrm{even}}\,$ and $\,(\textbf{1},\textbf{1})_{\textrm{odd}}\,$ irrep's respectively. This implies that there are two possible $\,\cN=1\,$ residual supersymmetry that can be preserved by the $\,1_{(s_{1},s_{2})}\,$ configurations. However, since the $\,\textrm{SU}(4)\,$ $R$-symmetry group of half-maximal supergravity is identified with $\,\textrm{SU}(4)_{\textrm{even}}\,$ and \textit{not} with $\,\textrm{SU}(4)_{\textrm{odd}}\,$, only those configurations preserving the $\,\cN=1\,$ supersymmetry associated to the $\,(\textbf{1},\textbf{1})_{\textrm{even}}\,$ irrep ($\,1_{(+,+)}\,$ and $\,1_{(-,+)}\,$) can still be truncated to half-maximal supergravity as ${\,\cN=1\,}$ supersymmetric solutions. In contrast, solutions preserving the ${\,\cN=1\,}$ supersymmetry associated to the $\,(\textbf{1},\textbf{1})_{\textrm{odd}}\,$ irrep ($\,1_{(-,-)}\,$ and $\,1_{(+,-)}\,$) appear as non-supersymmetric solutions when truncated to half-maximal supergravity. Nevertheless, they are fake supersymmetric in the sense that they are supersymmetric with respect to the ``wrong'' $R$-symmetry group, hence inheriting all the stability properties associated to supersymmetric solutions.

At this point, the nature of the two $\mathbb{Z}_{2}$ factors labelled by $(s_{1},s_{2})$ becomes clear. The first one, as already pointed out in ref.~\cite{Dibitetto:2011gm}, is a symmetry of the $\,\cN=4\,$ theory and hence it does not really label different solutions, whereas, at this level, the second $\mathbb{Z}_{2}$ seems to appear as an accidental symmetry forcing
the value of the energy and the mass spectra of inequivalent critical points to be identical. When lifting these solutions to maximal supergravity, the second $\mathbb{Z}_{2}$ becomes a symmetry as well: it corresponds precisely to the $\,\textrm{SU}(8)\,$ element interchanging $\,\textrm{SU}(4)_{\textrm{time-like}}\,$ with $\,\textrm{SU}(4)_{\textrm{space-like}}\,$, thus relating equivalent solutions. As a consequence, the number of inequivalent critical points reduces to four and they can be seen as different solutions of the same maximal gauged supergravity.

Let us now identify the gauge group underlying these type IIA geometric backgrounds in a maximal gauged supergravity context. Since we set $\,F_{M \dot{\mu}}\,=\,\Xi_{\a \b \mu}=0\,$, the $X_{\,\mathbb{MNP}}$ components in (\ref{Xfermionic}) do vanish. Then, the brackets (\ref{gauge_algebra}) of the gauge group $\,G_{0}\,$ take the simpler form
\be
\label{algebra_bos_general}
\begin{array}{ccclcc}
\left[ X_{\a M} , X_{\b N} \right] & = &  -  & {X_{\a M \b N}}^{\g P} \, X_{\g P} & , \\[2mm]
\left[ X_{\a M} , X_{\m} \right] & = &  - & {X_{\a M \m}}^{\r} \, X_{\r}  & , \\[2mm]
\left[ X_{\m} , X_{\n} \right] & = &  - & {X_{\m \n}}^{\g P} \, X_{\g P}  & .
\end{array}
\ee
The $12$-dimensional subgroup $\,G_{\textrm{bos}} \subset G_{0}\,$ spanned by the linearly independent\footnote{Only $12$ out the $24$ bosonic generators $X_{\a M}$ are linearly independent hence entering the gauging. Adopting the same choice that in ref.~\cite{Dibitetto:2011gm}, we decide to write the magnetic generators as a function of the electric ones, \emph{i.e.}  $\,X_{-M}(X_{+M})\,$.} $\,X_{\a M}\,$ bosonic generators in (\ref{algebra_IIA}) turns out to be
\be
\label{subalgebra_bos}
G_{\textrm{bos}}=\textrm{ISO}(3) \ltimes \textrm{U}(1)^6 \ .
\ee
This is the gauge group of the half-maximal theory in which fermionic generators $\,X_{\m}\,$ are projected out by the $\,\mathbb{Z}_{2}\,$ orientifold projection \cite{Dibitetto:2011gm}. The $16$ linearly independent fermionic generators extend $\,G_{\textrm{bos}}\,$ in (\ref{subalgebra_bos}) to the complete $28$-dimensional gauge group $\,G_{0}\,$ of maximal supergravity which is identified with
\be
\label{algebra_IIA}
G_{0}=\textrm{SO}(4) \ltimes \textrm{Nil}_{(22)} \ ,
\ee
for these type IIA geometric flux backgrounds. To be more concrete about the structure of the gauge group, let us split the $28$ linearly independent generators into $6$ generators $\,\{   T_{i}^{(0)} \, , \, T_{a}^{(0)} \} \,$ spanning the semisimple $\,\textrm{SO}(4) \sim \textrm{SU}(2)_{i}  \times \textrm{SU}(2)_{a}\,$ part in (\ref{algebra_IIA}) and $22$ generators 
\be
\left\{ T_{i}^{(1)} \, , \, T_{a}^{(1)} \, , \, T_{i}^{(2)} \, , \, T_{a}^{(2)} \, , \, T_{ia} \, , \,T \right\} 
\ee
associated to the nilpotent ideal $\,\textrm{Nil}_{(22)}\,$. The index structure of the generators, where $\,i,a=1,2,3\,$, reflects their transformation properties with respect to the semisimple part of the gauge group. In the appropriate basis, we can write the non-vanishing gauge brackets as
\be
[ \, T_{i}^{(0)} , T_{j}^{(p)} ] = \epsilon_{ijk} \, T_{k}^{(p)} \hspace{5mm} (p=0,1,2)
\hspace{15mm} \textrm{ and } \hspace{15mm}
[ \, T_{i}^{(0)} , T_{ja} ] = \epsilon_{ijk} \, T_{ka} \ ,
\ee
which involve the semisimple generators $\,T_{i}^{(0)}$, together with
\be
[ \, T_{i}^{(1)} , T_{j}^{(2)} ] = \delta_{ij} \, T
\hspace{5mm} \textrm{ , } \hspace{5mm}
[ \, T_{i}^{(1)} , T^{(1)}_{a} ] = - T_{ia} \ ,
\hspace{5mm} \textrm{ and } \hspace{5mm}
[ \, T_{i}^{(1)} , T_{ja} ] = \delta_{ij} \, T^{(2)}_{a} \ ,
\ee
involving generators in the nilpotent part. When non-equivalent, the above set of brackets must be supplemented with additional ones obtained by exchanging $\,T_{i}^{(p)} \leftrightarrow T_{a}^{(p)}\,$, $\,\epsilon_{ijk} \leftrightarrow \epsilon_{abc}\,$ and $\,\delta_{ij} \leftrightarrow \delta_{ab}\,$. By inspection of the above brackets, one finds that the $\,\textrm{Nil}_{(22)}\,$ piece is a nilpotent $22$-dimensional ideal of order three (four steps) with lower central series
\be
\left\lbrace   T_{i}^{(1)} \, , \, T_{a}^{(1)} \, , \, T_{i}^{(2)} \, , \, T_{a}^{(2)} \, , \, T_{ia} \, , \,T \right\rbrace
\supset
\left\lbrace T_{i}^{(2)} \, , \, T_{a}^{(2)} \, , \, T_{ia} \, , \,T \right\rbrace
\supset
\left\lbrace T_{i}^{(2)} \, , \, T_{a}^{(2)} \, , \,T  \right\rbrace
\supset
\left\lbrace T \right\rbrace
\supset
\left\lbrace 0 \right\rbrace \ . \nonumber
\ee

As an aside remark, we have taken the real realisation of gamma matrices (see appendix~\ref{App:spinors}) when building the structure constants of the gauge algebra in (\ref{algebra_bos_general}). Otherwise, if taking the $\,{\textrm{SU}(4) \times \textrm{SU}(4)}\,$ covariant realisation, the structure constants turn out to be complex an so the gauge generators in the adjoint representation. Thus, one still would have to impose a reality condition upon vectors when it comes to identify the gaugings.

Because of all the aforementioned, we conclude that the gauge group in (\ref{algebra_IIA}) gives rise to $\,\cN=1\,$ supersymmetric and non-supersymmetric AdS stable solutions of maximal supergravity at the origin of the moduli space which can be embedded in string theory as type IIA flux compactifications in the presence of geometric fluxes.

\chapter*{Conclusions and Outlook}
\markboth{Conclusions and Outlook}{Conclusions and Outlook}
\addcontentsline{toc}{chapter}{Conclusions and Outlook}
\label{Conclusions}
In this thesis we have analysed many aspects of flux compactifications in string theory preserving maximal and half-maximal supersymmetry. To this aim, we have exploited gauged supergravities, not only as lower-dimensional effective descriptions but also as a guideline to understand the role of string dualities in flux compactifications. The main reason for expecting such a relation is that supergravities in any dimension enjoy global symmetries which happen to exactly match the duality groups coming from their corresponding stringy origin (see tables~\ref{table:string_dualities} and \ref{table:max} -- \ref{table:half-max}). 

We have seen how gauged supergravities effectively describe compactifications of string theory preserving some supersymmetry in the presence of fluxes. Unfortunately, though, as we pointed out in figure~\ref{figure:higher-dim}, lower-dimensional supergravities allow for a wider set of deformations (\emph{i.e.} gaugings), some of which have no known higher-dimensional origin and therefore they are called non-geometric fluxes. 
From the viewpoint of the effective description, one realises that non-geometric fluxes are a very helpful (if not necessary!) ingredient for achieving moduli stabilisation. However, the open problem remains precisely that of providing a higher-dimensional origin for them.

We have used the so-called embedding tensor formalism in order to accomodate all the consistent deformations of half-maximal and maximal supergravities into irrep's of the aforementioned duality groups which appear in the compactified theories as actual symmetries. This implies as a straightforward consequence that embedding tensor configurations which are related to each other by a duality transformation describe the same physics. Thus, in order to understand whether or not non-geometric fluxes really bring new physics into the game, one realises the importance of classifying duality orbits of gaugings. Only in the case of new orbits with no geometric representative, one should actually worry about providing a higher-dimensional description thereof. This has been discussed in chapter~\ref{DFT}, where we have performed the orbit classification mentioned above starting from the highest dimensions ($D=9,\,8,\,7$) in both the maximal and half-maximal case. We found that all the considered maximal supergravities only admit theories which are geometric up to U-duality transformations. On the contrary, in the half-maximal case we found a number of non-geometric orbits of theories. Still, we were able to provide an uplift to Double Field Theory (DFT) for all of them by means of a twisted doubled torus reduction. 

DFT is a recently proposed construction which implements O($10,10$) invariance at the level of a field thoery living in $10+10$ dimensions. Its gauge invariance in the original background independent formulation was only proven by imposing an additional constraint on the dependence of the fields. Such a constraint is generally referred to as the strong constraint and, whenever satisfied, it implies the possibility of rotating away any dependence on doubled coordinates. Because of its general features, DFT seems to be a very natural framework for addressing the issue of uplifting non-geometric fluxes. An important point concerning our results is that such a successful uplift was found to require a more recent formulation of DFT in which the strong constraint turns out not to be strictly required by gauge invariance and hence can be relaxed. In reductions of DFT, in particular one could imagine of relaxing the strong constraint by allowing higher-dimensional fields to have a more general coordinate dependence along the internal directions and only requiring a single field theory description in the large dimensions, \emph{i.e.} after reduction. 

The interesting further steps in this research line would be to try to generalise the above results to lower dimensions ($D<7$). Unfortunately though, the duality groups become larger and larger; hence the problem of classifying duality orbits of gaugings ceases to be computationally accessible. The main interesting facts that one would like to confirm in full generality or disprove by finding a counterexample are the following:
\begin{itemize}
\item do U-duality orbits of gaugings in maximal supergravities always admit a geometric representative?

\item if not, does the above statement become true once the field equations are imposed?

\item do all T-duality orbits of gaugings in half-maximal supergravities allow for a DFT uplift?
\end{itemize}  

Concentrating now on $D=4$, the embedding tensor formalism has helped us in a somehow related but different type of analysis, that is building the dictionary between embedding tensor deformations in $\cN=D=4$ and geometric fluxes in orientifold compactifications of type II string theories. The conclusion in this context is that none of the semisimple gaugings giving rise to dS solutions in $\cN=4$ supergravity is accessible by geometric flux compactifications (see chapter~\ref{Half_Max}). Subsequently, we completed the dictionary of type II generalised fluxes corresponding with the full set of emebedding tensor deformations. Finally, we moved to the analysis of critical points. In order to reduce the computational complexity of the problem, we restricted to an interesting truncation admitting an $\cN=1$ description ($STU$-model). This truncation is very relevant in flux compactifications because it contains effective theories describing type IIA compactifications with O6/D6. We were able to analyse the full set of vacua of geometric type IIA compactifications. What we found was a number of AdS critical points, for which we could also compute the full mass spectrum for all the scalar fields. The most peculiar fact about these solutions is that they all required the absence of local sources, which suggested the possibility of uplifting such a solution to maximal supergravity.

Following this natural idea, we decided to exploit these solutions in order to understand how to embed flux compactifications into $\cN=8$ gauged supergravities in the case in which no branes are required in the construction. This has been discussed in chapter~\ref{Maximal}, where, as an intermediate step we worked out the truncation taking $\cN=8$ to $\cN=4$ in four dimensions. Subsequently, given an $\cN=4$ background related in a known way to fluxes (see previous chapter), we show which extra constraints the fluxes have to satisfy in order for it to admit an uplift to $\cN=8$. In the case of the geometric type IIA setup introduced above, we showed that these extra constraints can be interpreted as the absence of local sources. Surprisingly, the uplifted AdS solutions turn out to be critical points of a unique theory within $\cN=8$ with a non-semisimple gauging. Finally, by making use of the mass formula for the scalars given in the standard formulation of maximal supergravity, we were able to explicitely compute the full mass spectrum of the AdS critical points that we had just uplifted. We found the interesting presence of a non-supersymmetric and nevertheless fully stable vacuum. This represents a further example to be considered when discussing the issue of stability without supersymmetry in extended supergravities.

We would like to stress once more that the analysis done in chapter~\ref{Half_Max} in the context of geometric type IIA compactifications preserving half-maximal supersymmetry is exhaustive since it was performed by searching critical points only in the origin of moduli space but keeping all the embedding tensor components which are related among themselves by non-compact duality transformations. We have then shown that the critical points that we found can all be regarded as solutions of the maximal theory. However, they do not constitute a complete set of vacua of the maximal theory in that the set of included embedding tensor components is not as well closed with respect to more general non-compact U-dualities. This means that, in order to exhaustively study the full landscape of vacua of geometric type II compactifications without branes, one should also include geometric fluxes which are odd under the orientifold involution defining the truncation from maximal to half-maximal theory.

As we just argued, the analysis of the full set of critical points of geometric type II compactifications would require the inclusion of flux components which are odd under the orientifold projection, like \emph{e.g.} metric flux in type IIB with O3-planes. These odd fluxes sit in spinorial irrep's of SO($6,6$) and so far very few things are known about backgrounds including them. Therefore it would be extremely interesting to perform an exhaustive analysis able to scan the landscape of vacua of geometric type II compactifications in the absence of branes. The possibility of finding new interesting critical points makes it a very promising research line. Moreover, analysing these backgrounds might shed a light on the role of U-duality in string compactifications. Such a duality, which is indeed only a symmetry at the level of the effective supergravity description, could still constitute an organising principle at a more fundamental level. We hope to come back to these points in a future project.

\appendix
\chapter{Gaugings of $D=7,\,8$ Supergravities}
\label{appendix:D=7,8}
\renewcommand{\theequation}{A.\arabic{equation}}
\section{Different solvable and nilpotent gaugings}
\label{Appendix_A_Gaugings}

In section~\ref{sec:T_Dualitites} we have studied the T-duality orbits of gaugings in half-maximal $D=7$ supergravity and for each of them, we identified the gauge algebra and presented the results in table~\ref{orbits_halfmax7}. Since there is no exhaustive classification of non-semisimple algebras of dimension 6, we would like to explicitly give the form of the algebras appearing in table~\ref{orbits_halfmax7}.

\subsection*{Solvable algebras}

\subsubsection*{The CSO($2,0,2$) and CSO($1,1,2$) algebras}

The details about these algebras can be found in ref.~\cite{deRoo:2006ms}; we summarise here some relevant facts.

The six generators are labelled as $\{t_{0},\,t_{i},\,s_{i},\,z\}_{i=1,2}$, where $t_{0}$ generates SO($2$) (SO($1,1$)), under which $\{t_{i}\}$ and $\{s_{i}\}$ transform as doublets
\be
\begin{array}{cccc}
\left[t_{0},\,t_{i}\right]\,=\,{\epsilon_{i}}^{j}\,t_{j} & , &
\left[t_{0},\,s_{i}\right]\,=\,{\epsilon_{i}}^{j}\,s_{j} & ,
\end{array}
\ee
where the Levi-Civita symbol ${\epsilon_{i}}^{j}$ has one index lowered with the metric $\eta_{ij}\,=\,$diag$(\pm 1,1)$ depending on the two different signatures. $z$ is a central charge appearing in the following commutators
\be 
\left[t_{i},\,s_{j}\right]\,=\,\delta_{ij}\,z\ . 
\ee
The Cartan-Killing metric is diag($\mp 1, \underbrace{0, \cdots, 0}_{\textrm{6 times}}$), where the $\mp$ is again related to the two different signatures.

\subsubsection*{The $\mathfrak{f}_{1}$ and $\mathfrak{f}_{2}$ algebras}

These are of the form Solv$_{4}\,\times\,$U$(1)^{2}$. The 4 generators of Solv$_{4}$ are labeled by $\{t_{0},\,t_{i},\,z\}_{i=1,2}$, where $t_{0}$ generates SO($2$) (SO($1,1$)), under which $\{t_{i}\}$ transform as a doublet
\be 
\left[t_{0},\,t_{i}\right]\,=\,{\epsilon_{i}}^{j}\,t_{j}\ , 
\ee
\be 
\left[t_{i},\,t_{j}\right]\,=\,\epsilon_{ij}\,z\ . 
\ee
The Cartan-Killing metric is diag($\mp 1, \underbrace{0, \cdots, 0}_{\textrm{6 times}}$).

\subsubsection*{The $\mathfrak{h}_{1}$ and $\mathfrak{h}_{2}$ algebras}

The 6 generators are $\{t_{0},\,t_{i},\,s_{i},\,z\}_{i=1,2}$ and they satisfy the following commutation relations
\be
\begin{array}{lclc}
\left[t_{0},\,t_{i}\right]\,=\,{\epsilon_{i}}^{j}\,t_{j} & , &
\left[t_{0},\,s_{i}\right]\,=\,{\epsilon_{i}}^{j}\,s_{j}\,+\,t_{i} &
, \\[2mm]
\left[t_{i},\,s_{j}\right]\,=\,\delta_{ij}\,z & , &
\left[s_{i},\,s_{j}\right]\,=\,\epsilon_{ij}\,z & .
\end{array}
\ee
The Cartan-Killing metric is diag($\mp 1, \underbrace{0, \cdots, 0}_{\textrm{6 times}}$).

\subsubsection*{The $\mathfrak{g}_{0}$ algebra}

The 6 generators are $\{t_{0},\,t_{I},\,z\}_{I=1,\cdots,4}$, where $t_{0}$ transforms cyclically the $\{t_{I}\}$ amongst themselves such that
\be
\left[\bigg[\big[[t_{I},\,t_{0}],\,t_{0}\big],\,t_{0}\bigg],\,t_{0}\right]\,=\,t_{I}\ , 
\ee
and
\be 
\left[t_{1},\,t_{3}\right]\,=\,\left[t_{2},\,t_{4}\right]\,=\,z\ .
\ee
Note that this algebra is solvable and not nilpotent even though its Cartan-Killing metric is \emph{completely zero}.

\subsection*{Nilpotent algebras}

\subsubsection*{The CSO($1,0,3$) algebra}

The details about this algebra can be again found in ref.~\cite{deRoo:2006ms}; briefly summarizing, the 6 generators are given by  $\{t_{m},\,z^{m}\}_{m=1,2,3}$ and they satisfy the following commutation relations
\be 
\left[t_{m},\,t_{n}\right]\,=\,\epsilon_{mnp}\,z^{p}\ ,
\ee
with all the other brackets being vanishing. The order of nilpotency of this algebra is 2.

\subsubsection*{The $\mathfrak{l}$ algebra}

The 6 generators $\{t_{1},\cdots,\,t_{6}\}$ satisfy the following commutation relations
\be
\begin{array}{lclclc}
 \left[t_{1},\,t_{2}\right]\,=\,t_{4} & , &
 \left[t_{1},\,t_{4}\right]\,=\,t_{5} & , &
 \left[t_{2},\,t_{4}\right]\,=\,t_{6} & .
\end{array}
\ee
The corresponding central series reads
\be
\begin{array}{lclclcl}
 \left\{t_{1},\,t_{2},\,t_{3},\,t_{4},\,t_{5},\,t_{6}\right\} &\supset&
 \left\{t_{4},\,t_{5},\,t_{6}\right\} &\supset&
 \left\{t_{5},\,t_{6}\right\} &\supset& \left\{0\right\}\ ,
\end{array}
\ee
from which we can immediately conclude that its nilpotency order is 3.

\section{SO($2,2$) and SO($3,3$) 't Hooft symbols}
\label{Appendix_A_'tHooft}

In section~\ref{DFT_twists} we discuss the origin of a given flux configuration from DFT backgrounds specified by twist matrices $U$. The deformations of half-maximal supergravity in $D=10-d$ which can be interpreted as the gauging of a subgroup of the T-duality group O($d,d$) can be described by a 3-form of O($d,d$) $f_{ABC}$ which represents a certain (non-)geometric flux configuration.

In $D=8$ and $D=7$, the T-duality group happens to be isomorphic to SL($2)\,\times\,$SL($2$) and SL($4$) respectively. As a consequence, in order to explicitly relate flux configurations and embedding tensor orbits, we need to construct the mapping between T-duality irrep's and irrep's of SL($2)\,\times\,$SL($2$) and SL($4$) respectively.

\subsection*{From the $(\textbf{2},\textbf{2})$ of SL($2)\,\times\,$SL($2$) to the $\textbf{4}$ of SO($2,2$)}

The 't Hooft symbols $\left[G_A\right]^{\alpha i}$ are invariant tensors which map the fundamental representation of SO($2,2$) (here denoted by $A$), into the $(\textbf{2},\textbf{2})$ of SL($2)\,\times\,$SL($2$)
\be 
v^{\alpha i}\,=\,\left[G_A\right]^{\alpha i}\,v^{A}\ ,
\label{vec2vecvec}
\ee
where $v^{A}$ denotes a vector of SO($2,2$) and the indices $\alpha$ and $i$ are raised and lowered by means of $\epsilon_{\alpha\beta}$ and $\epsilon_{ij}$ respectively. $\left[G_A\right]^{\alpha i}$ and $\left[G_A\right]_{\alpha i}$ satisfy the following identities
\bea
&&\left[G_A\right]_{\alpha i}\,\left[G_B\right]^{\alpha i}\,=\,\eta_{AB}\ ,\\
&&\left[G_A\right]^{\alpha i}\,\left[G^{A}\right]^{\beta
j}\,=\,\epsilon^{\alpha \beta}\,\epsilon^{ij}\ , 
\eea
where $\eta_{AB}$ is the SO($2,2$) metric.

After choosing light-cone coordinates for SO($2,2$), our choice for the tensors $\left[G_A\right]^{\alpha i}$ is the following
\begin{align}
\left[G_1\right]^{\alpha i} &= \left[
  \begin{array}{cc}
  0 & 0 \\
  0 & 1
  \end{array}
  \right]\,\,,
& \left[G_2\right]^{\alpha i} &= \left[
  \begin{array}{cc}
  0 & 1 \\
  0 & 0
  \end{array}
  \right]\,\,,
\\
\left[G_{\bar{1}}\right]^{\alpha i} &= \left[
  \begin{array}{cc}
  1 & 0 \\
  0 & 0
  \end{array}
  \right]\,\,,
& \left[G_{\bar{2}}\right]^{\alpha i} &= \left[
  \begin{array}{cc}
  0 & 0 \\
  -1 & 0
  \end{array}
  \right]\,\,.
\end{align}
By making use of the mapping \eqref{vec2vecvec}, we can rewrite the structure constants $(X_{\alpha i})_{\beta j}{}^{\gamma k}$ as a 3-form of SO($2,2$) as follows:
\begin{align}
 f_{ABC}
&= (X_{\alpha i})_{\beta j}{}^{\gamma k} \left[G_{A}\right]^{\alpha i}
\left[G_{B}\right]^{\beta j} \left[G_{C}\right]_{\gamma k}\,\,. \label{X2f_D=8}
\end{align}

\subsection*{From the $\textbf{6}$ of SL($4$) to the $\textbf{6}$ of SO($3,3$)}

The 't Hooft symbols $\left[G_A\right]^{mn}$ are invariant tensors which map the fundamental representation of SO($3,3$), \emph{i.e.} the \textbf{6} into the anti-symmetric two-form of SL($4$)
\be 
v^{mn}\,=\,\left[G_A\right]^{mn}\,v^{A}\ ,
\label{vec2two-form}
\ee
where $v^{A}$ denotes a vector of SO($3,3$). The two-form irrep of SL($4$) is real due to the role of the Levi-Civita tensor relating $v^{mn}$ to $v_{mn}$
\be 
v_{mn}\,=\,\frac{1}{2}\,\epsilon_{mnpq}\,v^{pq}\ . 
\ee
The 't Hooft symbols with lower SL($4$) indices $\left[G_A\right]_{mn}$ carry out the inverse mapping of the one given in \eqref{vec2two-form}. The tensors $\left[G_A\right]^{mn}$ and $\left[G_A\right]_{mn}=\frac{1}{2}\,\epsilon_{mnpq}\,\left[G_A\right]^{pq}$ satisfy the following identities
\bea
&&\left[G_A\right]_{mn}\,\left[G_B\right]^{mn}\,=\,2\,\eta_{AB}\ ,\\
&&\left[G_{A}\right]_{mp}\,\left[G_{B}\right]^{pn}\,+\left[G_{B}\right]_{mp}\,\left[G_{A}\right]^{pn}\,=\,-\delta_m^{n}\,\eta_{AB}\ ,\\
&&\left[G_A\right]_{mp}\,\left[G_B\right]^{pq}\,\left[G_C\right]_{qr}\,\left[G_D\right]^{rs}\,\left[G_E\right]_{st}\,\left[G_F\right]^{tn}\,=\,\delta_m^{n}\,\epsilon_{ABCDEF}\ ,
\eea
where $\eta_{AB}$ and $\epsilon_{ABCDEF}$ are the SO($3,3$) metric and Levi-Civita tensor respectively.

After choosing light-cone coordinates for SO($3,3$) vectors, our choice of the 't Hooft symbols is
\begin{align}
\left[G_1\right]^{mn}
&=
\left[
  \begin{array}{cccc}
  0 & -1 & 0 & 0\\
  1 & 0 & 0 & 0\\
  0 & 0 & 0 & 0\\
  0 & 0 & 0 & 0
  \end{array}
  \right]\,\,,
&
\left[G_2\right]^{mn}
&=
\left[
  \begin{array}{cccc}
  0 & 0 & -1 & 0\\
  0 & 0 & 0 & 0\\
  1 & 0 & 0 & 0\\
  0 & 0 & 0 & 0
  \end{array}
  \right]\,\,,
\\
\left[G_3\right]^{mn}
&=
\left[
  \begin{array}{cccc}
  0 & 0 & 0 & -1\\
  0 & 0 & 0 & 0\\
  0 & 0 & 0 & 0\\
  1 & 0 & 0 & 0
  \end{array}
  \right]\,\,,
&
\left[G_{\bar1}\right]^{mn}
&=
\left[
  \begin{array}{cccc}
  0 & 0 & 0 & 0\\
  0 & 0 & 0 & 0\\
  0 & 0 & 0 & -1\\
  0 & 0 & 1 & 0
  \end{array}
  \right]\,\,,
\\
\left[G_{\bar2}\right]^{mn}
&=
\left[
  \begin{array}{cccc}
  0 & 0 & 0 & 0\\
  0 & 0 & 0 & -1\\
  0 & 0 & 0 & 0\\
  0 & 1 & 0 & 0
  \end{array}
  \right]\,\,,
&
\left[G_{\bar3}\right]^{mn}
&=
\left[
  \begin{array}{cccc}
  0 & 0 & 0 & 0\\
  0 & 0 & 1 & 0\\
  0 & -1 & 0 & 0\\
  0 & 0 & 0 & 0
  \end{array}
  \right]\,\,.
\end{align}

Thus, we can rewrite the structure constants in the \textbf{6}, $(X_{mn})_{pq}{}^{rs}$, arising from (\ref{gen_half-max}) as a 3-form of SO($3,3$) as follows:
\begin{align}
 f_{ABC} & = (X_{mn})_{pq}{}^{rs} \left[G_{A}\right]^{mn} \left[G_{B}\right]^{pq} \left[G_{C}\right]_{rs}\ .
\label{X2f_D=7}
\end{align}

\chapter{Gaugings and Superpotentials from Fluxes}
\label{appendix:Fluxes}
\renewcommand{\theequation}{B.\arabic{equation}}
\section{Type II fluxes and the embedding tensor $\,f_{\a MNP}$}
\label{App:fluxes}

In this appendix, we summarise the identification between embedding tensor components $\,f_{\a MNP}\,$ in the $\textbf{220}$ (alternatively $\, \Lambda_{\a ABC}\,$ as explained in section~\ref{FTheta_Dictionary}) and type II flux backgrounds for the $\,\cN=1\,$ supergravity theory.

\begin{figure}[h!]
\begin{center}
\scalebox{0.9}[0.9]{
\begin{tabular}{ccccc}
\includegraphics[scale=0.5,keepaspectratio=true]{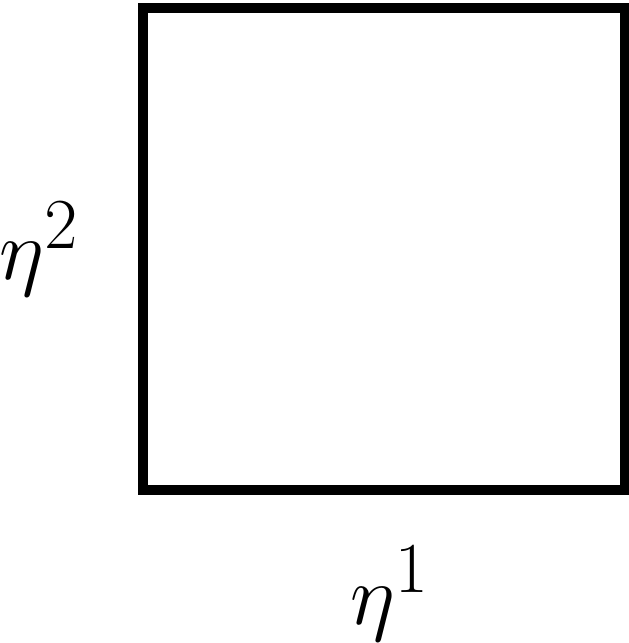} &    &  \includegraphics[scale=0.5,keepaspectratio=true]{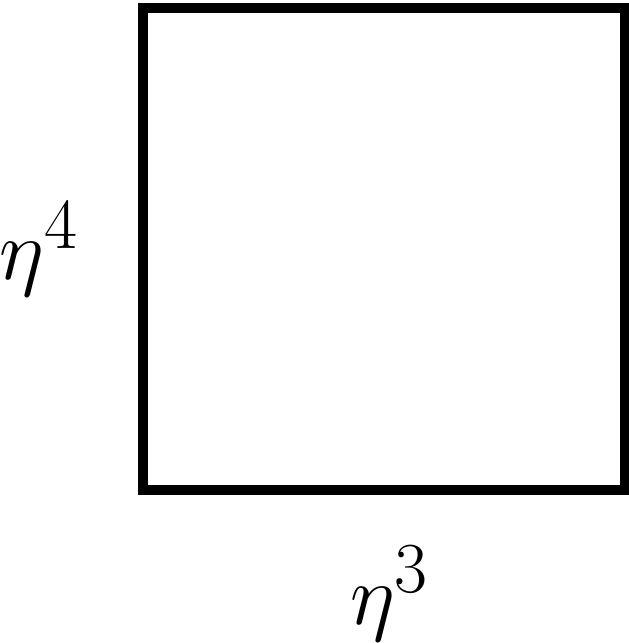}  &    & \includegraphics[scale=0.5,keepaspectratio=true]{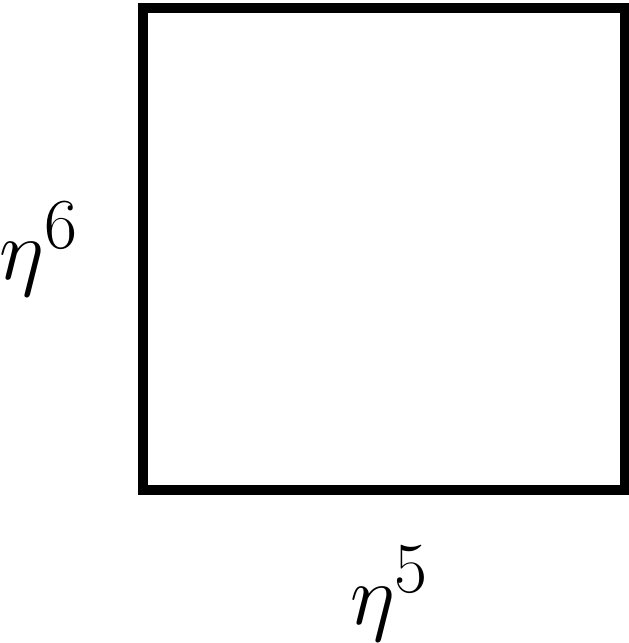} \\[-24mm]
     & \,\,\Large{$\times$} &        & \,\,\Large{$\times$} &
\end{tabular}
}
\end{center}
\vspace{9.5mm}
\caption{{\it $T^{6} =  T^{2}_{1} \times T_{2}^{2} \times T_{3}^{2}$ torus factorisation and the coordinate basis.}}
\label{fig:Torus_Factor1}
\end{figure}
In the following we will use early Latin indices $a,b,c$ for horizontal $\,``-"$ $x$-like directions $(\eta^{1},\eta^{3},\eta^{5})$ and late Latin indices $i,j,k$ for vertical $\,``\,|\,"$ $y$-like directions $(\eta^{2},\eta^{4},\eta^{6})$ in the 2-tori $\,T_{I}\,$ with $\,I=1,2,3$. This splitting of coordinates is in one-to-one correspondence with the SO($6,6$) index splitting of the embedding tensor components given in (\ref{defL}), where $A=(1,2,3,4) \equiv (a,i,\bar{a},\bar{i})\,$ refers to an SO($2,2$) fundamental index and $\epsilon_{IJK}$ denotes the usual totally antisymmetric tensor. The conventions adopted here are chosen such in a way that they match those ones introduced in section~\ref{Geom_Flux_Comp} in the context of $\cN=1$ supergravity models coming from fluxes.

\begin{table}[h!]
\renewcommand{\arraystretch}{1.25}
\begin{center}
\scalebox{0.84}[0.87]{
\begin{tabular}{ | c || c | c | c | c | c |}
\hline
couplings & SO($6,6$) & SO($2,2$) & type IIB & type IIA & fluxes\\
\hline
\hline
$1 $& $ -f_{+ \bar{a}\bar{b}\bar{c}} $  & $ - \Lambda_{+333} $& $ {F}_{ ijk} $& $F_{aibjck}$ & $  a_0 $\\
\hline
$U $& $f_{+ \bar{a}\bar{b}\bar{k}}$  &  $ \Lambda_{+334} $& ${F}_{ ij c} $& $F_{aibj}$ & $   a_1 $\\
\hline
$U^2 $& $ -f_{+ \bar{a}\bar{j}\bar{k}}$  & $ - \Lambda_{+344} $& ${F}_{i b c} $& $F_{ai}$ & $  a_2 $\\
\hline
$U^3 $& $f_{+ \bar{i}\bar{j}\bar{k}}$  & $ \Lambda_{+444} $& ${F}_{a b c} $& $F_{0}$ & $  a_3 $\\
\hline
\hline
$S $& $ -f_{- \bar{a}\bar{b}\bar{c}} $  &  $ - \Lambda_{-333} $& $ {H}_{ijk} $& $ {H}_{ijk} $  & $  - b_0$\\
\hline
$S \, U $& $f_{- \bar{a}\bar{b}\bar{k}}$  & $ \Lambda_{-334}$ & $ {H}_{ij c} $& ${\omega}^{c}_{ij}$ & $  - b_1 $\\
\hline
$S \, U^2 $& $ -f_{- \bar{a}\bar{j}\bar{k}}$   & $ - \Lambda_{-344} $& $ {H}_{ i b c}$ & $ {Q}_{ i }^{ b c}$  & $ - b_2 $\\
\hline
$S \, U^3 $& $f_{- \bar{i}\bar{j}\bar{k}}$   &  $ \Lambda_{-444} $& $ {H}_{a b c} $& $ {R}^{a b c} $ & $ - b_3 $\\
\hline
\hline
$T $& $f_{+ \bar{a}\bar{b}k}$   & $ \Lambda_{+233} $& $  Q^{a b}_k $&$ H_{a b k} $ & $  c_0 $\\
\hline
$T \, U $& $f_{+ \bar{a}\bar{j} k}=f_{+ \bar{i}\bar{b} k}\,\,\,,\,\,\,f_{+ a\bar{b}\bar{c}}$  &  $ \Lambda_{+234} \,\,\,,\,\,\, \Lambda_{+133} $& $ Q ^{a j}_k = Q^{i b}_k \,\,\,,\,\,\, Q^{b c}_a $& $ \omega^{j}_{k a} = \omega^{i}_{b k} \,\,\,,\,\,\, \omega_{b c}^a $  & $c_1 \,\,\,,\,\,\, \tilde {c}_1 $\\
\hline
$T \, U^2 $& $f_{+ \bar{i}\bar{b}c}=f_{+ \bar{a}\bar{j}c}\,\,\,,\,\,\,f_{+ \bar{i}\bar{j}k}$   &  $ \Lambda_{+134} \,\,\,,\,\,\, \Lambda_{+244} $& $ Q ^{ib}_c = Q^{a j}_c \,\,\,,\,\,\, Q^{ij}_k $& $ Q ^{ci}_b = Q^{j c}_a \,\,\,,\,\,\, Q^{ij}_k $ & $c_2 \,\,\,,\,\,\,\tilde{c}_2 $\\
\hline
$T \, U^3 $& $f_{+ \bar{i}\bar{j} c}$  &  $ \Lambda_{+144} $& $  Q^{ij}_{c} $& $  R^{ijc} $ & $c_3 $\\
\hline
\hline
$S \, T $& $f_{- \bar{a}\bar{b}k}$   &  $ \Lambda_{-233} $& $  P^{a b}_k $& & $  - d_0 $\\
\hline
$S \, T \, U $& $f_{- \bar{a}\bar{j} k}=f_{- \bar{i}\bar{b} k}\,\,\,,\,\,\,f_{- a\bar{b}\bar{c}}$   &  $ \Lambda_{-234} \,\,\,,\,\,\, \Lambda_{-133} $& $ P ^{a j}_k = P^{i b}_k \,\,\,,\,\,\, P^{b c}_a $&  & $- d_1 \,\,\,,\,\,\, - \tilde {d}_1 $\\
\hline
$S \, T \, U^2 $& $f_{- \bar{i}\bar{b}c}=f_{- \bar{a}\bar{j}c}\,\,\,,\,\,\,f_{- \bar{i}\bar{j}k}$   &  $ \Lambda_{-134} \,\,\,,\,\,\, \Lambda_{-244} $& $ P ^{ib}_c = P^{a j}_c \,\,\,,\,\,\, P^{ij}_k $&  & $ - d_2 \,\,\,,\,\,\, - \tilde{d}_2 $\\
\hline
$S \, T \, U^3 $& $f_{- \bar{i}\bar{j} c}$  &  $ \Lambda_{-144} $ & $  P^{ij}_{c} $&  & $ - d_3 $\\
\hline
\end{tabular}
}
\end{center}
\caption{{\it Mapping between unprimed fluxes, embedding tensor components and couplings in the flux-induced superpotential. We have made the index splitting $\,M=\{a,i,\bar{a},\bar{i}\}\,$ for SO($6,6$) light-cone coordinates.}}
\label{table:unprimed_fluxes4}
\end{table}
\noindent This identification\footnote{Notice that refs~\cite{Dibitetto:2010rg,Dibitetto:2011gm} use light-cone coordinates for $\,\textrm{SO}(6,6)\,$ fundamental indices. In this basis, the metric takes the form
$\,
\eta_{MN} =
{\scriptsize{
\left(
\begin{array}{cc}
0 & \mathds{1}_{6} \\
\mathds{1}_{6} & 0
\end{array}
\right)}}
\,$
which is related to the Lorentzian metric diag$(\underbrace{-1,\cdots,-1}_{6},\underbrace{+1,\cdots,+1}_{6})$ through an $\,\textrm{SO}(12)\,$ rotation of the form
\be
U=\frac{1}{\sqrt{2}} \,\left(
\begin{array}{cc}
-\mathds{1}_{6} & \mathds{1}_{6} \\
\mathds{1}_{6} & \mathds{1}_{6}
\end{array}
\right) \ .
\ee
}
was originally proposed in ref.~\cite{Dibitetto:2010rg} and further developed in ref.~\cite{Dibitetto:2011gm}. We include it here for the sake of completeness.
\begin{table}[h!]
\renewcommand{\arraystretch}{1.25}
\begin{center}
\scalebox{0.87}[0.87]{
\begin{tabular}{ | c || c | c |c | c | c |}
\hline
couplings & SO($6,6$) & SO($2,2$) & type IIB &  type IIA & fluxes\\
\hline
\hline
$T^3 \, U^3 $& $ -f_{+ abc} $ & $ - \Lambda_{+111} $& $ {F'}^{ijk} $&  & $  a_0' $\\
\hline
$T^3 \, U^2 $&  $f_{+ abk}$ &  $ \Lambda_{+112} $& ${F'}^{ ij c} $& &$   a_1' $\\
\hline
$T^3 \, U $& $ -f_{+ ajk}$  & $ - \Lambda_{+122} $& ${F'}^{i b c} $& &$  a_2' $\\
\hline
$ T^3 $& $f_{+ ijk}$ & $ \Lambda_{+222} $& ${F'}^{a b c} $& &$  a_3' $\\
\hline
\hline
$S \, T^3 \, U^3 $& $ -f_{- abc} $  & $ - \Lambda_{-111} $& $ {H'}^{ ijk} $& &$  - b_0'$\\
\hline
$S \, T^3 \, U^2 $& $f_{- abk}$  & $ \Lambda_{-112} $& $ {H'}^{i jc} $& &$ - b_1' $\\
\hline
$S \, T^3 \, U $& $ -f_{- ajk}$  & $ - \Lambda_{-122} $& $ {H'}^{ i b c} $& & $ - b_2' $\\
\hline
$S  \, T^3 $& $f_{- ijk}$   & $ \Lambda_{-222}$ & $ {H'}^{a b c} $& &$ - b_3' $\\
\hline
\hline
$T^2 \, U^3 $& $f_{+ ab\bar{k}}$ & $ \Lambda_{+114} $& $  {Q'}_{a b}^k $& &$  c_0' $\\
\hline
$T^2 \, U^2 $& $f_{+ aj\bar{k}}=f_{+ ib\bar{k}}\,\,\,,\,\,\,f_{+ \bar{a}bc}$ & $  \Lambda_{+124} \,\,\,,\,\,\, \Lambda_{+113} $ & $ {Q'}_{a j}^k = {Q'}_{i b}^k \,\,\,,\,\,\, {Q'}_{b c}^a $& &$c_1' \,\,\,,\,\,\, \tilde{c}_1' $\\
\hline
$T^2 \, U $& $f_{+ ib\bar{c}}=f_{+ aj\bar{c}}\,\,\,,\,\,\,f_{+ ij\bar{k}}$  & $ \Lambda_{+123} \,\,\,,\,\,\, \Lambda_{+224} $& $ {Q'}_{ib}^c = {Q'}_{a j}^c \,\,\,,\,\,\, {Q'}_{ij}^k $& &$c_2' \,\,\,,\,\,\,\tilde{c}_2' $\\
\hline
$T^2 $& $f_{+ ij\bar{c}}$  & $ \Lambda_{+223} $& $  {Q'}_{ij}^{c} $& &$c_3' $\\
\hline
\hline
$S \, T^2 \, U^3$& $f_{- ab\bar{k}}$  & $ \Lambda_{-114} $& $  {P'}_{a b}^k $& &$ - d_0' $\\
\hline
$S \, T^2 \, U^2 $& $f_{- aj\bar{k}}=f_{- ib\bar{k}}\,\,\,,\,\,\,f_{- \bar{a}bc}$ & $ \Lambda_{-124} \,\,\,,\,\,\, \Lambda_{-113} $& $ {P'}_{a j}^k = {P'}_{i b}^k \,\,\,,\,\,\, {P'}_{b c}^a $& &$ - d_1' \,\,\,,\,\,\, - \tilde {d}_1' $\\
\hline
$S \, T^2 \, U $& $f_{- ib\bar{c}}=f_{- aj\bar{c}}\,\,\,,\,\,\,f_{- ij\bar{k}}$  &$ \Lambda_{-123} \,\,\,,\,\,\, \Lambda_{-224} $ & $ {P'}_{ib}^c = {P'}_{a j}^c \,\,\,,\,\,\, {P'}_{ij}^k $& &$ - d_2' \,\,\,,\,\,\, - \tilde{d}_2' $\\
\hline
$S \, T^2  $& $f_{-ij\bar{c}}$ & $ \Lambda_{-223} $ & $  {P'}_{ij}^{c} $& &$ - d_3' $\\
\hline
\end{tabular}
}
\end{center}
\caption{{\it Mapping between primed fluxes, embedding tensor components and couplings in the flux-induced superpotential. We have made the index splitting $\,M=\{a,i,\bar{a},\bar{i}\}\,$ for SO($6,6$) light-cone coordinates.}}
\label{table:primed_fluxes4}
\end{table}

Irrespective of their IIA or IIB string theory interpretation, the above set of fluxes generates the following $\,\cN=1\,$ flux-induced superpotential
\be
\label{W_fluxes4}
W = (P_{F} + P_{H} \, S ) + 3 \, T \, (P_{Q} + P_{P} \, S ) + 3 \, T^2 \, (P_{Q'} + P_{P'} \, S ) + T^3 \, (P_{F'} + P_{H'} \, S ) \ ,
\ee
involving the three complex moduli $S$, $T$ and $U$ surviving the SO($3$) truncation introduced in section~\ref{FTheta_Dictionary}. However, just by a simple inspection of tables~\ref{table:unprimed_fluxes4} and \ref{table:primed_fluxes4}, it is clearly more convenient to adopt the terminology of the type IIB string theory when it comes to associate embedding tensor components to fluxes. In this picture, the superpotential in (\ref{W_fluxes4}) contains flux-induced polynomials depending on both electric and magnetic pairs -- schematically $\,(e,m)\,$ -- of gauge $(F_{3},H_{3})$ fluxes and non-geometric $(Q,P)$ fluxes,
\be
\begin{array}{lcll}
\label{Poly_unprim}
P_{F} = a_0 - 3 \, a_1 \, U + 3 \, a_2 \, U^2 - a_3 \, U^3 & \hspace{5mm},\hspace{5mm} & P_{H} = b_0 - 3 \, b_1 \, U + 3 \, b_2 \, U^2 - b_3 \, U^3 & ,  \\[2mm]
P_{Q} = c_0 + C_{1} \, U - C_{2} \, U^2 - c_3 \, U^3 & \hspace{5mm},\hspace{5mm} & P_{P} = d_0 + D_{1} \, U - D_{2} \, U^2 - d_3 \, U^3 & ,
\end{array}
\ee
as well as those induced by their less known primed counterparts $\,(F'_{3},H'_{3})\,$ and $\,(Q',P')\,$ fluxes,
\be
\begin{array}{lcll}
\label{Poly_prim}
P_{F'} = a_3' + 3 \, a_2' \, U + 3 \, a_1' \, U^2 + a_0' \, U^3 & \hspace{3mm},\hspace{3mm} &P_{H'} = b_3' + 3 \, b_2' \, U + 3 \, b_1' \, U^2 + b_0' \, U^3 & ,  \\[2mm]
P_{Q'} = -c_3' +  C'_{2} \, U + C'_{1} \, U^2 - c_0' \, U^3 & \hspace{3mm},\hspace{3mm} & P_{P'} = -d_3' + D'_{2} \, U + D'_{1} \, U^2 - d_0' \, U^3 & .
\end{array}
\ee

\noindent For the sake of clarity, we have introduced the flux combinations $\,C_{i} \equiv 2 \, c_i - \tilde{c}_{i}\,$, $\,D_{i} \equiv 2 \, d_i - \tilde{d}_{i}\,$, $\,C'_{i} \equiv 2 \, c'_i - \tilde{c}'_{i}\,$ and $\,D'_{i} \equiv 2 \, d'_i - \tilde{d}'_{i}\,$ entering the superpotential (\ref{W_fluxes4}), and hence also the scalar potential.

As we already saw in chapter~\ref{Fluxes}, these so-called primed fluxes have been conjectured in ref.~\cite{Aldazabal:2006up} to be needed in order to have a fully U-duality invariant flux background, but there is no further understanding of their physical role and of the types of sources coupling to them at the present stage. Still, those give a hint to understand the relation between doubled geometry and non-geometry as anticipated in the introduction. In the heterotic duality frame those two exactly coincide, in the sense that all the fluxes introduced by using doubled geometry happen to be interpretable as non-geometric fluxes. However, in such a duality frame it is impossible to introduce their magnetic dual counterparts. After performing an S-duality to go to type I (equivalent to type IIB with O9-planes) and subsequently a 6-tuple T-duality, we are in IIB with O3-planes. In such a duality frame, non-geometry and doubled geometry happen to give rise to two complementary generalised sets of fluxes (see figure~\ref{fig:DG_NG}), the second one consisting with these primed fluxes. Moreover, this particular frame is S-duality invariant and therefore such a flux background can be completed to a fully S-duality invariant one. This construction in the isotropic case allows us to at least formally\footnote{Primed fluxes do not have any well-defined string theory description, not even a local one, since they stem from some strongly coupled limit of the IIB theory.} describe all the embedding tensor components included in the SO($3$) truncation.

\section{Full $\,\mathcal{N}=1\,$ flux vacua of geometric type IIA}
\label{App:N1_vacua}

The techniques developed to analyse the vacua of the $\cN=4$ theory turn out to be powerful enough to also work out the complete set of solutions of type IIA geometric backgrounds compatible with minimal supersymmetry.
As we saw in section \ref{FTheta_Dictionary}, the SO($3$) truncation admits an
$\mathcal{N}=1$ superpotential formulation. In this context it
becomes natural to relax the QC in (\ref{N6_orth}) which can be
understood as the lack of D$6$-branes orthogonal to the O$6$-planes.
Namely,
\be
\label{QC_relax}
N_{6}^{\bot} = - a_{3}\, c_{0} - a_{2} \, (2\,
c_1 - \tilde{c}_{1}) \neq 0 \ .
\ee
After this, the theory no longer enjoys $\mathcal{N}=4$ supersymmetry but it
 still admits an $\mathcal{N}=1$ description\footnote{Nevertheless, any
solution of the $\mathcal{N}=1$ theory compatible with the absence
of such sources can be embedded into the $\mathcal{N}=4$ theory.}.
In this section we will explore its vacuum structure.

We will distinguish between two types of IIA geometric flux
backgrounds, namely, those having only gauge fluxes and those with
both gauge and metric fluxes.

\subsection*{Backgrounds only with gauge fluxes}

Let us start by fixing the components of the metric $\omega$ flux to zero, namely,
\be
b_{1}=c_{1}=\tilde{c}_{1}=0 \ .
\ee
Putting together the first and the second QC in (\ref{QC_IIA}) and the extremality conditions, and using again the GTZ algebraic method of prime decomposition (details explained
in section~\ref{Review_N=4}), we obtain a solution space consisting of two pieces:

\begin{table}[h!]
\renewcommand{\arraystretch}{1.80}
\begin{center}
\scalebox{0.84}[0.87]{
\begin{tabular}{ | c || c | c |c | c | c | c |c | c | c | c |}
\hline
\textrm{ID} & $a_{0}$ & $a_{1}$ & $a_{2}$ & $a_{3}$ & $b_{0}$ & $b_{1}$ & $c_{0}$ & $c_{1}=\tilde{c}_{1}$ & $V_{0}$ & BF \\[1mm]
\hline \hline
$1$ & $0$ & $\dfrac{3 \lambda}{2}$ & $0$ & $\dfrac{5 \lambda}{2}$ & $-\lambda$ & $0$ & $\lambda$ & $0$ & $-\dfrac{3 \lambda^{2}}{32}$ & $m^2=-\dfrac{2}{3}\rightarrow$ stable\\[1mm]
\hline
$2$ & $0$ & $-\dfrac{3 \lambda}{2}$ & $0$ & $\dfrac{5 \lambda}{2}$ & $-\lambda$ & $0$ & $\lambda$ & $0$ & $-\dfrac{3 \lambda^{2}}{32}$ & $m^2=-\dfrac{2}{3}\rightarrow$ stable\\[1mm]
\hline
$3$ & $0$ & $\sqrt{6} \lambda$ & $0$ & $5 \lambda$ & $-4 \lambda$ & $0$ & $\lambda$ & $0$ & $-\dfrac{\lambda^{2}}{4}$ & $\min$\\[1mm]
\hline
$4$ & $0$ & $-\sqrt{6} \lambda$ & $0$ & $5 \lambda$ & $-4 \lambda$ & $0$ & $\lambda$ & $0$ & $-\dfrac{\lambda^{2}}{4}$ & $\min$\\[1mm]
\hline
$5_{s_{1}}$ & $0$ & $s_{1}\,\lambda$ & $\lambda$ & $-2 \,s_{1}\, \lambda$ & $s_{1}\,\lambda$ & $0$ & $-s_{1}\,\lambda$ & $0$ & $-\dfrac{\lambda^{2}}{16}$ & $\min$\\[1mm]
\hline
$6_{s_{1}}$ & $0$ & $s_{1}\,\dfrac{7 \lambda}{3}$ & $-\dfrac{\lambda}{3}$ & $-s_{1}\,\dfrac{14 \lambda}{3}$ & $s_{1}\,\dfrac{11 \lambda}{3}$ & $0$ & $-s_{1}\,\lambda$ & $0$ & $-\dfrac{11 \lambda^{2}}{48}$ & $m^2=-0.14251 \rightarrow$ stable\\[1mm]
\hline
\end{tabular}
}
\end{center}
\caption{{\it The set of stable AdS$_{4}$ extrema of dimension $1$ in the $\mathcal{N}=1$ type IIA theory only with gauge fluxes.}}
\label{table:N=1_vacua_c11=0}
\end{table}

\begin{itemize}

\item[$i)$] The first piece has dimension $2\,$ and it is directly identified with the solution in (\ref{IIA_GKP}) of the $\,\mathcal{N}=4\,$ theory.

\item[$ii)$] The second piece consists of eight critical points of dimension $1$, all of them implying a non-vanishing tadpole for both
\be
N_{6}^{\bot} = - a_{3} \, c_{0} \neq 0 \hspace{10mm} \textrm{ and } \hspace{10mm} N_{6}^{||}= - a_{3} \, b_{0} \neq 0 \ ,
\ee
so they cannot be embedded into the previous $\,\mathcal{N}=4\,$ theory. These moduli solutions are stable AdS$_{4}$ vacua which are summarised in table~\ref{table:N=1_vacua_c11=0}. Finally, these solutions of the $\mathcal{N}=1$ theory are non-supersymmetric except that labelled with $\,1\,$ in table~\ref{table:N=1_vacua_c11=0} which turns out to preserve $\,\mathcal{N}=1\,$ supersymmetry. The scalar potential induced by the fluxes of solutions $2$ and $4$ is respectively  related to that one induced by the fluxes of $1$ and $3$ in table~\ref{table:N=1_vacua_c11=0} by the transformation
\be
\label{I_parity}
\begin{array}{cccccccc}
\alpha_{3}: & V(S,T,U) &=& - i \,\, V(\, i \, S, i \, T,-i \, U \,;\, -a_{1}, \,\,f_{i}) & ,
\end{array}
\ee
where $f_i$ refers to all the fluxes left invariant. Such a transformation can also be viewed at the level of the superpotential as $\,W(S,T,U) \rightarrow i \,W(S,T,U)$. Unlike those in the previous section, this transformation modifies the K\"ahler potential and, as a consequence, the mass spectrum for the solutions $1$ and $2$ (also $3$ and $4$) is different even when they share the lightest mass. They correspond to completely different solutions a they look quite similar each other.
\end{itemize}

\subsection*{Backgrounds with both gauge and metric fluxes}

Let us now allow for backgrounds with non-vanishing metric
fluxes. Putting again together the first and second QC
in (\ref{QC_IIA}) and the extremum conditions,
and running the GTZ method of prime decomposition, we obtain two
prime factors of dimension $2$ compatible with real fluxes:

\begin{itemize}

\item[$i)$] The first piece represents a branch of non-supersymmetric solutions which cannot be embedded into the $\,\mathcal{N}=4\,$ theory (all the solutions come out with $\,N_{6}^{\bot}\neq 0$). This piece implies $\,a_{0}=a_{1}=0\,$. Without loss of generality, we can set the global scale of $\,V\,$ by fixing $\tilde{c}_{1}=1$ in order to exhaustively explore the structure of extrema by varying the quantity $\,\delta \equiv \left| c_0\right|$. It is found to contain an unstable Minkowski solution \cite{deCarlos:2009qm} at the critical value $\,\delta_{c} \sim 2.69\,$ as well as unstable dS ones if going beyond this critical value (the region with $\,\delta > \delta_{c}\,$ presents an asymptotic behaviour). This is depicted in figure~\ref{fig:plot_dS}.

\begin{figure}[h!]
\centering
\includegraphics[width=7cm]{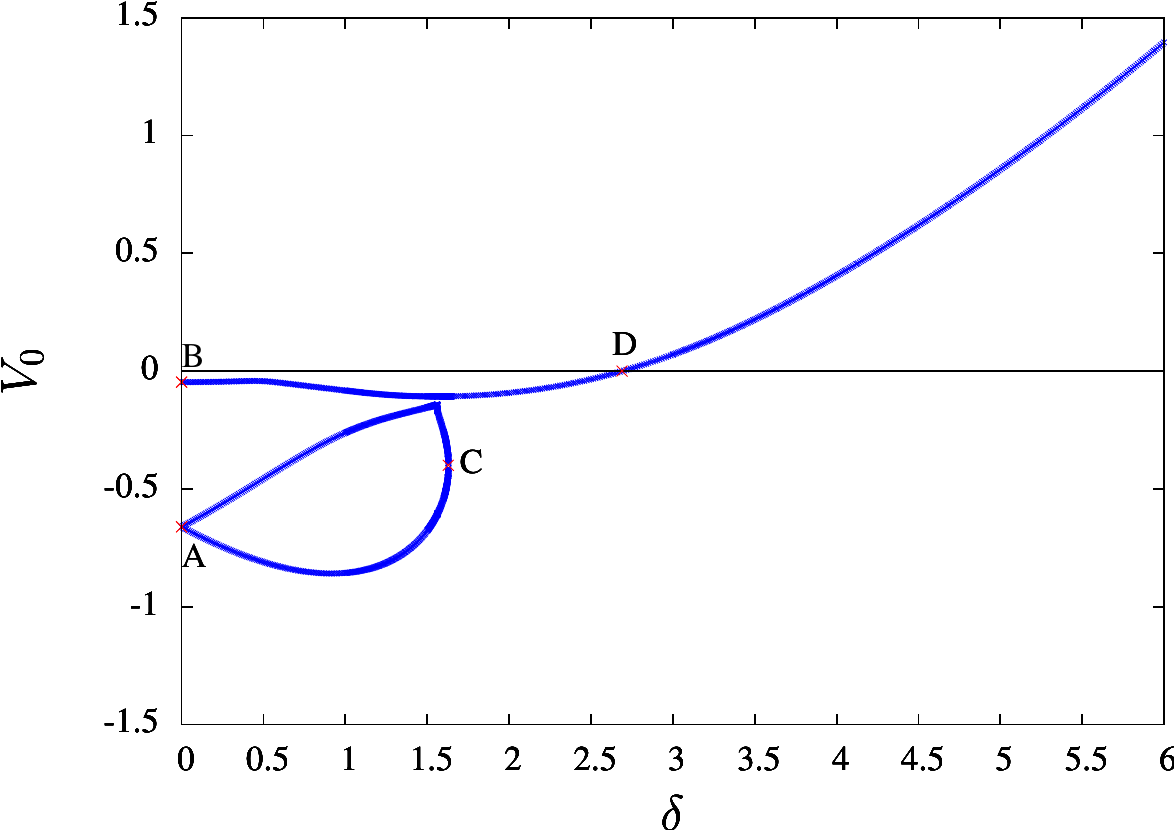}
\includegraphics[width=7cm]{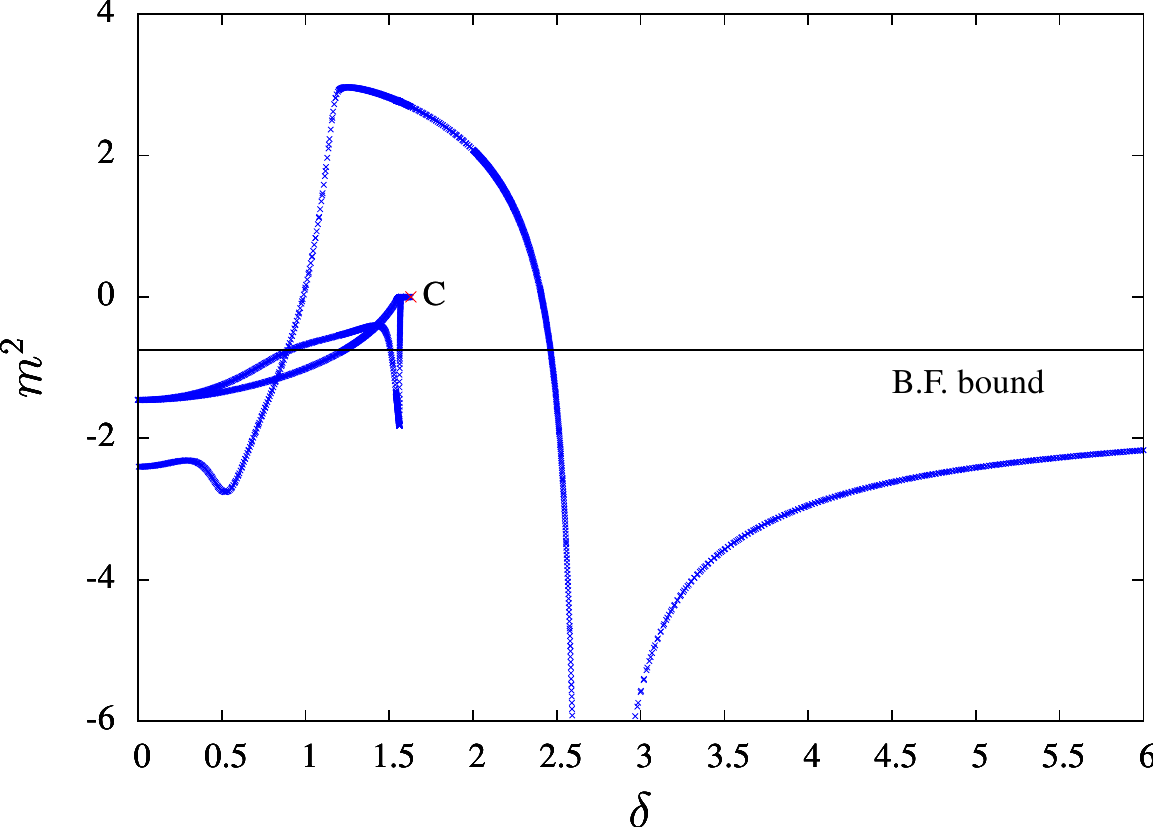}
\caption{{\it Left: Plot of the potential energy at the extrema,
$\,V_{0}\,$, as a function of the scanning parameter $\,\delta$: the
point A corresponds to two degenerate and unstable AdS$_{4}$
solutions; points B and C correspond to singular points; point D
associated to $\delta_{c}\sim 2.69$ is an unstable Minkowski
solution. Right: Plot of the lowest normalised mass in \eqref{mass} as a
function of the scanning parameter $\,\delta$. After reaching the dS
region, the system undergoes an asymptotic behaviour where
$\,m^{2}\rightarrow -\frac{4}{3}\,$ as long as $\,\delta
\rightarrow \infty$.}} \label{fig:plot_dS}
\end{figure}

\begin{figure}[h!]
\centering
\includegraphics[width=7cm]{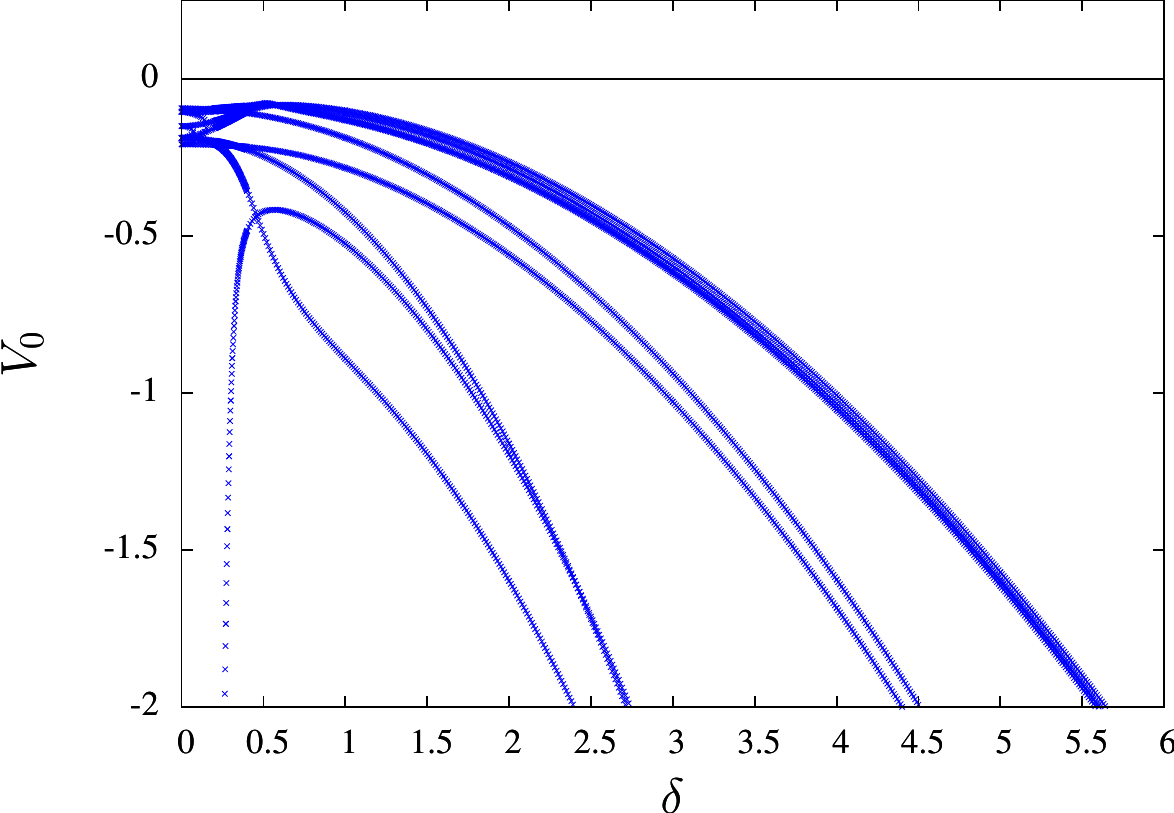}
\includegraphics[width=7cm]{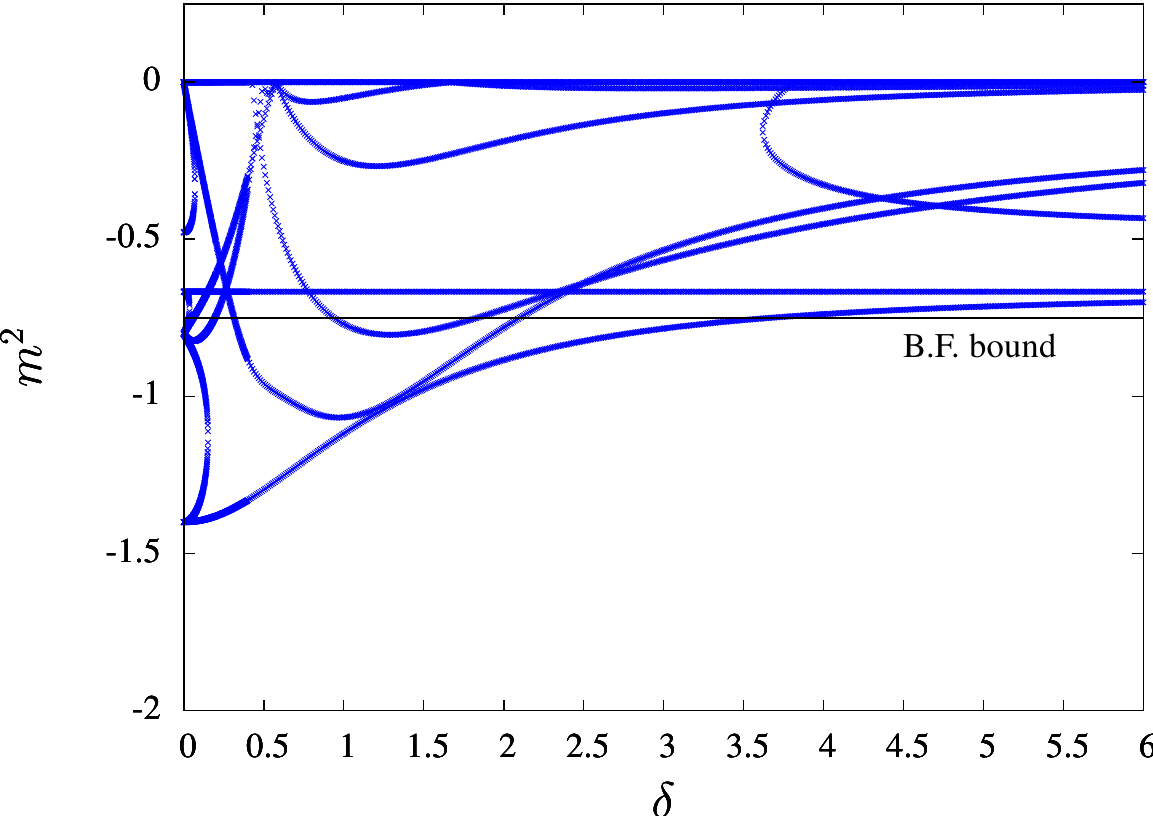}
\caption{{\it Left: Plot of the potential energy at the extrema,
$\,V_{0}\,$, as a function of the scanning parameter $\,\delta$.
Right: Plot of the lowest normalised mass as a function of
the scanning parameter $\,\delta$. As long as
$\,\delta\rightarrow\infty$, the system undergoes a four-fold
asymptotic behaviour with $\,m^{2}\,$ always above the BF
bound.}} \label{fig:plot_AdS}
\end{figure}

\item[$ii)$] The second piece can be also explored in terms of the quantity $\,\delta \equiv \left| c_0\right|$ after fixing again the global scale of $\,V\,$ by the choice $\,\tilde{c}_{1}=1$. It only contains AdS$_{4}$ solutions which are mostly non-supersymmetric\footnote{The $\,\mathcal{N}=4\,$ QC (after relaxing (\ref{QC_relax})) together with the vanishing of the F-terms imply $\,a_{0}=\frac{3}{2}\tilde{c}_{1}$, $\,a_{1}=\frac{3}{2}c_{0}$, $\,a_{2}=-\frac{1}{6}\tilde{c}_{1}$, $\,a_{3}=\frac{5}{2}c_{0}$, $\,b_{0}=-c_{0}$, $\,b_{1}=\frac{1}{3}\tilde{c}_{1}\,$ and $c_{1}=\tilde{c}_{1}$. As a result, for a given value of $\,(c_{0} \,,\, \tilde{c}_{1})\,$, one extremum is always supersymmetric whereas the others (solving $\partial V=0$) are not. At the supersymmetric extremum $\,N_{6}^{\bot}=N_{6}^{||}\,$ holds and $\,m^2=-\frac{2}{3}$. Furthermore, this supersymmetric extremum can be embedded into the $\,\mathcal{N}=4\,$ theory (even $\,\mathcal{N}=8$) when $\,\frac{c_0}{\tilde{c}_{1}}=\frac{1}{\sqrt{15}}\,$ since $\,N_{6}^{\bot}=N_{6}^{||}=0$.} and cannot be embedded into the $\,\mathcal{N}=4\,$ theory because of $\,N_{6}^{\bot}\neq 0$. Nevertheless, some special AdS$_{4}$ solutions with $\,N_{6}^{\bot} = 0\,$ do appear at the special values $\,\delta=0\,$, $\,\delta=1/\sqrt{15}\,$ and $\,\delta=1/\sqrt{3}$, hence being embeddable into the $\,\mathcal{N}=4\,$ theory. This is depicted in figure~\ref{fig:plot_AdS}.

\end{itemize}

\chapter{Different Formulations of $\cN=8$}
\label{appendix:Exceptional}
\renewcommand{\theequation}{C.\arabic{equation}}
\section{Summary of indices}
\label{App:indices}

All through the text in chapter~\ref{Maximal} we extensively make use of indices of different groups. Here we give a list of the notations retained here (conventions based on ref.~\cite{Dibitetto:2012ia})

\be
\label{Indices} 
\begin{array}{cccl}
\mathpzc{A}\,,\,\,\mathpzc{B}\,,\,\,\cdots & & & \textrm{adjoint of
E}_{7(7)}\\
\mathbb{M}\,,\,\,\mathbb{N}\,,\,\,\cdots & & & \textrm{fundamental
of
E}_{7(7)}\textrm{ (global)}\\
\underline{\mathbb{M}}\,,\,\,\underline{\mathbb{N}}\,,\,\,\cdots & &
& \textrm{fundamental of
E}_{7(7)}\textrm{ (local)}\\
\mathcal{I}\,,\,\,\mathcal{J}\,,\,\,\cdots & & & \textrm{fundamental
of SU}(8)\\
\a\,,\,\,\b\,,\,\,\cdots & & & \textrm{fundamental of SL}(2)\\
M\,,\,\,N\,,\,\,\cdots & & & \textrm{fundamental of SO}(6,6)\\
\m\,,\,\,\n\,,\,\,\cdots & & & \textrm{M-W spinor of
SO}(6,6)\textrm{
(L)}\\
\dot{\m}\,,\,\,\dot{\n}\,,\,\,\cdots & & & \textrm{M-W spinor of
SO}(6,6)\textrm{
(R)}\\
A\,,\,\,B\,,\,\,\cdots & & & \textrm{fundamental of SO}(2,2)\\
m\,,\,\,n\,,\,\,\cdots & & & \textrm{fundamental of SO}(6)_{\textrm{time-like}}\\
a\,,\,\,b\,,\,\,\cdots & & & \textrm{fundamental of
SO}(6)_{\textrm{space-like}}\\
i\,,\,\,j\,,\,\,\cdots & & & \textrm{fundamental of SU}(4)_{\textrm{time-like}}\\
\hat{i}\,,\,\,\hat{j}\,,\,\,\cdots & & & \textrm{fundamental of
SU}(4)_{\textrm{space-like}}\,\,\,.
\end{array} \ee

\section{Majorana-Weyl spinors of $\textrm{SO}(6,6)$}
\label{App:spinors}

Our starting point is a Majorana spinor in $6+6$ dimensions carrying $2^{6}=64$ real degrees of freedom. For Majorana spinors there exists a real
representation of the $\Gamma$-matrices $\{\Gamma_M\}_{M=1,\cdots,12}$
such that they satisfy
\bea\label{Clifford}
\left\{\Gamma_M,\Gamma_N\right\}=2\,\eta_{MN}\,\mathds{1}_{64}\,,
\eea
where $\,\eta_{MN}=\eta^{MN}\,$ is the $\,\textrm{SO}(6,6)\,$ invariant metric. We adopt a set of conventions in which Majorana spinors are naturally objects
of the form $\chi^{a}$ and hence $\Gamma$-matrices carry indices $\left[\Gamma_M\right]^{a}_{\phantom{a}b}\,$. In addition to the $\Gamma$-matrices, we
introduce two antisymmetric matrices, $\,\cC_{a b}\,$ and $\,\cC^{a b}\,$, which turn out to represent the components of the transposed charge conjugation
matrix $\,\cC\,$ and its inverse respectively. We will use these objects in order to raise and lower spinorial indices according to
the so-called SouthWest-NorthEast (SW-NE) conventions \cite{VanProeyen:1999ni}. This translates into the following rules
\bea
\chi^{a}=\chi_b \, \cC^{b a} \hspace{8mm},\hspace{8mm}
\chi_a=\cC_{a b} \, \chi^b \ ,
\eea
and the consistency of the two rules implies
\bea
\cC^{a b}\,\cC_{c
b}=\delta^{a}_{\phantom{a}c}\hspace{8mm}\textrm{and}\hspace{8mm}\cC_{b
a}\,\cC^{b c}=\delta_{a}^{\phantom{a}c} \ .
\eea
The charge conjugation matrix mentioned above relates $\Gamma$-matrices to their transpose in the following way
\bea
\label{GammaT}
\left(\Gamma_M\right)^T =-\, \cC\,\Gamma_M\,\cC^{-1} \ ,
\eea
whereas one can also define a conjugation matrix $\,B=-A^{-T} \, \mathcal{C}\,$, such that
\be
\label{B64}
B^{*} B = \mathds{1}_{64}   \hspace{10mm} \textrm{ and } \hspace{10mm} \Gamma_{M}^{*}= -B \, \Gamma_{M} \, B^{-1} \hspace{15mm} \textrm{with} \hspace{10mm} A = \Gamma_{1} ... \Gamma_{6} \ .
\ee

Majorana spinors live in the $\,\textbf{64}\,$ of $\,\textrm{SO}(6,6)\,$ which is not an irrep and can be decomposed in terms of left- and right-handed Majorana-Weyl (M-W) spinors. These are  related to the $\,\textbf{32}\,$ and $\,\textbf{32}^{\prime}\,$ irrep's, respectively. In a basis in which $\,\Gamma_{13}=\Gamma_1\cdots\Gamma_{12}\,$ takes the form $\Gamma_{13}=\textrm{diag}(+\mathds{1}_{32},-\mathds{1}_{32})\,$, one can introduce the so-called 2-component formalism such that
\bea \chi^a=\left(\begin{array}{c} \chi^{\m}\\
\chi_{\dm}\end{array}\right)\,, \eea
where the indices $\,\m\,$ and $\,\dm\,$ respectively denote left- and right-handed M-W spinors. Accordingly to this decomposition, the $\Gamma$-matrices split into $32\times 32$ blocks as follows
\bea
\label{Gamma_splitting}
\left[\Gamma_M\right]^{a}_{\phantom{a}b}=\left(\begin{array}{cc} 0 &
\left[\g_M\right]^{\m\dn}\\
\left[\bar{\g}_M\right]_{\dm\n} & 0
\end{array}\right) \ ,
\eea
and the charge conjugation and conjugation matrices become
\bea
\label{B&C_splitting}
\cC_{ab}=\left(
\begin{array}{cc}
\cC_{\m\n} & 0\\
0 & \cC^{\dm\dn}
\end{array}
\right)
\hspace{10mm} \textrm{and}  \hspace{10mm}
B_{ab}=\left(
\begin{array}{cc}
B_{\m\n} & 0\\
0 & B^{\dm\dn}
\end{array}
\right) \ .
\eea
In terms of these $\,32\times 32\,$ gamma matrices, the relations \eqref{Clifford} and \eqref{GammaT} can be respectively written as
\bea
\left[{\g}_{(M}\right]^{\m\dot{\rho}}\,\left[\bar{\g}_{N)}\right]_{\dot{\rho}\n}=\eta_{MN}\,\delta^{\m}_{\n}
\hspace{10mm}\textrm{and}\hspace{10mm}
\left[\bar{\g}_M\right]_{\dm\n} = \left[{\g}_M\right]_{\n\dm}= \cC_{\n \sigma}\left[{\g}_M\right]^{\sigma\dot{\rho}}\cC_{\dot{\rho} \dm}\ .
\eea

Antisymmetrised products of two gamma matrices can be defined both for left- and right-handed M-W representations as
\bea
[\g_{MN}]^{\m}_{\phantom{a}\n}\equiv[\g_{[M}]^{\m\dot{\rho}}\,[\bar{\g}_{N]}]_{\dot{\rho}\n}
\hspace{15mm} \textrm{and} \hspace{15mm}
[\g_{MN}]_{\dm}^{\phantom{a}\dn}\equiv[\bar{\g}_{[M}]_{\dm\rho}\,[\g_{N]}]^{\rho\dn} \ ,
\eea
and further extended to antisymmetrised products of an even number of gamma matrices. However, only those up to degree six are linearly independent since higher-degree ones (from $7$ to $12$) are related to them by Hodge duality\footnote{The limit case of the antisymmetrised product of six gamma matrices turns out to be anti-selfdual (ASD) when involving undotted indices and self-dual (SD) when involving dotted indices.}. After defining all the products of gamma matrices, one can make use of $\,\cC_{\m\n}$\,, $\,\cC^{\dm\dn}\,$ and their inverse transpose in order to rise and lower indices. As a result, antisymmetrised products of \emph{two} and \emph{six} gamma matrices are symmetric, whereas the ones with \emph{four} are antisymmetric.

\subsection*{$\textrm{SU}(4) \times \textrm{SU}(4)$ covariant formulation of M-W spinors}

Decomposing the vector and the left- and right-handed M-W spinor irrep's of $\,\textrm{SO}(6,6)\,$ under its maximal compact subgroup
\be
\label{A3A3_in_D6}
\textrm{SU}(4)_{\textrm{time-like}} \times \textrm{SU}(4)_{\textrm{space-like}} \sim \textrm{SO}(6)_{\textrm{time-like}} \times \textrm{SO}(6)_{\textrm{space-like}} \subset \textrm{SO}(6,6) \ ,
\ee
yields the following branching relations
\be
\begin{array}{ccccc}
\textrm{index} \,\,\,\,\,\,& \textrm{SO}(6,6) & \supset & \textrm{SU}(4) \times \textrm{SU}(4) & \\[2mm]
M \,\,\,\,\,\, & \textbf{12} & \rightarrow &  (\textbf{6},\textbf{1}) \oplus (\textbf{1},\textbf{6})  & \\[2mm]
\m \,\,\,\,\,\, & \textbf{32} & \rightarrow &  (\textbf{4},\textbf{4}) \oplus (\bar{\textbf{4}},\bar{\textbf{4}})  & \\[2mm]
\dot{\m} \,\,\,\,\,\, & \textbf{32'} & \rightarrow &  (\textbf{4},\bar{\textbf{4}}) \oplus (\bar{\textbf{4}},\textbf{4})  &  .
\end{array}
\ee
At the level of indices, this translates into the splittings $\,M=m \oplus a\,$, with $\,m,a=1,...,6\,$, together with $\,\mu = \left \lbrace _{i \hj} \,\oplus\, ^{i \hj} \right \rbrace\,$ and $\,\dot{\nu} = \left \lbrace {_{i}}^{\hj} \oplus {^{i}}_{\hj} \right \rbrace\,$ with $\,i,\hi=1,...,4\,$. The fundamental $\,\textrm{SO}(6)\,$ indices $\,m\,$ and $\,a\,$ respectively correspond to the time-like and space-like parts of the block-diagonal $\,\eta_{MN}=\eta^{MN}\,$ metric of $\,\textrm{SO}(6,6)\,$ in Lorentzian coordinates
\be
\label{eta_Lorentzian}
\eta_{MN} =
\left(
\begin{array}{c|c}
-\delta_{mn} & 0 \\[1mm]
\hline
\\[-4mm]
0 & \delta_{ab}
\end{array}
\right) \ .
\ee
When written in terms of $\,\textrm{SU}(4)\sim\textrm{SO}(6)\,$ invariant tensors, the antisymmetric charge conjugation matrices in (\ref{B&C_splitting}) take the block off-diagonal form
\be
\mathcal{C}_{\mu \nu} =
\left(
\begin{array}{c|c}
0 & {{\mathcal{C}}_{i \hj}}^{k \hl} = -i \, \delta_{i}^{k} \delta_{\hj}^{\hl}  \\[1mm]
\hline
\\[-4mm]
{\mathcal{C}^{i \hj}}_{k \hl} =  i \,  \delta^{i}_{k} \delta^{\hj}_{\hl} & 0
\end{array}
\right)
\hspace{1mm},\hspace{1mm}
{\mathcal{C}}^{\dot{\mu} \dot{\nu}} =
\left(
\begin{array}{c|c}
0 & \mathcal{C}^{i \phantom{\hj} \phantom{k} \hl}_{\phantom{i} \hj k \phantom{\hl}} = -i \,  \delta^{i}_{k} \delta^{\hl}_{\hj}  \\[1mm]
\hline
\\[-4mm]
\mathcal{C}^{\phantom{i} \hj k \phantom{\hl}}_{i \phantom{\hj} \phantom{k} \hl}  =  i \, \delta_{i}^{k} \delta_{\hl}^{\hj} & 0
\end{array}
\right) \ ,
\ee
whereas the gamma matrices in (\ref{Gamma_splitting}) split into a set of time-like matrices with a block-diagonal structure
\be
\label{gamma_1}
[\gamma_{m}]^{\mu \dot{\nu}} =
\left(
\begin{array}{c|c}
[\gamma_{m}]^{i \hj k \phantom{\hl}}_{\phantom{i} \phantom{\hj} \phantom{k} \hl} = [G_{m}]^{ik}  \delta_{\hl}^{\hj} & 0 \\[1mm]
\hline
\\[-4mm]
0 & [\gamma_{m}]^{\phantom{i} \phantom{\hj} \phantom{k} \hl}_{i \hj k \phantom{\hl}}  = [G_{m}]_{ik}  \delta_{\hj}^{\hl}
\end{array}
\right) \ ,
\ee
\be
\label{gamma_2}
[\bar{\gamma}_{m}]_{\dot{\mu} \nu} =
\left(
\begin{array}{c|c}
[\bar{\gamma}_{m}]^{\phantom{i}  \hj \phantom{k} \phantom{\hl}}_{i \phantom{\hj} k \hl} = [G_{m}]_{ik}  \delta_{\hl}^{\hj} & 0 \\[1mm]
\hline
\\[-4mm]
0 & [\bar{\gamma}_{m}]^{i \phantom{\hj} k \hl}_{\phantom{i} \hj \phantom{k} \phantom{\hl}} = [G_{m}]^{ik}  \delta_{\hj}^{\hl}
\end{array}
\right) \ ,
\ee
and a set of space-like ones
\be
\label{gamma_3}
[\gamma_{a}]^{\mu \dot{\nu}} =
\left(
\begin{array}{c|c}
0 & [\gamma_{a}]^{i \hj \phantom{k} \hl}_{\phantom{i} \phantom{\hj} k \phantom{\hl}}  =  [G_{a}]^{\hj \hl}  \delta_{k}^{i}  \\[1mm]
\hline
\\[-4mm]
[\gamma_{a}]^{\phantom{i} \phantom{\hj} k \phantom{\hl}}_{i \hj \phantom{k} \hl} =  [G_{a}]_{\hj \hl}  \delta_{i}^{k}  & 0
\end{array}
\right) \ ,
\ee
\be
\label{gamma_4}
[\bar{\gamma}_{a}]_{\dot{\mu} \nu}  =
\left(
\begin{array}{c|c}
0 & [\bar{\gamma}_{a}]^{\phantom{i} \hj k \hl}_{i \phantom{\hj} \phantom{k} \phantom{\hl}} = - [G_{a}]^{\hj \hl}  \delta^{k}_{i} \\[1mm]
\hline
\\[-4mm]
[\bar{\gamma}_{a}]^{i \phantom{\hj} \phantom{k} \phantom{\hl}}_{\phantom{i} \hj k \hl}  =  - [G_{a}]_{\hj \hl}  \delta^{i}_{k}  & 0
\end{array}
\right) \ ,
\ee
with a block off-diagonal structure. The invariant tensors $\,G_{m}=[G_{m}]^{ij}\,$ and $\,G_{a}=[G_{a}]^{\hi \hj}\,$ are defined with upper indices and correspond to the gamma matrices for each of the $\,\textrm{SO}(6) \sim \textrm{SU}(4)\,$ factors in (\ref{A3A3_in_D6}). Often they are also called 't Hooft symbols and we take them to satisfy the (anti-)self-duality conditions\footnote{The non-vanishing parts of the scalar vielbeins $\,{\mathcal{V}_{M}}^{ij}\,$ and $\,{\mathcal{V}_{M}}^{\hi \hj}\,$ in (\ref{fermi_mass_N=4_new}) and (\ref{fermi_mass_N=4_extension}) reduce to $\,\frac{1}{2}[G_{m}]^{ij}\,$ and $\,\frac{1}{2}[G_{a}]^{\hi \hj}\,$ when evaluated at the origin of the moduli space, so them both must square to the identity in order to satisfy $\,\mathcal{V} \, \mathcal{V}^{T}=\mathds{1}\,$ at the origin.}
\be
[G_{m}]_{ij} = - \frac{1}{2} \,\, \epsilon_{ijkl} \,\, [G_{m}]^{kl}
\hspace{10mm} \textrm{and} \hspace{10mm}
[G_{a}]_{\hi \hj} = \frac{1}{2} \,\, \epsilon_{\hi \hj \hk \hl} \,\, [G_{a}]^{\hk \hl} \ ,
\ee
where $\,[G_{m}]_{ij} = ([G_{m}]^{ij})^{*}\,$ and $\,[G_{a}]_{\hi \hj} = ([G_{a}]^{\hi \hj})^{*}\,$. All along the present work, we have used the following explicit realisation of anti-self-dual $\,G_{m}\,$ 't Hooft symbols
\begin{equation}
\begin{array}{c}
[G_{1}]=
{\scriptsize{\left[
\begin{array}{cccc}
 0 & 1 & 0 & 0 \\
 -1 & 0 & 0 & 0 \\
 0 & 0 & 0 & -1 \\
 0 & 0 & 1 & 0
\end{array}
\right]}}
\hspace{2mm} , \hspace{2mm}
[G_{3}]=
{\scriptsize{\left[
\begin{array}{cccc}
 0 & 0 & 1 & 0 \\
 0 & 0 & 0 & 1 \\
 -1 & 0 & 0 & 0 \\
 0 & -1 & 0 & 0
\end{array}
\right]}}
\hspace{2mm} , \hspace{2mm}
[G_{5}]=
{\scriptsize{\left[
\begin{array}{cccc}
 0 & 0 & 0 & 1 \\
 0 & 0 & -1 & 0 \\
 0 & 1 & 0 & 0 \\
 -1 & 0 & 0 & 0
\end{array}
\right]}} \ ,

\\[8mm]

[G_{2}]=
{\scriptsize{\left[
\begin{array}{cccc}
 0 & i & 0 & 0 \\
 -i & 0 & 0 & 0 \\
 0 & 0 & 0 & i \\
 0 & 0 & -i & 0
\end{array}
\right]}}
\hspace{2mm} , \hspace{2mm}
[G_{4}]=
{\scriptsize{\left[
\begin{array}{cccc}
 0 & 0 & i & 0 \\
 0 & 0 & 0 & -i \\
 -i & 0 & 0 & 0 \\
 0 & i & 0 & 0
\end{array}
\right]}}
\hspace{2mm} , \hspace{2mm}
[G_{6}]=
{\scriptsize{\left[
\begin{array}{cccc}
 0 & 0 & 0 & i \\
 0 & 0 & i & 0 \\
 0 & -i & 0 & 0 \\
 -i & 0 & 0 & 0
\end{array}
\right]}} \ ,
\end{array}
\label{'tHooft_timelike}
\end{equation}
together with the self-dual $\,G_{a}\,$ ones
\begin{equation}
\begin{array}{c}
[G_{1}]=
{\scriptsize{\left[
\begin{array}{cccc}
 0 & 1 & 0 & 0 \\
 -1 & 0 & 0 & 0 \\
 0 & 0 & 0 & 1 \\
 0 & 0 & -1 & 0
\end{array}
\right]}}
\hspace{2mm} , \hspace{2mm}
[G_{3}]=
{\scriptsize{\left[
\begin{array}{cccc}
 0 & 0 & 1 & 0 \\
 0 & 0 & 0 & -1 \\
 -1 & 0 & 0 & 0 \\
 0 & 1 & 0 & 0
\end{array}
\right]}}
\hspace{2mm} , \hspace{2mm}
[G_{5}]=
{\scriptsize{\left[
\begin{array}{cccc}
 0 & 0 & 0 & 1 \\
 0 & 0 & 1 & 0 \\
 0 & -1 & 0 & 0 \\
 -1 & 0 & 0 & 0
\end{array}
\right]}} \ ,

\\[8mm]

[G_{2}]=
{\scriptsize{\left[
\begin{array}{cccc}
 0 & i & 0 & 0 \\
 -i & 0 & 0 & 0 \\
 0 & 0 & 0 & -i \\
 0 & 0 & i & 0
\end{array}
\right]}}
\hspace{2mm} , \hspace{2mm}
[G_{4}]=
{\scriptsize{\left[
\begin{array}{cccc}
 0 & 0 & i & 0 \\
 0 & 0 & 0 & i \\
 -i & 0 & 0 & 0 \\
 0 & -i & 0 & 0
\end{array}
\right]}}
\hspace{2mm} , \hspace{2mm}
[G_{6}]=
{\scriptsize{\left[
\begin{array}{cccc}
 0 & 0 & 0 & i \\
 0 & 0 & -i & 0 \\
 0 & i & 0 & 0 \\
 -i & 0 & 0 & 0
\end{array}
\right]}} \ .
\end{array}
\label{'tHooft_spacelike}
\end{equation}
Notice that they are complex matrices and then will lead to a complex representation of gamma matrices in (\ref{gamma_1})-(\ref{gamma_4}). This is related to the fact that $\,\textrm{SO}(6)\,$ does not admit \mbox{M-W} spinors (it is $\,\textrm{SO}(3,3)\,$ which does), so a real representation of $\,\textrm{SO}(6,6)\,$ gamma matrices is no longer possible when moving to an $\,\textrm{SU}(4) \times \textrm{SU}(4)\,$ covariant formulation.

When arranged into $\,64 \times 64\,$ matrices according to the splittings (\ref{Gamma_splitting}) and (\ref{B&C_splitting}), one verifies that the defining relations (\ref{Clifford}) and (\ref{GammaT}) hold and also that $\,{\Gamma_{13}=\textrm{diag}(+\mathds{1}_{32},-\mathds{1}_{32})}\,$. Finally, the conjugation matrices (\ref{B&C_splitting}) entering the definition of the scalar matrix in (\ref{M_origin}) take the form
\be
\label{B_A3A3}
B_{\mu \nu} =
\left(
\begin{array}{c|c}
0 & \mathds{1}_{16}  \\[1mm]
\hline
\\[-4mm]
\mathds{1}_{16} & 0
\end{array}
\right)
\hspace{5mm} \textrm{ , } \hspace{5mm}
B^{\dot{\mu} \dot{\nu}} =
\left(
\begin{array}{c|c}
0 & - \mathds{1}_{16}  \\[1mm]
\hline
\\[-4mm]
- \mathds{1}_{16} & 0
\end{array}
\right) \ ,
\ee
producing a non-standard definition of the origin of the moduli space, as discussed in detail in the main text.

\subsection*{Real formulation of M-W spinors}

In addition to the $\,\textrm{SU}(4) \times \textrm{SU}(4)\,$ covariant formulation of M-W spinors described above, we can adopt another realisation such that: $i)$ it is a real realisation of M-W spinors $ii)$ it is compatible with the standard choice of (\ref{M_origin_choice}) as the origin of the moduli space.

We build our real $\,64 \times 64\,$ $\Gamma$-matrices in a Majorana representation out of the $\,2 \times 2\,$ Pauli matrices $\,\sigma_{1,2,3}\,$ in the following way
\be
\label{Gamma_Real}
\begin{array}{lcc}
\Gamma_{1} & = & i \, \s_{2} \otimes \mathds{1}_{2} \otimes \mathds{1}_{2}\otimes \mathds{1}_{2}\otimes \mathds{1}_{2} \otimes \mathds{1}_{2} \\
\Gamma_{2} & = & i \, \s_{3} \otimes \s_{2}         \otimes \mathds{1}_{2}\otimes \mathds{1}_{2}\otimes \mathds{1}_{2} \otimes \mathds{1}_{2} \\
\Gamma_{3} & = & i \, \s_{3} \otimes \s_{3}         \otimes \s_{2}        \otimes \mathds{1}_{2}\otimes \mathds{1}_{2} \otimes \mathds{1}_{2} \\
\Gamma_{4} & = & i \, \s_{3} \otimes \s_{3}         \otimes \s_{3}        \otimes\s_{2}         \otimes \mathds{1}_{2} \otimes \mathds{1}_{2} \\
\Gamma_{5} & = & i \, \s_{3} \otimes \s_{3}         \otimes \s_{3}        \otimes \s_{3}        \otimes \s_{2}         \otimes \mathds{1}_{2} \\
\Gamma_{6} & = & i \, \s_{3} \otimes \s_{3}         \otimes \s_{3}        \otimes \s_{3}        \otimes \s_{3}         \otimes \s_{2}
\end{array}
\hspace{3mm} , \hspace{3mm}
\begin{array}{lcc}
\Gamma_{7} & = & \s_{1} \otimes \mathds{1}_{2} \otimes \mathds{1}_{2}\otimes \mathds{1}_{2}\otimes \mathds{1}_{2} \otimes \mathds{1}_{2} \\
\Gamma_{8} & = & \s_{3} \otimes \s_{1}         \otimes \mathds{1}_{2}\otimes \mathds{1}_{2}\otimes \mathds{1}_{2} \otimes \mathds{1}_{2} \\
\Gamma_{9} & = & \s_{3} \otimes \s_{3}         \otimes \s_{1}        \otimes \mathds{1}_{2}\otimes \mathds{1}_{2} \otimes \mathds{1}_{2} \\
\Gamma_{10} & = & \s_{3} \otimes \s_{3}         \otimes \s_{3}        \otimes\s_{1}         \otimes \mathds{1}_{2} \otimes \mathds{1}_{2} \\
\Gamma_{11} & = & \s_{3} \otimes \s_{3}         \otimes \s_{3}        \otimes \s_{3}        \otimes \s_{1}         \otimes \mathds{1}_{2} \\
\Gamma_{12} & = & \s_{3} \otimes \s_{3}         \otimes \s_{3}        \otimes \s_{3}        \otimes \s_{3}         \otimes \s_{1}
\end{array}
\ee
where we decide to use a set of Pauli matrices satisfying $\,[\s_{i} , \s_{j}] = 2 \,  i \,\epsilon_{ijk} \, \s_{k}\,$. This corresponds to the choice
\be
\s_{1} =
\left(
\begin{array}{cc}
0 & 1  \\
1 & 0
\end{array}
\right)
\hspace{5mm} \textrm{ , } \hspace{5mm}
\s_{2} =
\left(
\begin{array}{cc}
0 & -i  \\
i & 0
\end{array}
\right)
\hspace{5mm} \textrm{ , } \hspace{5mm}
\s_{3} =
\left(
\begin{array}{cc}
1 & 0  \\
0 & -1
\end{array}
\right) \ .
\ee
Building the $\,64 \times 64\,$ charge conjugation matrix as
\be
\label{C_Real}
\begin{array}{lcc}
\mathcal{C} & = & -i\, \s_{2} \otimes \s_{1} \otimes \s_{2} \otimes \s_{1} \otimes \s_{2} \otimes \s_{1} \ ,
\end{array}
\ee
one can easily check that (\ref{Gamma_Real}) and (\ref{C_Real}) automatically satisfy the conditions in (\ref{Clifford}) and (\ref{GammaT}) with the $\,\eta_{MN}\,$ metric given in (\ref{eta_Lorentzian}). By applying an $\,\textrm{SO}(64)\,$ rotation taking $\,{\Gamma_{13}=\textrm{diag}(+\mathds{1}_{32},-\mathds{1}_{32})}\,$, we go to a real M-W basis according to the splittings (\ref{Gamma_splitting}) and (\ref{B&C_splitting}). In this basis, the conjugation matrix in (\ref{M_origin}) happens to be $\,B_{\m \n}=-B^{\dm \dn}=\mathds{1}_{32}\,$, hence being compatible with the standard choice for the origin of the moduli space in (\ref{M_origin_choice}). We will use this real representation of M-W spinors when it comes to identify gaugings associated to critical points at the origin of the moduli space.

\section{$X_{\mathbb{M} \mathbb{N} \mathbb{P}}\,$ in the $\,\textrm{SL}(2) \times \textrm{SO}(6,6)\,$ formulation}
\label{App:X-tensor_components}

In this appendix we derive the explicit form of the components of the $\,X_{\mathbb{M} \mathbb{N} \mathbb{P}}\,$ tensor given in (\ref{Xbosonic}) and (\ref{Xfermionic}). Let us first start by giving the explicit form of the $\,\textrm{E}_{7(7)}\,$ symmetric generators $\,[t_{\mathpzc{A}}]_{\mathbb{MN}}\,$ in the fundamental representation following the conventions in ref.~\cite{Dibitetto:2011eu}. By virtue of the index splittings $\,{\mathbb{M}=\a M \oplus \mu}\,$ and $\,\mathpzc{A}=\a M \b N \oplus \g \dot{\mu}\,$ associated to the branching of the $\textbf{56}$ and $\textbf{133}$ irrep's of $\,\textrm{E}_{7(7)}\,$ under $\,\textrm{SL}(2) \times \textrm{SO}(6,6)$, they are given by
\be
\label{E7_gen}
\begin{array}{cclc}
\left[t_{\a M \b N}\right]_{\g P \d Q} & = & \epsilon_{\a \b} \, \epsilon_{\g \d} \, \left[t_{MN}\right]_{PQ} + \eta_{MN}\, \, \eta_{PQ} \, \left[t_{\a \b}\right]_{\g \d} & , \\[3mm]
\left[t_{\a M \b N}\right]_{\m \n} & = &\dfrac{1}{4} \, \epsilon_{\a \b} \, \left[\g_{MN}\right]_{\m \n} & , \\[3mm]
\left[t_{\a \dm}\right]_{\b N \n} = \left[t_{\a \dm}\right]_{\n \b N} & = & \epsilon_{\a \b} \, \left[\bg_{N}\right]_{\dm \n} = \epsilon_{\a \b} \, \left[\g_{N}\right]_{\n \dm}& ,
\end{array}
\ee
where $\,[t_{\a \b}]^{\g \d} = \delta_{\a}^{(\g}\delta_{\b}^{\d)}\,$ and $\,[t_{M N}]^{P Q} = \delta_{MN}^{PQ}\,$ are the generators of $\,\textrm{SL}(2)\,$ and $\,\textrm{SO}(6,6)\,$, respectively.

On the other hand, we need the $\Theta$-components of the embedding tensor $\,{\Theta_{\mathbb{M}}}^{\mathpzc{A}}\,$ in order to compute $\,X_{\mathbb{MNP}}\,$. These can be split into those components involving an even number of fermionic indices
\be
\label{emb_tens_comp_old}
\begin{array}{cclc}
{\Theta_{\a M}}^{\b N \g P} & = & - \dfrac{1}{2} \, \epsilon^{\b \g} \, {f_{\a M}}^{NP} \, -  \, \dfrac{1}{2} \, \epsilon^{\b \g}  \, \delta_{M}^{[N} \, \xi_{\a}^{\,\,\,\,P]} \, + \,  \dfrac{1}{12} \,  \delta_{\a}^{(\b} \, \xi^{\g)}_{\,\,\,\,M} \, \eta^{N P} & , \\[4mm]
{\Theta_{\m}}^{\a \dn} & = & \dfrac{1}{24} \, \epsilon^{\a \b} \, f_{\b M N P} \, \left[  \g^{MNP} \right]_{\mu}^{\,\,\,\,\,\,\dn} \, - \, \dfrac{1}{8} \, \epsilon^{\a \b} \, \xi_{\b M} \left[ \gamma^{M}\right]_{\mu}^{\,\,\,\,\,\,\dn} & ,
\end{array}
\ee
which were already derived in ref.~\cite{Dibitetto:2011eu}, together with a set of additional ones involving an odd number of fermionic indices. The most general ansatz for the latter according to the symmetry is given by
\be
\label{emb_tens_comp_new}
\begin{array}{cclc}
{\Theta_{\a M}}^{\b \dot{\m}} & = & h_{1} \, \d_{\a}^{\b} \, {F_{M}}^{\dot{\m}} \, +  \, h_{2} \, \epsilon^{\b \g}  \, \Xi_{\a\g\n} \, [\g_{M}]^{\n \dot{\m}} & , \\[4mm]
{\Theta_{\m}}^{\a M \b N} & = & h_{3} \, \epsilon^{\a \b} \, {F^{[M}}_{\dot{\n}} \, \left[  \g^{N]} \right]_{\mu}^{\,\,\,\,\,\dn} \, + \, h_{4} \,\, {\Xi^{\a \b}}_{\m} \,\, \eta^{MN} & ,
\end{array}
\ee
where $\,h_{i=1,2,3,4}\,$ are constant coefficients to be fixed as follows: $i)$ Following the definition in (\ref{gauge_algebra}) and using the form of the $\,\Omega_{\mathbb{MN}}\,$ matrix in (\ref{Omega_matrix}), we can build the $\,X_{\mathbb{MNP}}\,$ tensor as $\,X_{\mathbb{M} \mathbb{N} \mathbb{P}} = {\Theta_{\mathbb{M}}}^{\mathpzc{A}} \, {[t_{\mathpzc{A}}]_{\mathbb{N}}}^{\mathbb{R}} \, \Omega_{\mathbb{RP}}\,$.  $ii)$ By requiring the $\,X_{\mathbb{MNP}}\,$ tensor to live in the $\,\textbf{912}\,$ irrep of $\,\textrm{E}_{7(7)}\,$ we still have to impose the LC in (\ref{linear_const}). This imposes the relations $\,h_{3}=h_{1}\,$ and $\,h_{4}=-\frac{1}{6} \, h_{2}\,$ on the coefficients. $iii)$ Finally we set the remaining free parameters to the values $\,h_{1}=-1\,$ and $\,h_{2}=1\,$, what fixes the relative normalisation between $\,F_{M \dot{\m}}\,$ and $\,\Xi_{\a \b \m}\,$. The final expression is then given by
\be
\label{emb_tens_comp_new}
\begin{array}{cclc}
{\Theta_{\a M}}^{\b \dot{\m}} & = & - \d_{\a}^{\b} \, {F_{M}}^{\dot{\m}} \, +  \,  \epsilon^{\b \g}  \, \Xi_{\a\g\n} \, [\g_{M}]^{\n \dot{\m}} & , \\[4mm]
{\Theta_{\m}}^{\a M \b N} & = & - \epsilon^{\a \b} \, {F^{[M}}_{\dot{\n}} \, \left[  \g^{N]} \right]_{\mu}^{\,\,\,\,\,\dn} \, -\, \dfrac{1}{6} \,\, {\Xi^{\a \b}}_{\m} \,\, \eta^{MN} & ,
\end{array}
\ee
and after some algebra, the set of independent components of the $\,X_{\mathbb{MNP}}\,$ tensor are those given in (\ref{Xbosonic}) and (\ref{Xfermionic}).

\section{The vielbein in the origin of the moduli space}
\label{App:vielbein}

As stated in section~\ref{sec:connecting_A7&A1D6}, the vielbein $\,\mathcal{V}_{\mathbb{M}}^{\phantom{\mathbb{M}}\underline{\mathbb{N}}}\,$ is the fundamental object connecting the $\,\,\textrm{SU}(8)\,$ and $\,\textrm{SL}(2) \times \textrm{SO}(6,6)\,$ formulations of maximal supergravity. Recalling the relation (\ref{X(T)})
\be
\label{X(T)_2}
X_{\mathbb{MNP}} = 2 \, \mathcal{V}_{\mathbb{M}}^{\phantom{\mathbb{M}}\underline{\mathbb{Q}}} \, \mathcal{V}_{\mathbb{N}}^{\phantom{\mathbb{N}}\underline{\mathbb{R}}} \, \mathcal{V}_{\mathbb{P}}^{\phantom{\mathbb{P}}\underline{\mathbb{S}}}\,\,\, T_{\underline{\mathbb{QRS}}} \ ,
\ee
one concludes that the vielbein $\,\mathcal{V}_{\mathbb{M}}^{\phantom{\mathbb{M}}\underline{\mathbb{Q}}}\,$ has the row index $\,\mathbb{M}\,$ in the $\,\textrm{SL}(2)\times \textrm{SO}(6,6)\,$ basis and the column index $\,\underline{\mathbb{Q}}\,$ in the $\,\textrm{SU}(8)\,$ one. In order to determine the form of $\,\mathcal{V}_{\mathbb{M}}^{\phantom{\mathbb{M}}\underline{\mathbb{Q}}}\,$ we must go to a common basis where to simultaneously describe $\,\textrm{SL}(2)\times \textrm{SO}(6,6)\,$ and $\,\textrm{SU}(8)\,$ indices. This common basis turns out to be
\be
\begin{array}{ccc}
\textrm{SO}(2) \times \textrm{SO}(6)_{\textrm{time}} \times \textrm{SO}(6)_{\textrm{space}} &\sim& \textrm{U}(1) \times \textrm{SU}(4)_{\textrm{time}} \times \textrm{SU}(4)_{\textrm{space}} \ ,
\end{array}
\ee
since it is the only maximal subgroup being shared by them both. Furthermore, it coincides with their maximal compact subgroup. All our conventions related to the $\textrm{SU}(4) \times \textrm{SU}(4)\,$ covariant formulation of $\,\textrm{SO}(6,6)\,$ spinors, gamma matrices, etc. are summarised in appendix~\ref{App:spinors}.

In what follows we will make an extensive use of two different decompositions of an $\,\textrm{E}_{7(7)}\,$ fundamental index:
\begin{itemize}

\item[$i)$] the decomposition with respect to $\,\textrm{SL}(2)\times \textrm{SO}(6,6)\,$
\be
\label{A1D6_decomp}
\begin{array}{ccccc}
\textrm{E}_{7(7)} & \supset &  \textrm{SL}(2)\times \textrm{SO}(6,6) & \supset  & \textrm{SL}(2) \,\times\, \textrm{SO}(6)_{\textrm{time}} \,\times\, \textrm{SO}(6)_{\textrm{space}} \\[2mm]
\textbf{56} & \rightarrow &  (\textbf{2},\textbf{12}) \oplus (\textbf{1},\textbf{32}) &  \rightarrow  & (\textbf{2},\textbf{6},\textbf{1}) \oplus (\textbf{2},\textbf{1},\textbf{6}) \oplus (\textbf{1},\textbf{4},\textbf{4}) \oplus (\textbf{1},\bar{\textbf{4}},\bar{\textbf{4}})\\[2mm]
\mathbb{M} & = &  \a M \oplus \mu & =  & \a \, m \,\,\, \oplus \,\,\, \a \,a \,\,\, \oplus \,\,\, i \, \hat{j} \,\,\, \oplus \,\,\, \overline{i} \, \overline{\hat{j}}\\[2mm]
\end{array}
\ee

\item[$ii)$] the decomposition with respect to $\,\textrm{SU}(8)\,$
\be
\label{A7_decomp}
\begin{array}{ccccc}
\textrm{E}_{7(7)} & \supset &  \textrm{SU}(8) & \supset  & \textrm{U}(1) \,\times\, \textrm{SU}(4)_{\textrm{time}} \,\times\, \textrm{SU}(4)_{\textrm{space}} \\[2mm]
\textbf{56} & \rightarrow &  \textbf{28} \oplus \overline{\textbf{28}} &  \rightarrow  & (\textbf{6},\textbf{1})_{(-2)} \oplus (\textbf{1},\textbf{6})_{(2)} \oplus (\textbf{4},\textbf{4})_{(0)} \,\,\, \oplus \,\,\, \textrm{c.c.}  \\[2mm]
\underline{\mathbb{M}} & = &  [\mathcal{I} \, \mathcal{J}] \, \oplus \, \textrm{c.c.} & =  & [i\, j] \,\,\, \oplus \,\,\, [\hat{i} \,\hat{j}] \,\,\, \oplus \,\,\,  i \, \hat{j} \,\,\, \oplus \,\,\, \textrm{c.c.}
\end{array}
\ee
\end{itemize}
where $\,_{(q)}\,$ in (\ref{A7_decomp}) denotes the $\,\textrm{U}(1)\,$ charge of the $\,\textrm{SU}(4) \,\times\, \textrm{SU}(4)\,$ irrep's. By comparing the decompositions in (\ref{A1D6_decomp}) and (\ref{A7_decomp}), the non-vanishing components of the vielbein\footnote{As discussed in section~\ref{sec:connecting_A7&A1D6}, we are setting all the ``fermionic'' scalars to the origin of the moduli space.} will correspond to
\be
\label{vielbein_components}
\begin{array}{ccl}
\mathcal{V}_{\mathbb{M}}^{\phantom{\mathbb{M}}\underline{\mathbb{Q}}} & = & \left\lbrace \, {\mathcal{V}_{\a M}}^{[\mathcal{I}\mathcal{J}]} \,\, , \,\, \mathcal{V}_{\a M [\mathcal{I} \mathcal{J}]}  \,\, , \,\, {\mathcal{V}_{\mu}}^{[\mathcal{I} \mathcal{J}]} \,\, , \,\, \mathcal{V}_{\mu [\mathcal{I}\mathcal{J}]} \, \right\rbrace \\[2mm]
&  =  & \left\lbrace \, {\mathcal{V}_{\a m}}^{[ij]} \,\, , \,\, \mathcal{V}_{\a m [ij]} \,\, , \,\,  {\mathcal{V}_{\a a}}^{[\hat{i} \hat{j}]}  \,\, , \,\,  \mathcal{V}_{\a a [\hat{i} \hat{j}]} \,\, , \,\, {\mathcal{V}_{i \hat{j}}}^{k \hat{l}} \,\, , \,\, {\mathcal{V}^{i \hat{j}}}_{k \hat{l}} \, \right\rbrace \ ,
\end{array}
\ee
where $\,\a=+,-\,$ is an $\,\textrm{SL}(2)\,$ index raised and lowered by $\,\epsilon_{\a \b}\,$ and where $\,[ij]\,$, $\,[\hat{i} \hat{j}]\,$ and $\,i\hat{j}\,$ are pairs of fundamental $\,\textrm{SU}(4)\,$ indices with $\,i,\hat{i}=1,...,4\,$. The fundamental $\,\textrm{SO}(6)\,$ indices $\,m\,$ and $\,a\,$ correspond to the time-like and space-like parts of the diagonal metric $\,\eta_{MN}=(\underbrace{-1,...,-1}_{\textrm{$6$ times}},\underbrace{1,...,1}_{\textrm{$6$ times}})\,$ of $\,\textrm{SO}(6,6)\,$ in Lorentzian coordinates.
\\[-1mm]

If setting all the scalar fields to zero, \emph{i.e.} moving to the origin of the moduli space, the set of vielbein components in (\ref{vielbein_components}) must reduce to the product of a constant $\,\textrm{SL}(2)\,$ complexified vielbein $\,L_{\a}\equiv\mathcal{V}_{\a}|_{\textrm{origin}} =\left( i \,,\, 1\right)\,$ satisfying
\be
\label{LLstar}
L_{\a} \, L^{*}_{\b} = \d_{\a \b} + i \, \epsilon_{\a \b} \ ,
\ee
with a set of $\,\textrm{SO}(6) \sim \textrm{SU}(4)\,$ invariant tensors. These are the \text{'t Hooft symbols} $\,{\mathcal{V}_{m}}^{ij}|_{\textrm{origin}}=\frac{1}{2} \, [G_{m}]^{ij}\,$, $\,{\mathcal{V}_{a}}^{\hi \hj}|_{\textrm{origin}}=\frac{1}{2} \, [G_{a}]^{\hat{i}\hat{j}}\,$ (see appendix~\ref{App:spinors}) and the Kronecker deltas $\,\delta_{i}^{j}\,$ and $\,\delta_{\hat{i}}^{\hat{j}}\,$. The non-vanishing components of the vielbein $\,{\mathcal{V}_{\mathbb{M}}}^{\underline{\mathbb{N}}}\,$ are given by
\be
\begin{array}{lclclc}
{\mathcal{V}_{\a m}}^{ij} = \dfrac{-i}{2\sqrt{2}} \, (L_{\a})^{*} \, [G_{m}]^{ij} & \hspace{0mm} , \hspace{0mm} & \mathcal{V}_{\a m \, ij} = \dfrac{i}{2\sqrt{2}} \, L_{\a} \, [G_{m}]_{ij}
& \hspace{0mm} , \hspace{0mm} &   {\mathcal{V}_{i \hj}}^{k \hl} = \dfrac{(1+i)}{2} \, \d^{k}_{i} \, \d^{\hl}_{\hj} & , \\[5mm]
{\mathcal{V}_{\a a}}^{\hi \hj} = \dfrac{-1}{2\sqrt{2}} \, L_{\a} \, [G_{a}]^{\hi \hj} & \hspace{0mm} , \hspace{0mm} & \mathcal{V}_{\a a \, \hi \hj} = \dfrac{-1}{2\sqrt{2}} \, (L_{\a})^{*} \, [G_{a}]_{\hi \hj} & \hspace{0mm} , \hspace{0mm} & {\mathcal{V}^{i \hj}}_{k \hl} = \dfrac{(1-i)}{2} \, \d^{i}_{k} \, \d^{\hj}_{\hl} & ,
\end{array}
\ee
whereas those of the inverse vielbein $\,{\mathcal{V}^{\mathbb{M}}}_{\underline{\mathbb{N}}}\,$ read
\be
\begin{array}{lclclc}
{\mathcal{V}^{\a m}}_{ij} = \dfrac{-1}{2\sqrt{2}} \, L^{\a} \, [G^{m}]_{ij} & \hspace{0mm} , \hspace{0mm} & \mathcal{V}^{\a m \, ij} = \dfrac{-1}{2\sqrt{2}} \, (L^{\a})^{*} \, [G^{m}]^{ij}
& \hspace{0mm} , \hspace{0mm} &   {\mathcal{V}^{i \hj}}_{k \hl} = \dfrac{(1-i)}{2} \, \d_{k}^{i} \, \d_{\hl}^{\hj} & ,\\[5mm]
{\mathcal{V}^{\a a}}_{\hi \hj} = \dfrac{-i}{2\sqrt{2}} \, (L^{\a})^{*} \, [G^{a}]_{\hi \hj} & \hspace{0mm} , \hspace{0mm} & \mathcal{V}^{\a a \, \hi \hj} = \dfrac{i}{2\sqrt{2}} \, L^{\a} \, [G^{a}]^{\hi \hj} & \hspace{0mm} , \hspace{0mm} & {\mathcal{V}_{i \hj}}^{k \hl} = \dfrac{(1+i)}{2} \, \d_{i}^{k} \, \d_{\hj}^{\hl} & ,
\end{array}
\ee
and completely specify the relations (\ref{A's_fluxes}). One can check that the vielbein $\,{\mathcal{V}_{\mathbb{M}}}^{\underline{\mathbb{N}}}\,$ satisfies the normalisation conditions \cite{deWit:2007mt}
\be
\begin{array}{rcll}
{\mathcal{V}_{\mathbb{M}}}^{\mathcal{IJ}} \, {\mathcal{V}_{\mathbb{N}}}_{\mathcal{IJ}}  - {\mathcal{V}_{\mathbb{M}}}_{\mathcal{IJ}} \, {\mathcal{V}_{\mathbb{N}}}^{\mathcal{IJ}} & = & i \, \Omega_{\mathbb{M} \mathbb{N}} & , \\[2mm]
\Omega^{\mathbb{M} \mathbb{N}} \, {\mathcal{V}_{\mathbb{M}}}^{\mathcal{IJ}} \, {\mathcal{V}_{\mathbb{N}}}_{\mathcal{KL}} & = & i \, \delta^{\mathcal{IJ}}_{\mathcal{KL}} & ,\\[2mm]
\Omega^{\mathbb{M} \mathbb{N}} \, {\mathcal{V}_{\mathbb{M}}}^{\mathcal{IJ}} \, {\mathcal{V}_{\mathbb{N}}}^{\mathcal{KL}} & = & 0 & .
\end{array}
\ee

\cleardoublepage 
\addcontentsline{toc}{chapter}{Bibliography}
\bibliography{thesis.bbl}
\bibliographystyle{utphys}

\end{document}